\documentclass[useAMS,usenatbib]{mn2e}

\bibpunct{(}{)}{;}{a}{}{,}

\include{graphicx}
\defcitealias{norton01}{NGZZ}
\title[The kinematics and stellar populations in E+A galaxies]{The kinematics and spatial distribution of stellar populations in E+A galaxies}
\author[Michael B. Pracy et al.]{
\parbox[t]{\textwidth}{Michael B.~Pracy$^{1}$\thanks{E-mail:mpracy@mso.anu.edu.au}, Harald Kuntschner$^{2}$, Warrick J.~Couch$^{3}$, Chris Blake$^{3}$, \\
Kenji Bekki$^{4}$ and Frank Briggs$^{1}$}
\vspace*{6pt}\\
$^{1}$Research School of Astronomy \& Astrophysics, The Australian National University, Weston Creek, ACT 2611, Australia \\
$^{2}$Space Telescope European Coordinating Facility, European Southern Observatory, Karl-Schwarzschild Strasse 2, 85748, Garching, Germany \\
$^{3}$Centre for Astrophysics and Supercomputing, Swinburne University of Technology, P.O. Box 218, Hawthorn, VIC, 3122, Australia \\
$^{4}$School of Physics, University of New South Wales, Sydney NSW 2052, Australia \\
}

\begin{document}

\date{Received 0000; Accepted 0000}

\pagerange{\pageref{firstpage}--\pageref{lastpage}} \pubyear{2006}

\maketitle

\label{firstpage}

\begin{abstract}
We have used the GMOS instrument on the 8.1-m Gemini--South telescope to obtain spatially--resolved two--colour imaging and
integral field unit (IFU) spectroscopy of a sample of ten nearby ($z=0.04$--$0.20$) ``E+A'' galaxies selected from the Two Degree Field
Galaxy Redshift Survey. These galaxies have been selected to lie in a variety of environments from isolated systems to rich clusters. 
Surface brightness profiles measured using our imaging data show the isophotal profiles of our sample are generally $r^{1/4}$--like,
consistent with a sample dominated by early--type galaxies. Only one galaxy in our sample has an obvious exponential (`disk--like')
component in the isophotal profile. This is further underscored by all galaxies having early Hubble type morphological classifications, and
showing a behaviour in the central velocity dispersion--absolute magnitude plane that is consistent with the
Faber-Jackson relation, once the transitory brightening that occurs in the E+A phase is corrected for. In addition, 
two-thirds of our sample shows clear evidence of either ongoing or recent tidal interactions/mergers, as 
evidenced by the presence of tidal tails and disturbed morphologies. While all the galaxies
in our sample have total integrated colours that are relatively blue (in keeping with their E+A status), they
show a diversity of colour gradients, possessing central core regions that are either redder, bluer, or indistinct in colour 
relative to their outer regions. The E+A spectra are well fitted by that of a young stellar population, the light 
from which is so dominant that it is impossible to quantify the presence of the underlying old stellar population. 
Consistent with other recent findings, there is little evidence for radial gradients in the Balmer absorption 
line equivalent widths over the central few kiloparsecs ($< 4$\,kpc), although we are unable to search for the previously reported
radial gradients at larger galacto--centric radii due to the limited spatial extent of our IFU data.
Kinematically, the most striking property is the significant and unambiguous rotation that
is seen in all our E+A galaxies, with it being generally aligned close to the photometric major axis. This is contrary to the 
findings of \citet{norton01}, who found little or no evidence for rotation in a very similar sample of nearby E+A galaxies.
We also clearly demonstrate that our E+A galaxies are, in all but one case, consistent with being
"fast rotators" \citep{emsellem07}, based on their internal angular momentum per unit mass measured
as a function of radius and ellipticity. We argue that the combination of disturbed morphologies and significant rotation 
in these galaxies supports their production via gas--rich galaxy mergers and interactions. The large fraction of fast rotators argues
against equal mass mergers being the dominant progenitor to the E+A population.
\end{abstract}

\begin{keywords}
galaxies: evolution -- galaxies: formation -- galaxies: stellar content
\end{keywords}

\section{Introduction}
Galaxy mergers and interactions are a fundamental driving force of galaxy evolution. `E+A' galaxies are
a crucial part of this picture. They are conspicuous by their unusual spectra \citep{dressler83}: strong hydrogen
Balmer absorption lines (implying a relatively young `A'-type stellar population) superimposed upon an 
elliptical (`E') galaxy spectrum with no optical emission lines (implying
no ongoing star formation). The A-stars must have been produced by a powerful recent episode of star formation, which has been 
suddenly truncated in the past 1\,Gyr \citep{couch87,poggianti99}. The spectroscopic data are also consistent with the uniform truncation of star 
formation in a disk, without necessarily requiring a starburst \citep[e.g][]{Shioya04}. Observational evidence 
indicates that the E+A phenomenon may, in part, mark the transitory phase between star-forming disk galaxies and 
quiescent spheroidal systems (e.g. \citealt{caldwell96,zabludoff96,norton01}, hereafter NGZZ).

Understanding the origin of the E+A phase is complicated by the fact that E+A galaxies 
inhabit a wide range of environments, in a manner that depends upon both redshift and luminosity. 
In summary: luminous E+A galaxies are commonplace in intermediate redshift
clusters, where they were first identified and studied \citep{dressler83}. In the low-redshift universe
cluster E+A's are still frequent, but their luminosities are much lower than their higher-z counterparts \citep{poggianti04}.
Most luminous E+A galaxies at low redshift are located in the field, although they represent a very low fraction of the 
overall field galaxy population \citep{zabludoff96}.

The physical mechanism(s) responsible in triggering intense star-formation and subsequently rapidly quenching it remains
the key question to better understand the E+A galaxy phase and the role it plays in galaxy evolution. There have been several
mechanisms suggested which could give rise to the E+A spectral signature. These include major mergers \citep{mihos96,bekki05}, 
unequal mass mergers \citep{bekki01} and galaxy interactions \citep{bekki05}. For E+As galaxies residing in clusters, other plausible 
mechanisms exist, such as: interaction with the strongly varying global cluster tidal field \citep{bekki99}, galaxy harassment 
\citep{moore98} or interaction with the hot intra-cluster gas \citep{gunn72,dressler83,bothun86}. 

The continuous accretion of cold gas onto galaxies is predicted in galaxy formation models 
\citep[e.g.][]{birnboim03,keres05,semelin05,dekel06}and could also be a driver of galaxy
evolution. Such a mechanism, however,  
should not induce a starburst, but rather contribute to slow and continuous levels of star formation. Given the large and rapid changes 
needed to produce the spectral properties of E+A galaxies, and the
evidence that a high fraction of these systems are morphologically  disturbed, accretion of cold gas from large scale 
structure is unlikely to be an important mechanism in E+A formation.

There is evidence that the E+As in the low-redshift field are the result of merging or tidal interactions between galaxies.
Ground--based imaging of a sample of 21 E+A galaxies drawn from the Las Campanas Redshift Survey (LCRS) revealed an increased
incidence of tidal features associated with these galaxies implying galaxy interactions or galaxy mergers had taken place 
\citep{zabludoff96}. Their conclusions were later confirmed using high resolution Hubble Space Telescope imaging of the LCRS sample 
\citep{yang08}, which found a high incidence of tidal features consistent with  mergers or interactions.
Samples of E+A galaxies constructed from larger redshift surveys reinforce these ideas. Using a robust sample of
E+A galaxies selected from the Two Degree Field Galaxy Redshift Survey \citep[2dFGRS;][]{Colless01}, \citet{blake04} concluded that their sample was 
consistent with a major merger origin based on their morphologies, incidence of tidal disruption, and luminosity function.
In addition, Goto's (2005) study of a sample of low-redshift E+A galaxies selected from the Sloan Digital Sky Survey \citep[SDSS;][]{abazajian04}
revealed an excess in the projected galaxy density on small scales surrounding the E+A galaxies providing a strong hint of 
galaxy-galaxy interactions as a formation mechanism.

A more powerful and direct method of inferring the physical mechanism(s) responsible for the E+A galaxy phase is to study the internal 
properties of individual galaxies.  The kinematics and spatial distribution of the stellar populations of a galaxy in the E+A phase 
holds a wealth of information about the history of its formation and is critical information for discerning which of the candidate 
mechanisms is responsible. Mergers and tidal interaction are expected to give rise to star formation which is centrally concentrated
as gas is funnelled to the galactic center \citep{noguchi88,barnes91,mihos92,mihos96,bekki05,hopkins09}. 
For an equal mass merger (e.g. two massive spirals),  
the remnant should settle quickly to the virial plane and be dynamically pressure supported \citep[NGZZ;][]{bekki05}.
Such mergers can also produce rotating remnants a small fraction of the time, but this requires specific merger configurations 
\citep{bekki05,bournaud08}. In the case of unequal mass mergers
and tidal interactions, rotation of the young stellar populations is generally expected \citep{bekki05}. In contrast, if star formation is
simply truncated abruptly in a `normal' star-forming disk, the young stellar population should be spread throughout the extent of the 
galaxy and rotation should be present in both the young and old stellar populations \citep[NGZZ;][]{bekki05,pracy05}.

Spatially resolved spectroscopic studies of E+A galaxies residing in clusters have revealed a young stellar population which is 
widely spread throughout the galaxy and not confined to the galaxy core; observations have also shown evidence for strong 
rotation in these galaxies
 \citep{caldwell96,franx93}. \citet{pracy05} found diversity in the spatial distribution of the young population in the E+A galaxy 
population in the intermediate redshift cluster AC114; some E+As have a centrally concentrated young stellar population consistent with a tidal 
or merger origin and others revealed a more distributed young star component consistent with the truncation of a spiral disk.
The small number of examples of spatially resolved spectroscopy of E+A galaxies in the nearby field have generally revealed a central 
concentration of the young stellar population \citep[NGZZ;][]{goto08} although there exist exceptions \citep{swinbank05}.
\citetalias{norton01} used long slit spectroscopy to study the internal kinematics of field E+A galaxies over the central $\sim 2$--4\,kpc.
They found little difference in the kinematics of the young and old stellar populations and that in all but a small fraction of cases the
E+As showed little or no rotation. They interpret their results as being consistent with E+As being in the midst of a transformation
from  gas rich, rotationally supported disks into  gas poor, pressure supported early--type galaxies. Yamauchi \& Goto's (2005) 
investigation of the two--dimensional color distributions of E+A galaxies found an excess of galaxies with blue cores (relative to normal early--type galaxies) 
and they interpreted their results as being consistent with the galaxies having undergone a centralized star-burst caused by a merger or interaction.
\citet{yang08}, using HST imaging found diversity in the internal color distributions of the LCRS sample, but also found
a tendency toward bluer centres which they interpreted as evidence for merger/interaction--induced star formation.

In this paper we present high quality imaging and spatially-resolved two-dimensional spectroscopy of a sample of robustly selected 
E+A galaxies from the 2dFGRS \citep{blake04}. These observations consist of deep two color imaging with GMOS on the 8.1-m 
Gemini South Telescope to investigate internal color gradients and to look for faint tidal tails or disturbances 
which could be symptomatic of a recent interaction or merger. We have also obtained Integral Field Unit (IFU) observations with GMOS in order 
to study the internal kinematics and line-strength distribution. Throughout we adopt an $\Omega_{M}=0.3,\Omega_{\Lambda}=0.7$ 
and $H_{0}=70 \;{\rm kms^{-1} Mpc^{-1}}$ cosmology.

\section{ Observations and Data reduction}

\subsection{Sample}
We have selected 10 relatively bright ($b_{J}\sim 18.4$) and nearby ($z<0.2$) E+A galaxies from the 2dFGRS E+A catalogue
compiled by \citet{blake04} for follow up with high resolution imaging and spatially resolved spectroscopic 
observations. The selection criteria used for the \citet{blake04} sample was based on that of \citet{zabludoff96}, specifically
a galaxy was required to have [OII] equivalent width of less than 2.5\,\AA\, in emission and a mean Balmer absorption line strength
(based on a weighted combination of the H$\delta$, H$\gamma$ and H$\beta$ lines) of greater than 5.5\,\AA\, in absorption.
These galaxies have already had their morphologies and external 
environments investigated by \citet{blake04} and we have selected our sample to span a wide range of environments.
Specifically, four galaxies are `isolated', four are located in groups (i.e. linked to 
other survey objects by a percolation algorithm, \citealt{eke04}) and two galaxies inhabit
cluster environments. The details of our target galaxies are summarized in Table \ref{tab:targets}.
\begin{table*}
\caption{\label{tab:targets}List of target galaxies}
\begin{tabular}{|c|c|c|c|c|c|c|c|c|c|c|} \hline
Name        & 2dFGRS ID    & RA         & Dec   &   z   &  $g$  &  $r$  & $g-r$   & $M_R$   &  $R_{e}$ (arsec) & environment \\ \hline
E+A\_1	& TGS439Z075 & 00 29 10.97 & -32 42 34.2 & 0.108 &  17.25 & 16.29 & 0.96  & -22.34  &  1.44       & Isolated  \\  
E+A\_2	& TGS271Z130 & 23 41 08.90 & -28 55 25.4 & 0.082 &  17.83 & 16.75 & 1.08  & -21.32  &  1.65          & Group     \\
E+A\_3	& TGS519Z227 & 02 33 10.60 & -33 52 24.4 & 0.070 &   N/A  & 15.70 & N/A   & -21.96  &  1.63          & Cluster  \\
E+A\_4	& TGS520Z261 & 02 40 24.27 & -33 25 50.6 & 0.035 &  17.64 & 16.61 & 1.03  & -19.38  &  2.57          & Group     \\
E+A\_5	& TGS480Z208 & 22 18 22.99 & -33 02 36.7 & 0.101 &  17.80 & 16.77 & 1.13  & -21.66  &  1.70          & Isolated \\
E+A\_6	& TGS358Z179 & 23 59 29.87 & -30 16 21.9 & 0.120 &  17.24 & 16.36 & 0.88  & -21.45  &  1.39          & Group     \\
E+A\_7	& TGS387Z032 & 02 15 40.52 & -30 50 54.5 & 0.092 &  17.13 & 16.12 & 1.01  & -22.17  &  1.60          & Group     \\
E+A\_8	& TGS539Z123 & 23 15 25.13 & -35 12 59.1 & 0.196 &  17.71 & 16.75 & 0.96  & -23.27  &  1.52          & Isolated  \\
E+A\_9	& TGS266Z090 & 23 10 46.57 & -28 31 49.7 & 0.088 &  17.21 & 16.24 & 0.97  & -21.99  &  1.64          & Cluster   \\
E+A\_10	& TGS350Z150 & 23 26 36.76 & -30 19 27.4 & 0.158 &  17.86 & 16.97 & 0.89  & -22.52  &  1.23         & Isolated  \\ \hline
\end{tabular}\\
\begin{flushleft}
Notes: Listed are the basic parameters of our E+A sample. Column 1 is an arbitrary ID and Column 2 is the 2dfGRS ID. The
remainder of the columns are (from left to right): target right ascension and declination (J2000), redshift, $g$--band
magnitude, $r$  magnitude, $g-r$ colour, absolute R magnitude inclusive of a k-correction \citep{wild04}, the effective radius measured
from the imaging using the \sc{IRAF ellipse} task, and galaxy 
environmental classification from \citet{blake04}.
\end{flushleft}
\end{table*}

\subsection{Imaging}

\subsubsection{Observations}
For each galaxy in our sample we obtained $g$-- and $r$--band imaging
using GMOS on Gemini-South. The imaging was  obtained in queue mode between
2005 September 1 and 2005 December 4. The imaging consisted of
3 $\times$ 420.5\,s exposures for each object, in each band, giving
a total integration time of 1261.5\,s. Longer integrations were 
obtained for E+A\_4 and E+A\_5. The details of the imaging observations
are given in Table \ref{tab:observations}.   The imaging was taken in seeing
of $\sim 1$\arcsec\, and never worse than 1.2\arcsec, and was sampled with 0.145\arcsec\, pixels (see Table \ref{tab:observations}).
\begin{table*}
\caption{\label{tab:observations} Summary of GMOS observations}
\begin{tabular}{|c|c|c|c|c|c|c|c|c|} \hline
Name    & g exptime (s) & FWHM (arcsec) & r exptime (s) & FWHM (arcsec)  & spec exptime (s) & observed $\lambda$\,(\AA)    & rest $\lambda$\, (\AA)  \\  \hline
E+A\_1  & 1262      &  0.88  & 1262      & 0.76    &  4082   &     4122--5380     &  3720--4855     \\ 
E+A\_2  & 1262      &  0.92  & 1262      & 0.79    &  4082   &     4122--5380     &  3810--4972     \\
E+A\_3  & 1262      &  0.57  & 1262      & 1.15    &  4082   &     4122--5380     &  3852--5028     \\ 
E+A\_4  & 2524      &  1.14  & 2524      & 1.08    &  N/A     &        N/A          &      N/A         \\
E+A\_5  & 2103      &  1.16  & 1262      & 1.23    &  4082   &     4122--5380     &  3744--4886     \\ 
E+A\_6  & 1262      &  0.98  & 1262      & 1.02    &  4082   &     4122--5380     &  3680--4803     \\ 
E+A\_7  & 1262      &  0.95  & 1262      & 0.96    &  4082   &     4122--5380     &  3775--4926     \\  
E+A\_8  & 1682      &  0.95  & 1262      & 0.83    &  N/A     &        N/A          &     N/A          \\ 
E+A\_9  & 1262      &  1.02  & 1262      & 0.99    &  4082   &     4122--5380     &  3788--4945     \\  
E+A\_10 & 1262      &  1.05  & 1262      & 1.00    &  4082   &     4122--5380     &  3559--4646     \\ \hline 
\end{tabular}
\begin{flushleft}
Notes:  Column 1 is galaxy ID. Columns 2 and 3 are the $g$--band exposure time
and seeing. Columns 4 and 5 are the $r$--band exposure times and the corresponding seeing. Column 6, 7 and 8 are the spectroscopic
exposure time, observed spectral wavelength coverage and rest-frame spectral wavelength coverage.
\end{flushleft}
\end{table*}
The final reduced $g$--band images of each target are shown in the first column
of Figure \ref{fig:images}. 
\setcounter{figure}{0}
\begin{figure*}
   \begin{center}
    \begin{minipage}{0.65\textwidth}
         \includegraphics[width=2.1cm, angle=-90, trim=0 0 0 0]{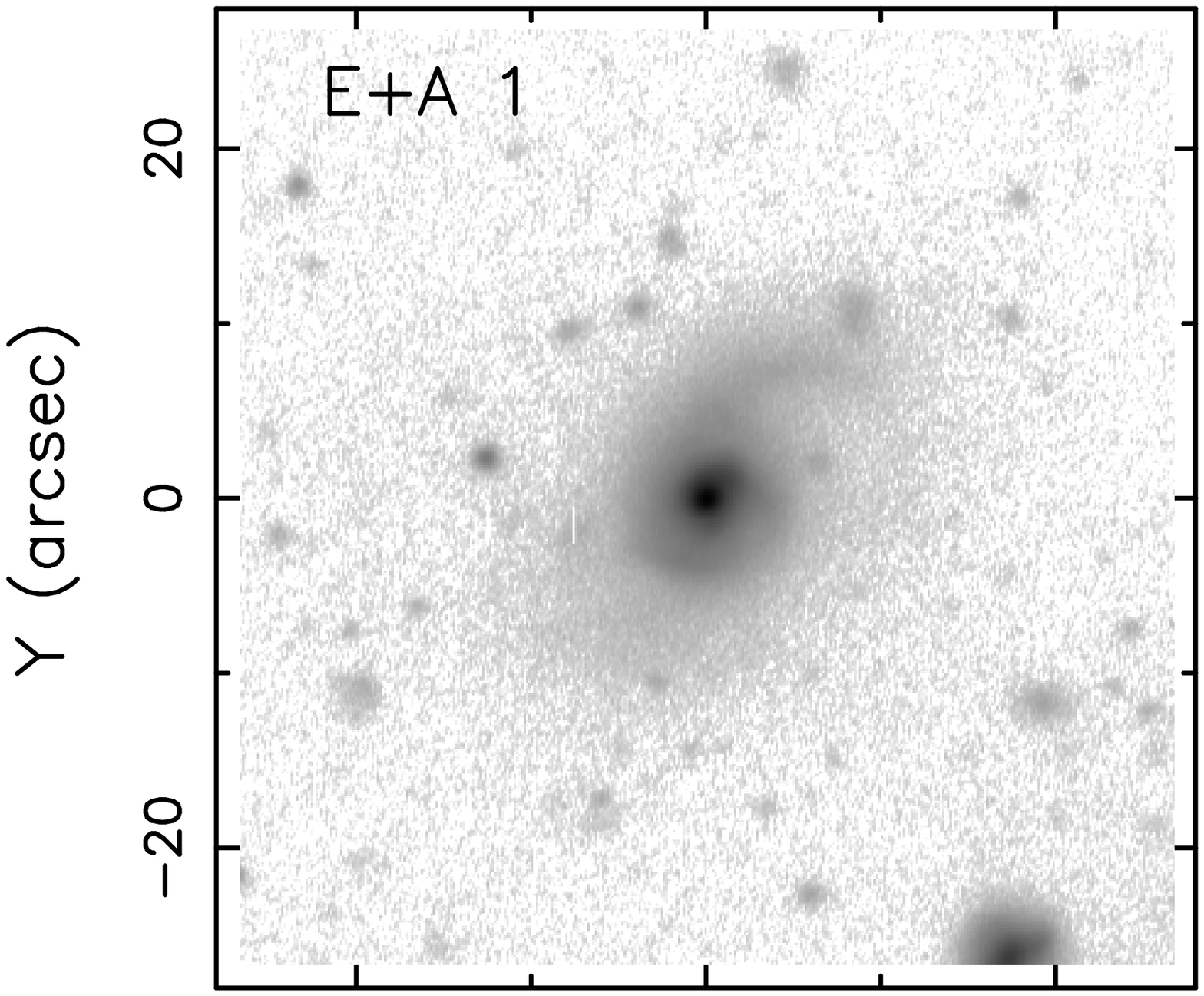}
          \includegraphics[width=2.1cm, angle=-90, trim=0 0 0 0]{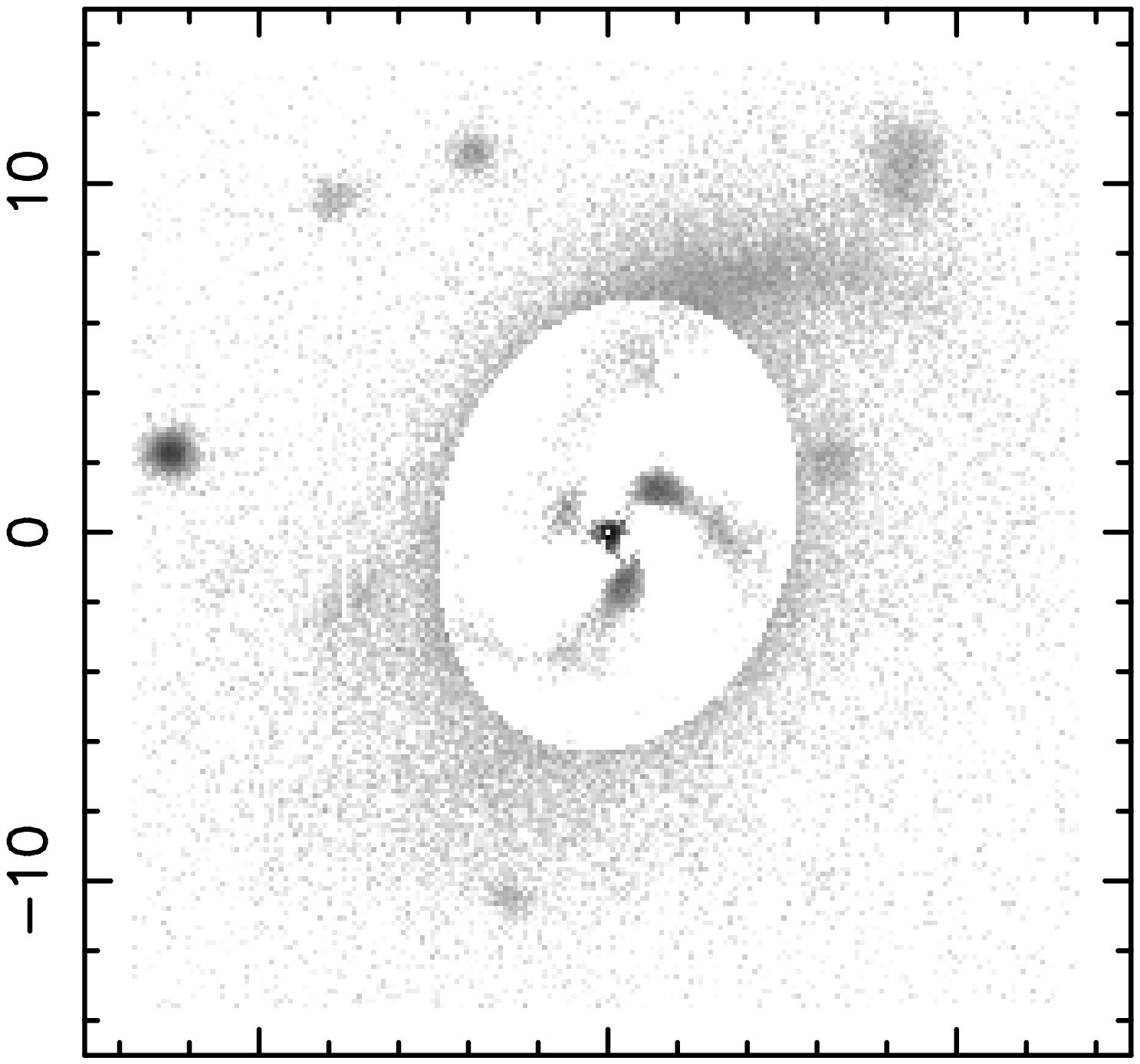}
         \includegraphics[height=2.62cm, angle=-90, trim=0 0 0 0]{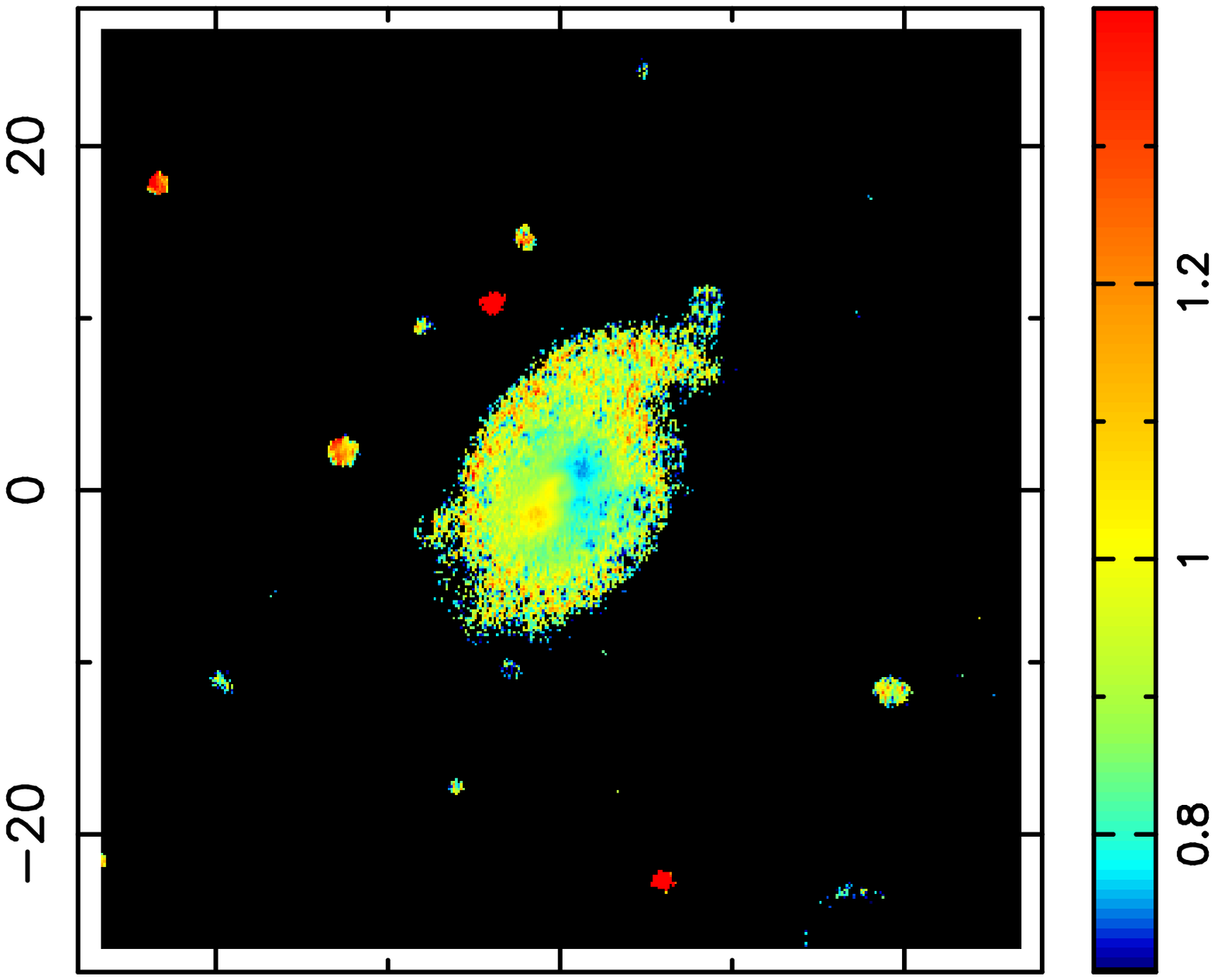}
         \includegraphics[height=2.62cm, angle=-90, trim=0 0 0 0]{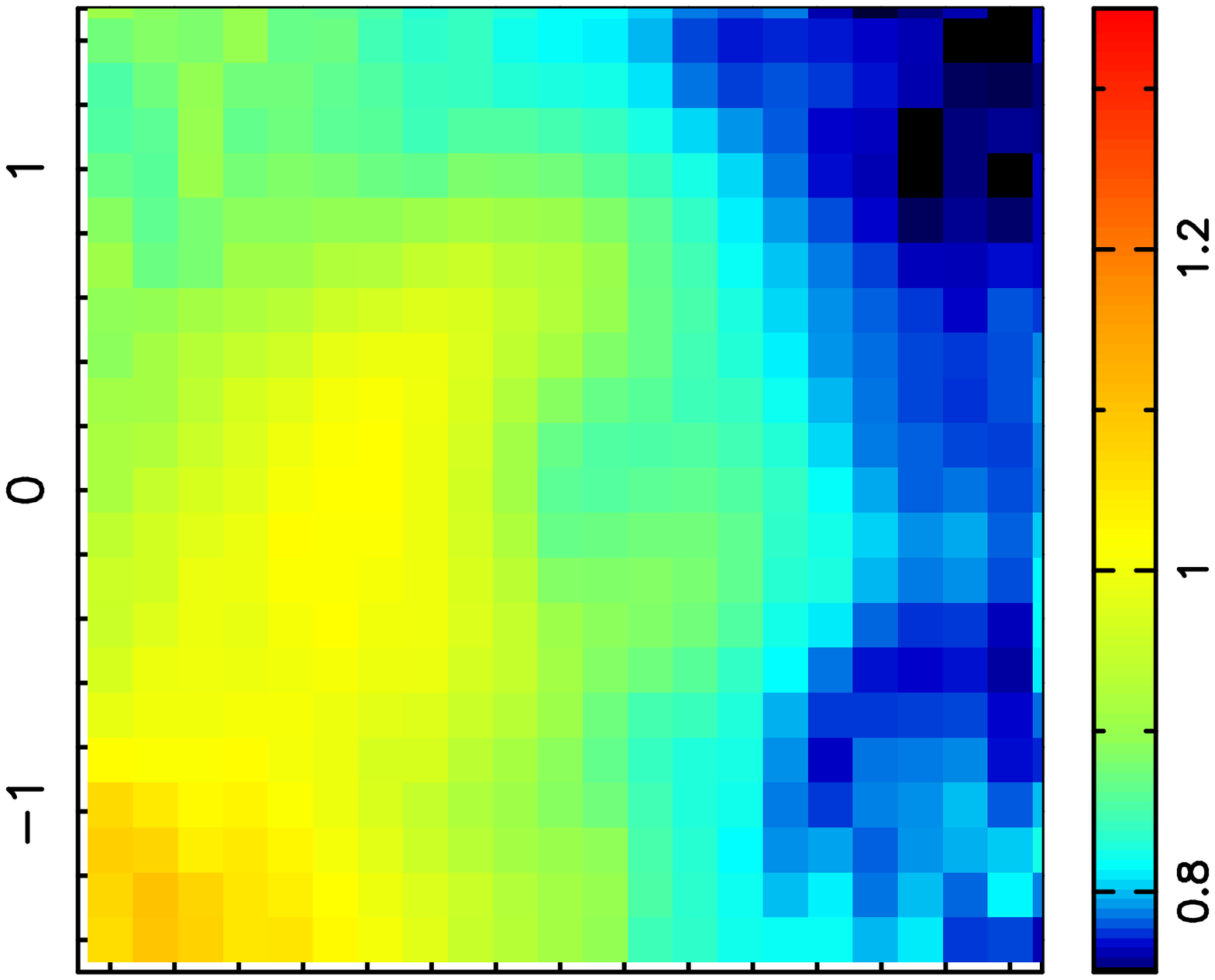}
     \end{minipage}
    \begin{minipage}{0.65\textwidth}
         \includegraphics[width=2.1cm, angle=-90, trim=0 0 0 0]{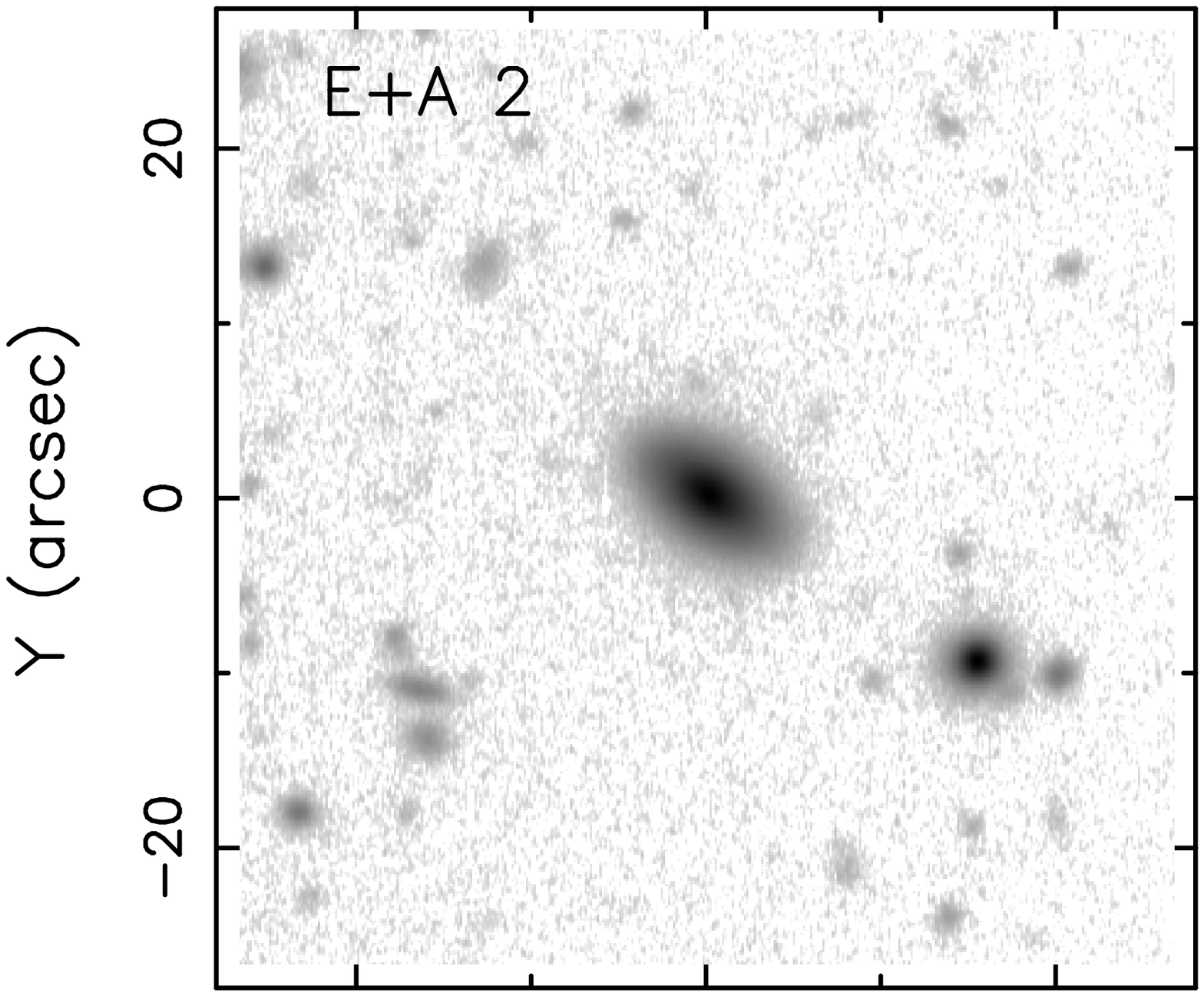}
          \includegraphics[width=2.1cm, angle=-90, trim=0 0 0 0]{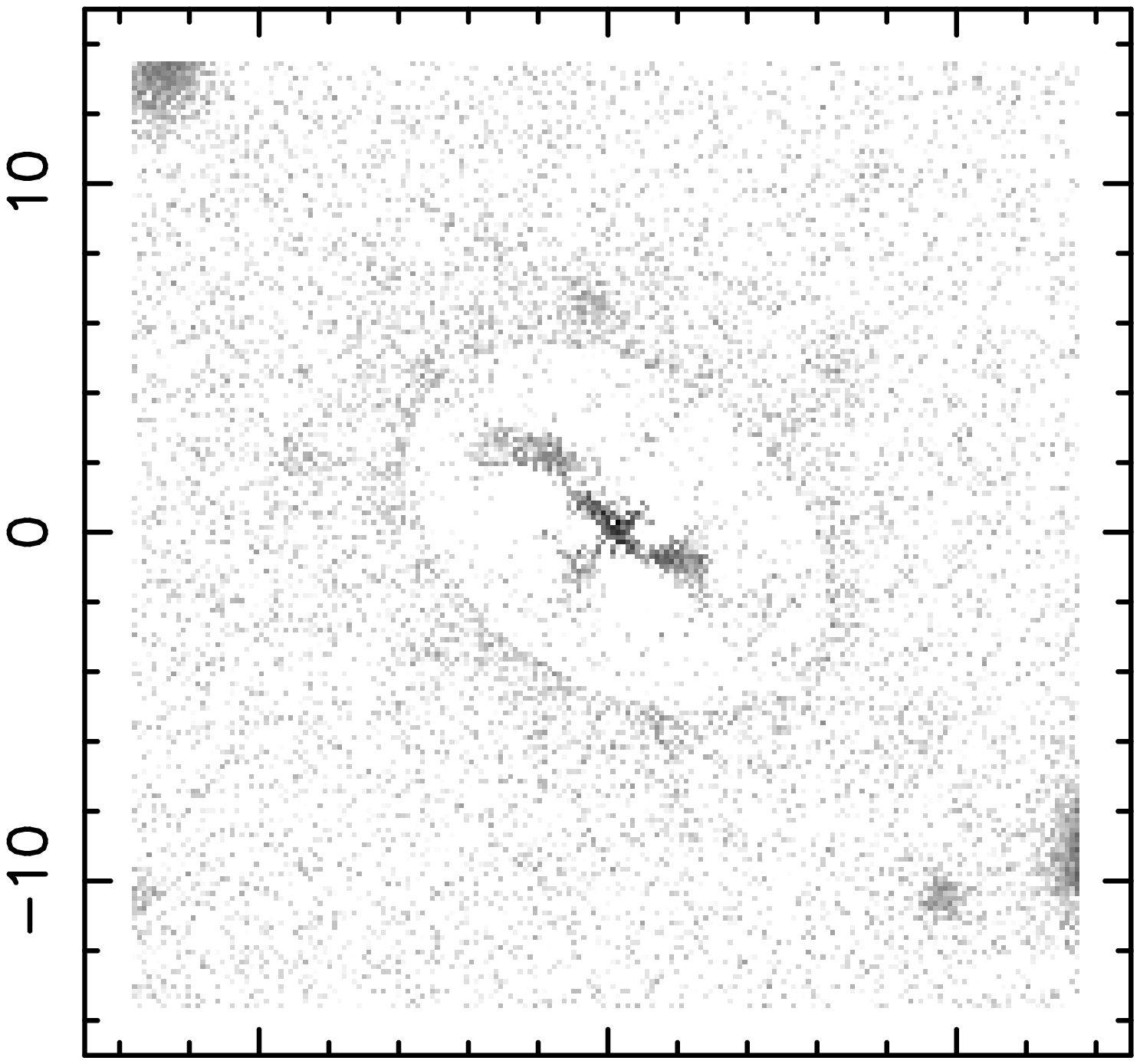}
         \includegraphics[height=2.62cm, angle=-90, trim=0 0 0 0]{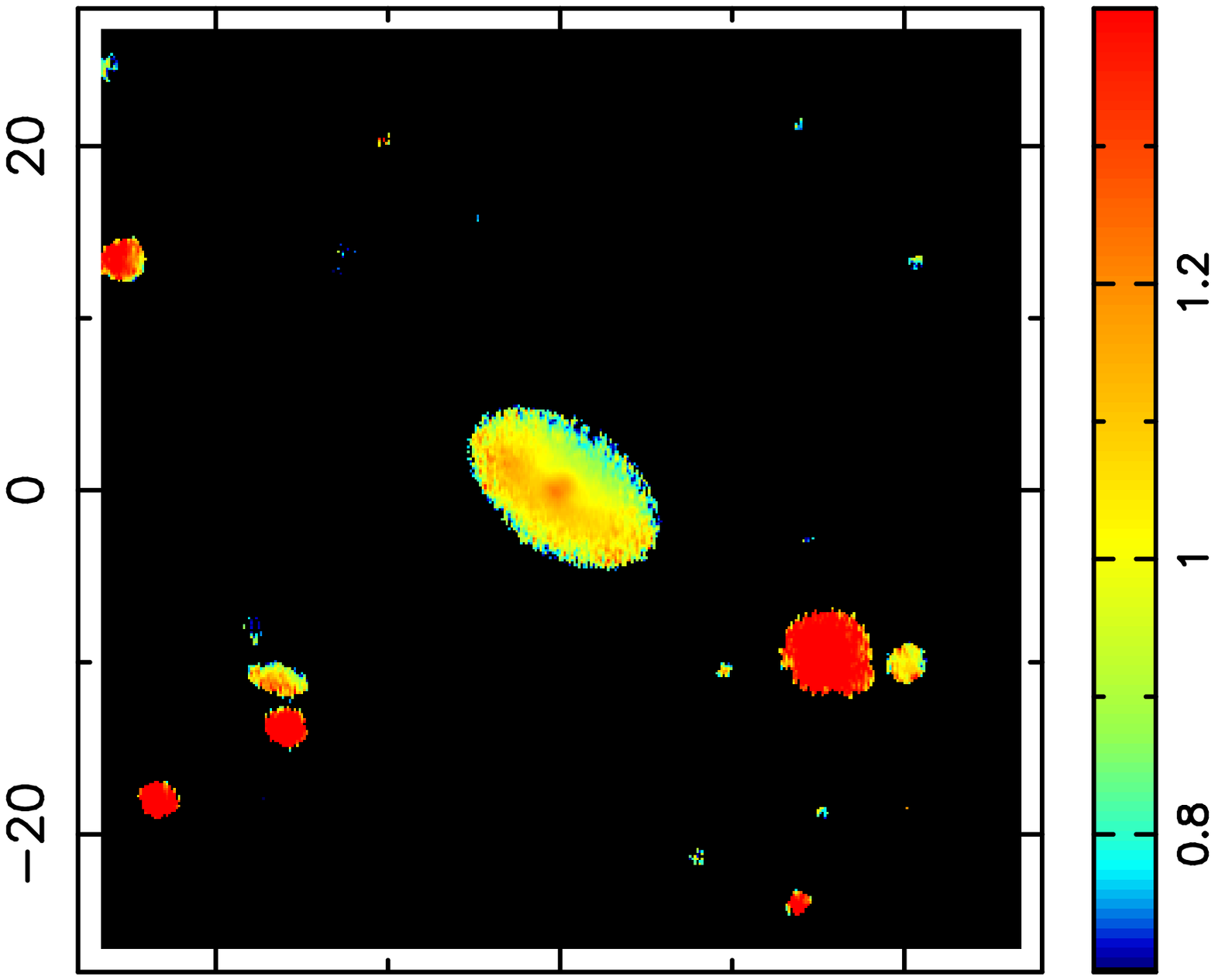}
         \includegraphics[height=2.62cm, angle=-90, trim=0 0 0 0]{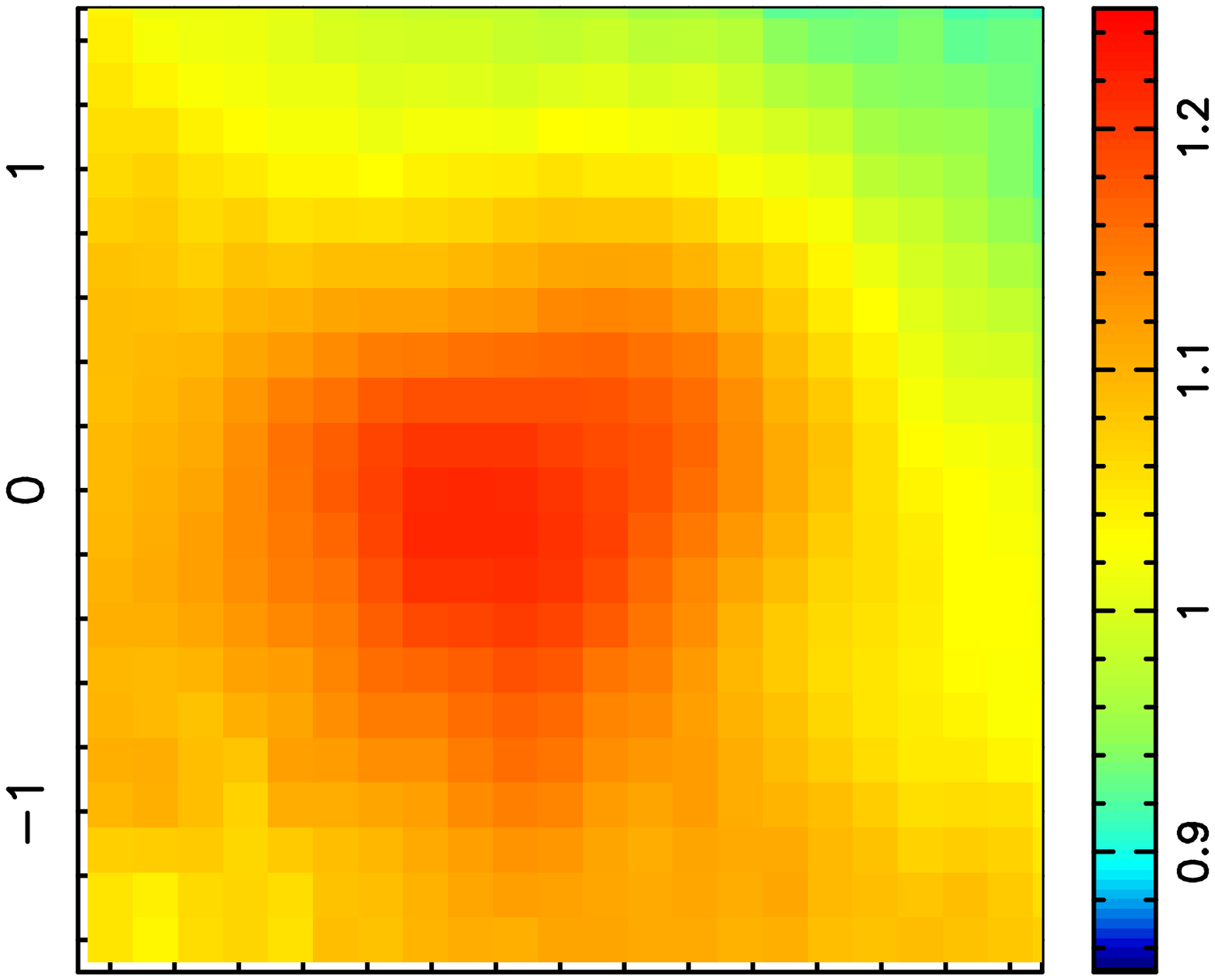}
     \end{minipage}
    \begin{minipage}{0.65\textwidth}
         \includegraphics[width=2.1cm, angle=-90, trim=0 0 0 0]{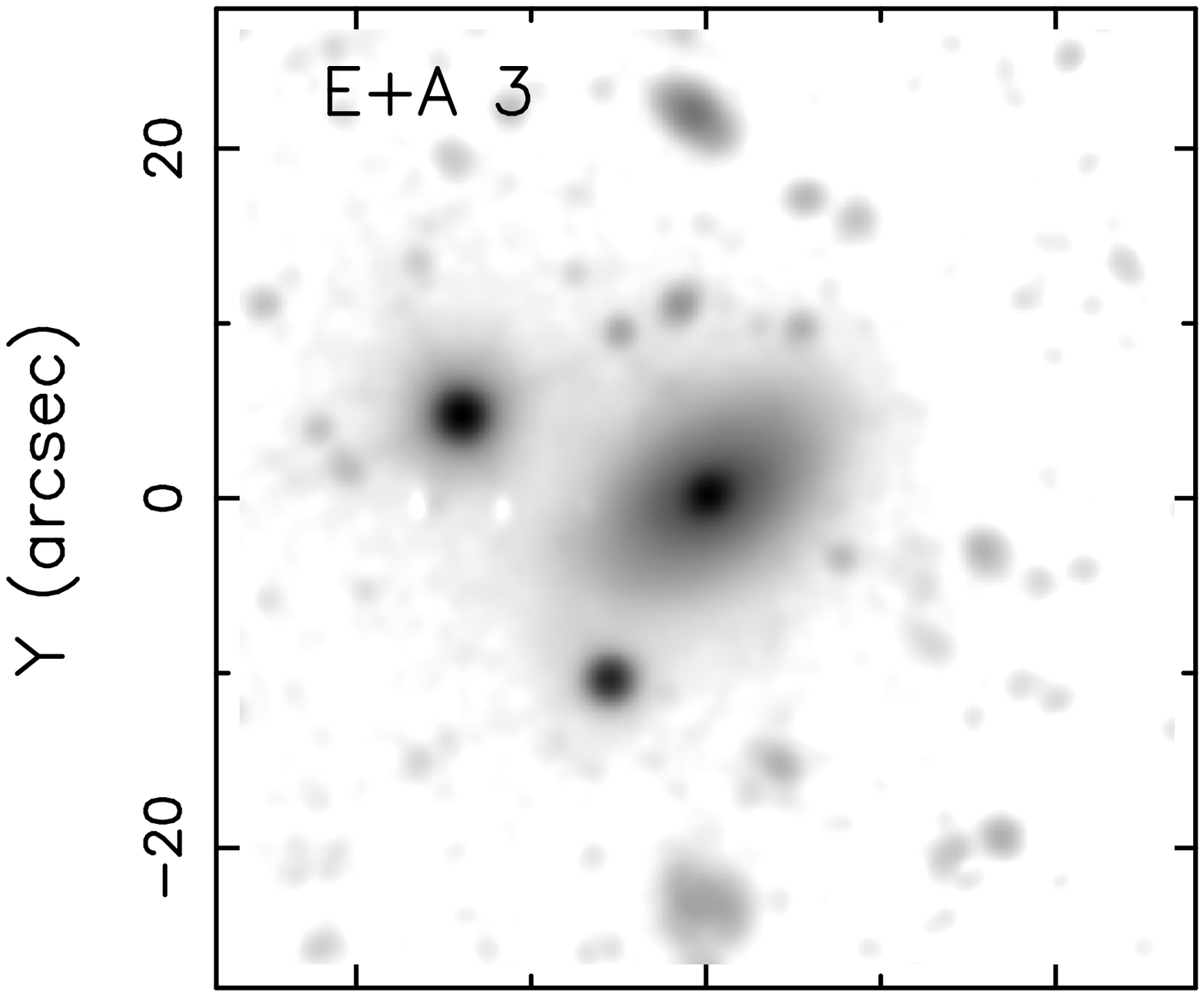}
         \includegraphics[width=2.1cm, angle=-90, trim=0 0 0 0]{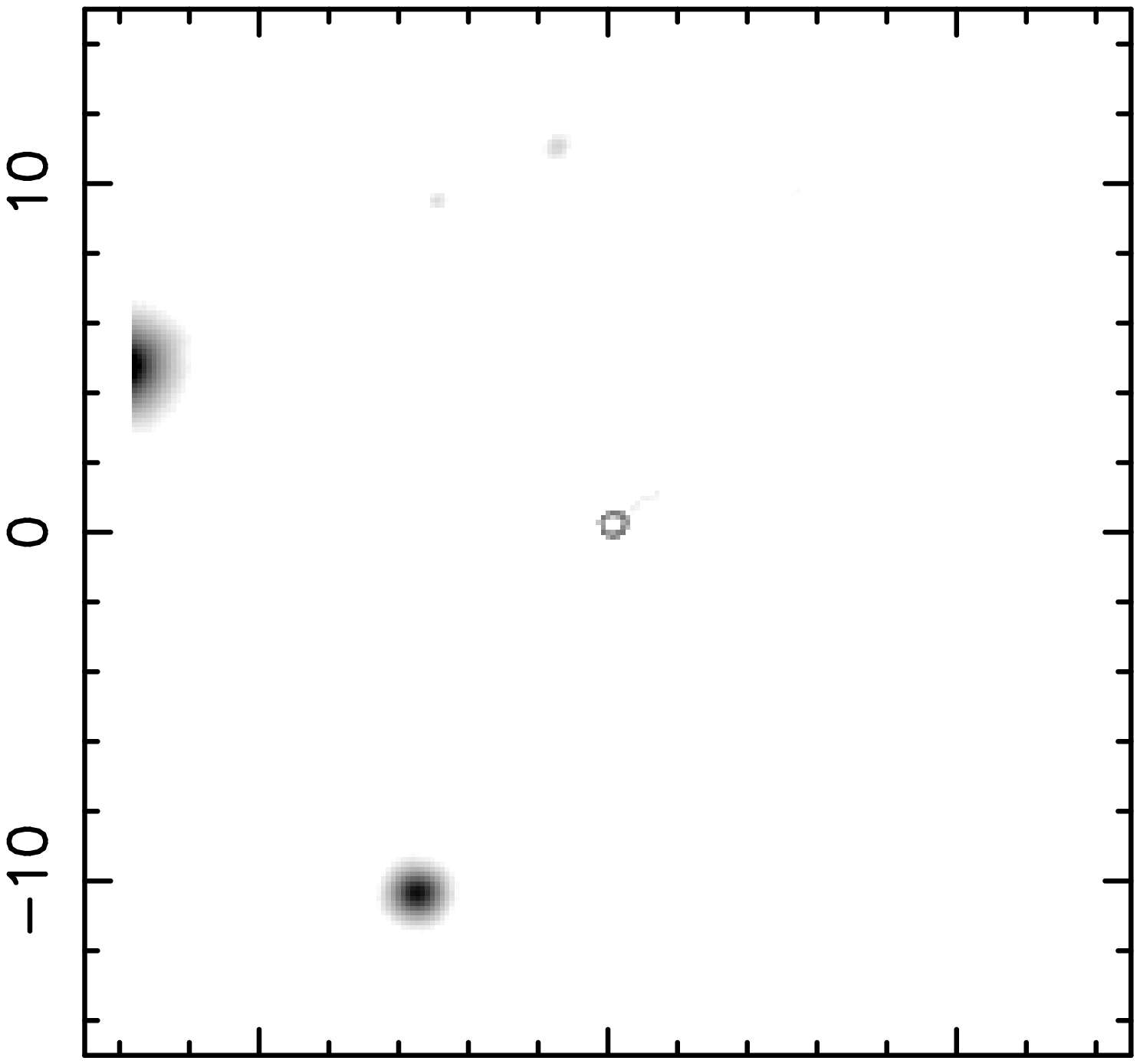}
     \end{minipage}
    \begin{minipage}{0.65\textwidth}
         \includegraphics[width=2.1cm, angle=-90, trim=0 0 0 0]{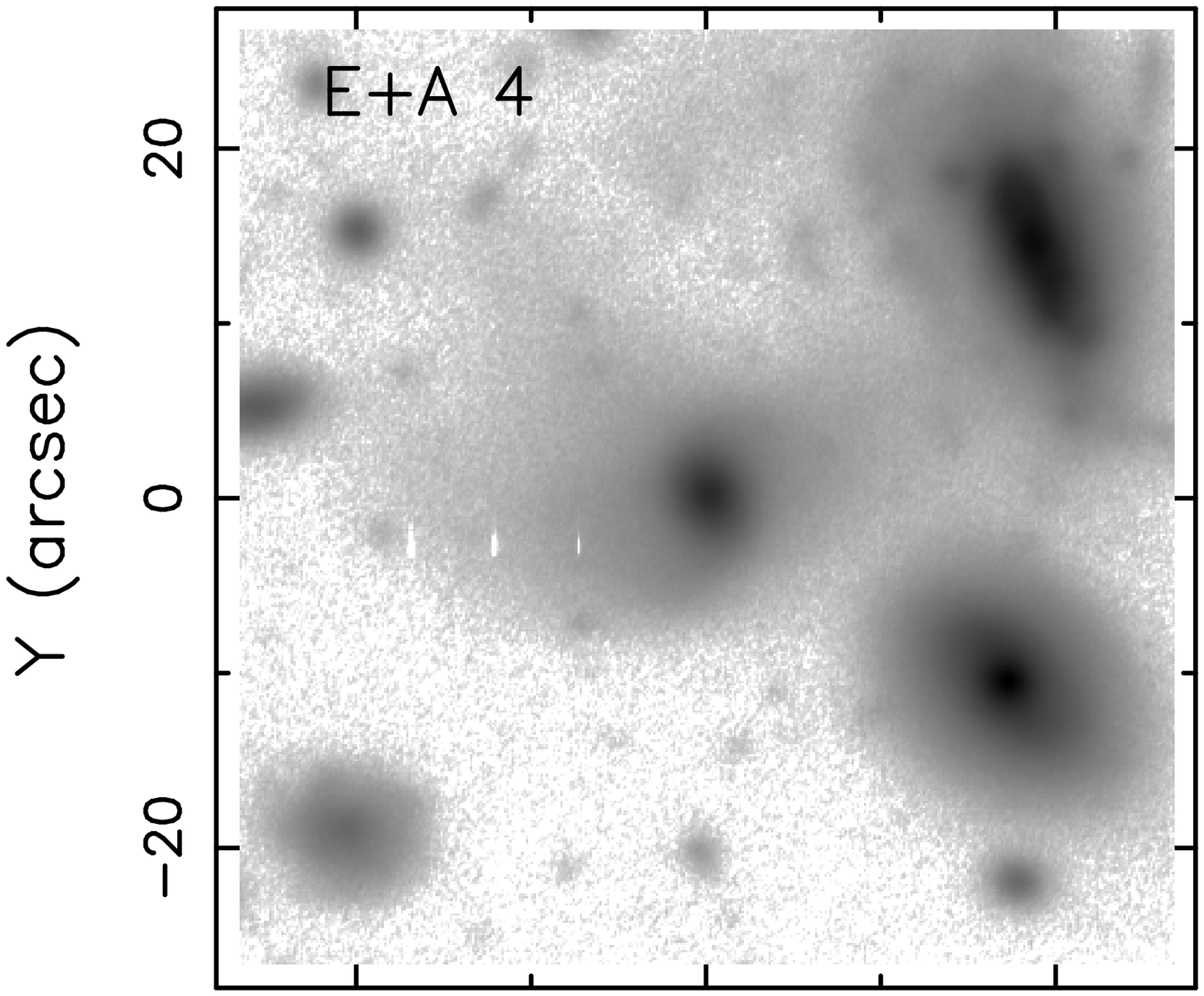}
         \includegraphics[width=2.1cm, angle=-90, trim=0 0 0 0]{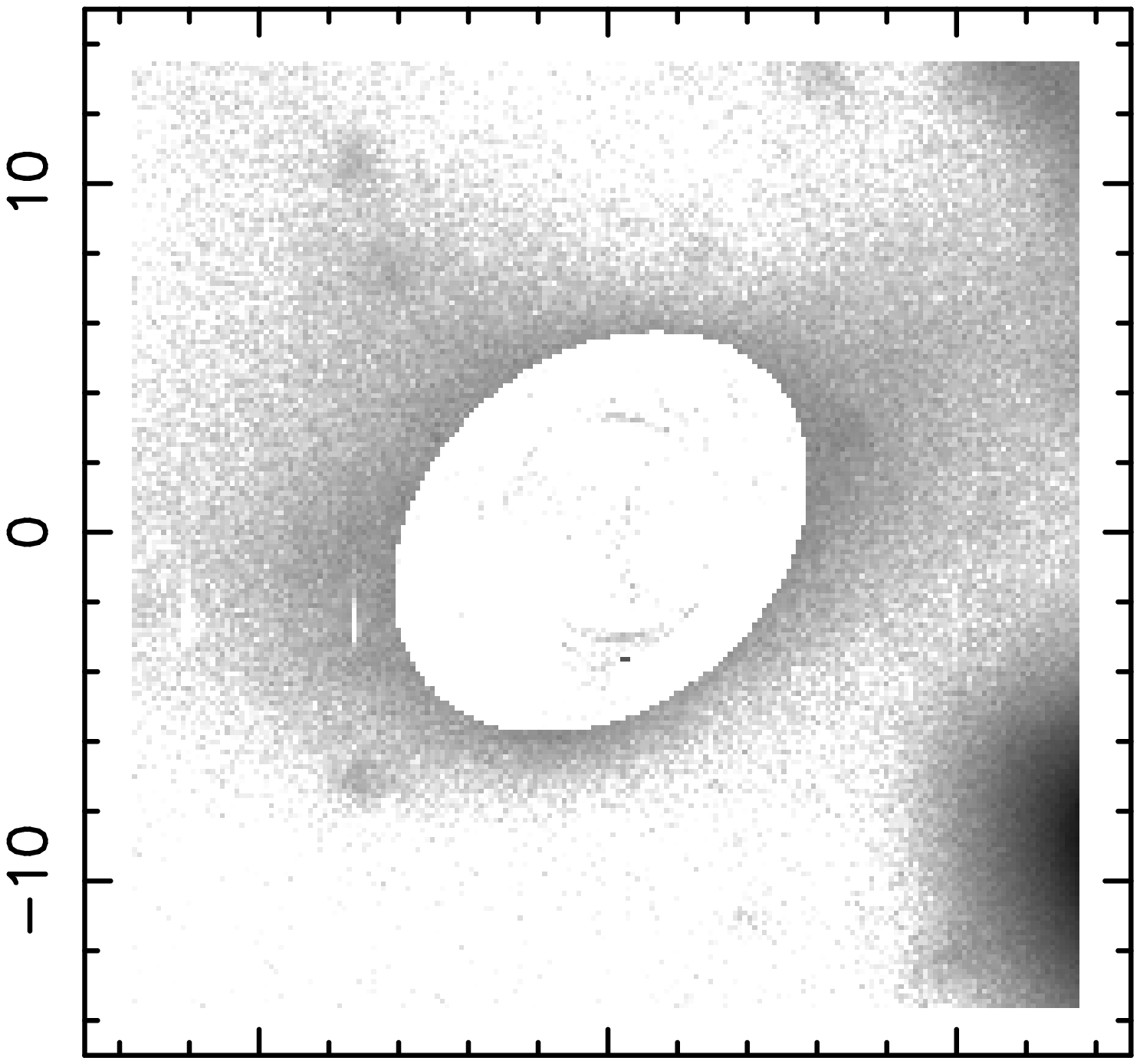}
         \includegraphics[height=2.62cm, angle=-90, trim=0 0 0 0]{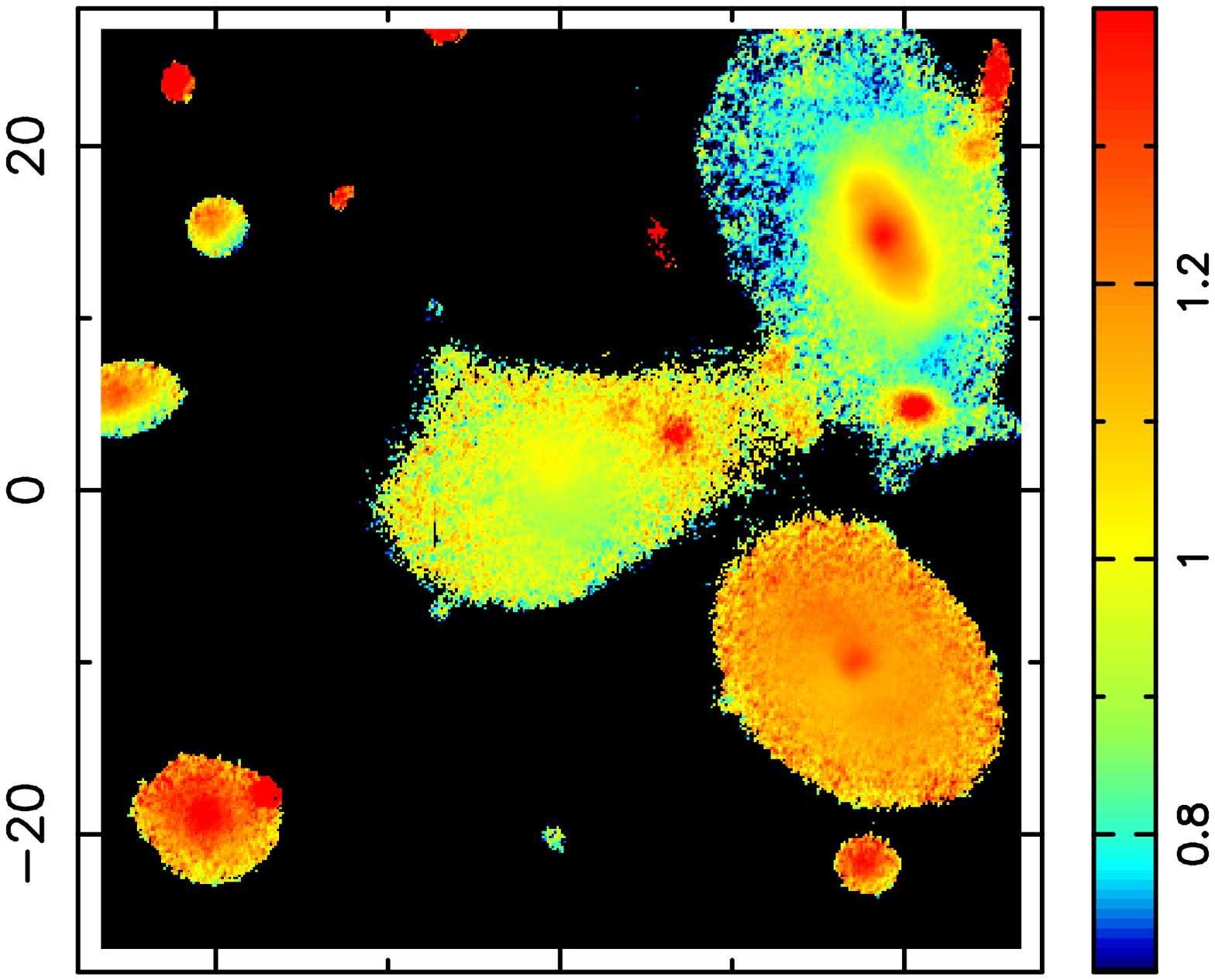}
         \includegraphics[height=2.62cm, angle=-90, trim=0 0 0 0]{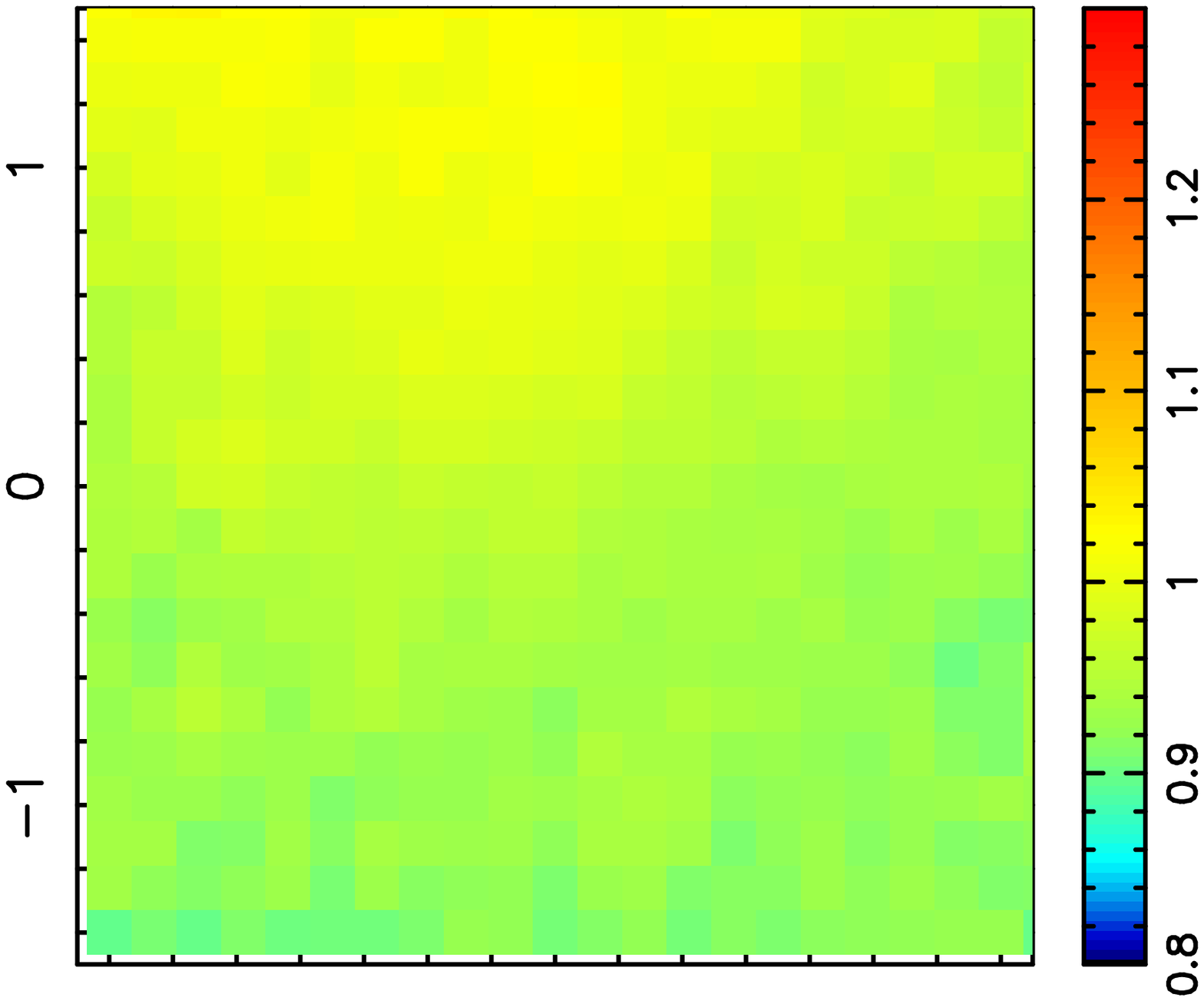}
     \end{minipage}
    \begin{minipage}{0.65\textwidth}
         \includegraphics[width=2.1cm, angle=-90, trim=0 0 0 0]{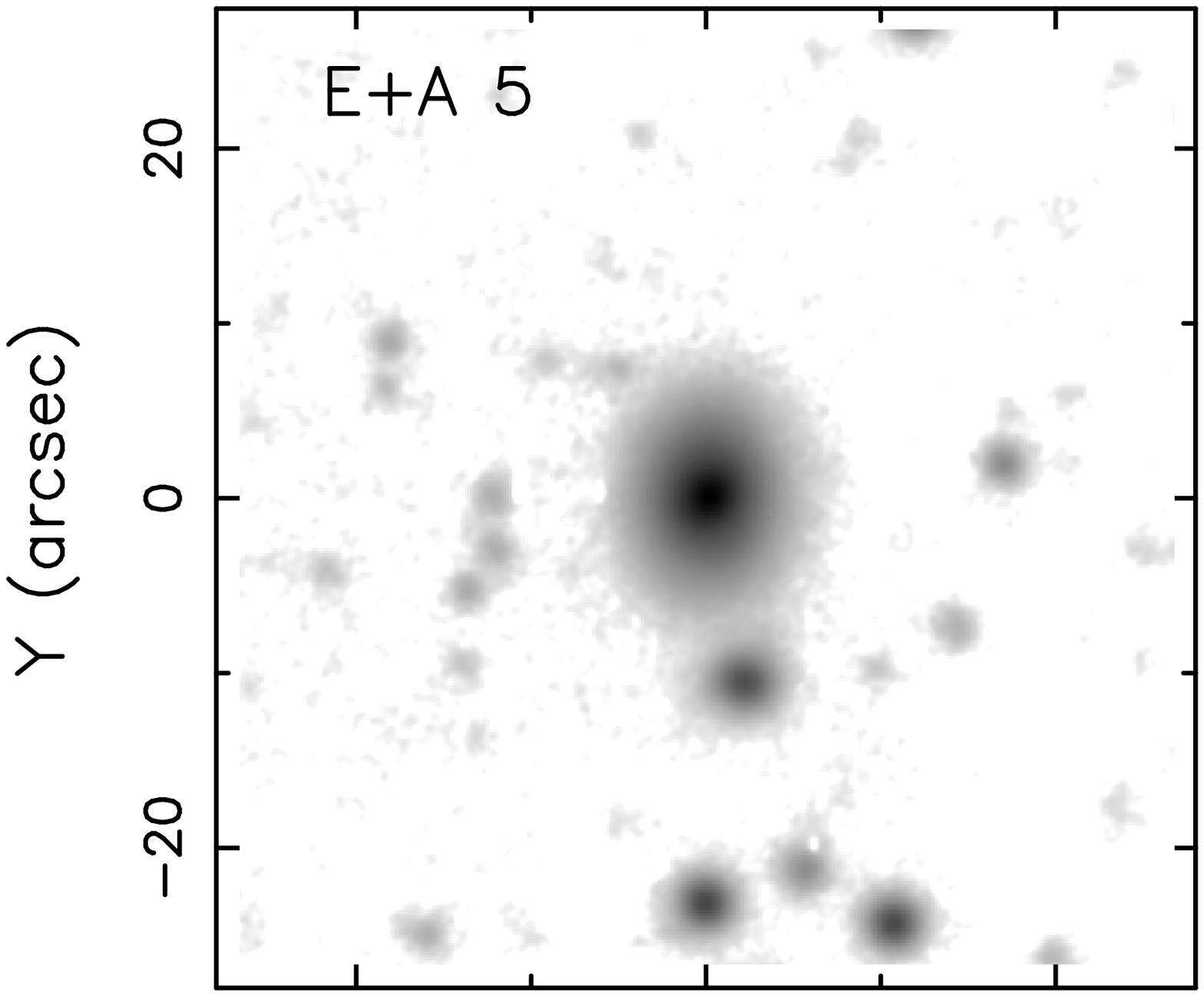}
         \includegraphics[width=2.1cm, angle=-90, trim=0 0 0 0]{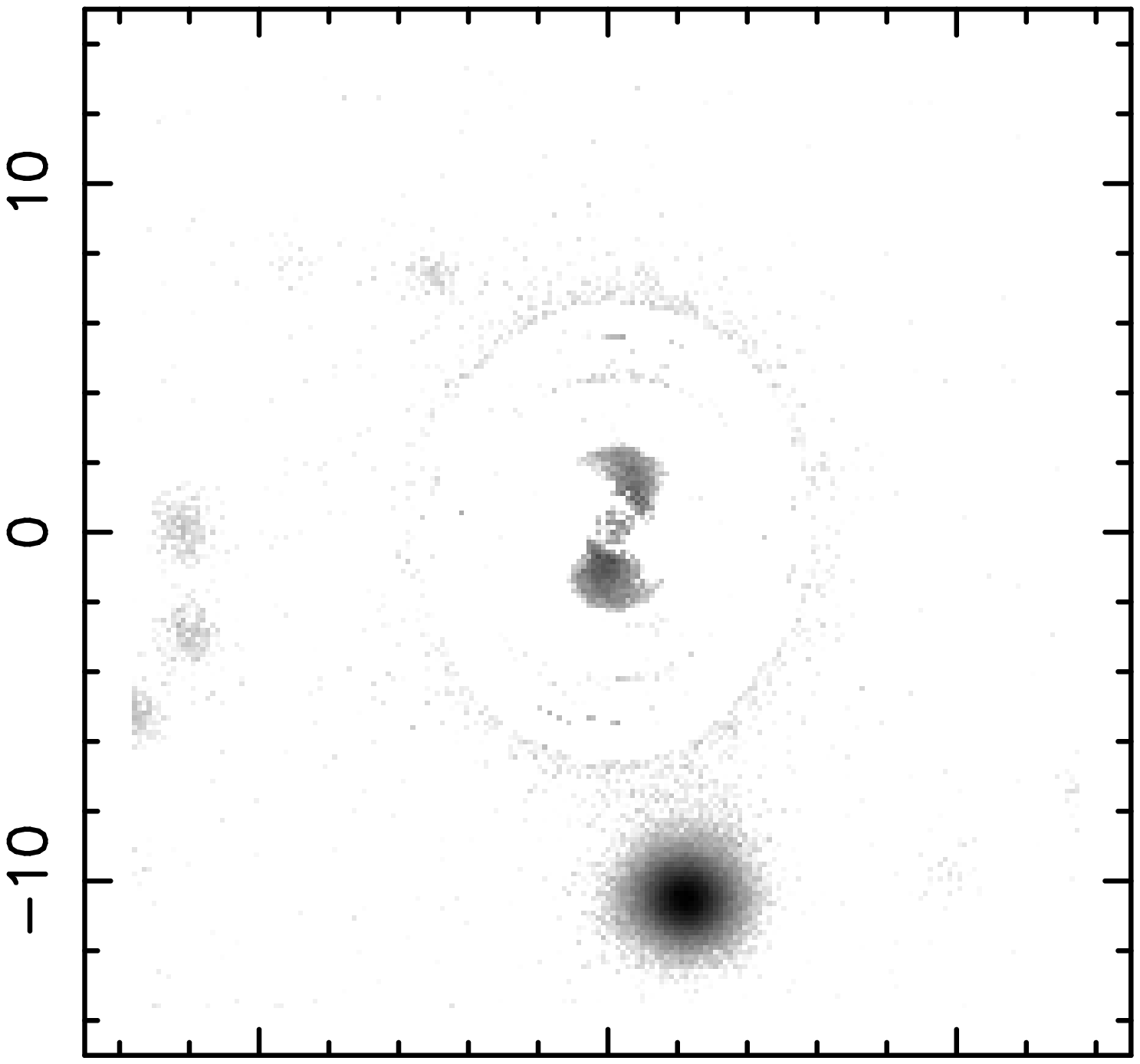}
         \includegraphics[height=2.62cm, angle=-90, trim=0 0 0 0]{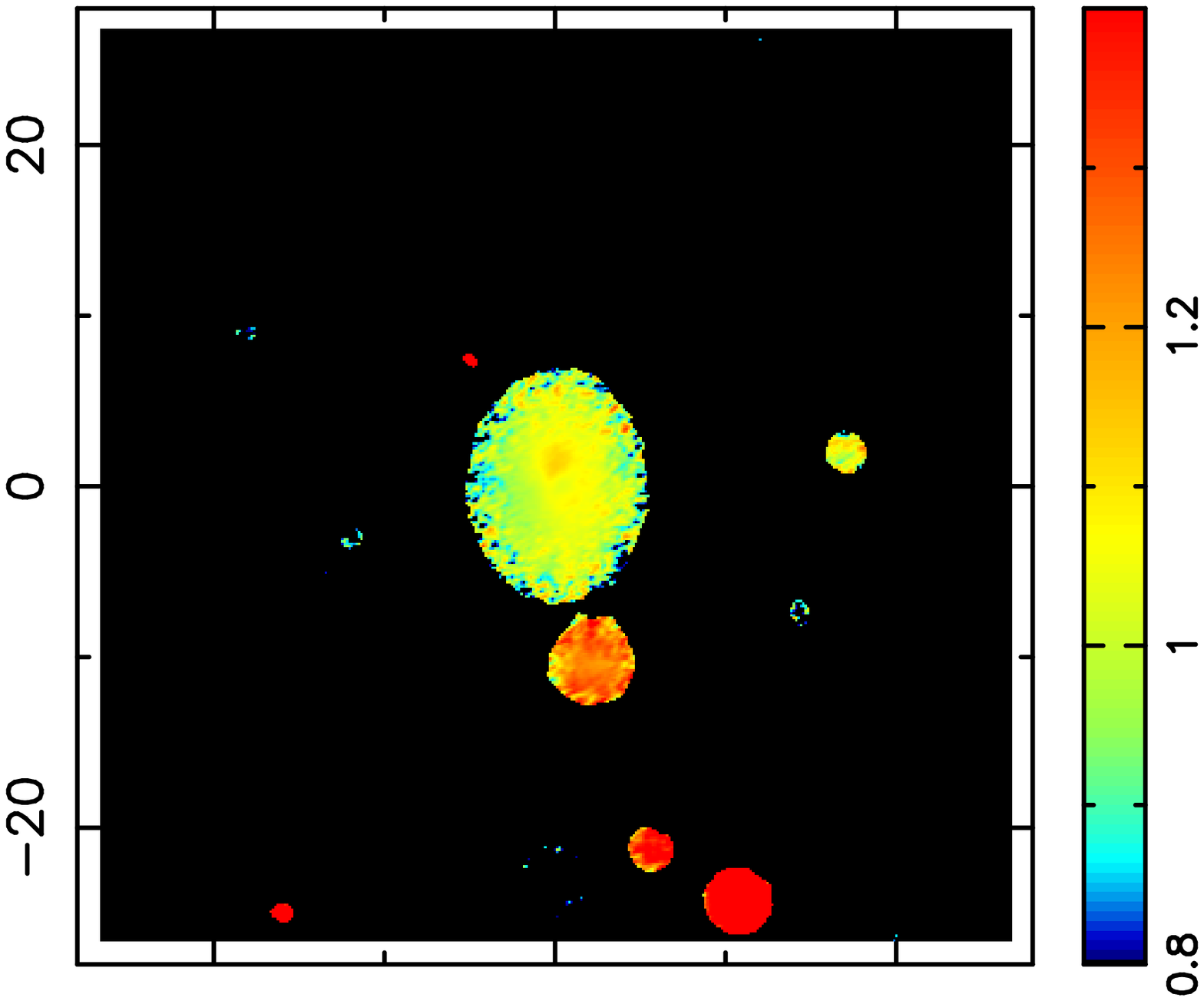}
         \includegraphics[height=2.62cm, angle=-90, trim=0 0 0 0]{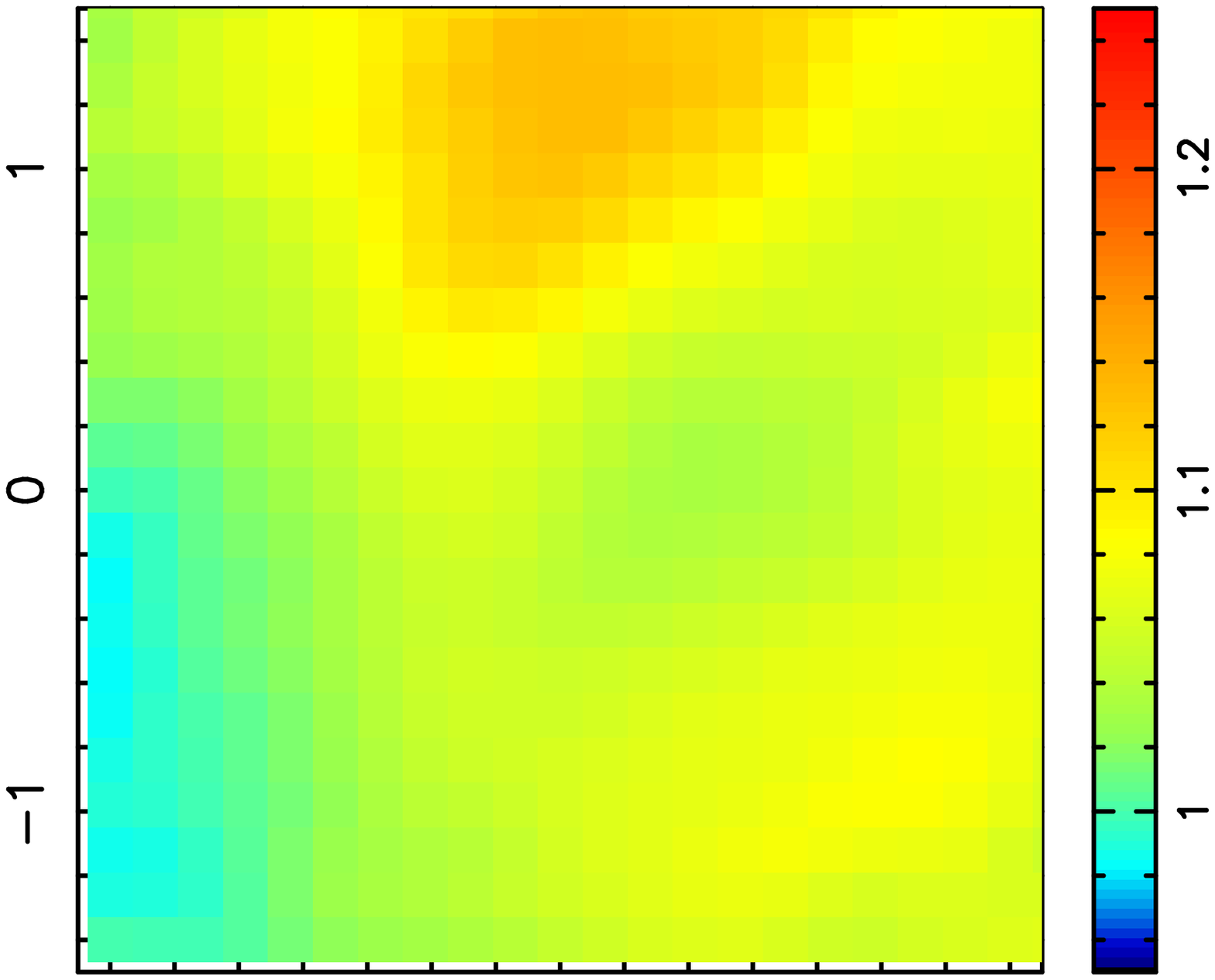}
     \end{minipage}
    \begin{minipage}{0.65\textwidth}
         \includegraphics[width=2.1cm, angle=-90, trim=0 0 0 0]{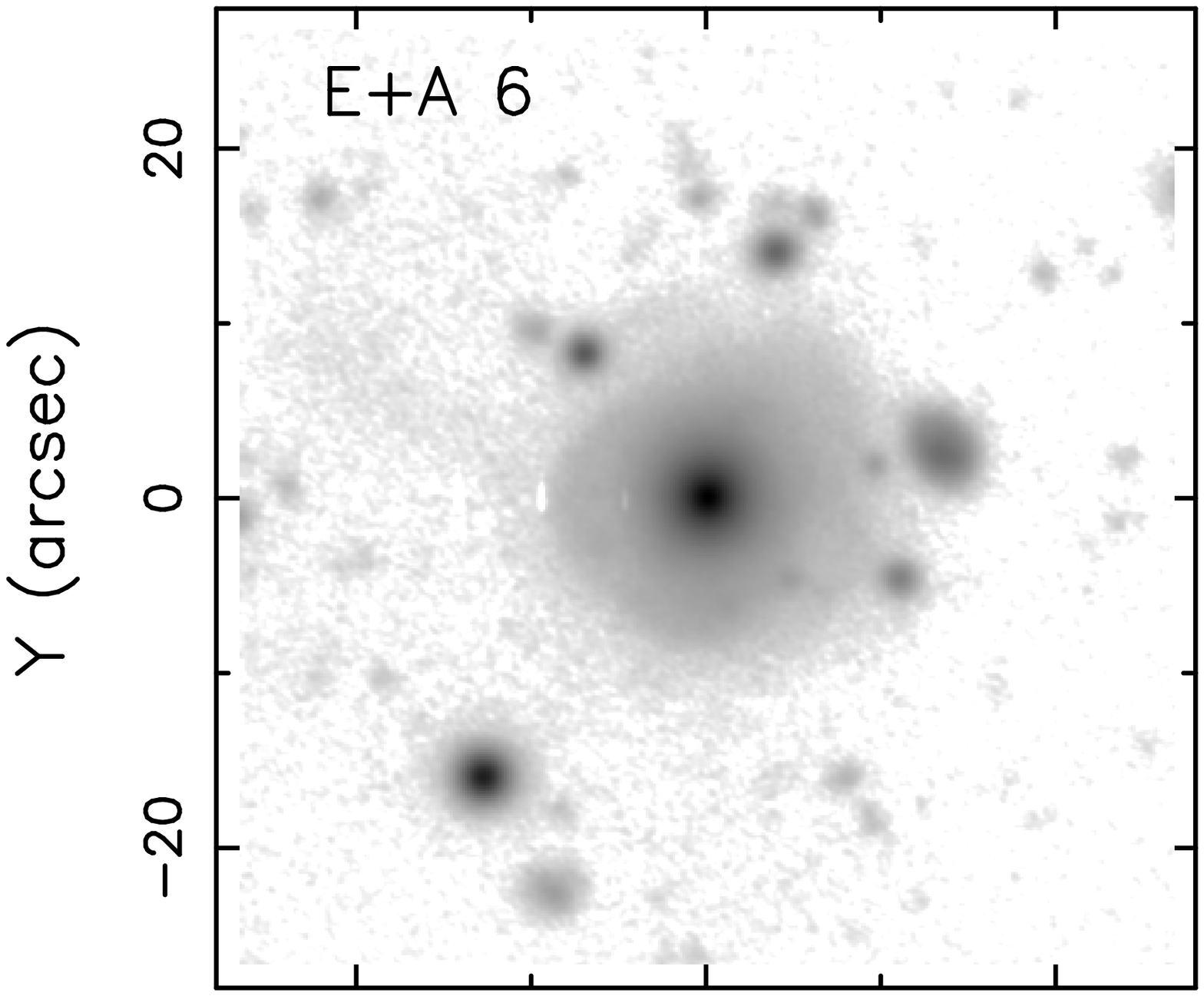}
         \includegraphics[width=2.1cm, angle=-90, trim=0 0 0 0]{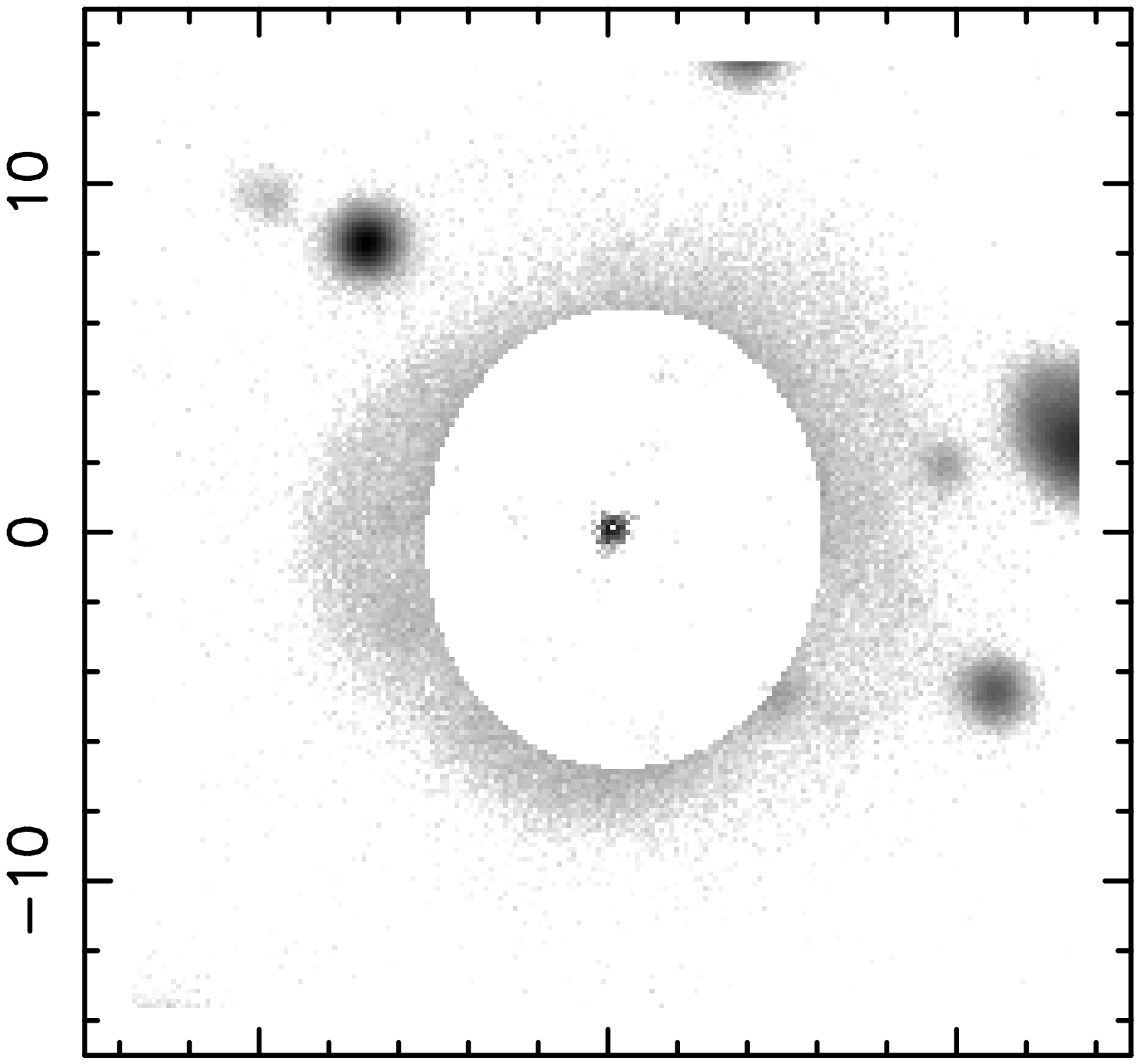}
         \includegraphics[height=2.62cm, angle=-90, trim=0 0 0 0]{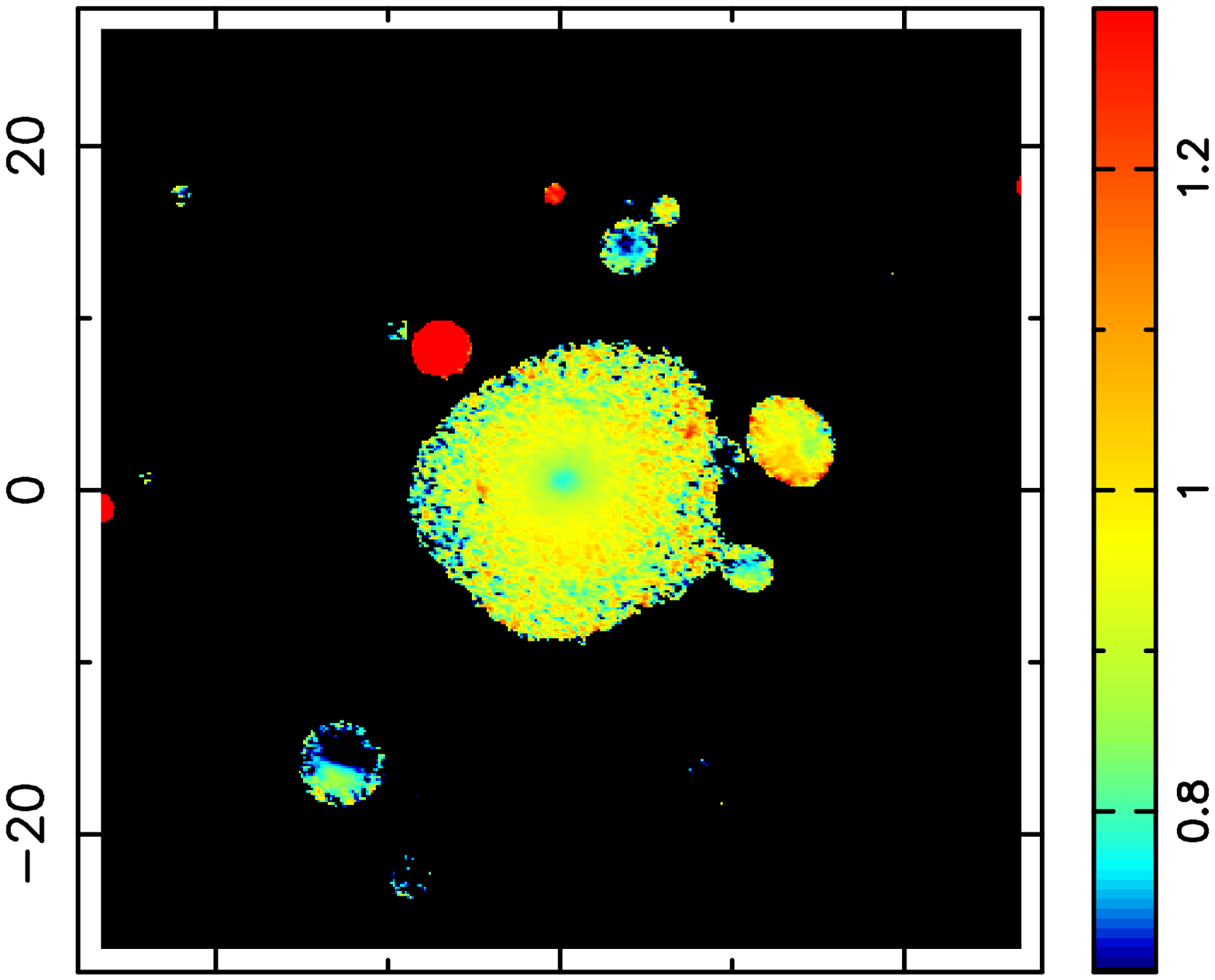}
         \includegraphics[height=2.62cm, angle=-90, trim=0 0 0 0]{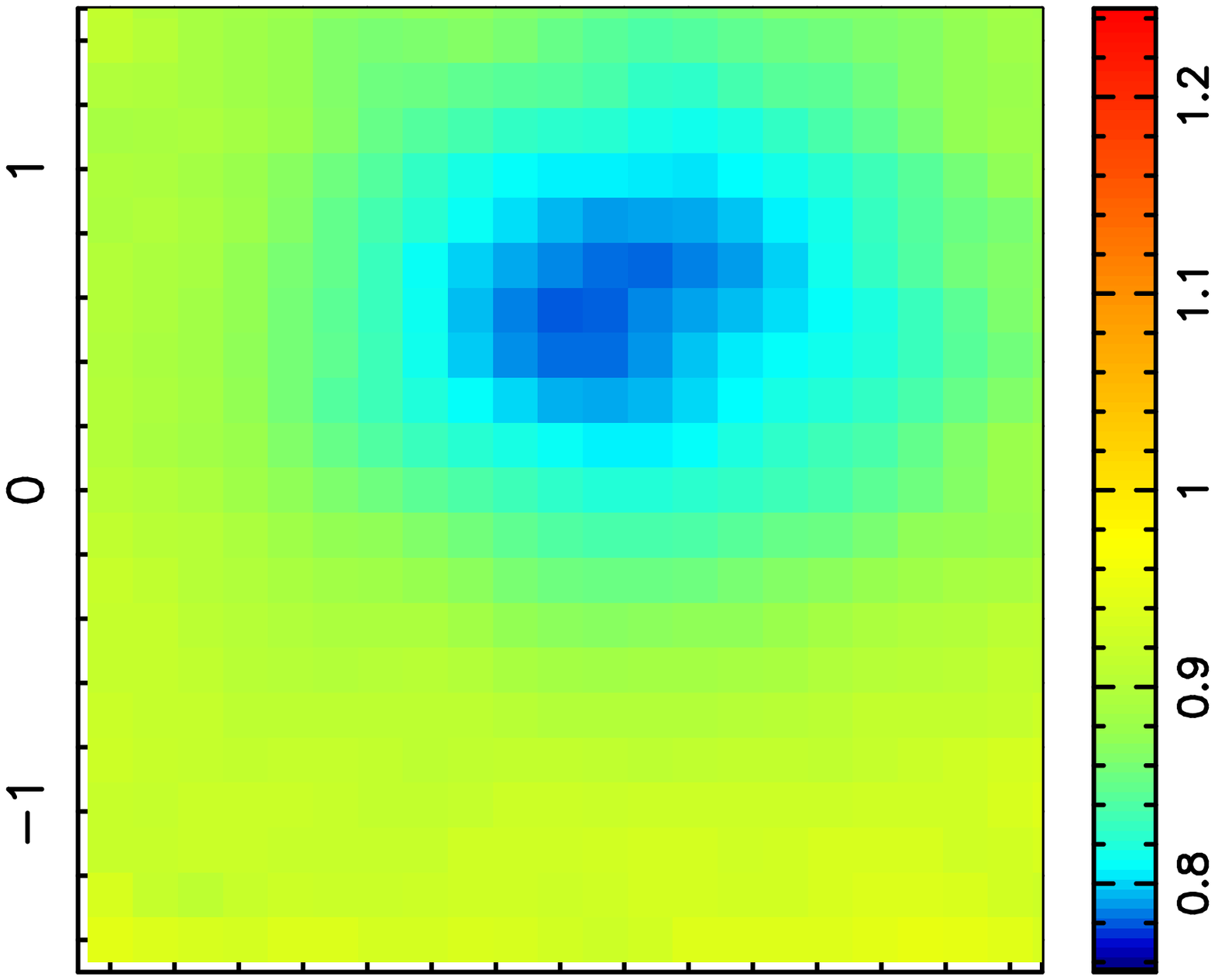}
     \end{minipage}
    \begin{minipage}{0.65\textwidth}
         \includegraphics[width=2.1cm, angle=-90, trim=0 0 0 0]{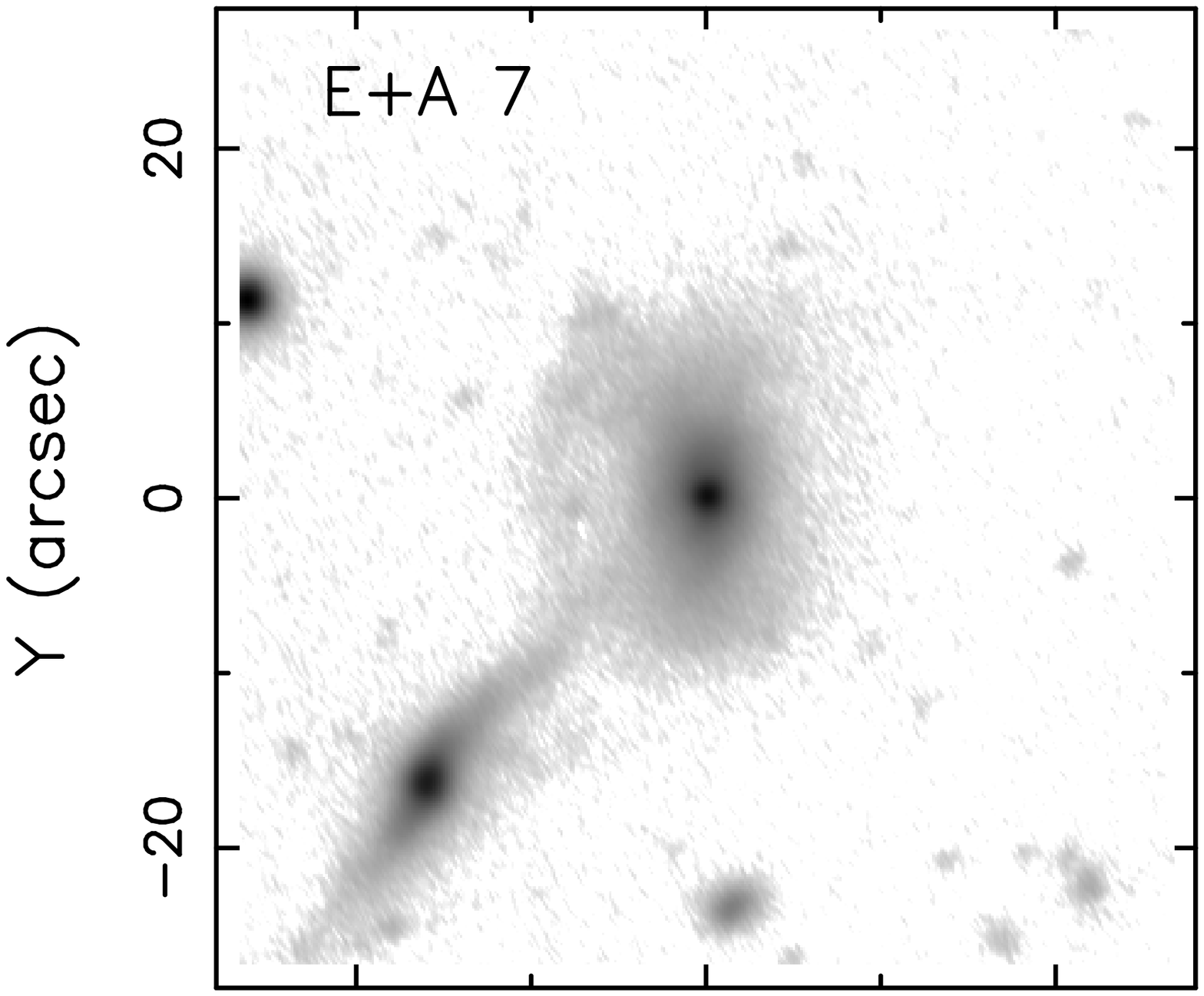}
         \includegraphics[width=2.1cm, angle=-90, trim=0 0 0 0]{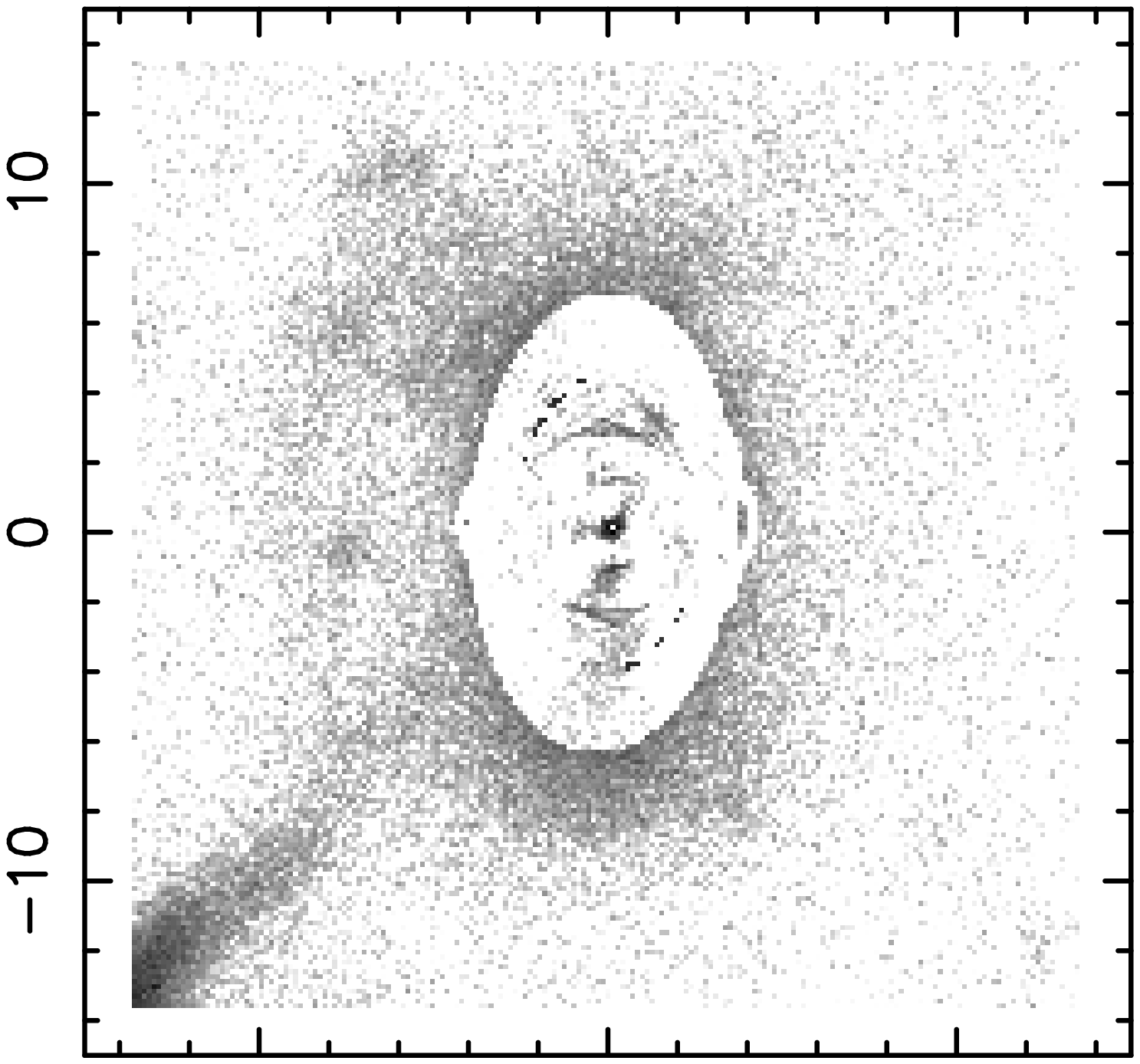}
         \includegraphics[height=2.62cm, angle=-90, trim=0 0 0 0]{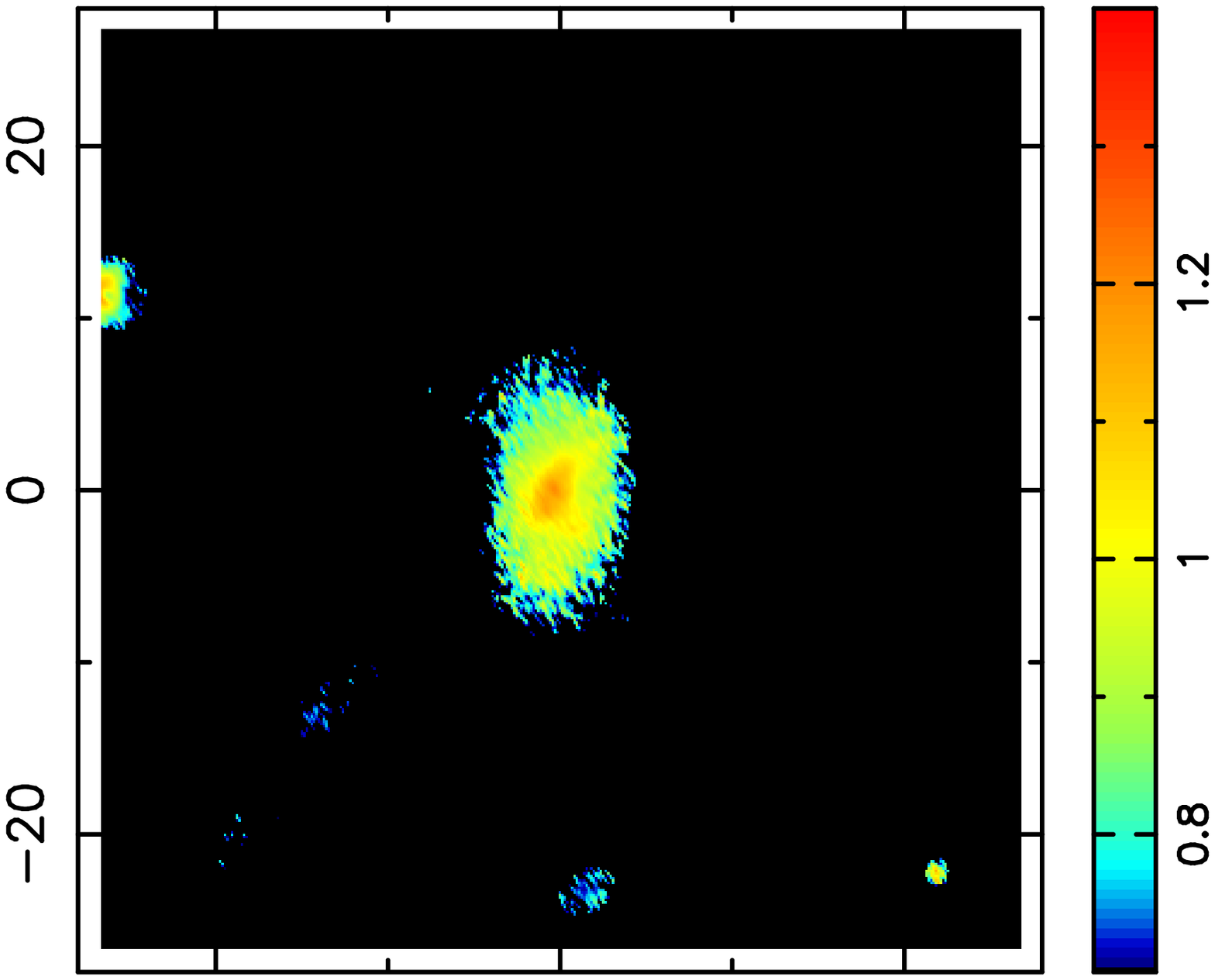}
         \includegraphics[height=2.62cm, angle=-90, trim=0 0 0 0]{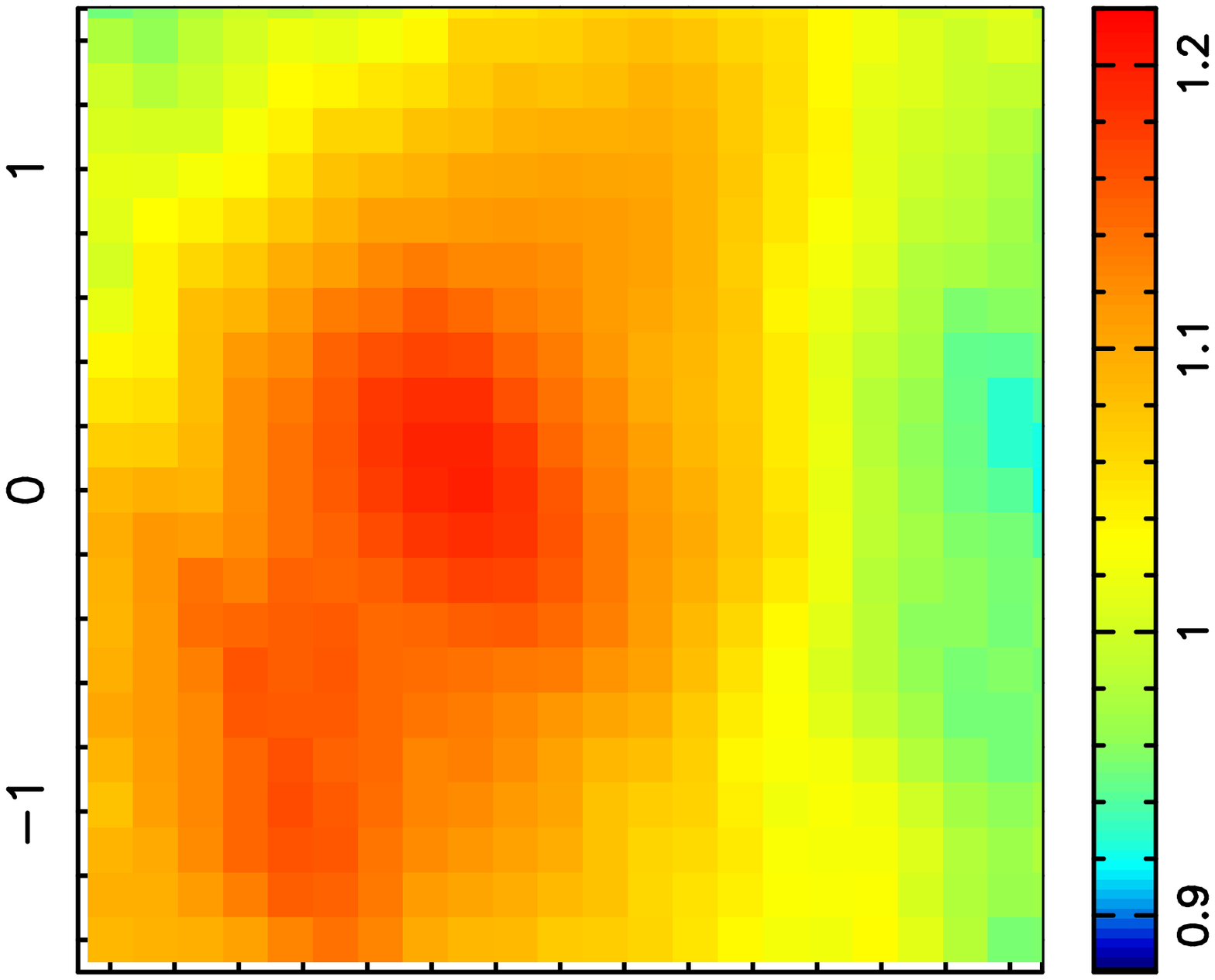}
     \end{minipage}
    \begin{minipage}{0.65\textwidth}
         \includegraphics[width=2.1cm, angle=-90, trim=0 0 0 0]{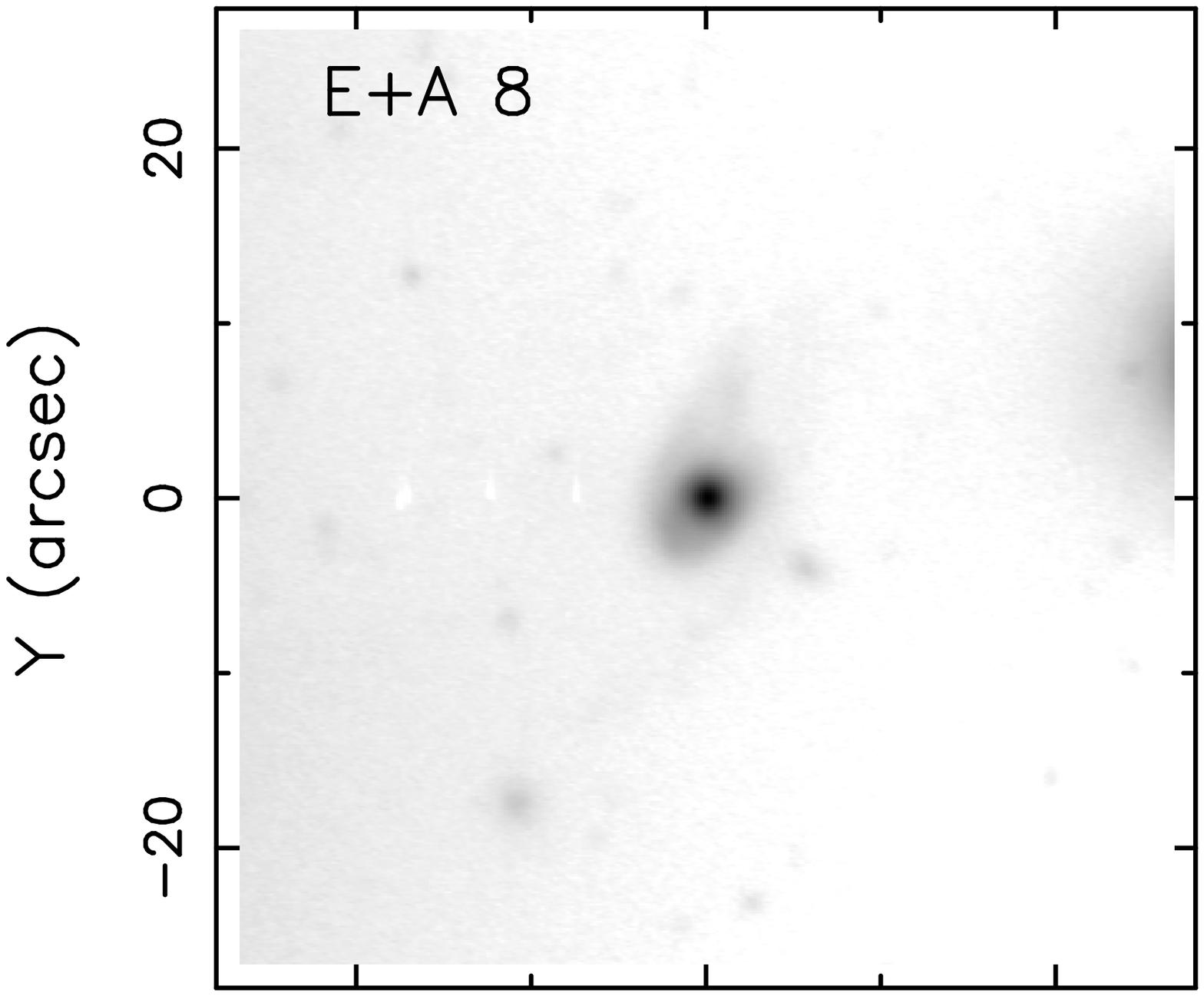}
         \includegraphics[width=2.1cm, angle=-90, trim=0 0 0 0]{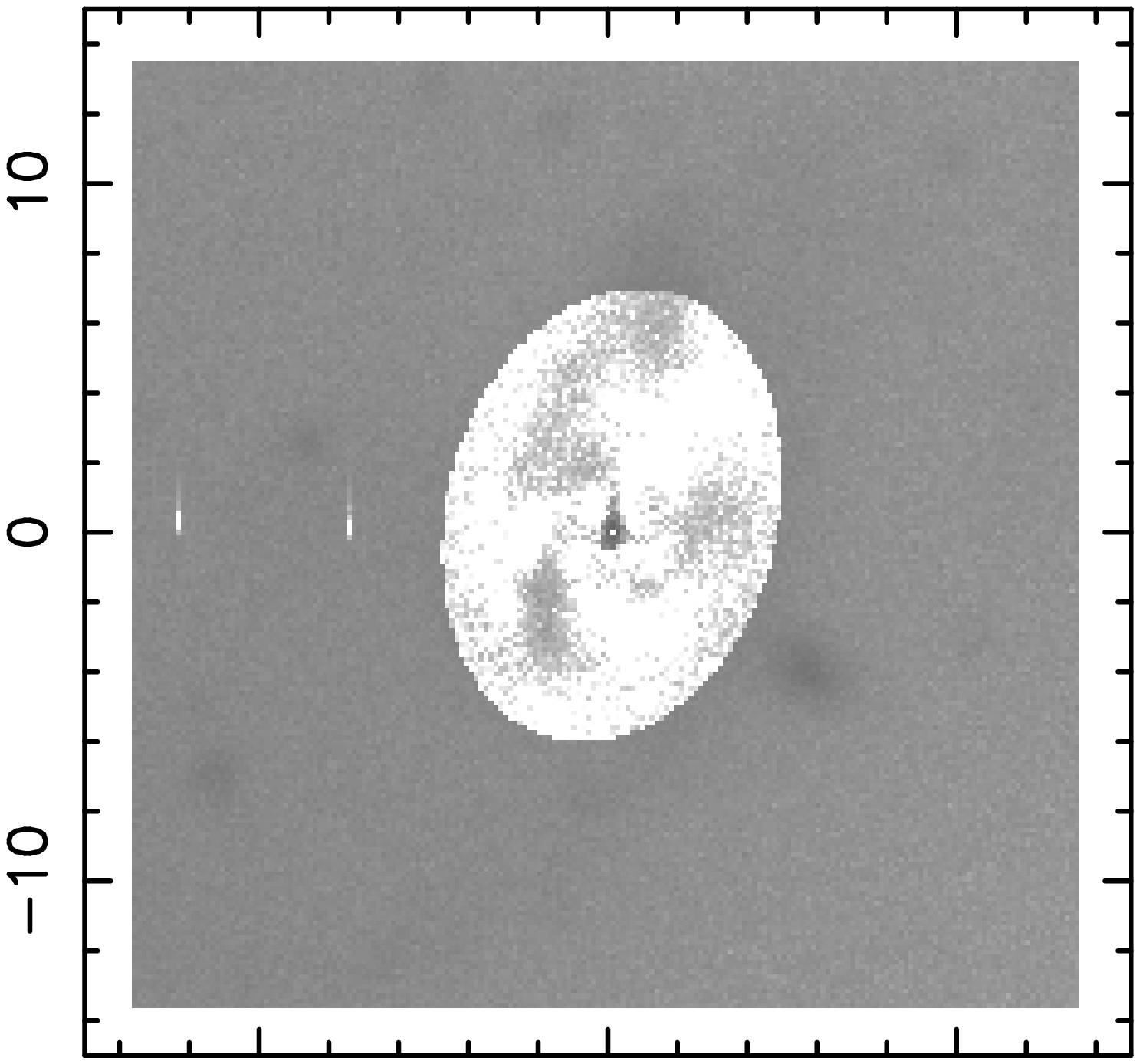}
         \includegraphics[height=2.62cm, angle=-90, trim=0 0 0 0]{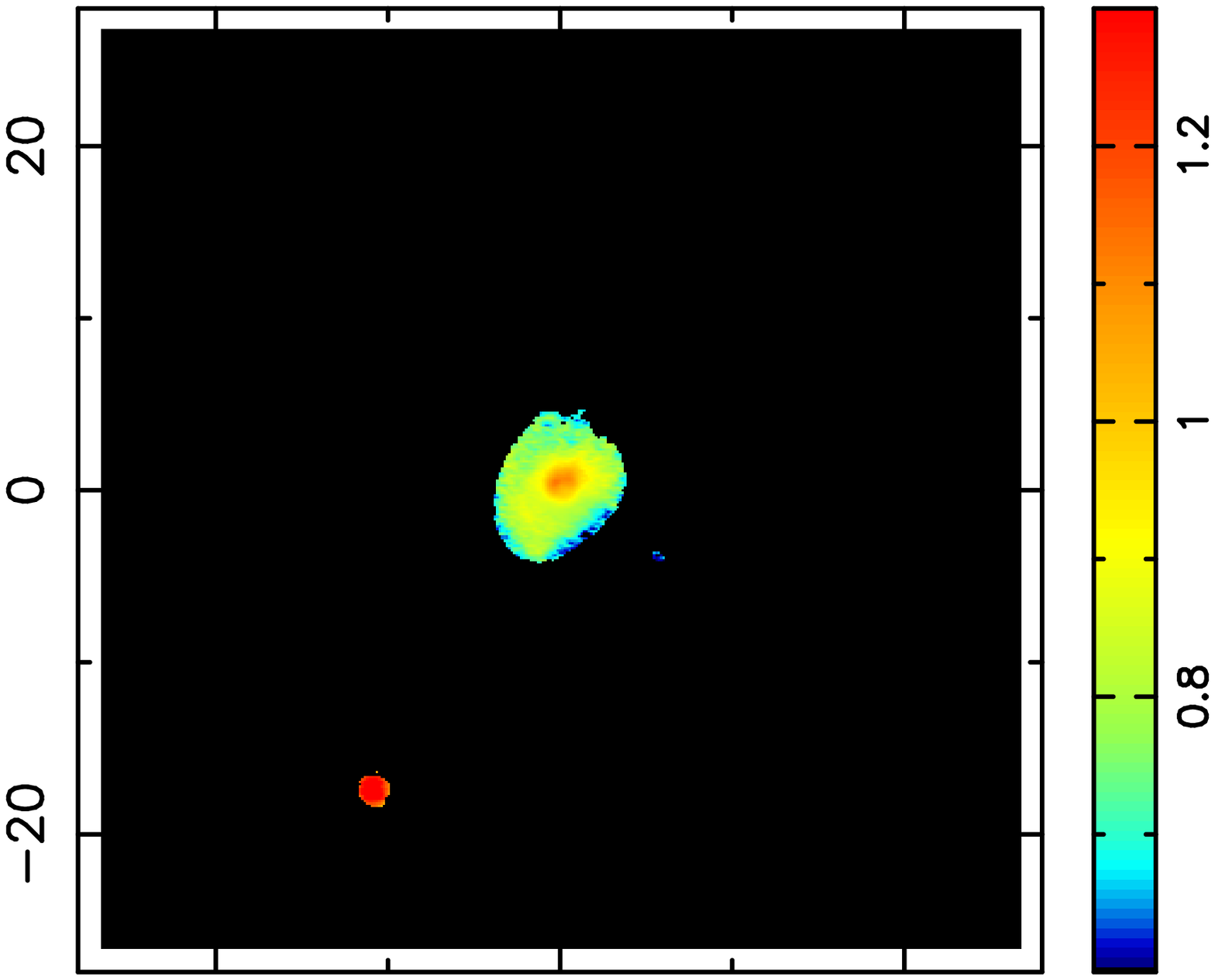}
         \includegraphics[height=2.62cm, angle=-90, trim=0 0 0 0]{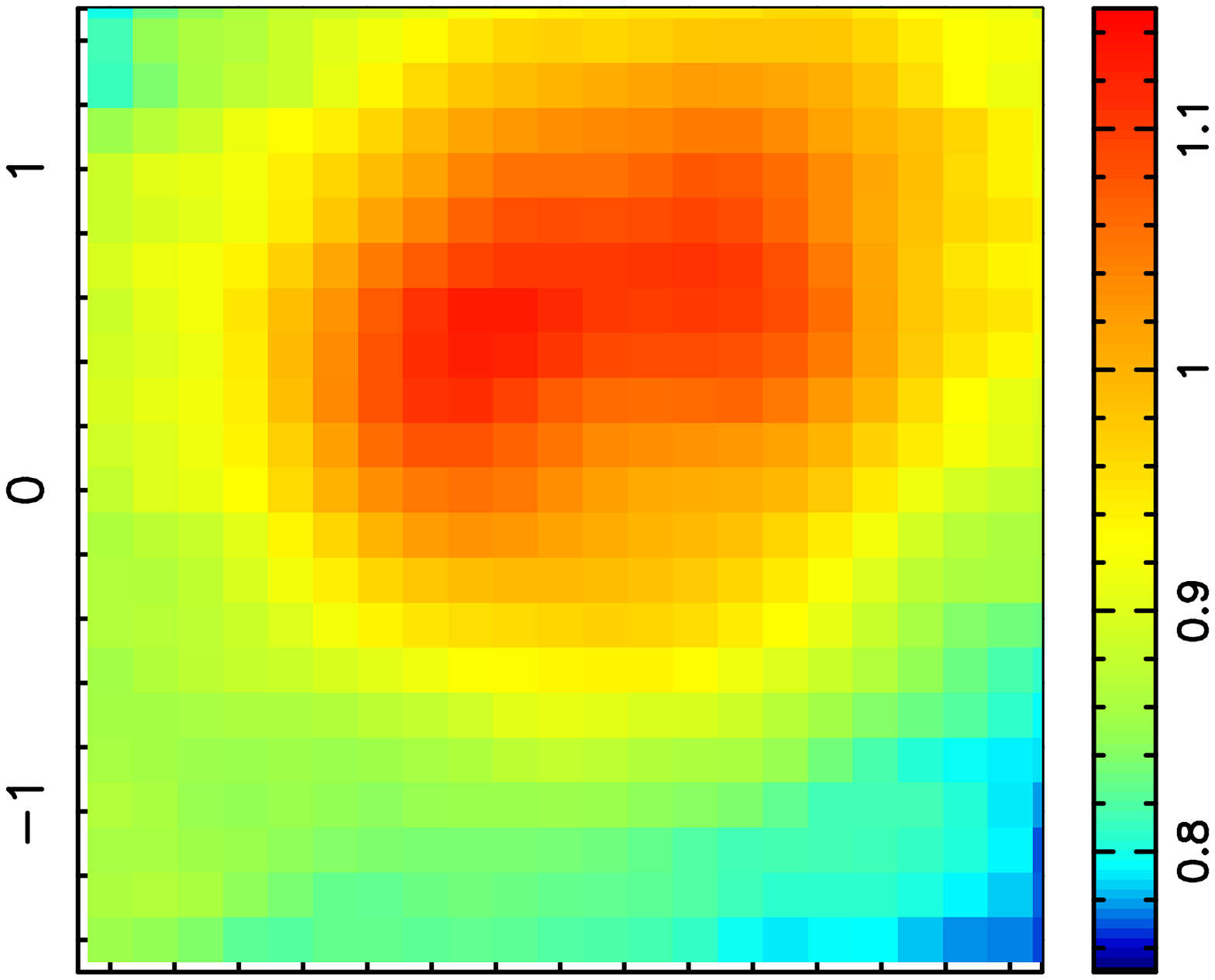}
     \end{minipage}
    \begin{minipage}{0.65\textwidth}
         \includegraphics[width=2.1cm, angle=-90, trim=0 0 0 0]{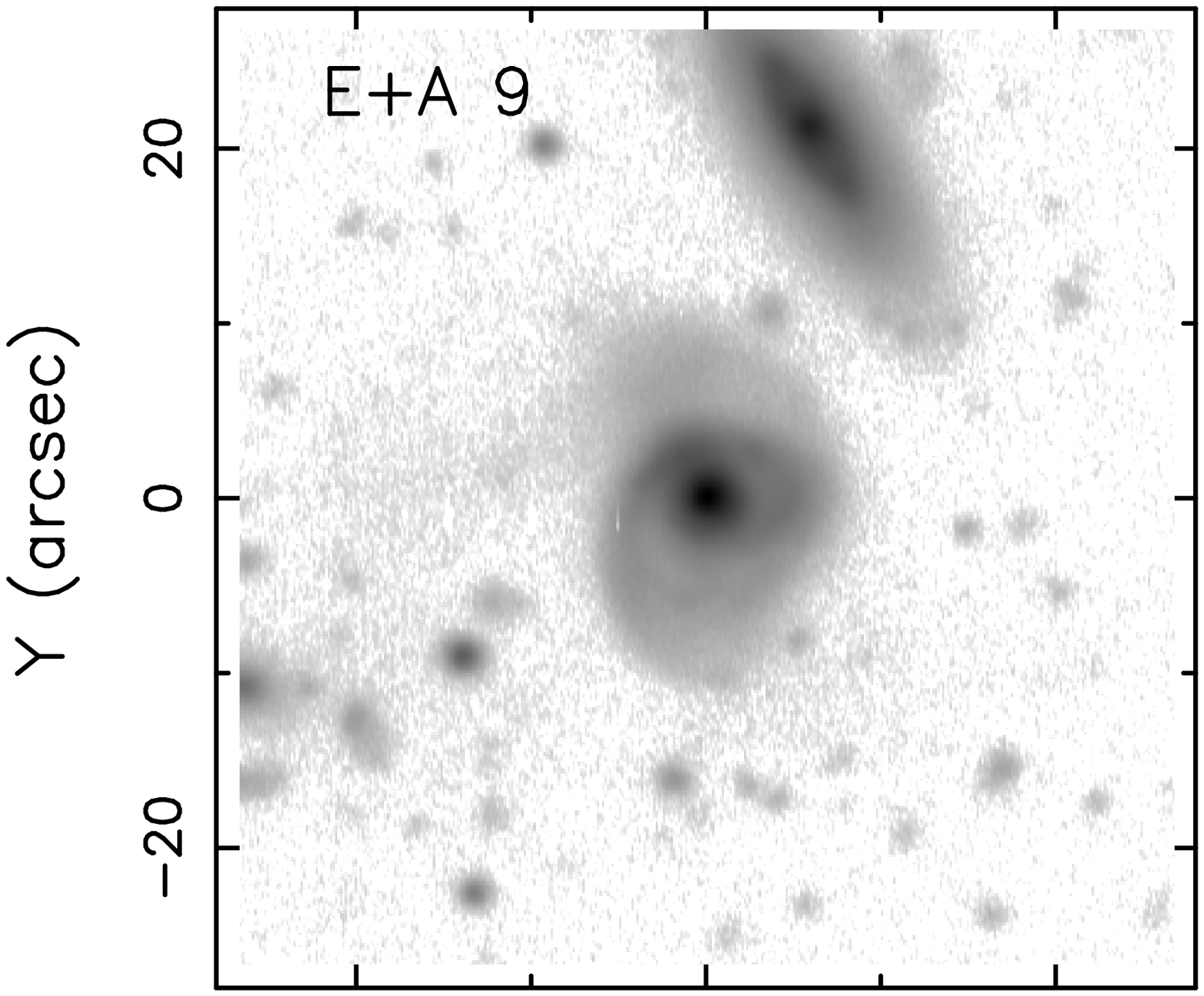}
         \includegraphics[width=2.1cm, angle=-90, trim=0 0 0 0]{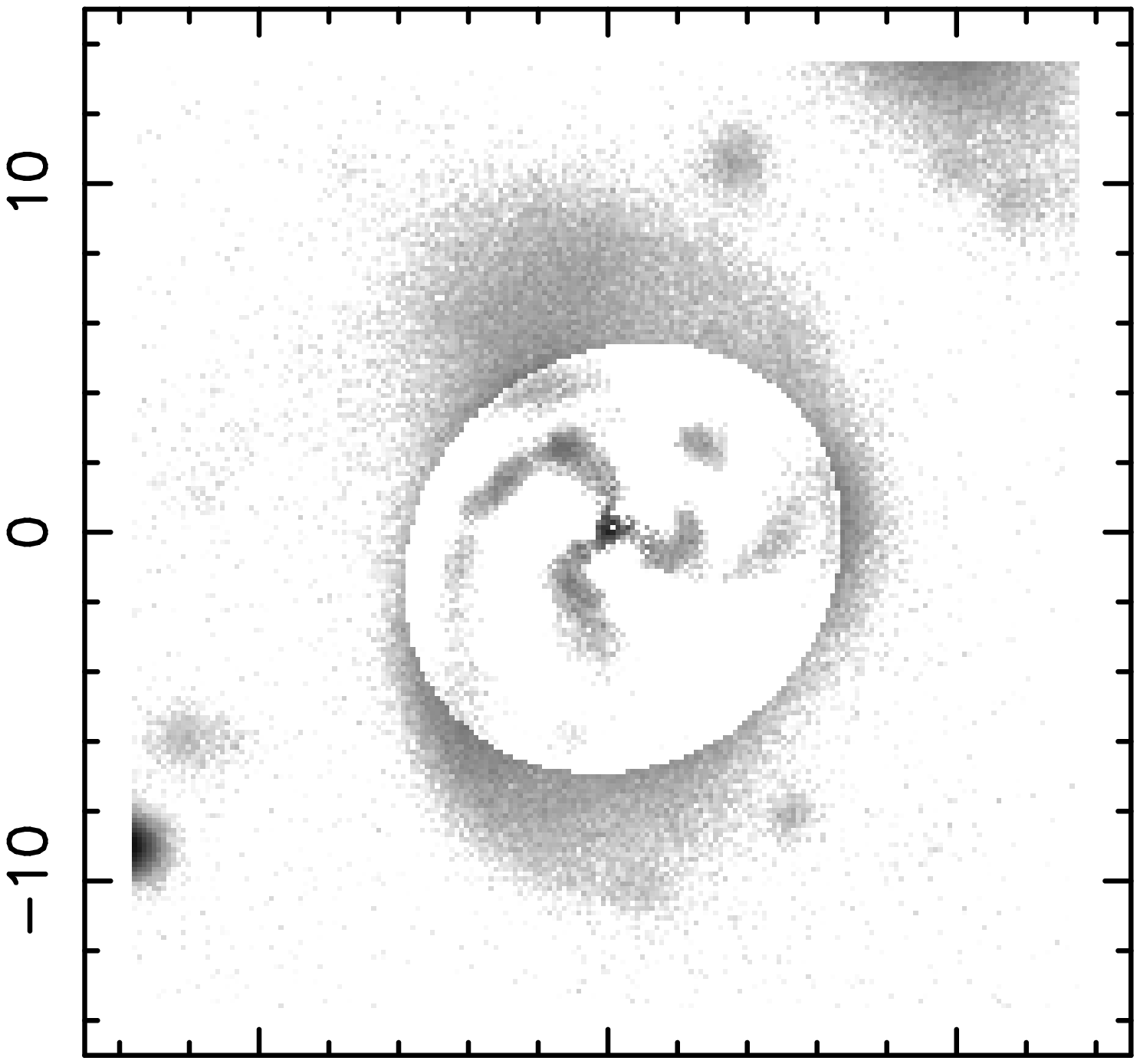}
         \includegraphics[height=2.62cm, angle=-90, trim=0 0 0 0]{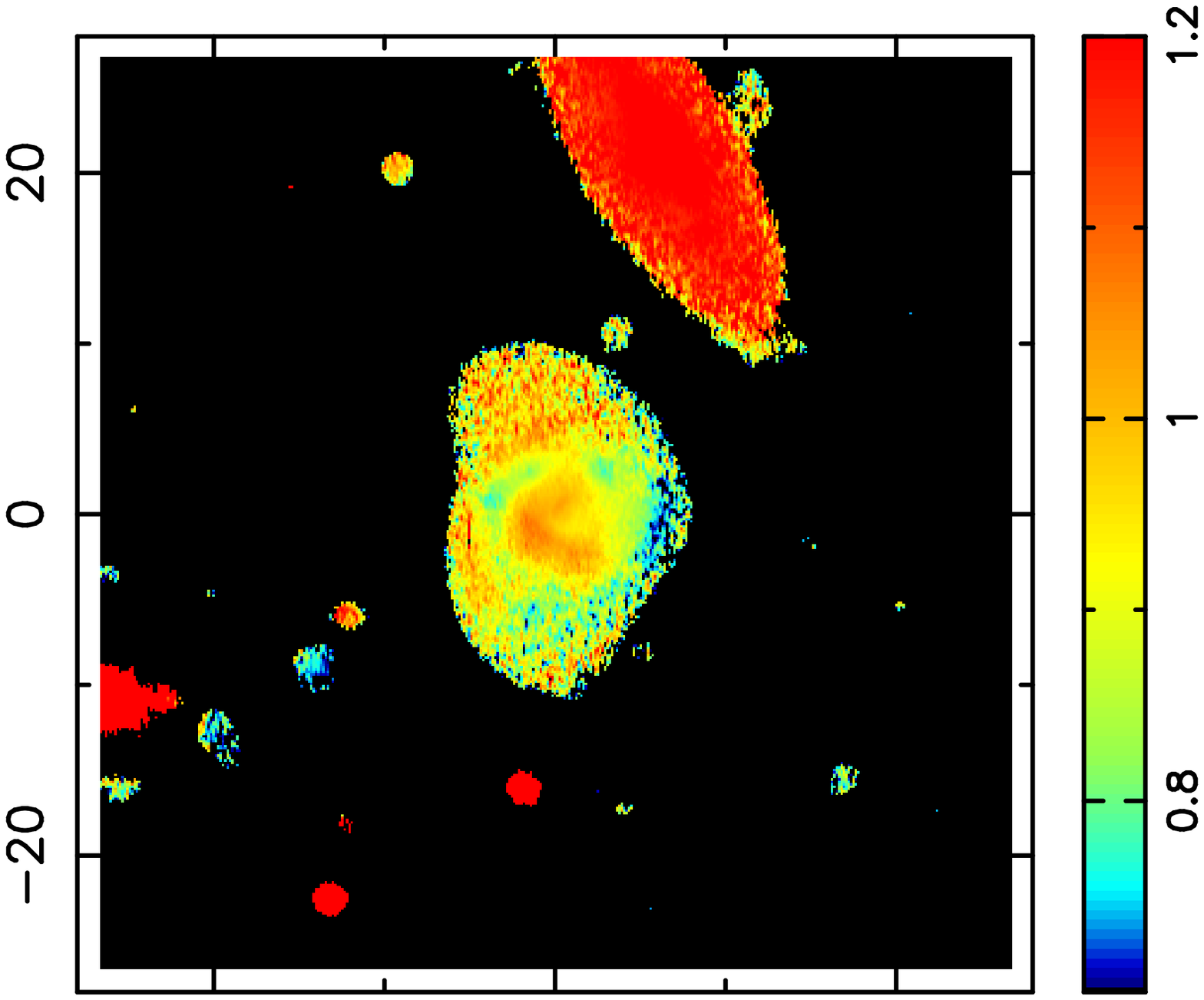}
         \includegraphics[height=2.62cm, angle=-90, trim=0 0 0 0]{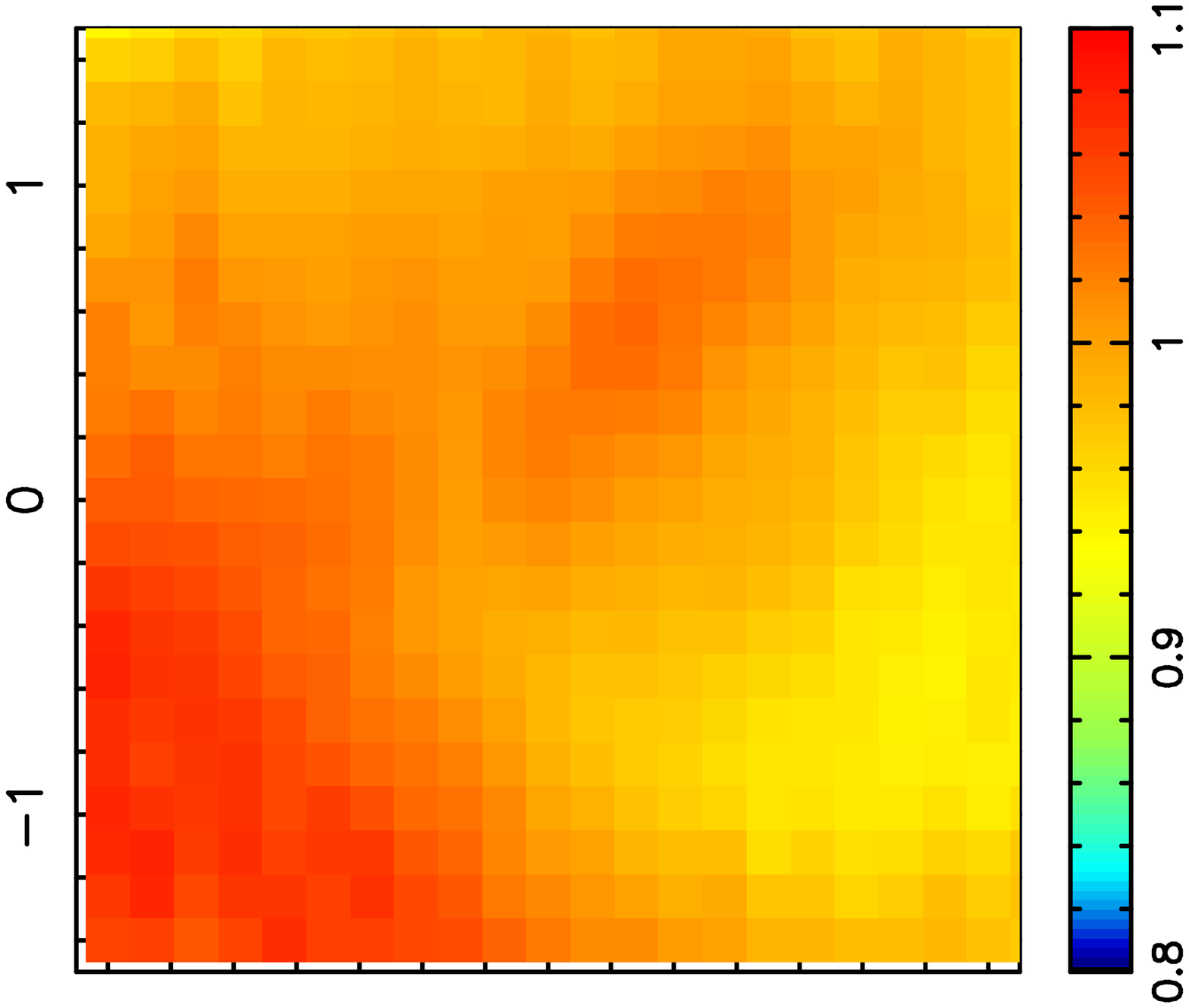}
     \end{minipage}
    \begin{minipage}{0.65\textwidth}
         \includegraphics[width=2.45cm, angle=-90, trim=0 0 0 0]{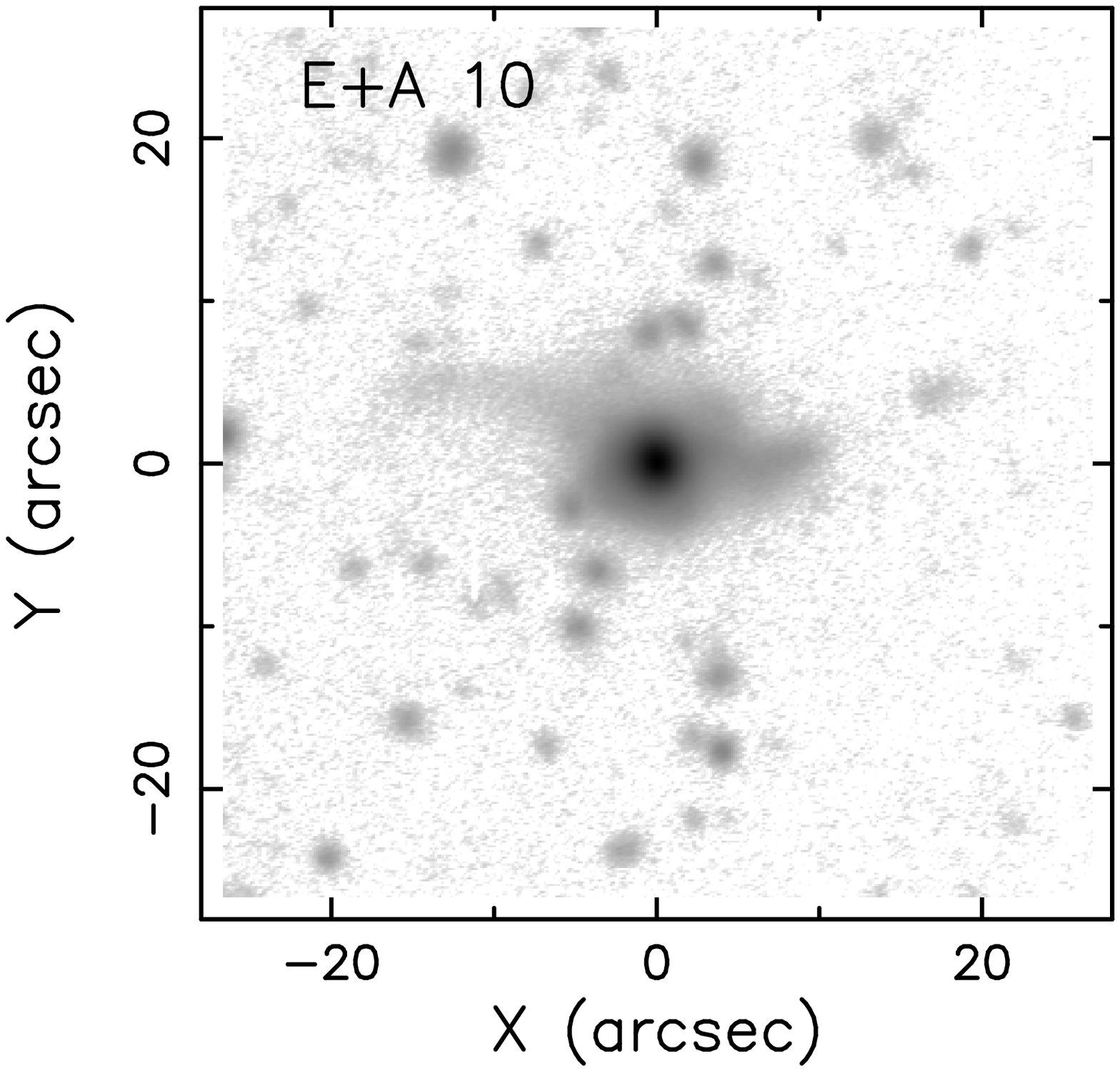}
         \includegraphics[width=2.45cm, angle=-90, trim=0 0 0 0]{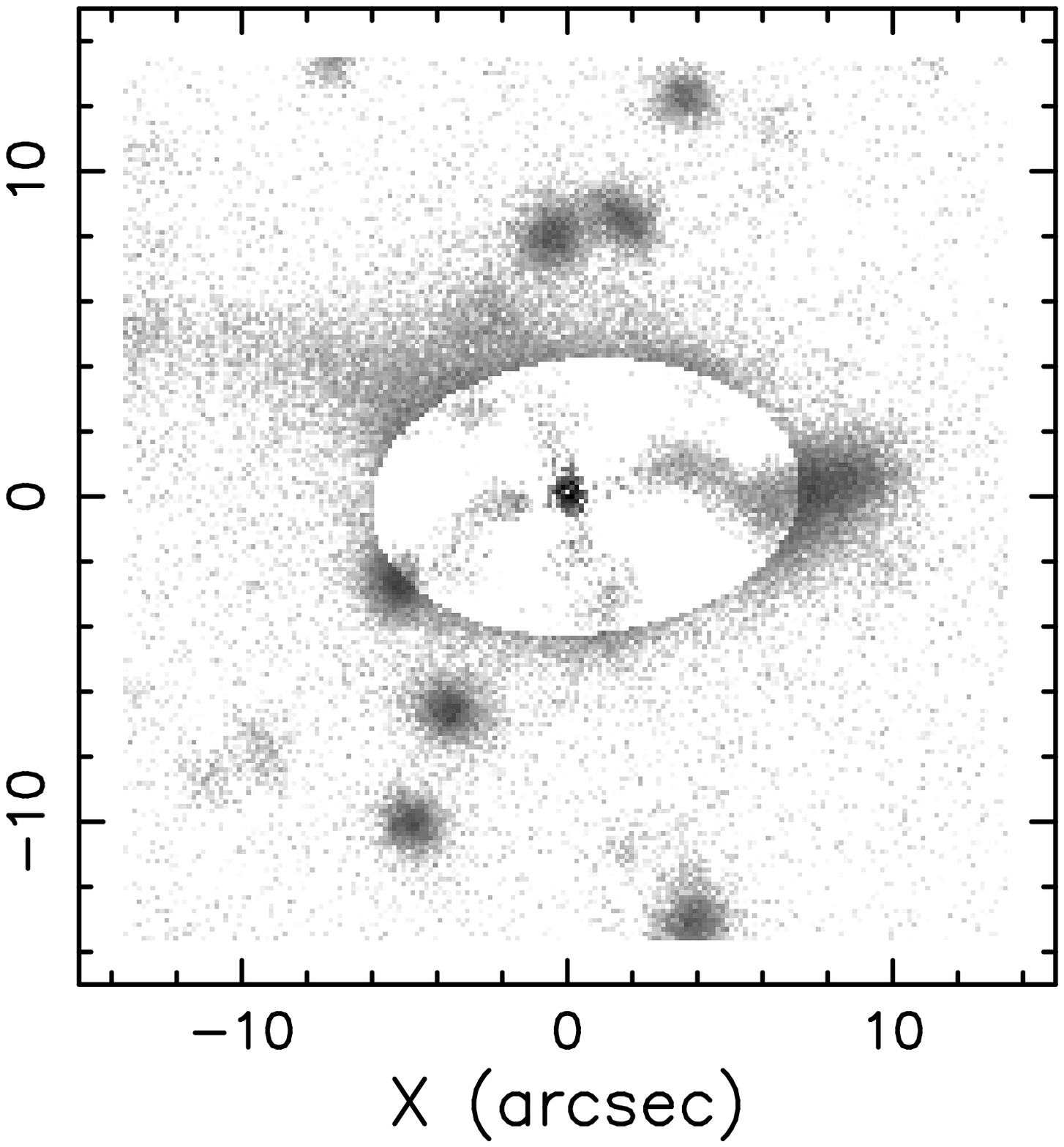}
         \includegraphics[height=2.59cm, angle=-90, trim=0 0 0 0]{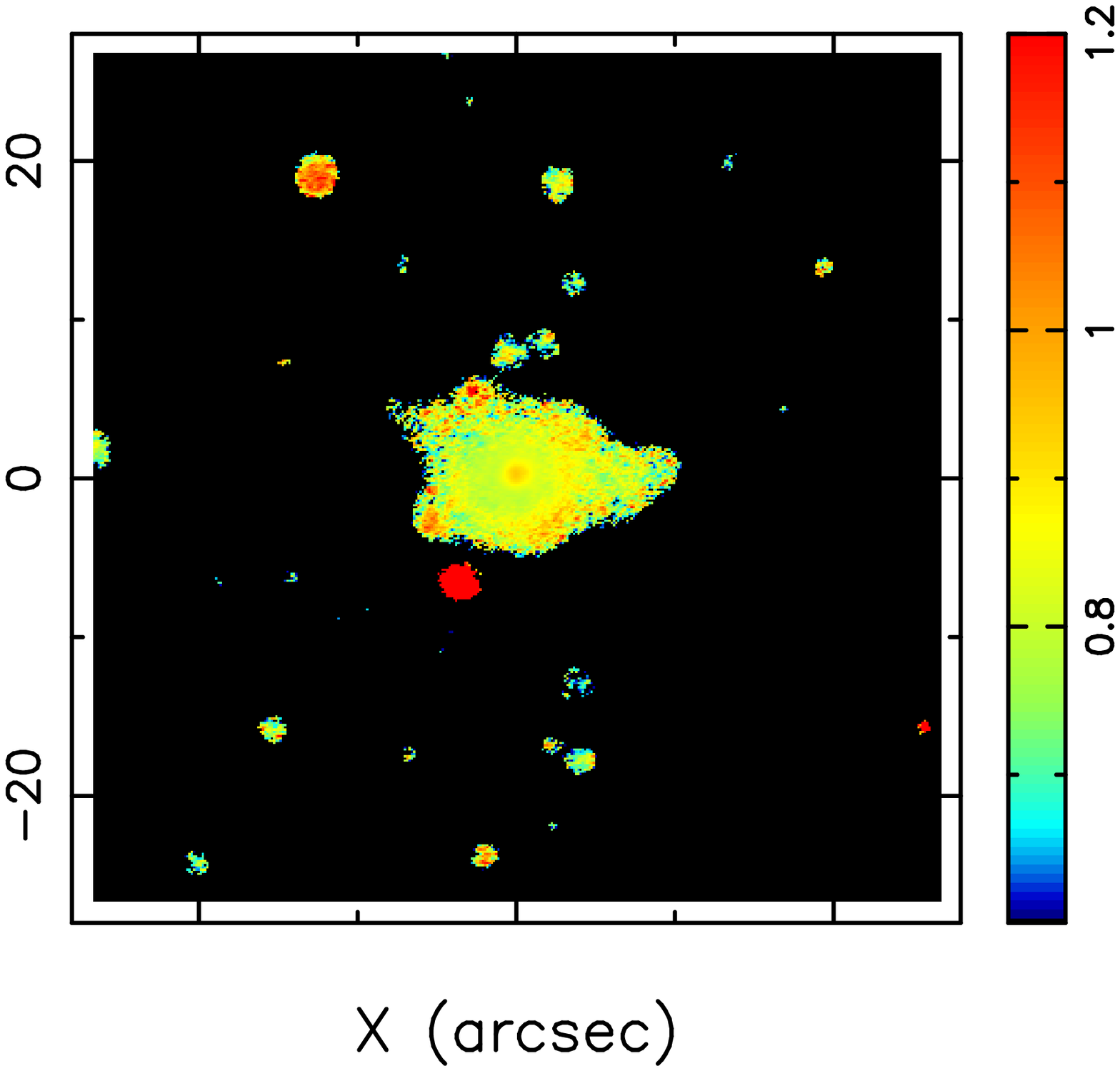}
         \includegraphics[height=2.59cm, angle=-90, trim=0 0 0 0]{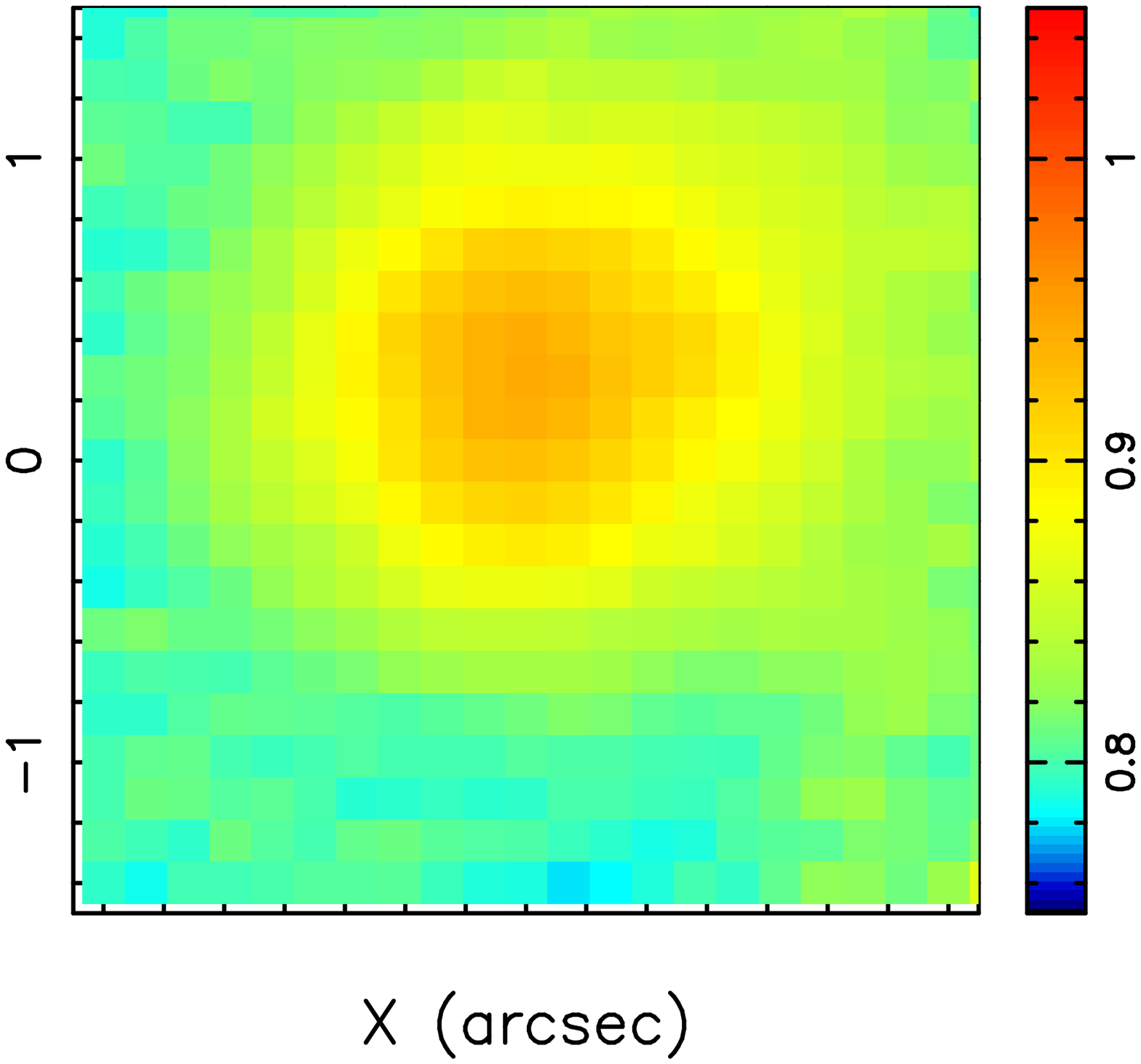}
     \end{minipage}
\end{center}
\caption{\label{fig:images}{\it Top to bottom:} E+A\_1 to E+A\_10. {\it left column:} $g$--band image; {\it 2nd column:}
residual image after subtraction of a model elliptical profile; {\it 3rd column:} $g-r$ colour image; {\it right column:} colour image of the central 3\arcsec. There is no colour image for E+A\_3 because of saturation in the galaxy core in the $g$--band image as well as poor image quality in the $r$--band. Note: the spatial and colour scales vary between columns.}
\end{figure*}

\subsubsection{Data reduction}
The imaging was reduced in the standard manner using the {\sc iraf} Gemini package
task {\sc gireduce} to perform bias and flat field corrections and 
remove the overscan region. The individual exposures were then combined
using the task {\sc gemcombine}. Since no standard star calibrations
were obtained, photometric zero points were calculated by matching
stars in the field to the SuperCosmos catalogue. The SuperCosmos magnitudes were
converted from $b_{J}$ and $r_{F}$ to $g$ and $r$ using the filter conversions given
in \citet{cross04} and references therein. For each GMOS observation we had between
four and ten stars that were suitable for use in calibration (isolated within a suitable magnitude
range). It is known that a small fraction of objects in the SuperCosmos catalogue have spurious magnitudes
with errors of order $\sim 0.5$\,mags. We therefore excluded the few obvious outliers in calculating
the zero--point values. The rms scatter in the derived zero point for all images was
$\sim 0.05$\,mags. This calibration is more than adequate for our purposes since 
the most important information is contained in the relative colour differences 
across individual objects rather than their absolute colours.

The point spread function (PSF) for each image was measured from stars near to the target 
galaxy and modelled as an elliptical Gaussian (since the delivered images can have a slight asymmetry). 
We created colour images by carefully 
aligning the $g$-- and $r$--band images and matching the PSFs by convolving each image with
an elliptical Gaussian such that the final PSFs of both images were circular
Gaussians with final image qualities slightly worse than the worst original image.
i.e.
\begin{equation}
\sigma_{\rm final}^{x,y}=\sqrt{\left(\sigma_{\rm original}^{x,y}\right)^2 + \left(\sigma_{\rm smooth}^{x,y}\right)^2}
\end{equation}
where $\sigma_{\rm final}^{x,y}$ is chosen to be slightly larger than the sigma of the 
semi--major axis of the worst seeing image. These matched images were then divided by 
one another to produce a colour image. Colour images of each galaxy are shown in column 3 of Figure \ref{fig:images}, 
with a zoom-in of the central $3\arcsec \times 3\arcsec$ region shown in the following column. Note that the colour
scales vary between the two columns.
We could not construct a colour map of E+A\_3 because the 
$g$--band image is saturated.

\subsection{Spectroscopy}

\subsubsection{Observations}
Our spectroscopic observation were obtained with the Gemini-South GMOS
spectrograph in IFU mode. The observations were conducted in queue mode 
between 2005 September 2 and 2005 December 6. We used the B600 grating in combination with
the $g$--filter resulting  in a spectral coverage of $\sim$\,4100--5380\,\AA\, with a resolution of $\sim 1.9$\,\AA.
We used the IFU in two--slit mode, which gives a rectangular field-of-view of 
5\arcsec $\times$ 7\arcsec\, sampled by 1000 $\times$ 0.2\arcsec\, individual lenslets. 
For each galaxy we obtained $4\times 1020.5$\,s dithered exposures, resulting in a total 
exposure time for each target of 4082\,s. Only 8 of our 10 targets were observed spectroscopically.
A summary of the spectroscopic observations is given in Table \ref{tab:observations}.

\subsubsection{Data reduction} 
The spectroscopic data were reduced using standard {\sc iraf} routines.
Firstly, the flat field spectra were overscan--subtracted and trimmed using
the {\sc iraf gemini} package task {\sc gireduce}. The flat field fiber spectra were 
traced, extracted and the individual CCDs mosaiced using the {\sc gfextract} task.
The task {\sc gfresponse} was used to calculate the relative fiber throughputs from
the flat field spectra. Arc images were then matched to their temporally nearest flat-field image
and reduced in a similar manner but using the flat-field to define the spectroscopic
apertures and traces. Following this, the wavelength solution was determined interactively using
the {\sc gswavelength} task. The science spectra were overscan subtracted, trimmed, extracted using
the trace of the nearest flat-field image, wavelength calibrated using the appropriate wavelength
solution and corrected for variations in the relative fiber throughput using their corresponding response images.

At this point, inspection of the extracted 2-D spectra revealed several problems. There 
were throughput discontinuities in the wavelength direction where the different CCDs had been mosaiced 
together. There was also a discontinuity in the 'spatial' direction of the 2-D spectra corresponding 
to the cross over between the 2 slits. Further, looking at the spectra from sky fibers  (and fibers corresponding to
the outer part of the science IFU which are expected to have little contribution from the target galaxy)
revealed a gradient in throughput with aperture number (spatial axis of the 2-D spectrum).
This gradient appeared (and had the same sense) in both halves of the 2-D spectra corresponding to the 
two separate slits. These gradients were fitted, in each individual exposure,  with a linear 
function on either side of each discontinuity 
and then divided out to give an approximately uniform level in the sky spectra. The sky spectra corresponding
to each slit were then averaged and sky subtraction was performed on all spaxels\footnote{A spaxel is the 
SPAtial piXture ELement of the instrument; in this case the IFU lenslet array}. The accuracy and 
systematics in the sky subtraction were checked by examining both the sky-subtracted sky spectra and also
(importantly) the sky-subtracted spectra from the outer regions of the target galaxy IFU, which have
essentially no signal from the galaxy, and hence are a good indication of how well our reduction and sky subtraction
has worked for our galaxy spectra. In general this procedure worked well, however, small residual offsets and gradients
in the sky-subtracted sky spectra remained. At this point data cubes were constructed for each IFU observation
using the task GFCUBE. The data were re-sampled spatially by this procedure and the resulting data cube had 
spatial pixels which were square with 0.2\arcsec\, sides.  The data cubes corresponding to individual dithered exposures for each target were shifted and averaged 
into a single data cube. The original observations were performed with a square dither pattern with each individual exposure offset by 1\arcsec.
This dither pattern between exposures was checked a posteriori by collapsing the cubes of individual exposures along the spectral direction and
comparing the resulting images. This confirmed the accuracy of the dither pattern to sub-spaxel accuracy and means any smearing of the image 
quality during combination will be small compared to the seeing.
Therefore we were able to combine the individual data cubes by shifting an integer number of spaxels (i.e. 5 $\times$ 0.2\arcsec\, spaxels) 
before combining the spectra using the {\sc iraf} task {\sc scombine}.  Each spectrum was then 
cleaned of any remaining cosmic rays using the {\sc iraf} task {\sc lineclean}.
The residual offsets described above could be seen in the final data cube as additive offsets between different 
columns of IFU elements and these were removed by subtracting off the average spectrum from the two outermost 
lenslets from all other lenslets in that column. After this procedure, examination of those spectra in the data cube, which 
had no discernible signal from the galaxy, revealed a well behaved sky subtraction with little sign of systematics present.

The systematics in the sky subtraction can be quantified by measuring the rms of the mean flux in spaxels expected
to have no contribution from the object. This can then be compared with the random noise by measuring the mean of the rms
scatter in those spaxels. For our data the ratio of these is in the range 19 to 26\,per cent, implying that the
systematics in the sky subtraction are small compared with the random errors.
A relative flux calibration was performed using observations of a 
flux standard star taken with the same set up as the science observations, which are done as part of the standard 
calibrations at Gemini Observatory.

\section{Photometric characteristics}

\subsection{Isophotal profiles}
\begin{figure*}
   \begin{minipage}{0.75\textwidth}
         \includegraphics[width=4.2cm, angle=90, trim=0 0 0 0]{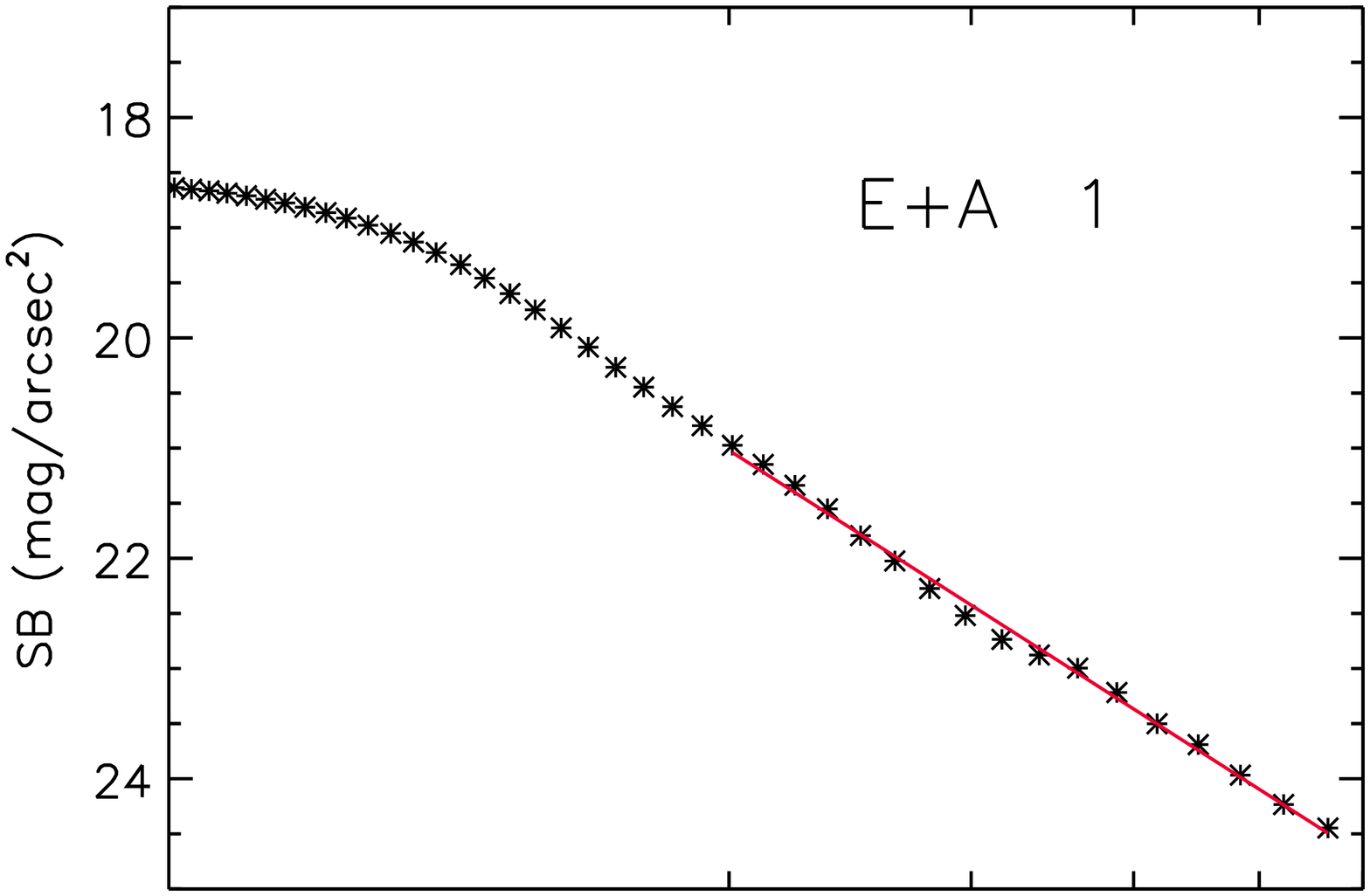}
         \includegraphics[width=4.2cm, angle=90, trim=0 0 0 0]{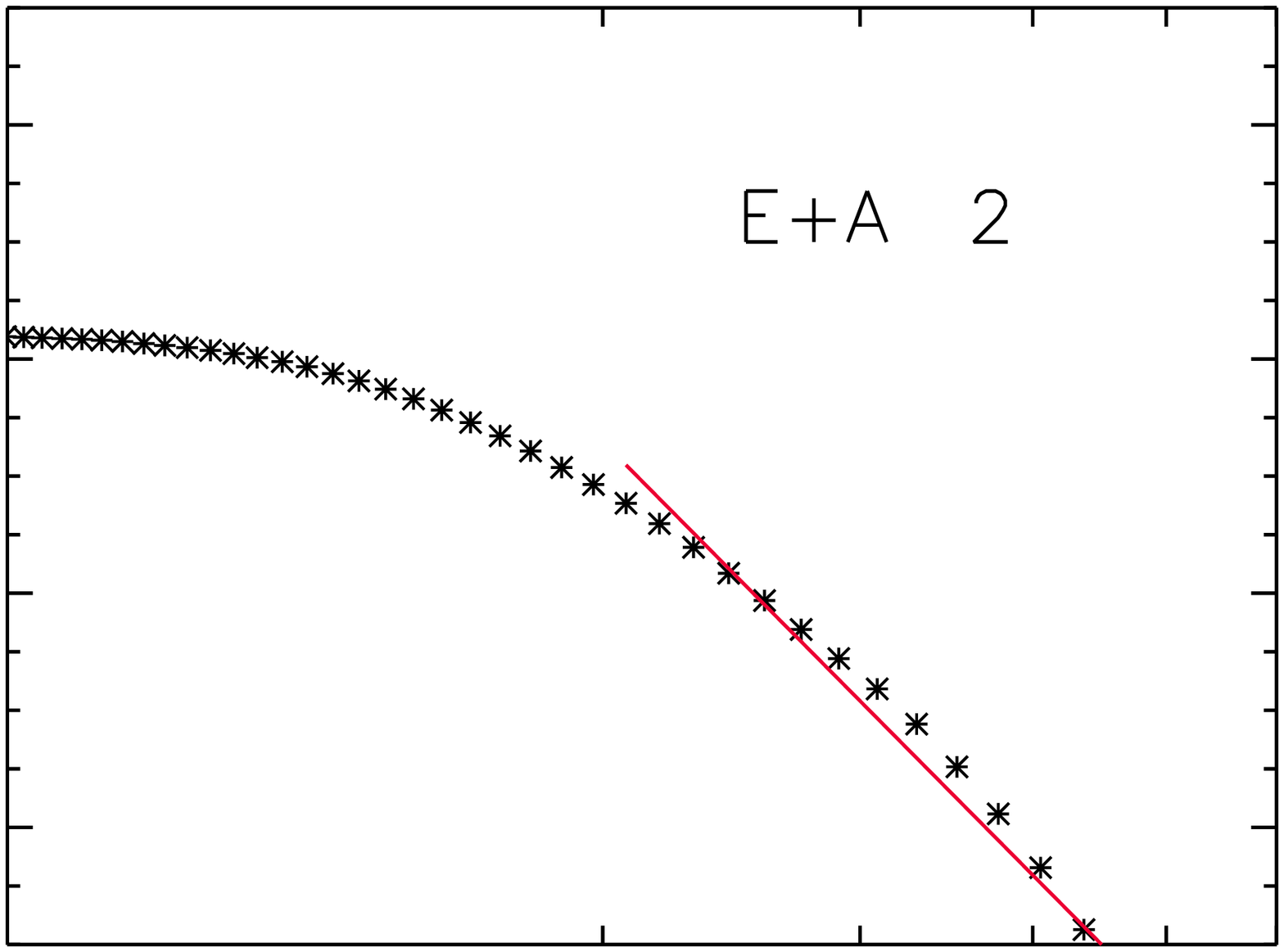}
\vspace{-0.7cm}
     \end{minipage}
    \begin{minipage}{0.75\textwidth}
         \includegraphics[width=4.2cm, angle=90, trim=0 0 0 0]{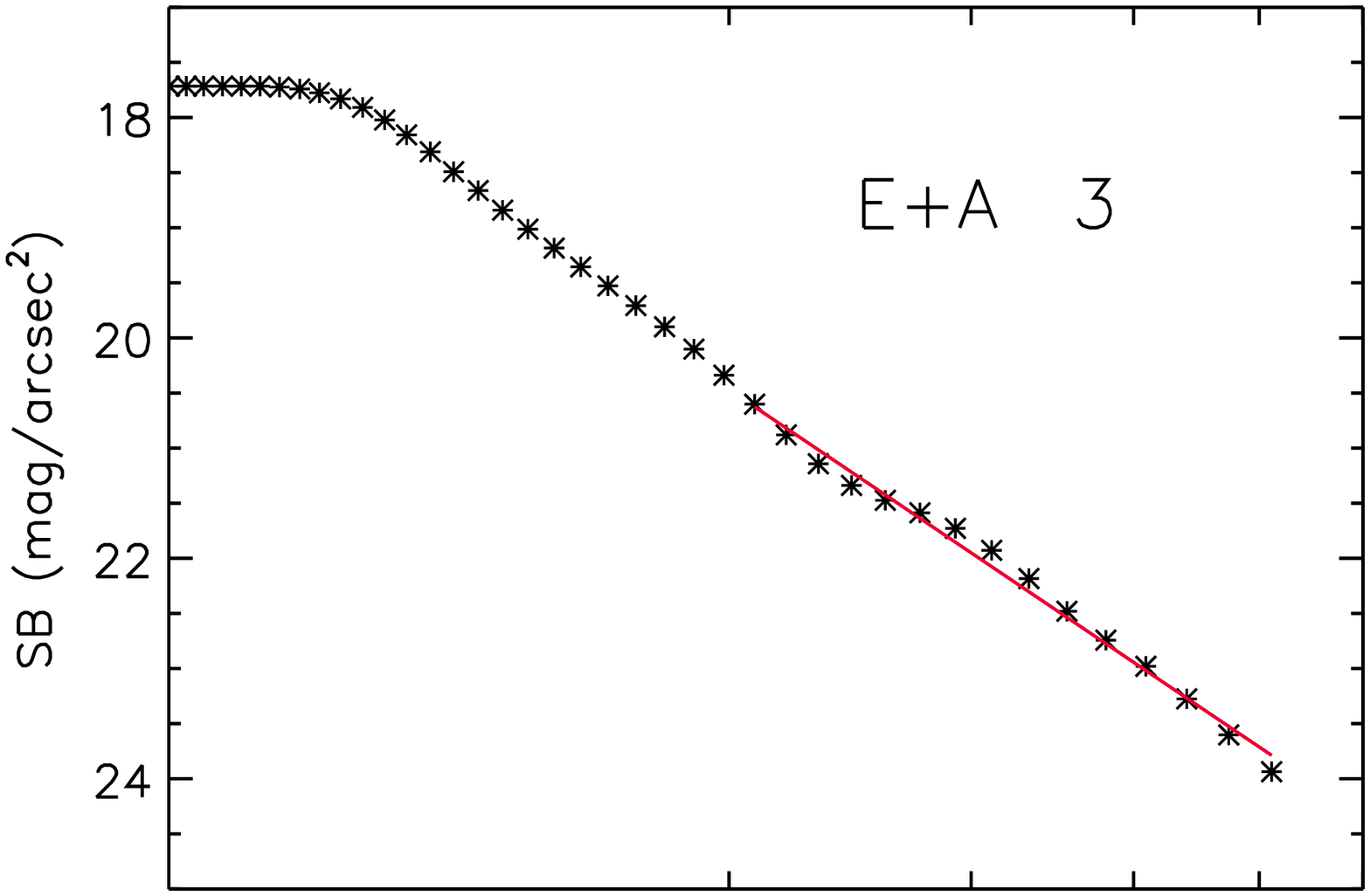}
         \includegraphics[width=4.2cm, angle=90, trim=0 0 0 0]{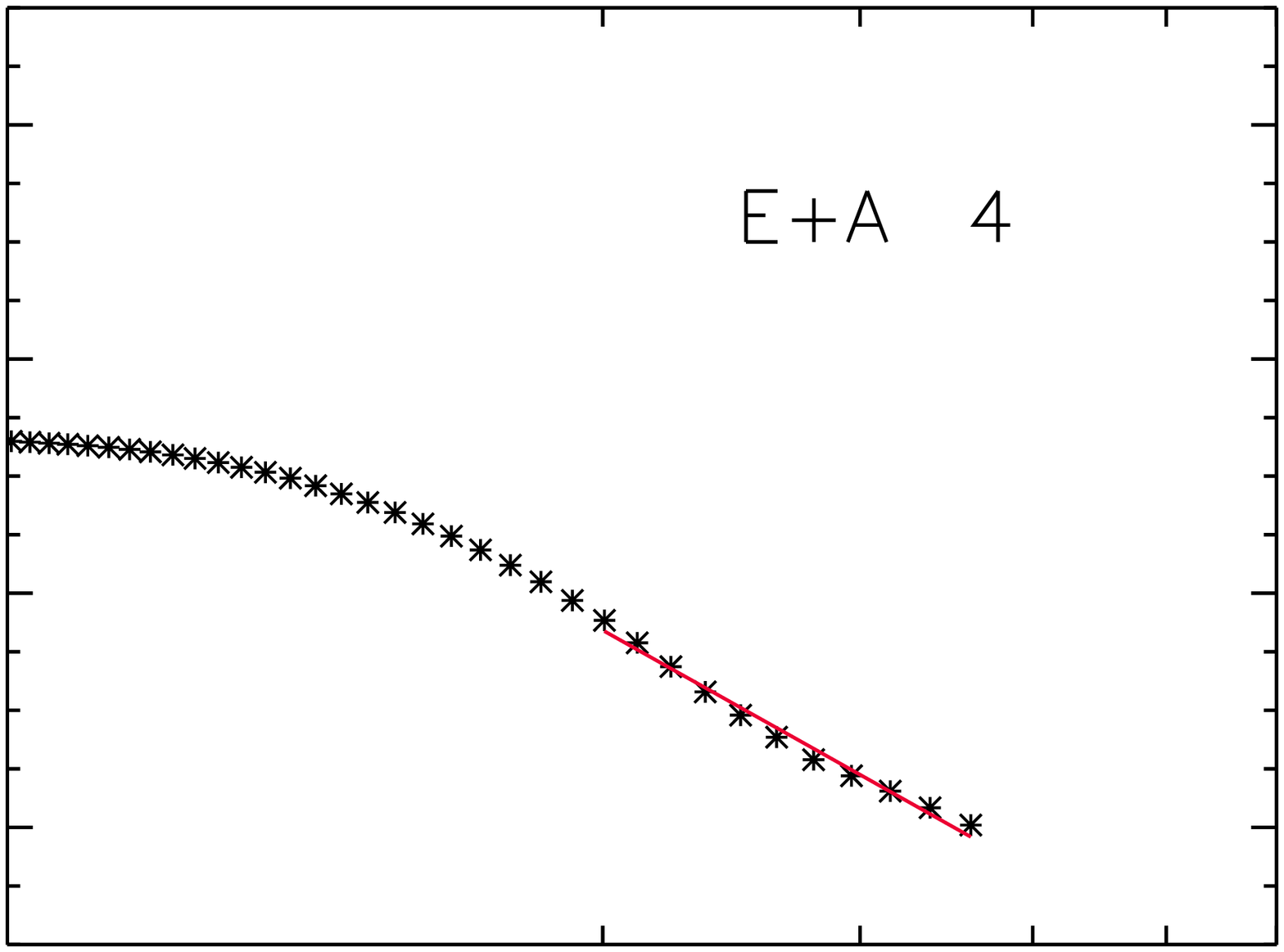}
\vspace{-0.7cm}
     \end{minipage}
    \begin{minipage}{0.75\textwidth}
         \includegraphics[width=4.2cm, angle=90, trim=0 0 0 0]{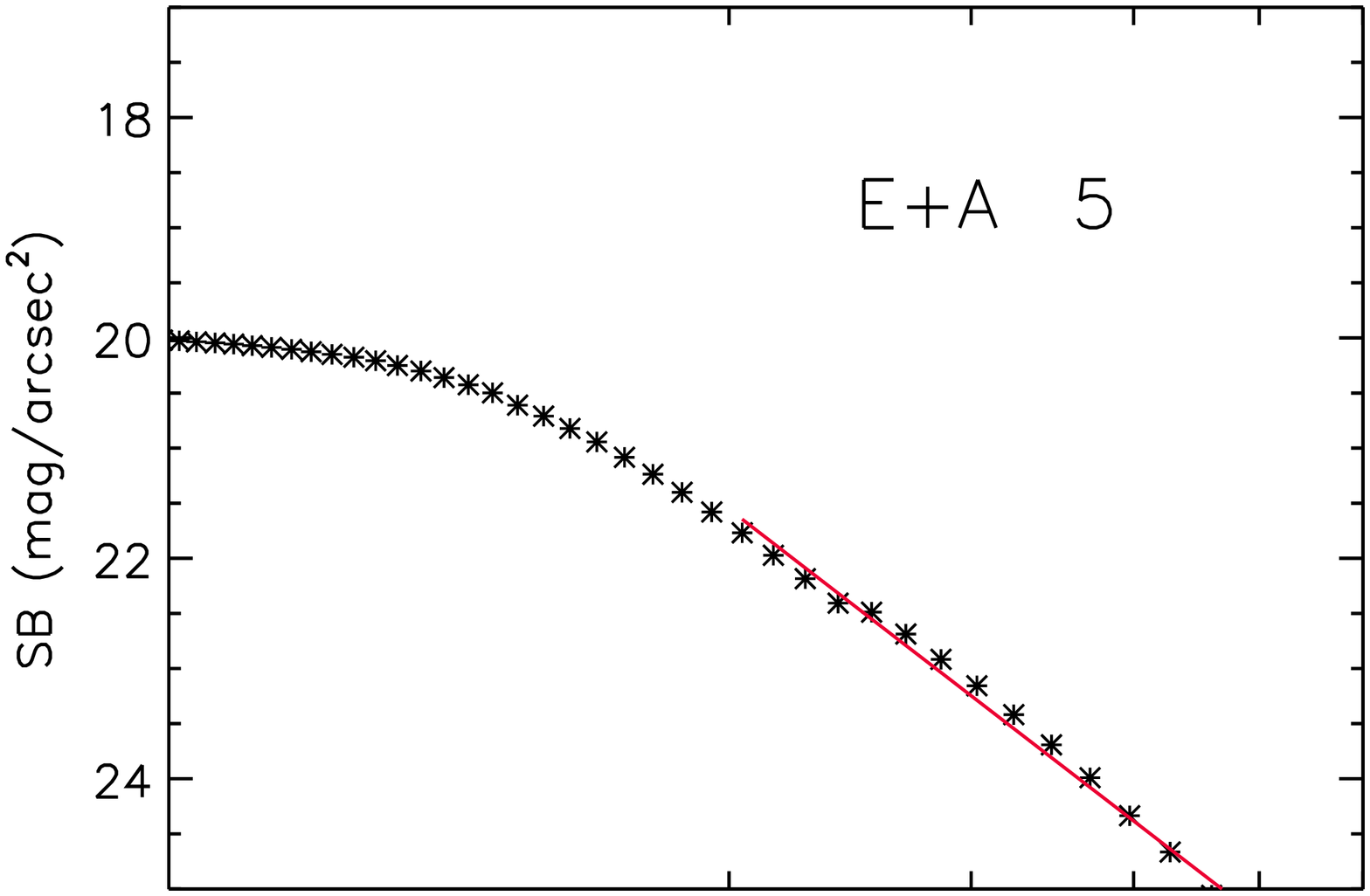}
         \includegraphics[width=4.2cm, angle=90, trim=0 0 0 0]{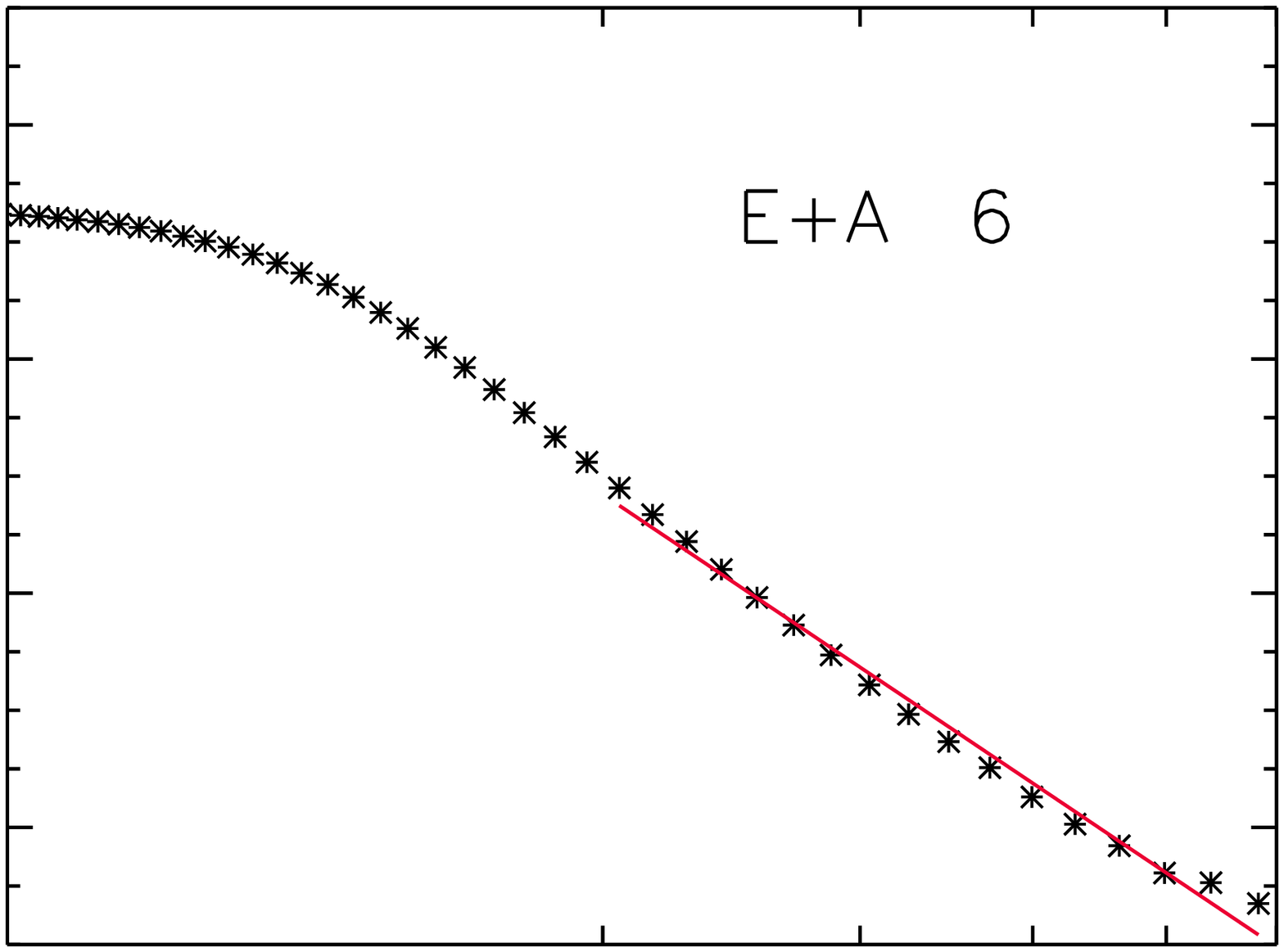}
\vspace{-0.7cm}
     \end{minipage}
    \begin{minipage}{0.75\textwidth}
         \includegraphics[width=4.2cm, angle=90, trim=0 0 0 0]{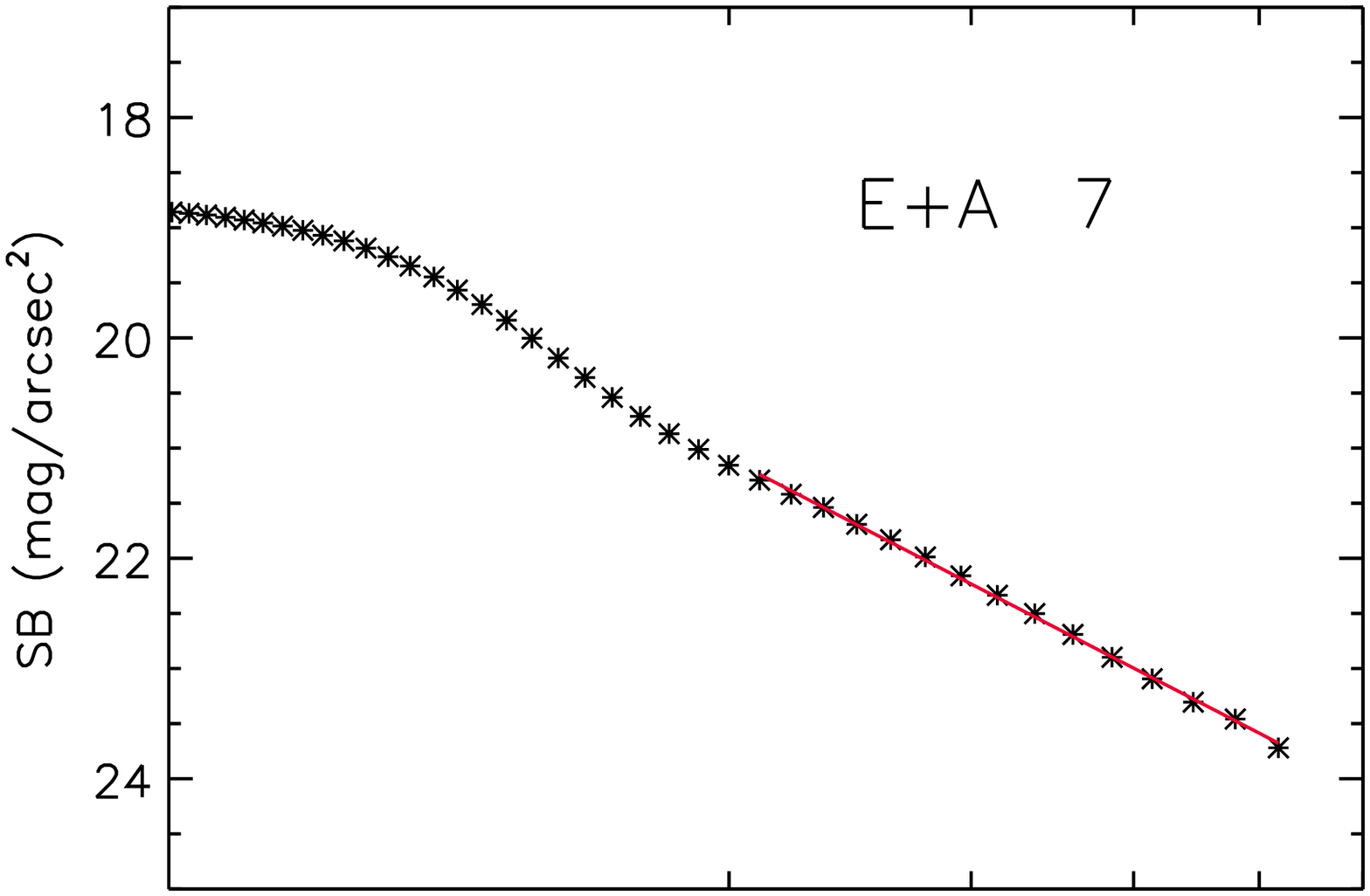}
         \includegraphics[width=4.2cm, angle=90, trim=0 0 0 0]{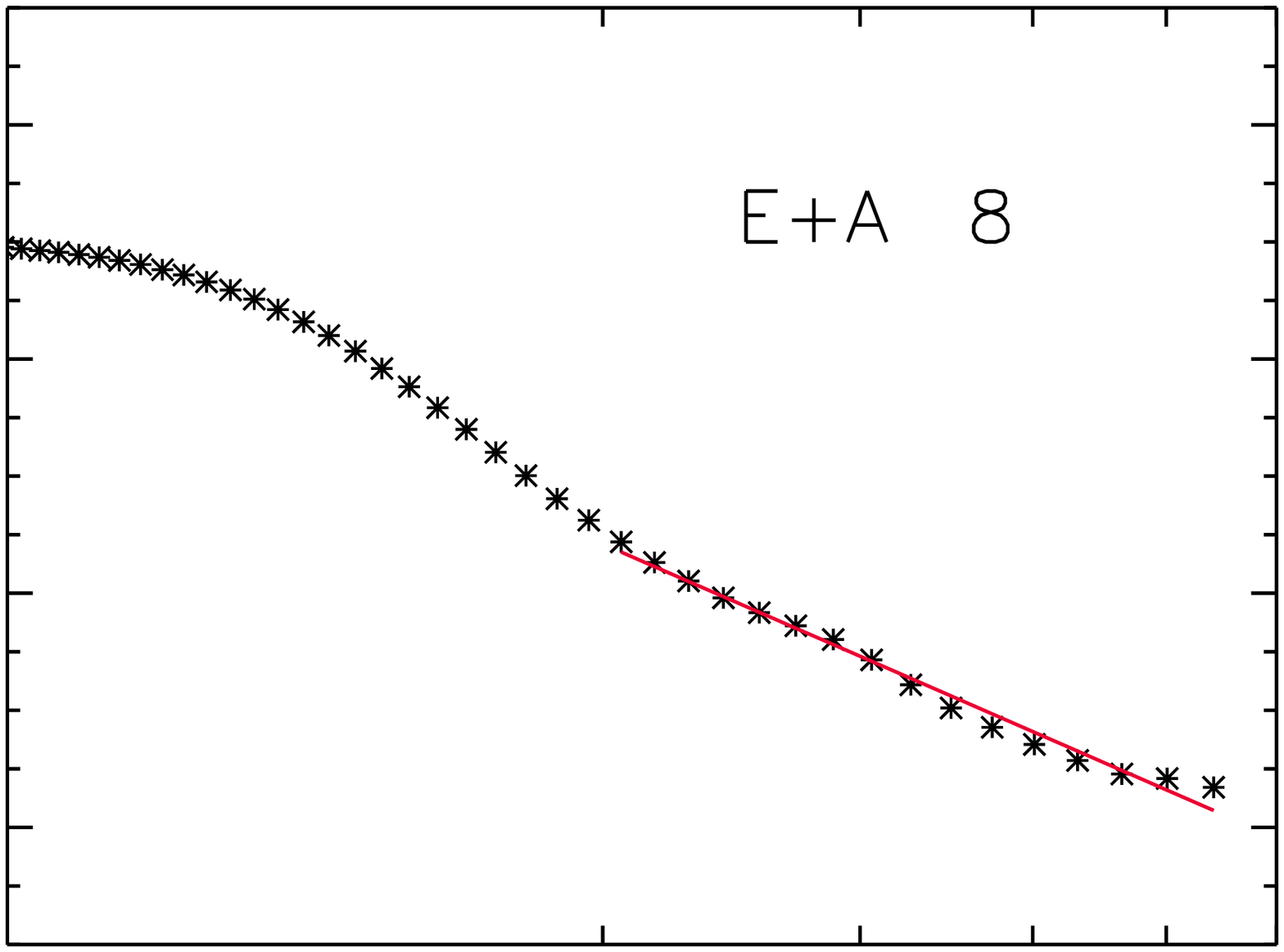}
\vspace{-0.7cm}
     \end{minipage}
    \begin{minipage}{0.75\textwidth}
         \includegraphics[width=4.2cm, angle=90, trim=0 0 0 0]{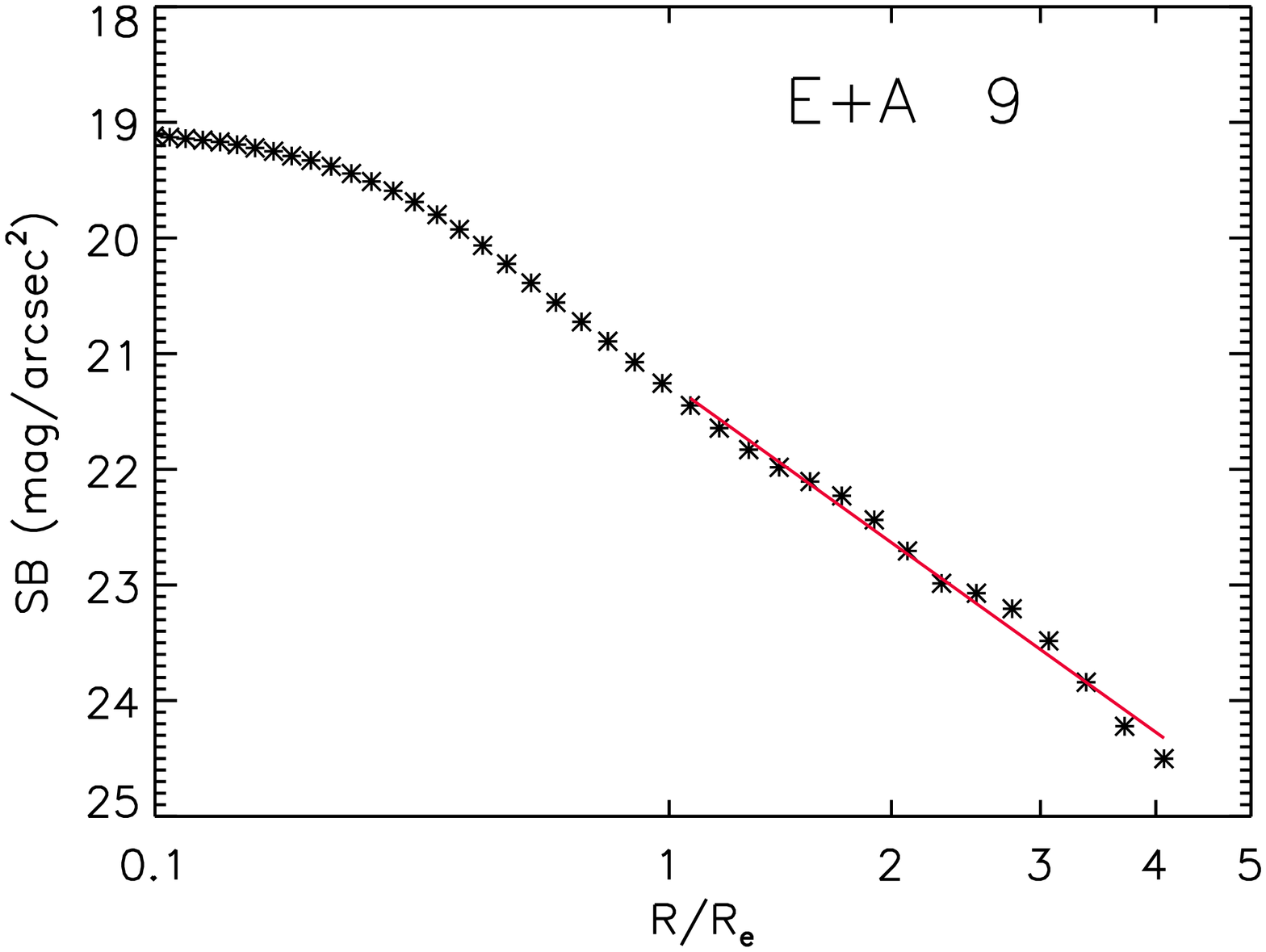}
         \includegraphics[width=4.2cm, angle=90, trim=0 0 0 0]{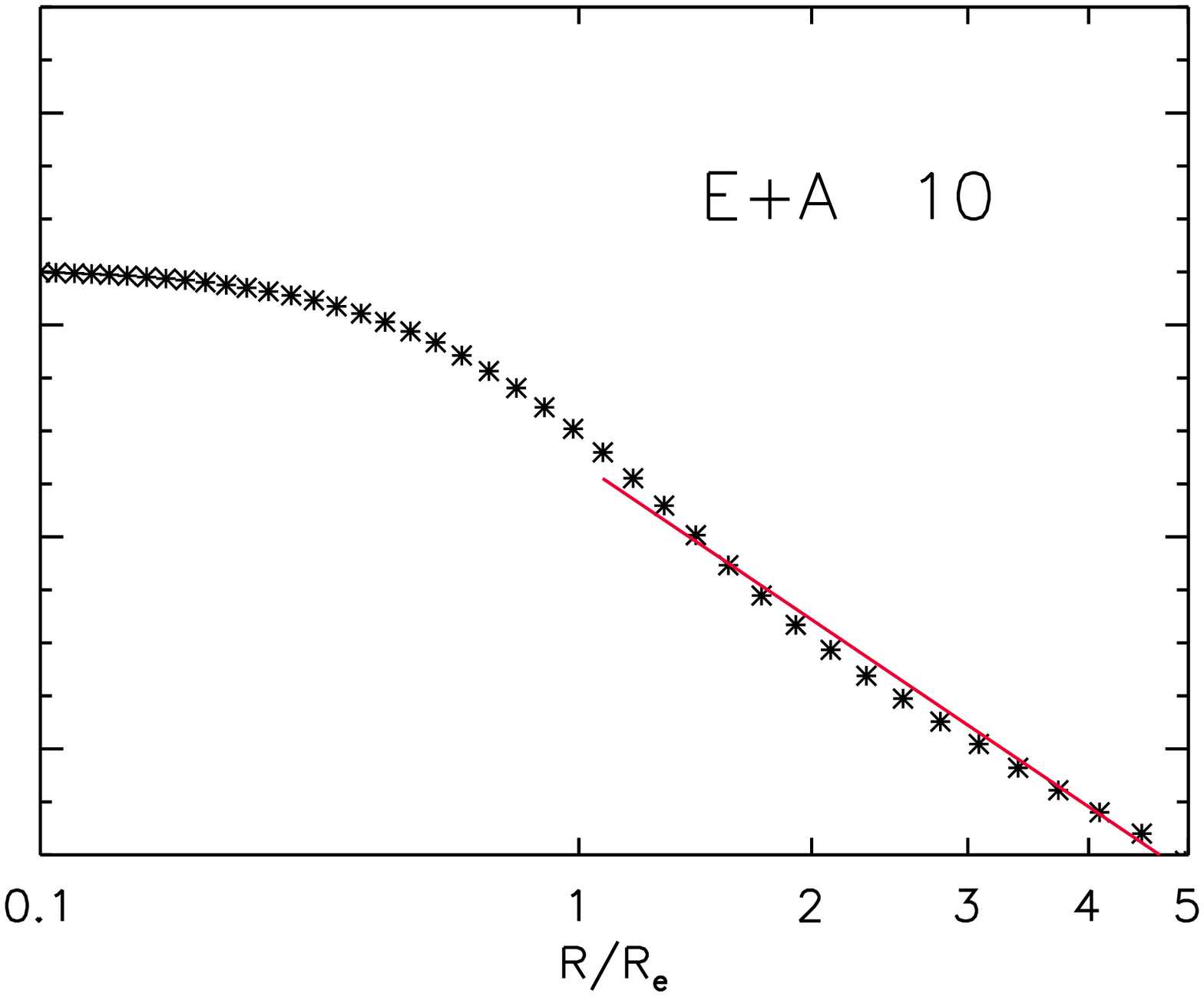}
     \end{minipage}
\caption{\label{fig:iso} $g$--band surface brightness profiles of all ten galaxies in the sample plotted on an $r^{{1 \over 4}}$ abscissa and given in
units of the effective radius. The profiles are flattened in the inner $\sim 1$\arcsec\, by convolution 
with the seeing disk -- the saturation in the centre of E+A\_3 is also evident. At larger radii they are generally 
consistent with an $r^{{1 \over 4}}$ law elliptical profile -- overploted as a {\it red line}. 
There are some deviations from $r^{{1 \over 4}}$ profiles; E+A\_2 in particular has a more exponential (disk--like) profile.}
\end{figure*}

We constructed isophotal surface brightness profiles using the {\sc iraf} task {\sc ellipse}. The $g$--band
surface brightness profiles are shown in Figure \ref{fig:iso}. The $r$--band profiles (not shown) are qualitatively similar.
The inner parts of the profiles out to $\sim 0.8$\arcsec\, are flattened by convolution with the seeing disk.  
At larger galacto--centric radii, the profiles appear like typical early--type galaxies with $r^{1 \over 4}$--like profiles 
(linear on a $r^{1 \over 4}$ horizontal axis). The exception is the isophotal profile
of E+A\_2 (and to a lesser extent E+A\_5) which appears more typical of a disk i.e. a strong downward concavity in the surface brightness 
profile. An $r^{1 \over 4}$ law fit to the data beyond $1 R_{e}$ is shown as the red line in Figure \ref{fig:iso}.
The profiles display some irregular structure expected for objects which have disturbed morphologies (see Figure \ref{fig:images}),
which is also evidenced by the position angles and ellipticities of the best fitting ellipse changing with semi--major axis distance.
These results are generally consistent with the HST study of \citet{yang08} which found E+A galaxies from 
the \citet{zabludoff96} LCRS sample to be predominantly early--type systems but with a greater than normal 
level of asymmetries. 

\subsection{Morphological properties}
Several previous studies have found E+A samples to have a large fraction of morphologically 
disturbed members \citep{zabludoff96,yang04,blake04} and nearby companion galaxies \citep{goto05}. A primary aim of acquiring deep imaging
with an 8-m class telescope of our sample was to search for the signatures of interactions and mergers in 
the form of faint tidal tails and debris. Inspection of the $g$--band images (see column 1, Fig. 1), as 
indeed the $r$--band images of our sample reveals that six out of ten galaxies in the sample 
(E+A\_1, E+A\_4, E+A\_7, E+A\_8, E+A\_9, E+A\_10) show either tidal bridges or tails
or clearly  disturbed morphologies  with at least three of these
apparently currently undergoing an interaction with a companion (E+A\_4, E+A\_7, E+A\_9). The remaining four galaxies
appear like undisturbed early--type systems. 

Table \ref{tab:imagesum} contains a summary of the morphological properties 
of the sample. The Hubble classifications listed in column 2 of the 
Table were determined visually from our $r$-band GMOS images.
This was done by one of us (W.J.C.) using the 
same approach and system described in \citet{blake04}, and which 
is based on the methodology adopted by the ``Morphs'' 
collaboration in their HST-based study of the morphologies 
of the galaxy populations in distant clusters \citep{smail97}. 
One small modification was required here and that was to
adopt a classification denoted by "S0a", to indicate galaxies
where it was impossible to distinguish them between an S0 and
an Sa galaxy, due to them showing tentative but not convincing
evidence for harbouring spiral arm structure. As can be seen in
Table 3, 40\,per cent of our sample have an S0a classification, and 
indeed our E+A sample is quite striking for its very narrow
range in Hubble type (S0-S0a-Sa).

In column 2 of Figure \ref{fig:images} we show the results of subtracting a smooth elliptical model from the images in an attempt
to reveal any faint sub--structure more clearly. The elliptical model was produced with the {\sc iraf ellipse} and {\sc bmodel} 
tasks and we allowed the position angle and ellipticity  of each isophotal ellipse to vary freely. In several cases 
the tidal/debris structures can be seen more clearly and further toward the galaxy centers. E+A\_2 and E+A\_5 reveal disk--like 
structures along their major axes consistent with the shapes of their isophotal profiles.

We find an even higher rate of morphological disturbance than the original \citet{zabludoff96} study of the
LRCS which found 5 out of 21 ($\sim 24$\,per cent) galaxies showed signs of disturbance. This difference is likely due to the limited depth of the Digital Sky Survey (DSS) imaging used. Indeed many of the tidal features in the imaging presented here are of too low surface brightness
to be detected in the DSS which was one of the motivations for acquiring deeper imaging.
\begin{table*}
\caption{\label{tab:imagesum} Morphological properties}
\begin{tabular}{|c|c|c|c|c|c|c|c|c|c|c|} \\ \hline
Name          & Morph & Neighbour & Disturbed/Tidal & centre col  &   col grad  & Environment       & Comment                                     \\  \hline
E+A\_1        &     Sa       &    No     & Yes     &    Irregular &    P         & Isolated    &   Tidal arm or bridge                       \\ 
E+A\_2        &     S0a       &    No     & No      &    Red       &   F         & Group   &   Normal/isolated                           \\ 
E+A\_3        &     S0       &    No     & No      &    N/A       &    F         & Cluster      &   Normal/isolated                           \\ 
E+A\_4        &     Sa       &    Yes    & Yes     &    None      &    P         & Group       &   Compact group/Lots of diffuse light       \\ 
E+A\_5        &     S0a       &    No     & No      &    Irregular &   F         & Isolated   &   Normal/isolated                           \\ 
E+A\_6        &     S0a       &    No     & No      &    Blue      &   P         & Group      &   Normal/isolated      \\ 
E+A\_7        &     S0a       &    Yes    & Yes     &    Red       &   N     & Group      &   Neighbour/tidal bridge      \\ 
E+A\_8        &     S0       &    No     & Yes     &    Red       &    N      & Isolated   &   Tidal tail      \\ 
E+A\_9        &     Sa      &    Yes    & Yes      &   None      &     F         & Cluster  &  Neighbour/disturbed     \\ 
E+A\_10       &     S0       &    No     & Yes     &    Red       &    F          & Isolated   &   Tidal tail      \\ \hline
\end{tabular}
\begin{flushleft}
Notes: Listed are the morphological and colour properties of our sample. Column 1 is galaxy ID, Column 2 is a
by eye morphological classification, Columns 3 and 4 indicate the presence or otherwise of near neighbours and
tidal features. Columns 5 and 6 are the core colour and colour gradient classifications. Column 7 is the environmental
classification from \citet{blake04} and column 8 is a comment on specific photometric properties.
\end{flushleft}
\end{table*}

\subsection{Colour morphologies}
We constructed two dimensional $g-r$ colour maps as outlined in Section 2.2.2. These maps
are shown in the two right--most columns in Figure \ref{fig:images} and show a 60\arcsec\, $\times$\, 60\arcsec\, map
(3rd column) and a 3\arcsec\, $\times$\, 3\arcsec\, map of the galaxy core (4th column). The colour distributions in
the galaxy cores are varied with the most common property being the presence of a red nuclear
core. The cores are red only in a relative sense with respect to the outer parts of the galaxy -- the absolute 
colours of the galaxies are blue in agreement with their E+A status. 
A red core is evident in 4 of the 9 galaxies (E+A\_2, E+A\_7, E+A\_8, E+A\_10).
Two galaxies have irregular color structure in their centres (E+A\_1,E+A\_5) and two  little structure
at all (E+A\_4, E+A\_9) whilst one galaxy in our sample shows a clear blue core (E+A\_6).
This is in contrast to the recent study of \citet{yang08} using HST imaging of 21 E+A galaxies of which
6 had compact blue cores with very steep profiles. These structures, however, are on scales much smaller 
than our resolution element. Some of these profiles have complicated core color structures 
and it is unclear how these color morphologies would appear at our resolution given the flux and area variation with radius,
or even whether the cores would be red or blue (see e.g. EA12, EA02 and EA09 in Figure 6 of \citet{yang08}). While both samples display
diversity in colour morphology with examples of blue, red and irregular cores our sample does appear to have a higher fraction of 
red core galaxies even if direct comparison is difficult.

\subsection{Colour gradients}
\begin{figure*}
   \begin{minipage}{0.75\textwidth}
         \includegraphics[width=4.2cm, angle=90, trim=0 0 0 0]{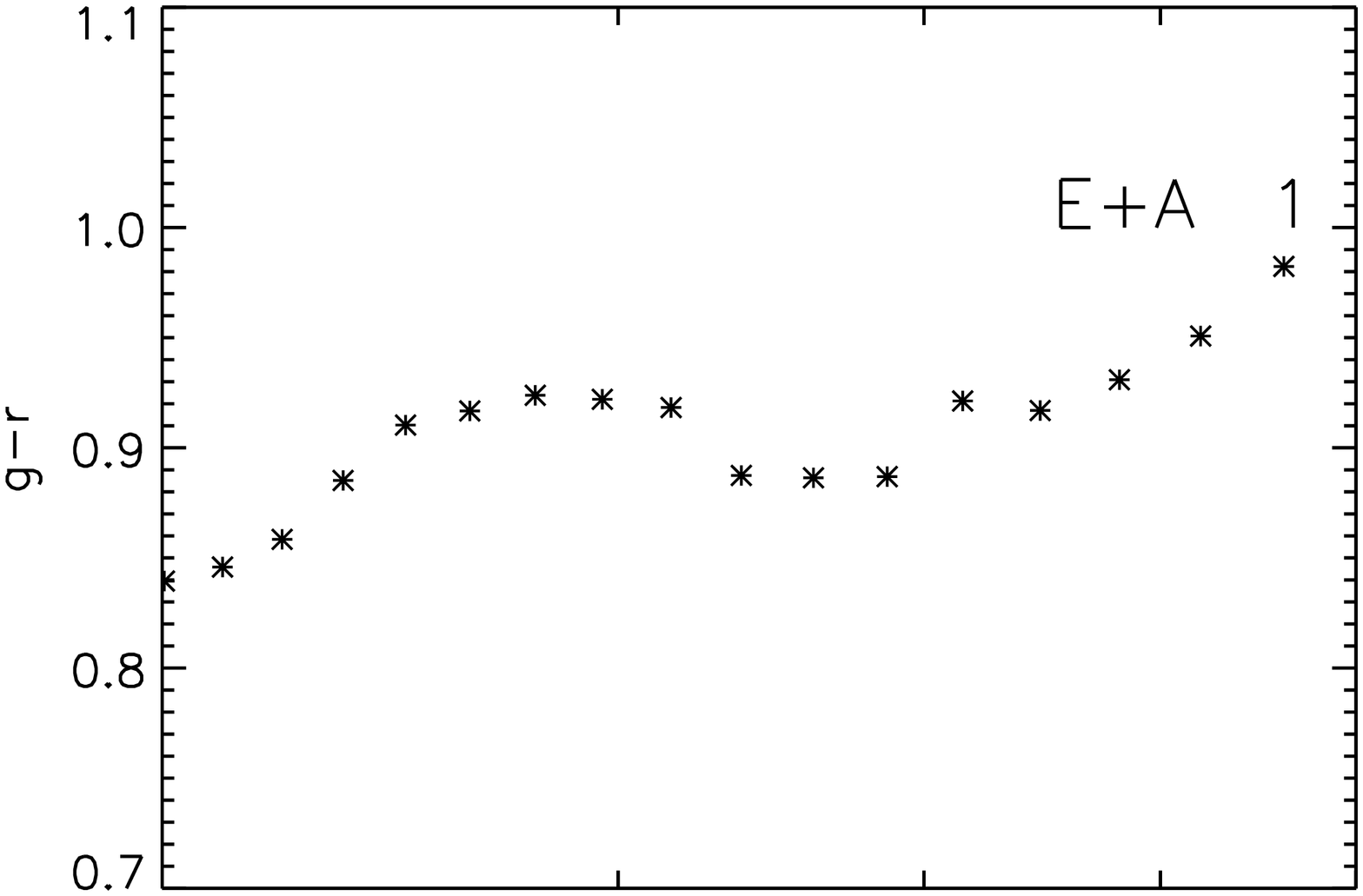}
         \includegraphics[width=4.2cm, angle=90, trim=0 0 0 0]{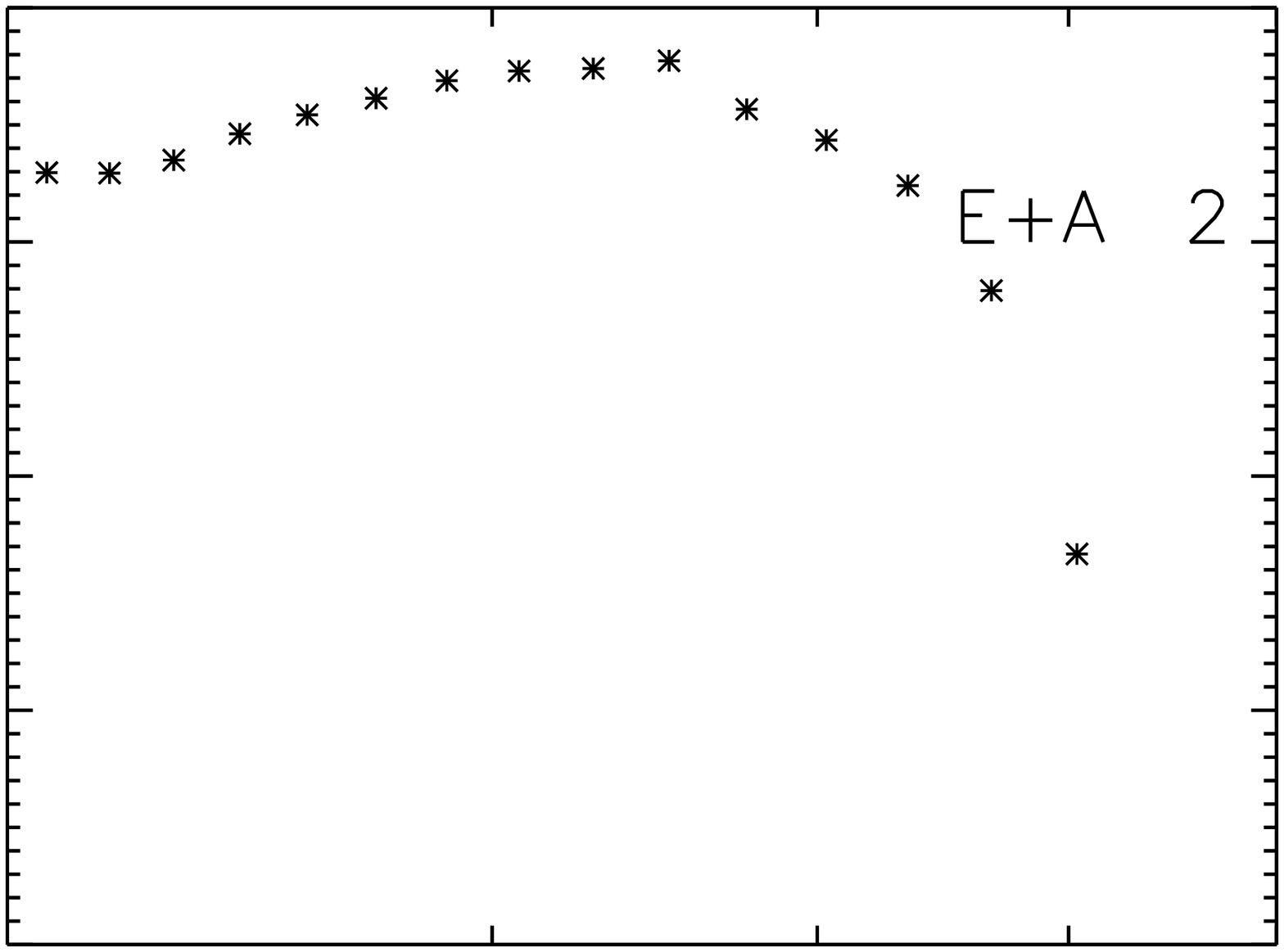}
\vspace{-0.7cm}
     \end{minipage}
    \begin{minipage}{0.75\textwidth}
         \includegraphics[width=4.2cm, angle=90, trim=0 0 0 0]{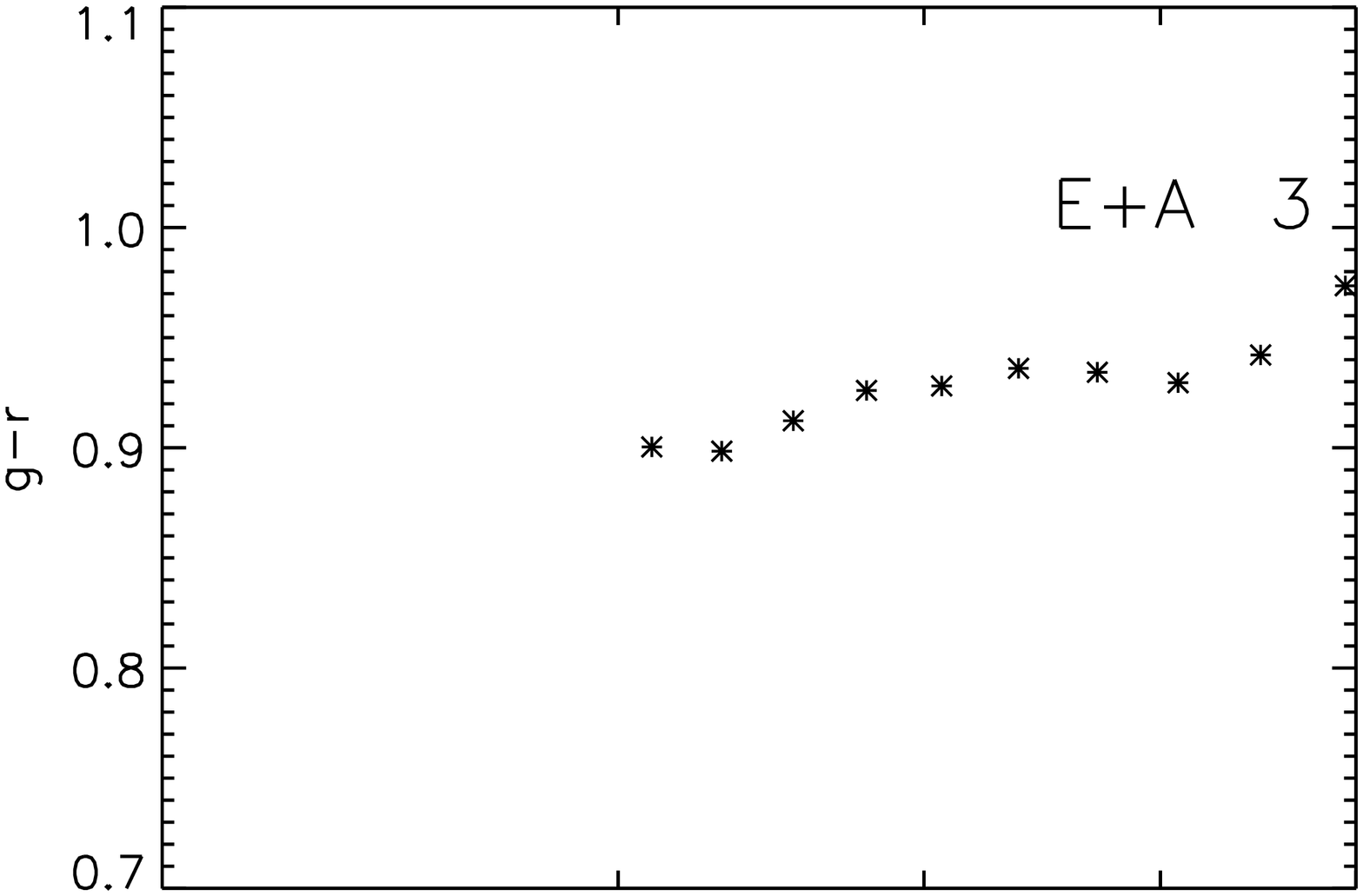}
         \includegraphics[width=4.2cm, angle=90, trim=0 0 0 0]{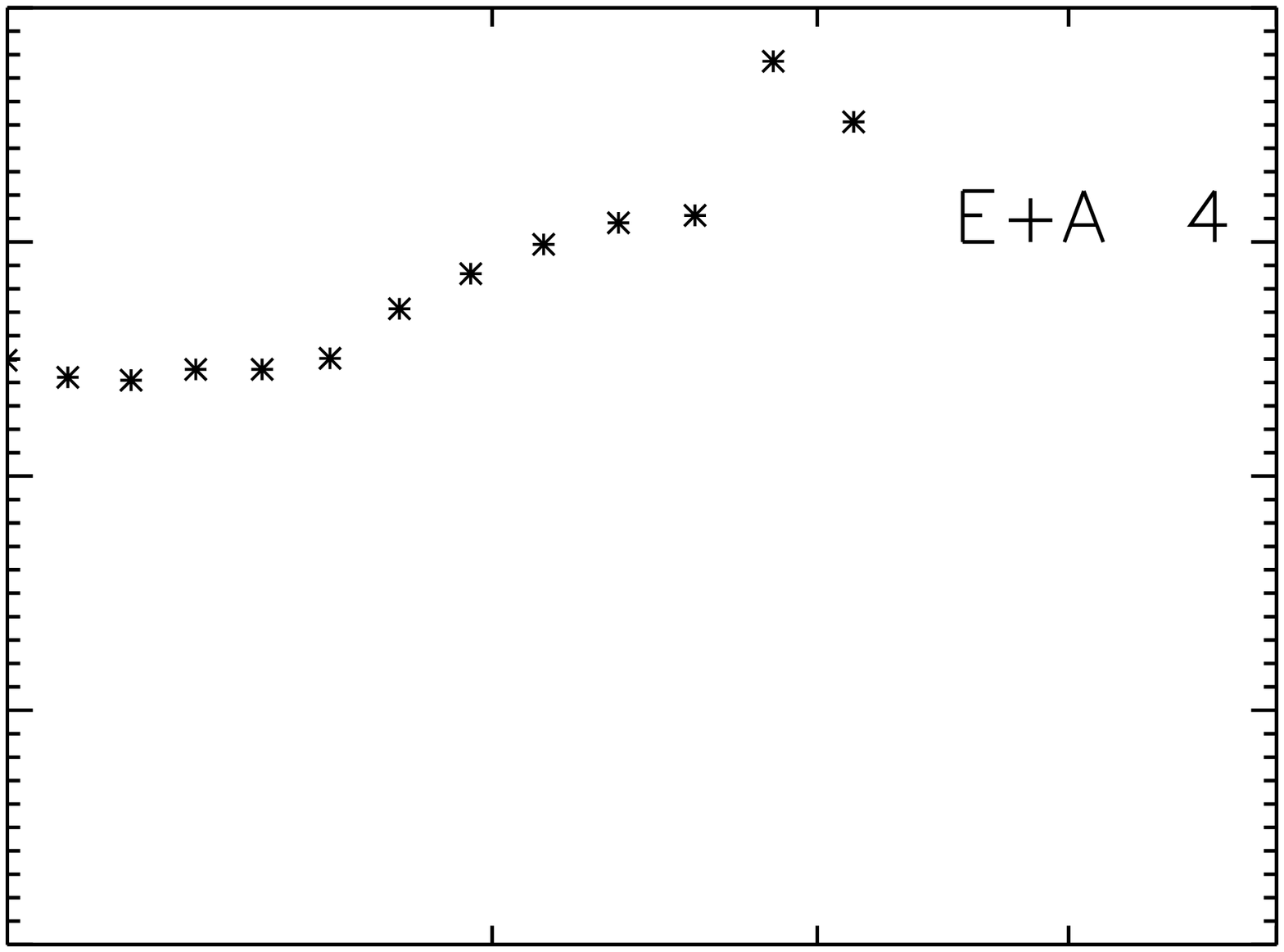}
\vspace{-0.7cm}
     \end{minipage}
    \begin{minipage}{0.75\textwidth}
         \includegraphics[width=4.2cm, angle=90, trim=0 0 0 0]{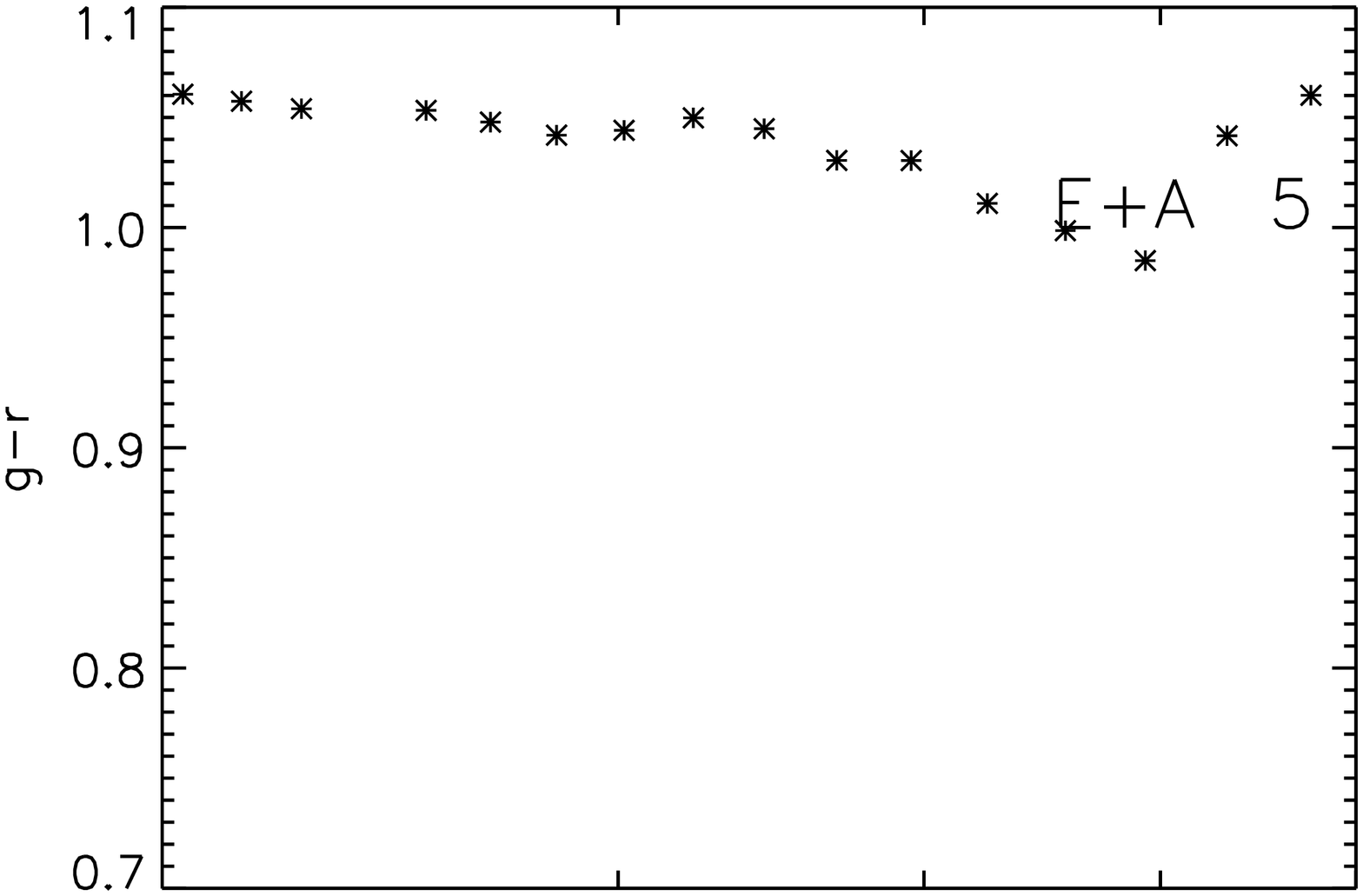}
         \includegraphics[width=4.2cm, angle=90, trim=0 0 0 0]{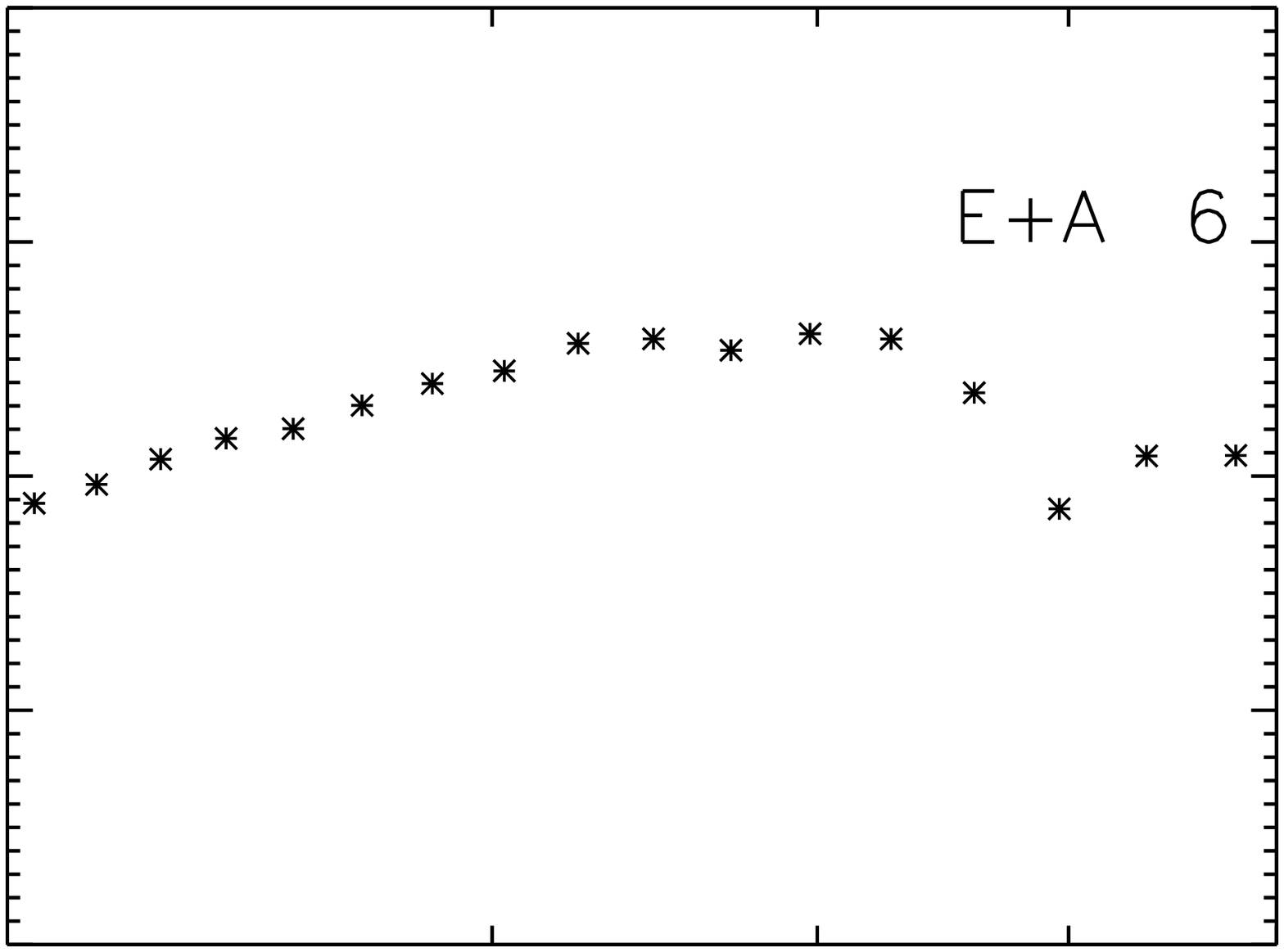}
\vspace{-0.7cm}
     \end{minipage}
    \begin{minipage}{0.75\textwidth}
         \includegraphics[width=4.2cm, angle=90, trim=0 0 0 0]{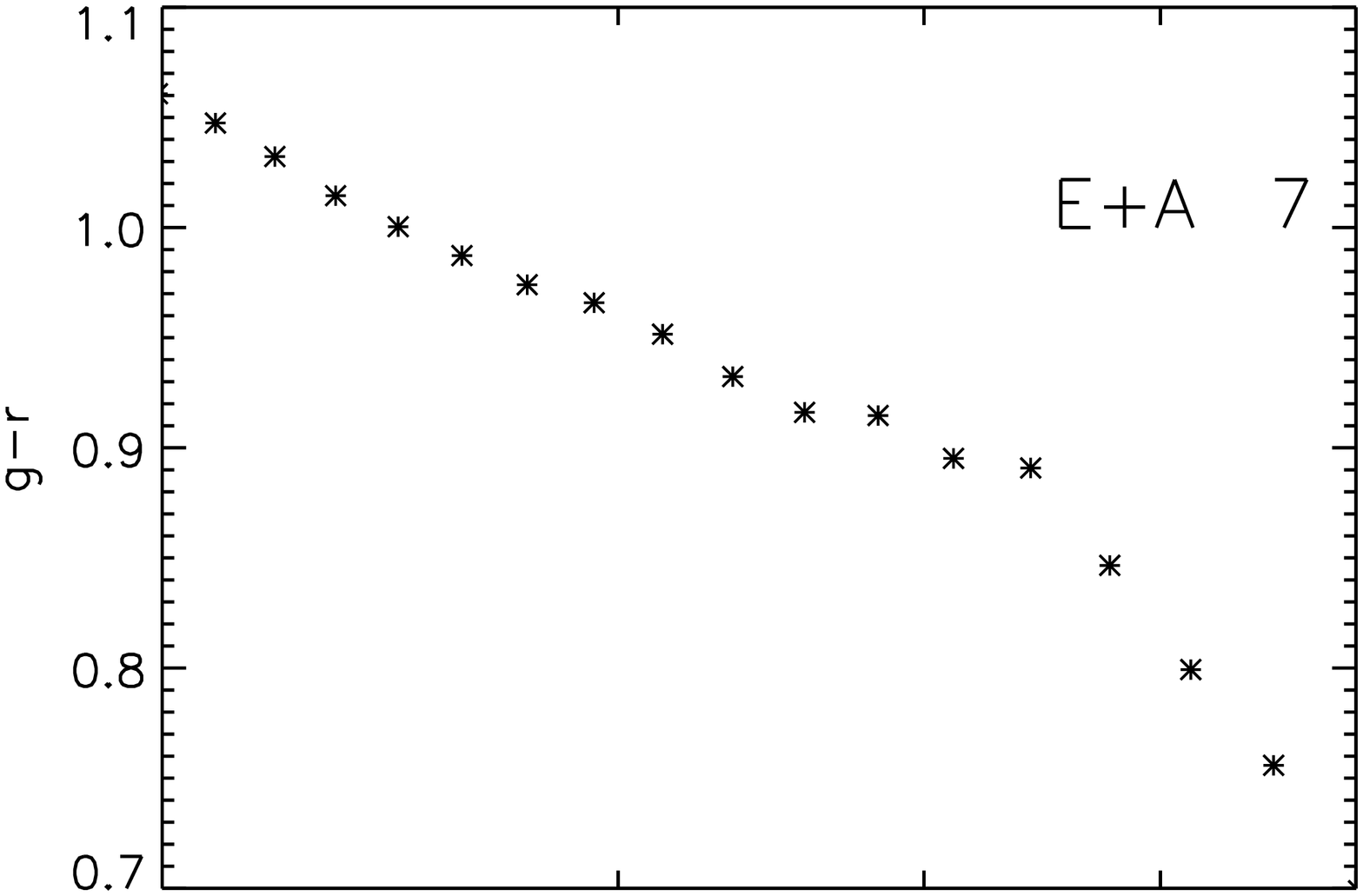}
         \includegraphics[width=4.2cm, angle=90, trim=0 0 0 0]{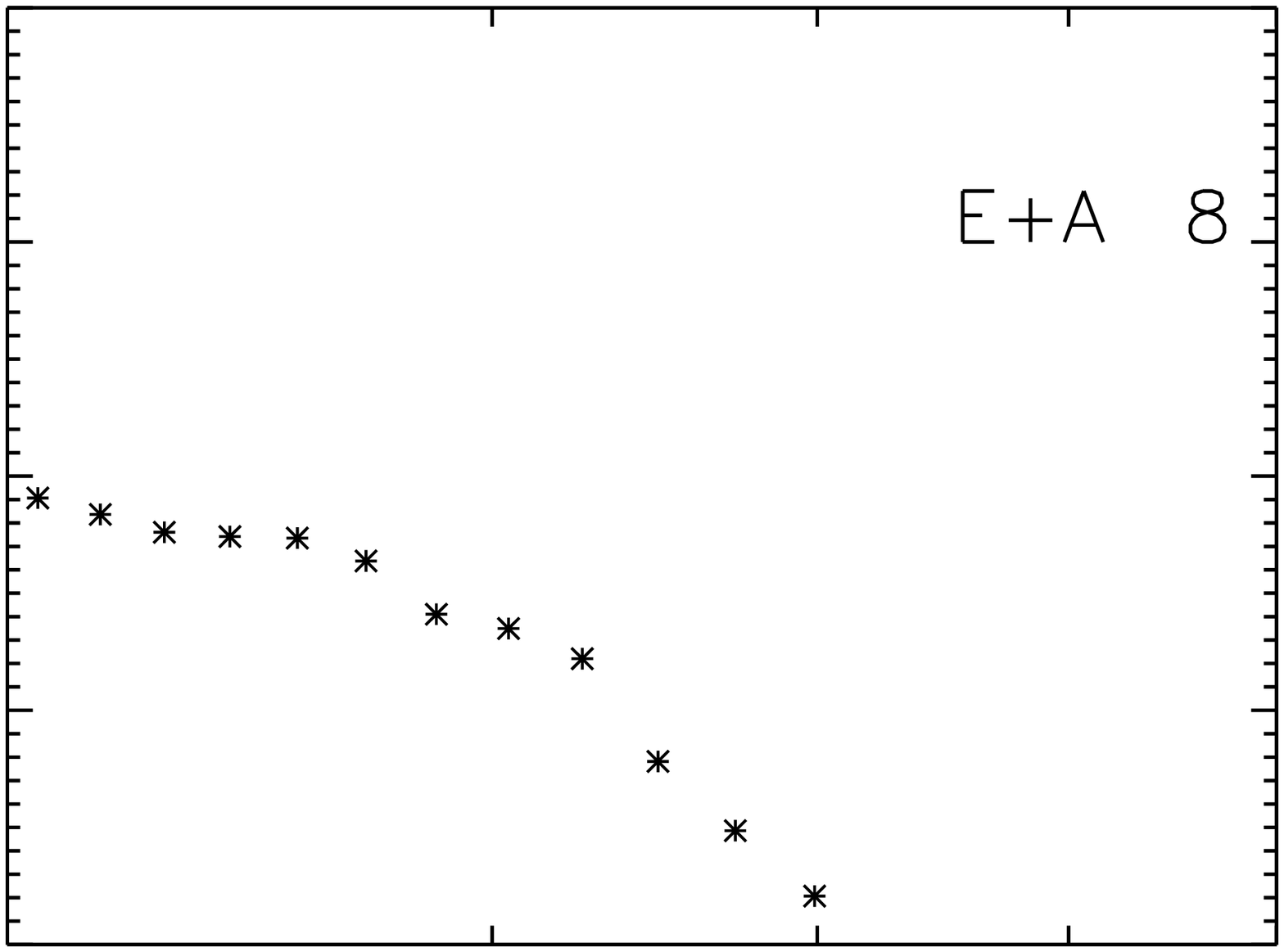}
\vspace{-0.7cm}
     \end{minipage}
    \begin{minipage}{0.75\textwidth}
         \includegraphics[width=4.2cm, angle=90, trim=0 0 0 0]{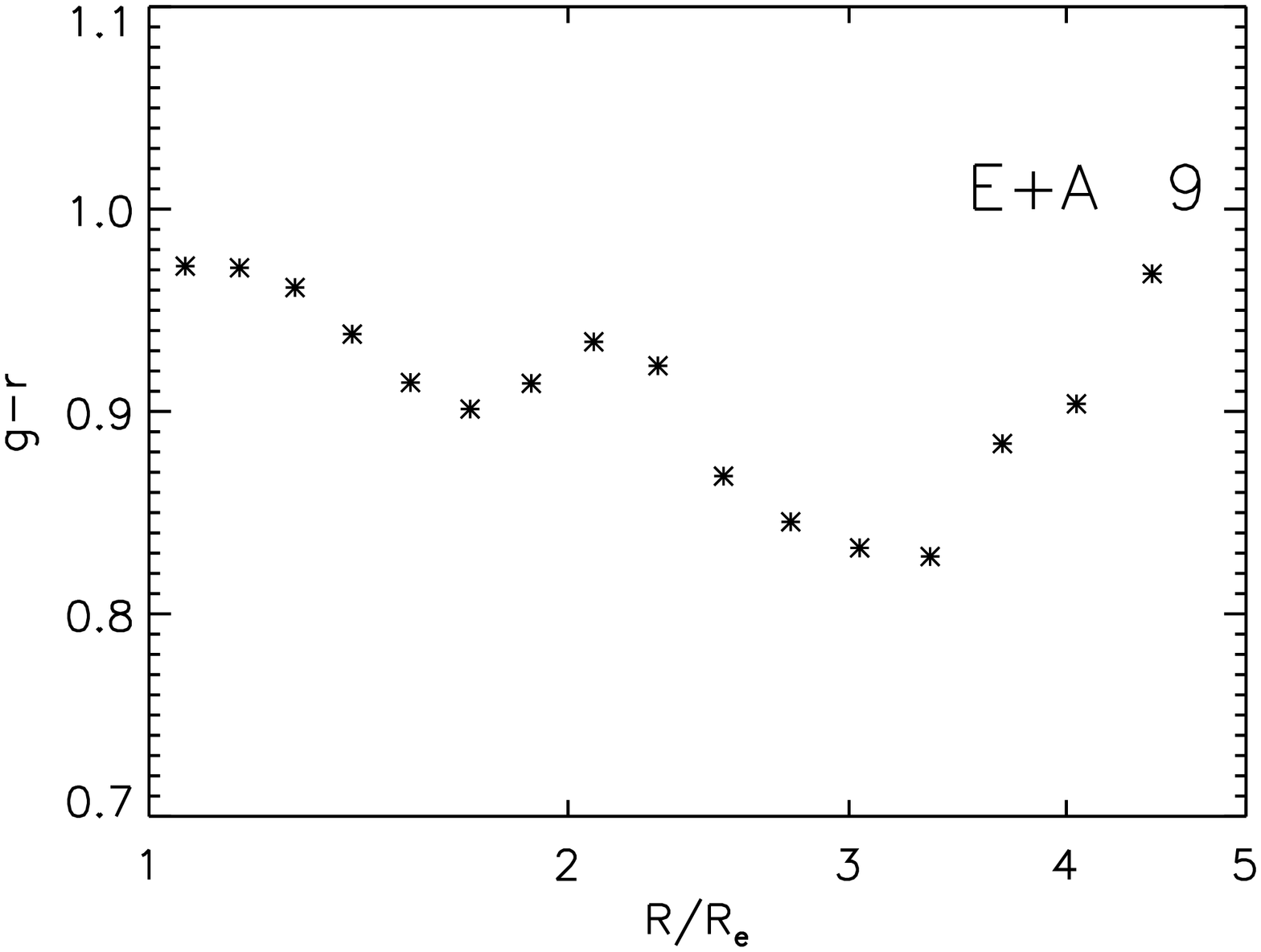}
         \includegraphics[width=4.2cm, angle=90, trim=0 0 0 0]{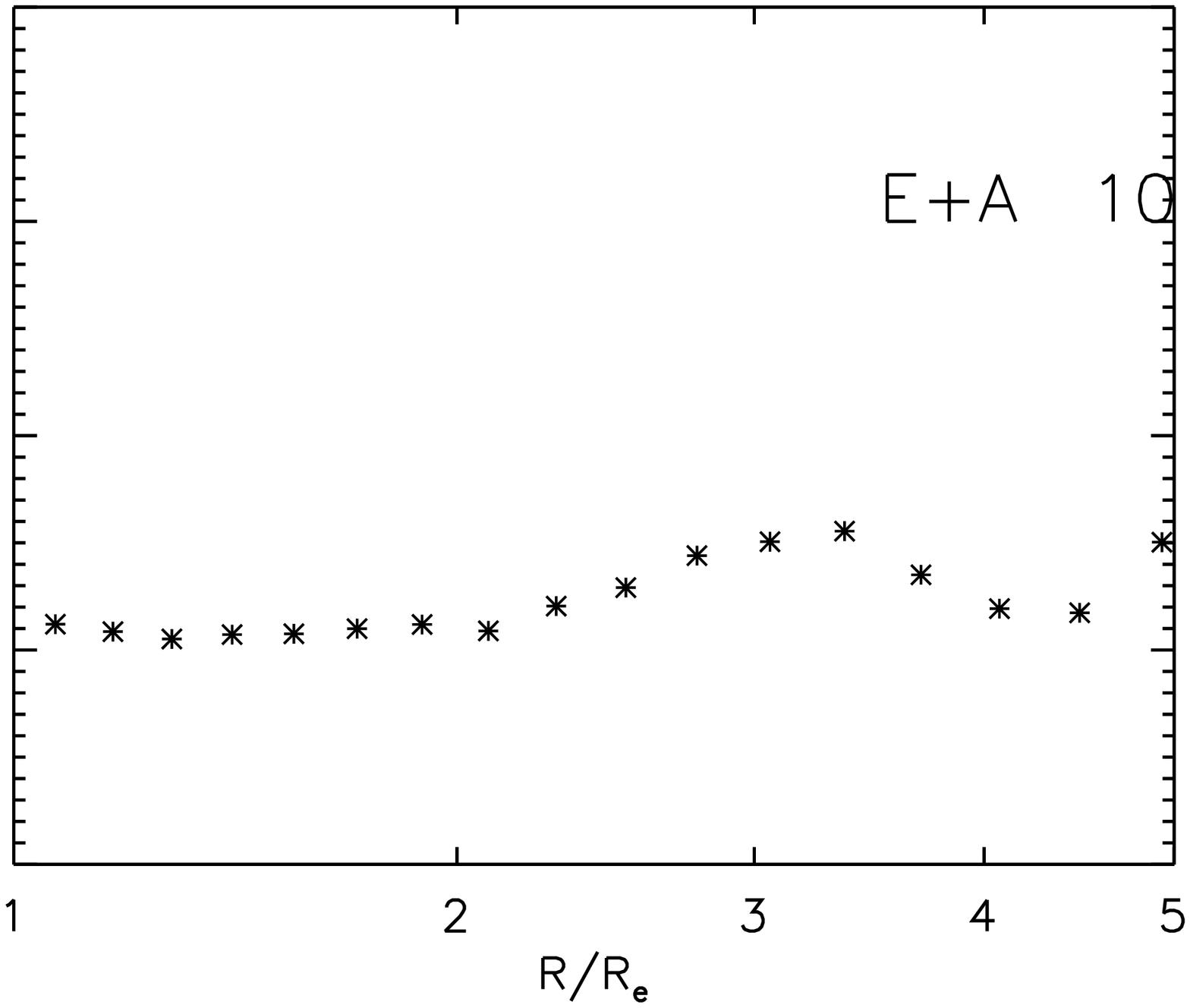}
     \end{minipage}
\caption{\label{fig:gradients} $g-r$ colour gradients along the semi-major axis. Gradients are shown
beyond 1\,$R_{e}$ except for E+A\_3 which is shown only beyond 2\,$R_{e}$ due to saturation in the galaxy centre
and poor seeing in the r--band image.}
\end{figure*}

On larger scales, outside the core, we still see a diversity in behaviour. In Figure \ref{fig:images} several galaxies have
negative colour gradients becoming bluer with increasing galacto--centric radius (e.g. E+A\_7) whilst others
have little sign of a gradient at all (e.g. E+A\_10) or have large scale patchiness in their colour distribution (e.g. E+A\_9).
We use the {\sc iraf  ellipse} task to examine the nature of the  large scale radial (semi-major axis) colour gradients. This is done by 
fitting elliptical isophotes to the PSF matched $g$-- and $r$--band images at regularly spaced semi-major axis lengths. The results
for our sample are shown in Figure \ref{fig:gradients}.
On these scales we again see variation in the radial color profile shapes. We use the same classification scheme that
 \citet{yang08} use to classify their large scale colour gradients by categorizing profiles as positive, negative or flat/variable.
This colour gradient classification for each galaxy is listed in Table \ref{tab:imagesum}. 
We classify two galaxies as having negative colour gradients (20\,per cent), five to be  flat/variable (50\,per cent) and three to be 
positive (30\,per cent). These can be compared with \citet{yang08} who find 29, 19, and 51\,per cent of their sample to be negative, 
flat/variable and positive, respectively. Note, however,  while the positive gradient galaxies have mild slopes and are not too dissimilar 
to the flat gradient galaxies the two most striking gradients (E+A\_7 and E+A\_8) are both negative. This is true for many examples 
in the \citet{yang08} sample as well, where the slope of the profile outside the core could easily be classified as flat rather
than positive. Qualitatively, and taken as a whole, the large scale colour gradient behaviour in the two samples is  quite
similar (c.f. Figure \ref{fig:gradients} with their Figure 6). Likewise, \citet{yamauchi05} find the majority of their E+A sample to 
have positive colour gradients -- again with slopes that are generally quite shallow.

\section{Spectroscopic characteristics}

\subsection{IFU data cubes}
The data reduction procedure outlined in Section 2 produces for each target
galaxy a data-cube with spatial dimensions of 7\arcsec\, $\times$\, 5\arcsec\,
with 0.2\,\arcsec\, spaxels and a wavelength coverage of 
$\sim  4122 < \lambda < 5380$\,\AA\, sampled by 1410 spectral pixels. 

There is little galaxy signal in the outer parts of the IFU and even spatially binning over large areas 
in the very outskirts is unable to produce a reasonable quality spectrum. We therefore restrict
our analysis to the central 3\arcsec\, region where the signal--to--noise ratio is sufficient for a robust
analysis.

\subsection{Integrated spectra}
\begin{figure*}
   \begin{center}
     \begin{minipage}{0.95\textwidth}
         \includegraphics[width=5.6cm, angle=90, trim=0 0 0 0]{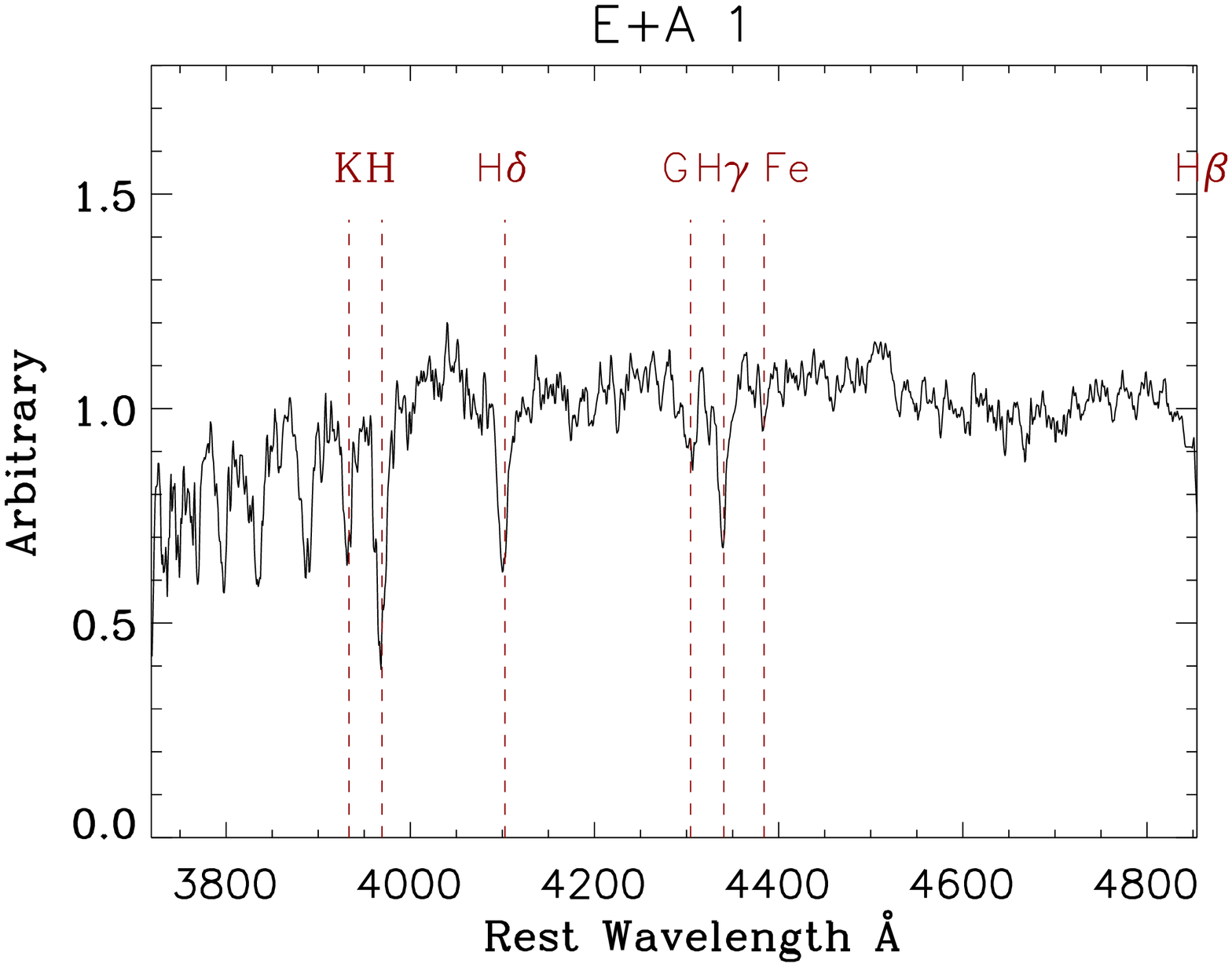}
         \includegraphics[width=5.6cm, angle=90, trim=0 0 0 0]{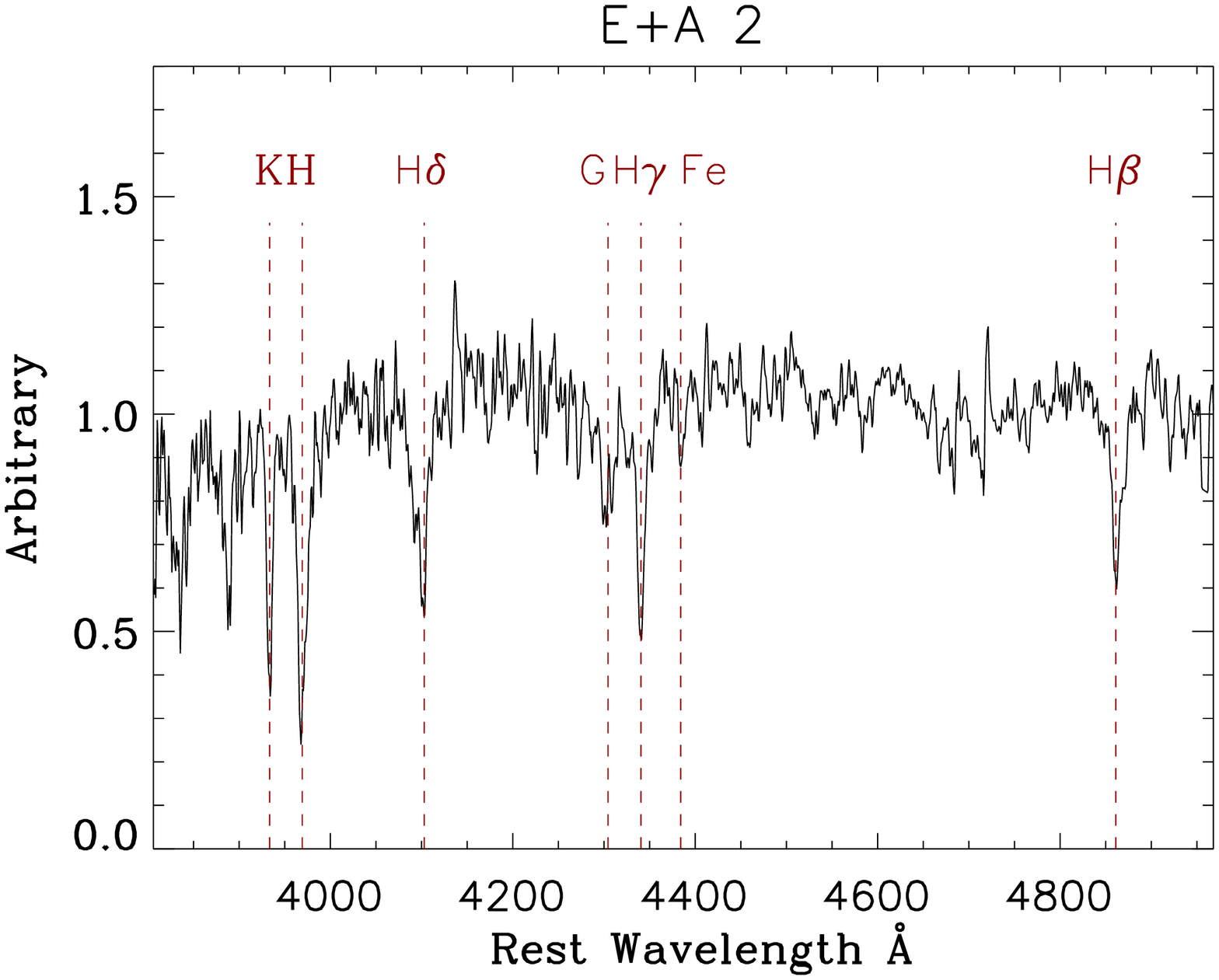}
      \end{minipage}
    \begin{minipage}{0.95\textwidth}
        \includegraphics[width=5.6cm, angle=90, trim=0 0 0 0]{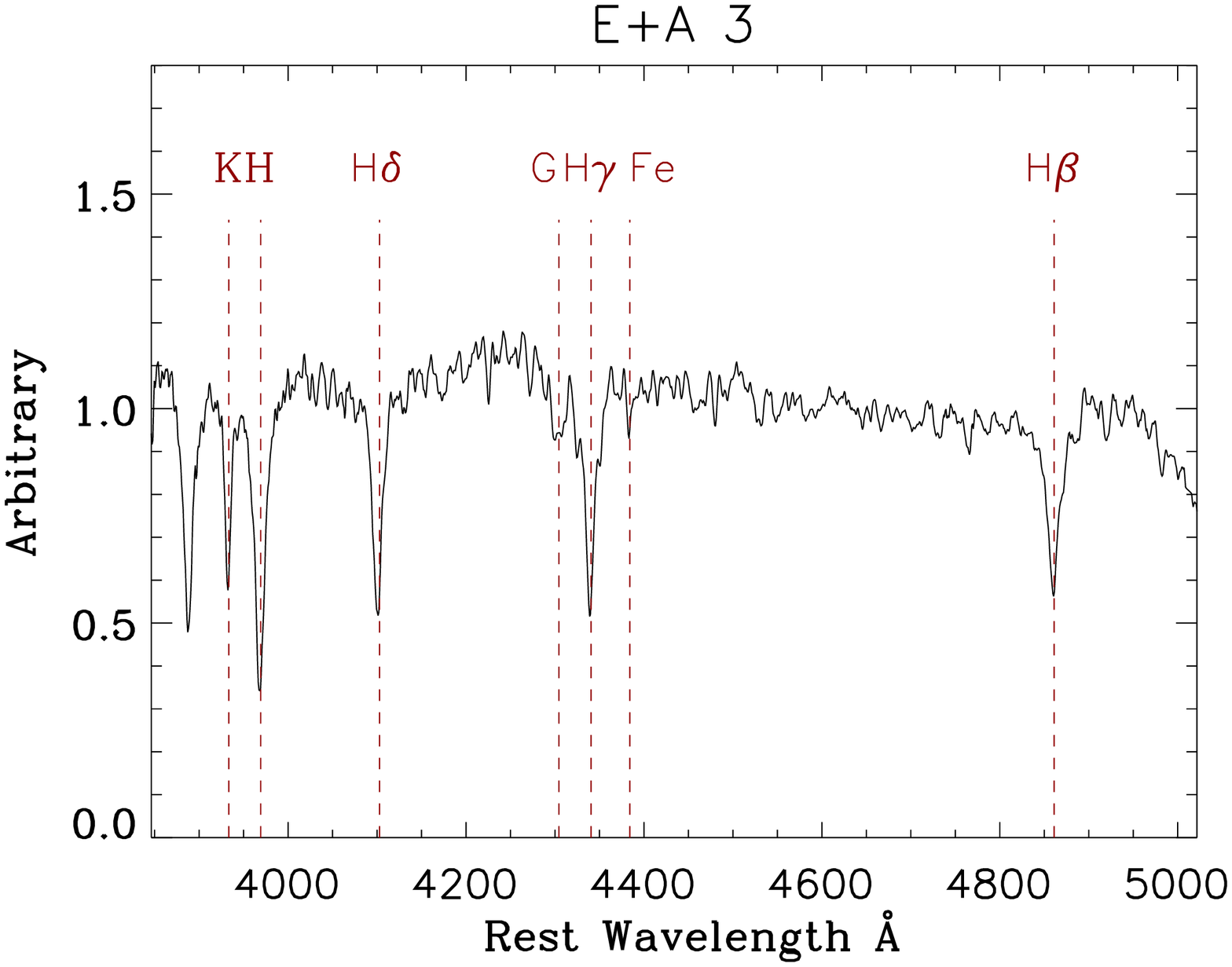}
        \includegraphics[width=5.6cm, angle=90, trim=0 0 0 0]{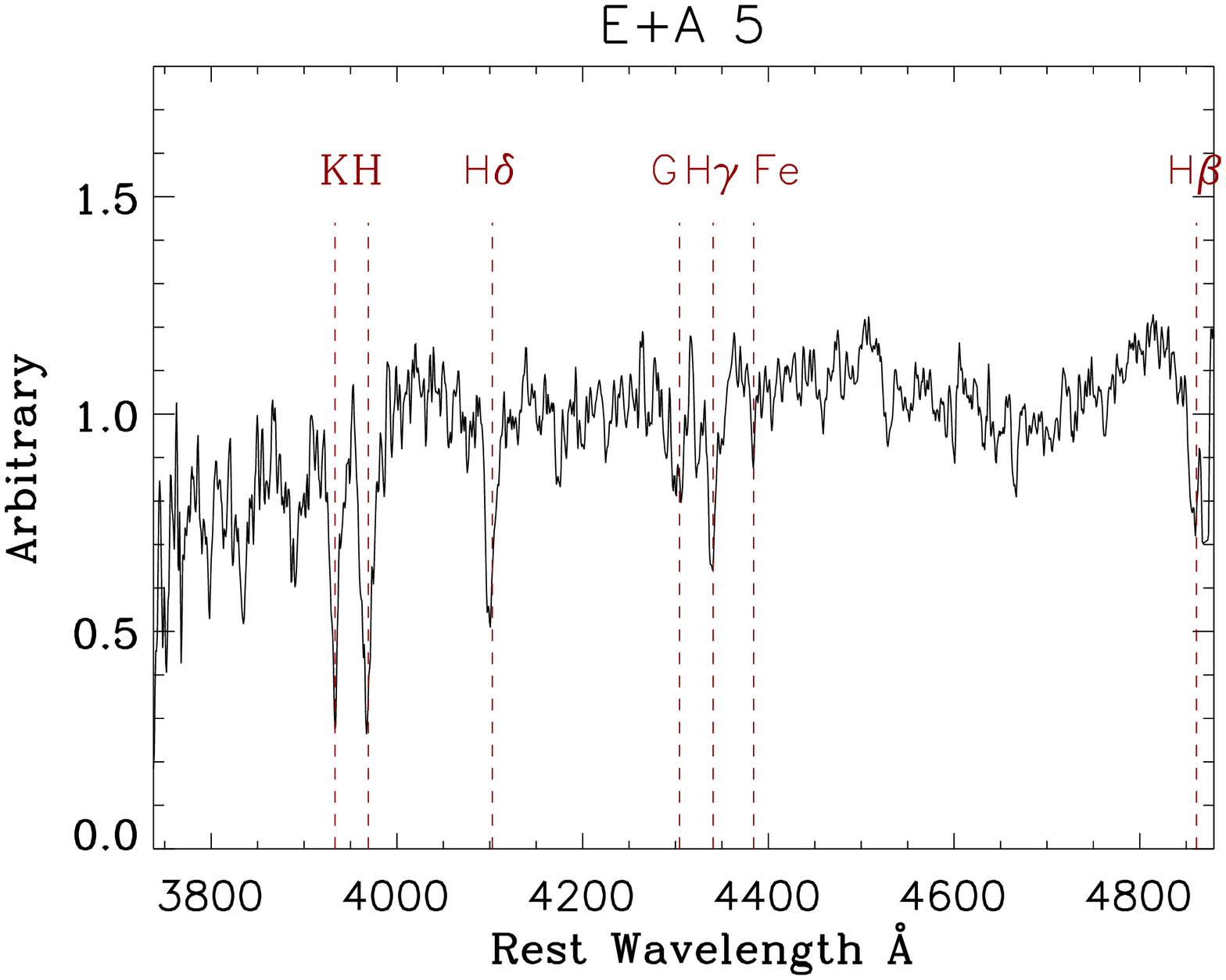}
     \end{minipage}
     \begin{minipage}{0.95\textwidth}
         \includegraphics[width=5.6cm, angle=90, trim=0 0 0 0]{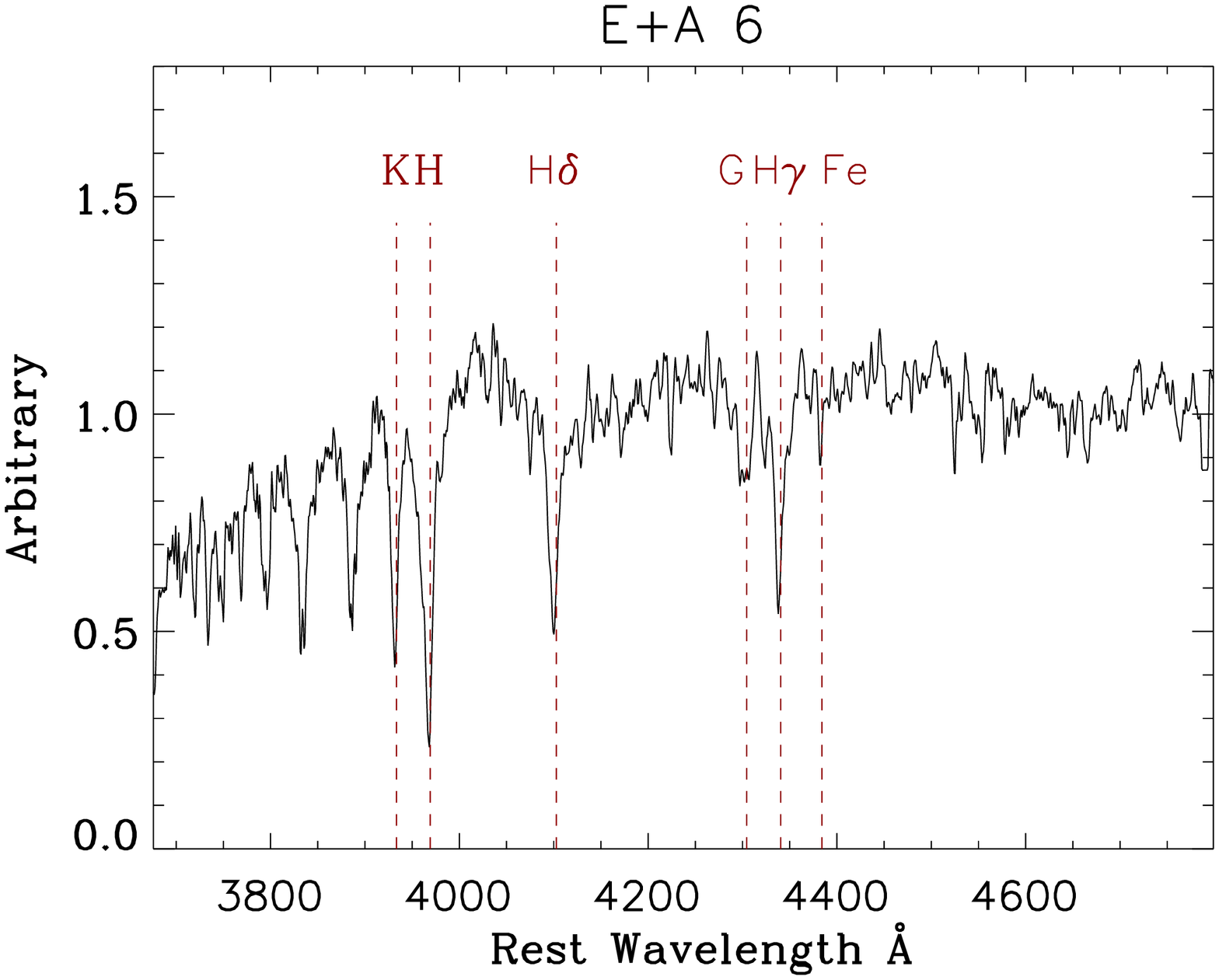}
         \includegraphics[width=5.6cm, angle=90, trim=0 0 0 0]{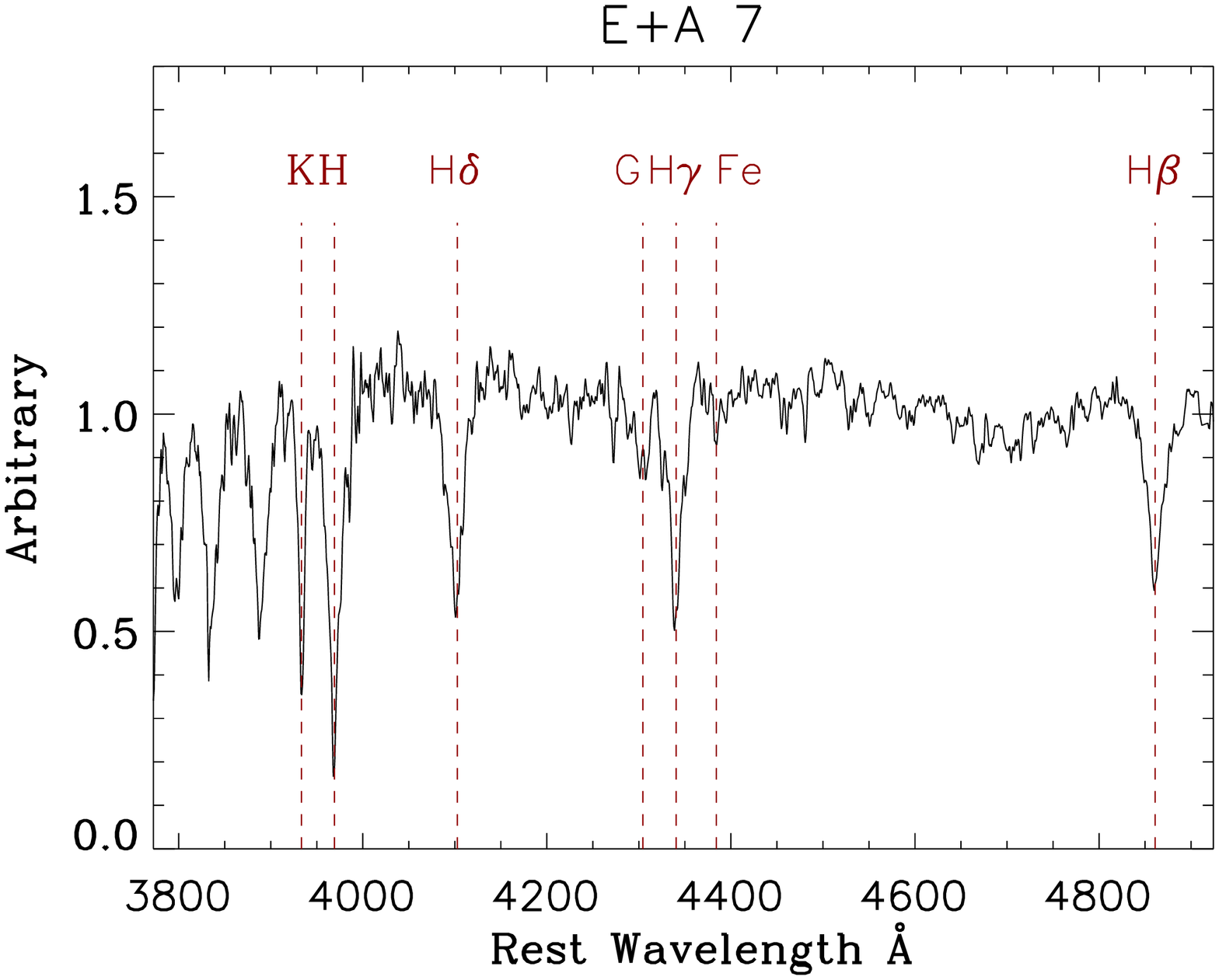}
      \end{minipage}
    \begin{minipage}{0.95\textwidth}
        \includegraphics[width=5.6cm, angle=90, trim=0 0 0 0]{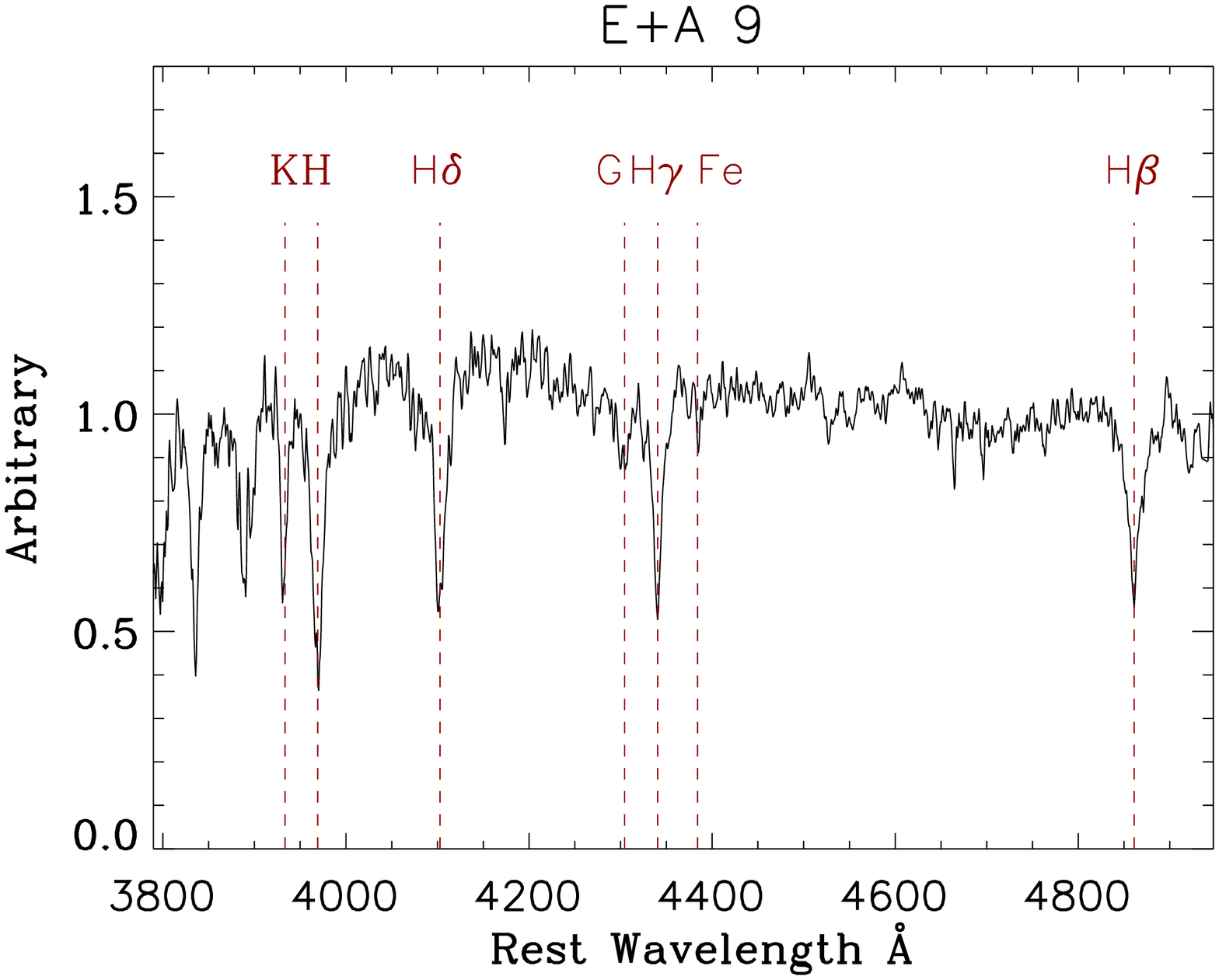}
        \includegraphics[width=5.6cm, angle=90, trim=0 0 0 0]{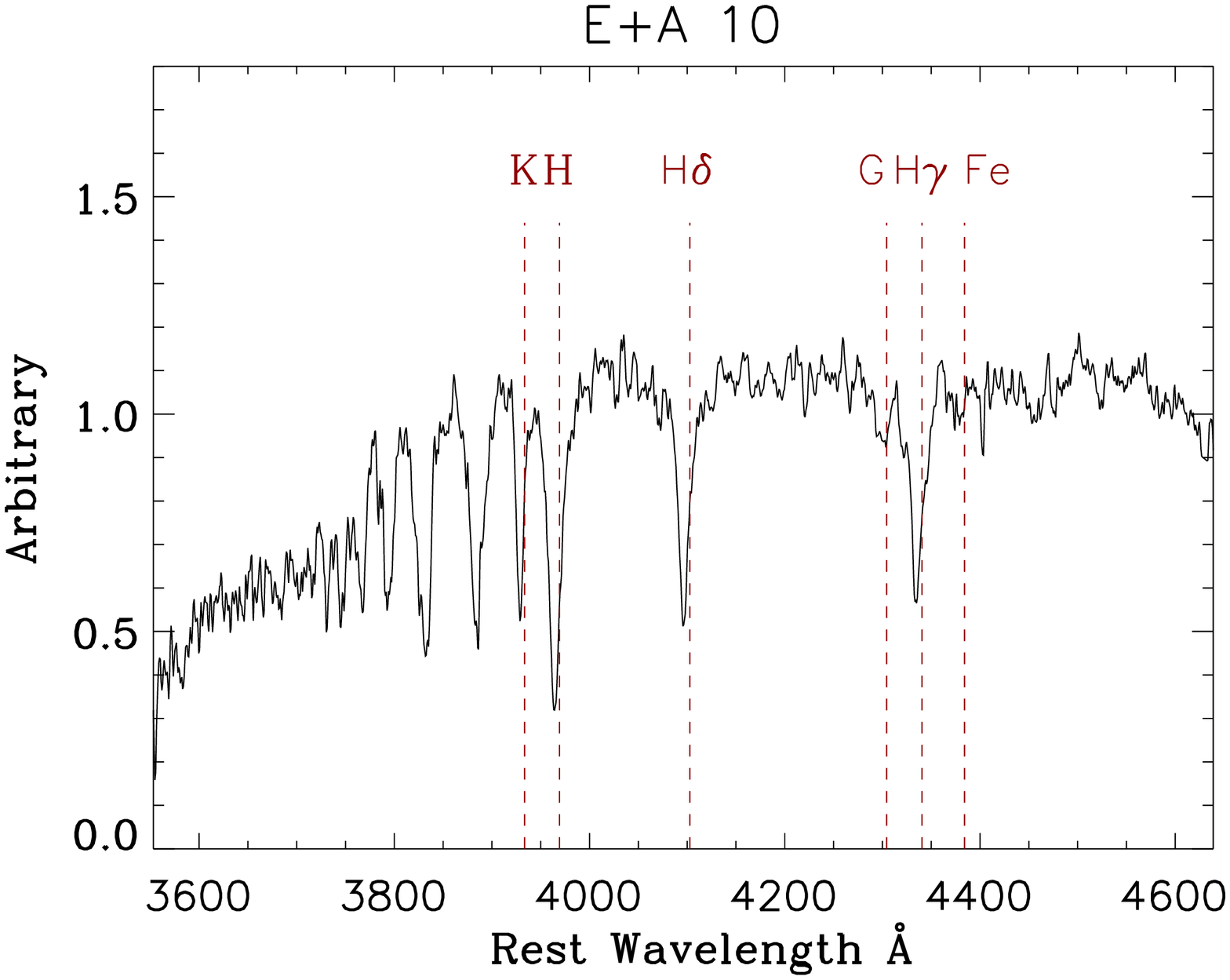}
     \end{minipage}
\end{center}
\caption{\label{fig:intspec} Integrated spectra made by combining the IFU data spatially. Galaxy names are shown as
the title in each panel.}
\end{figure*}

We first produce a high quality integrated spectrum for each of our target galaxies by collapsing the
data cube in the spatial directions. That is, we co-added all of the individual spaxels together
weighted by their variance. These spectra are shown in Figure \ref{fig:intspec} where they have been
adjusted to their rest-frame by division of the wavelength scale by $1+z$.

\subsection{Line strength indices}
We measure line strength equivalent widths on the Lick/IDS system \citep{worthey97,trager98}. This system uses
a flux summing technique whereby a central bandpass containing the line is flanked by
two bandpasses, one each to the red and blue side of the spectral line. The mean `continuum'
level in each of the bands adjacent to the line is measured and a straight line extrapolation
between the two is used as a measure of the interpolated continuum level in the central bandpass.
The ratio of the summed counts in the central bandpass to the expected height of the continuum 
is used as the estimate of the line equivalent width. Prior to measuring the indices the 
spectra were broadened to Lick resolution by convolution with a wavelength dependent Gaussian such that the 
quadrature sum of the Gaussian width and the instrumental resolution were equal to 
the Lick resolution ($\sim 9$\,\AA\, depending on the precise wavelength). The line index strengths were corrected for smearing effects 
due to the intrinsic velocity dispersion of the galaxies following the prescription of
\citet{kuntschner04}. We also correct for differences in the measured equivalent widths 
and the Lick system using the known offsets for data with flux calibrated continuum \citep{norris06}.

\subsection{Global spectral properties}
\begin{figure}
         \includegraphics[width=5.8cm, angle=90, trim=0 0 0 0]{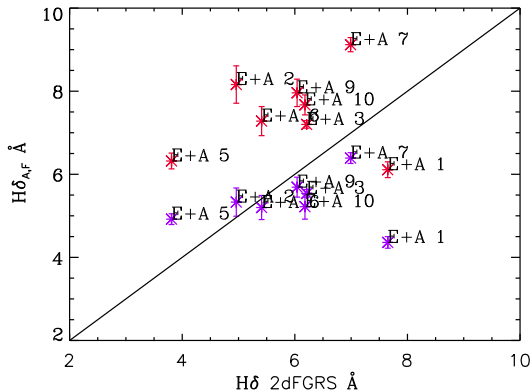}
\caption{\label{fig:indcomp} The H$\delta$ equivalent width measurements from our spatially integrated  GMOS spectroscopy
 compared with the original measurements that were made from the 2dFGRS spectra using Gaussian line fitting. The 
{\it red and blue points} show the H$\delta_{\rm A}$ and H$\delta_{\rm F}$ measurements, respectively. The {\it black line} shows
the one-to-one relationship. The H$\delta_{\rm A}$ index returns larger equivalent widths which appear like a constant
offset whilst the H$\delta_{\rm F}$ index more closely resembles the line-fit values. } 
\end{figure}

We first re-examine the key spectral line diagnostic for our galaxy sample to confirm
their original classification from the 2dFGRS spectroscopy.  All the galaxies studied are clearly dominated
by young stellar populations evidenced by the strong Balmer series absorption clearly visible 
in Figure \ref{fig:intspec}. The defining spectral line 
for E+A galaxies is the H$\delta$ absorption line. In Figure \ref{fig:indcomp} we compare the previously
measured H$\delta$ absorption line equivalent widths from 2dFGRS spectra using a line-fitting
technique \citep{lewis02} with the values derived from our spatially integrated spectra using the line-summing
technique described above. We utilize the two commonly applied index definitions for H$\delta$; the 
H$\delta_{\rm F}$ and H$\delta_{\rm A}$ indices \citep{worthey97}. The H$\delta_{\rm F}$ definition has a narrow
central bandpass designed as a probe of F-star populations whilst the  H$\delta_{\rm A}$ definition has
a wider central bandpass designed to include the broader A-star light -- seemingly a more natural 
definition for E+A galaxies. 

In Figure \ref{fig:indcomp} we compare our new index measurements with those
previously derived from the 2dFGRS spectra \citep{lewis02,blake04}. The H$\delta_{\rm A}$ index generally results
in larger equivalent widths than those originally derived via line-fitting but the relative line strengths between objects
are similar between the two different sets of observations and measurement techniques. The H$\delta_{\rm F}$ index returns
values similar to the original line-fit results. Our selection from the 2dFGRS sample of \citet{blake04} is 100\,per cent
successful in selecting E+A galaxies. 

In Figure 6 we show some age-metallicity diagnostic
diagrams using Lick index measurements of the Balmer
absorption lines H$\delta$, H$\gamma$ and H$\beta$ along with the metallicity
indices Fe4383 and C$_2$4668. The over-plotted predictions of single stellar population (SSP)
models are taken from \citet{thomas03,thomas04} and assume solar abundance ratios. The position of our target
galaxies with respect to the grids confirm their young luminosity--weighted ages 
(scattered around $\sim 0.5$\,Gyr) consistent with
their E+A classification. Our selection criteria of strong H$\delta$ absorption without the
presence of emission lines basically select galaxies where the Balmer absorption strength
is maximized (model prediction $\sim 0.4$\,Gyr). Our sample shows a range of luminosity weighted metallicities
from about solar to several times solar metallicity. Due to the sparsity of suitable Lick
indices in our rest wavelength range, attempts to obtain a meaningful estimate of the abundance
ratios failed and hence we assume solar values.
\begin{figure*}
         \includegraphics[width=15.8cm, angle=0, trim=0 0 0 0]{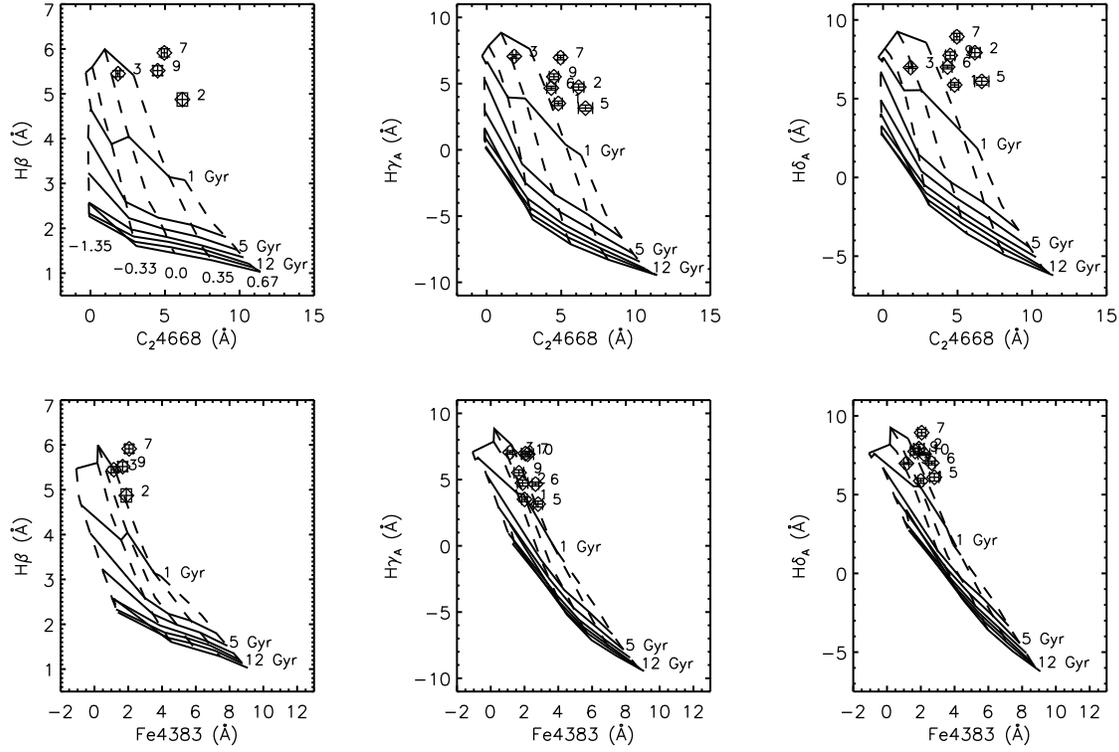}
\caption{\label{fig:grid} Age metallicity diagnostic diagrams confirming the young luminosity
weighted ages of our E+A sample. Galaxy names are indicated by numbers.
Over-plotted are models by \citet{thomas03,thomas04}
with solar abundance ratios.  The models span a range in
metallicity [Z/H] = $-1.35$ to $+0.67$ (dashed lines), and solid lines indicate ages from top to
bottom of 0.4, 1, 3, 5, 8, 10, 12 and 15\,Gyr, respectively.
}
\end{figure*}

\subsection{Adaptive spatial binning}
There is large variation in the signal--to--noise ratio of the individual spectra
within each data cube. For the most part the signal--to--noise ratio decreases rapidly
with galacto--centric radius and in general the signal--to--noise ratio of an individual 
spaxel is insufficient for quantitative analysis. As a result, we need to bin the data-cube spatially to increase 
the signal--to--noise ratio in each spectrum. To do this we use the Voronoi spatial binning
method of \citet{cappellari03}. In this scheme the data are adaptively binned in order 
to achieve a predefined signal--to--noise ratio across the full area of the IFU. In practice, this means
binning over large areas where the signal is low -- usually the galaxy outskirts -- while using 
small bins where the signal is high -- usually in the centres of the galaxies. 
We choose different final signal--to--noise ratios
for each galaxy depending on the intrinsic quality of the spectra for that galaxy. The chosen final, per
spatially binned element, signal--to--noise ratios range from 5\,\AA$^{-1}$\, for the lowest quality data
to 15\,\AA$^{-1}$\, for our brightest target. 

\subsection{Spatially resolved line index measurements}
\begin{figure*}
   \begin{center}
     \begin{minipage}{0.95\textwidth}
         \includegraphics[width=5.6cm, angle=90, trim=0 0 0 0]{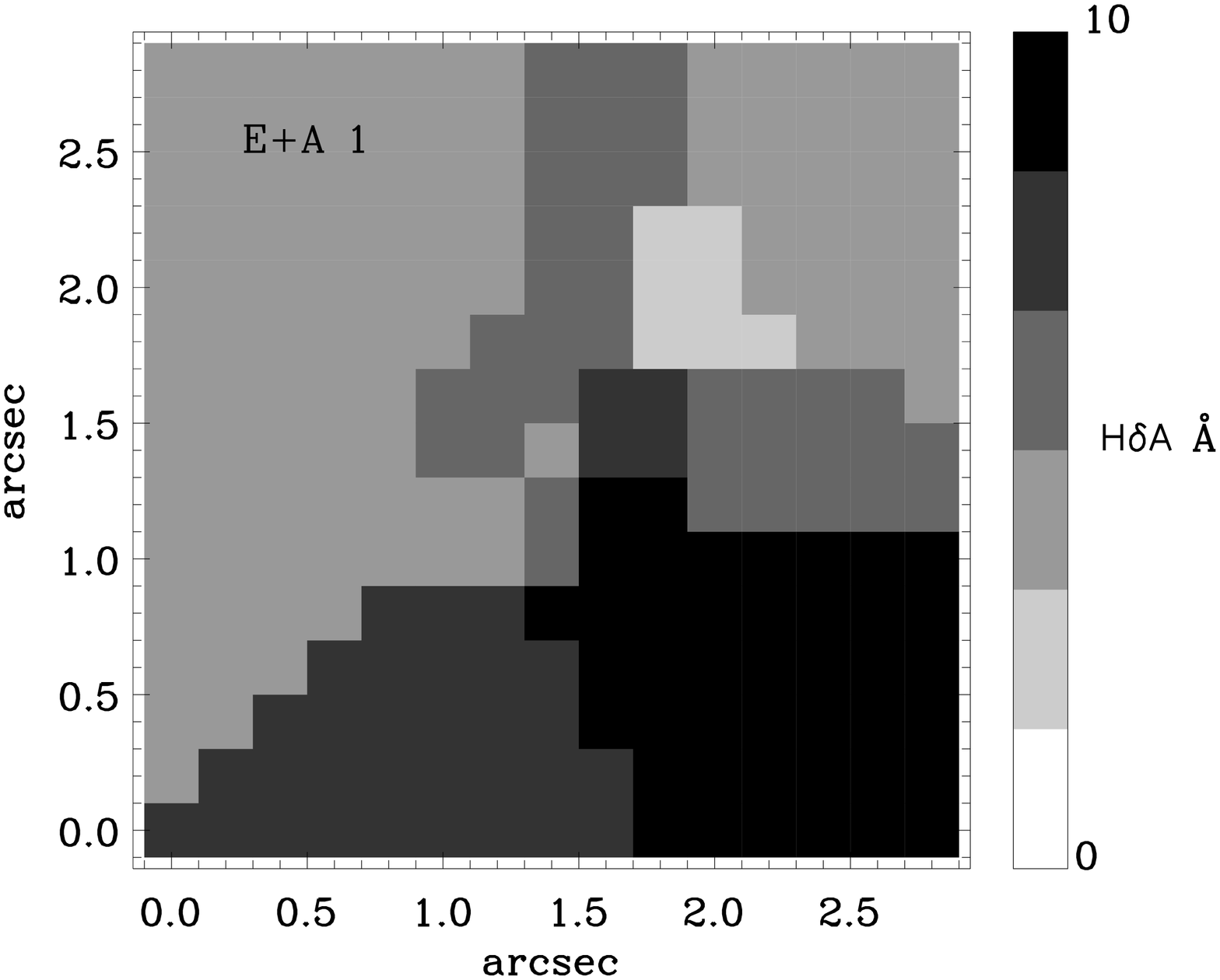}
         \includegraphics[width=5.6cm, angle=90, trim=0 0 0 0]{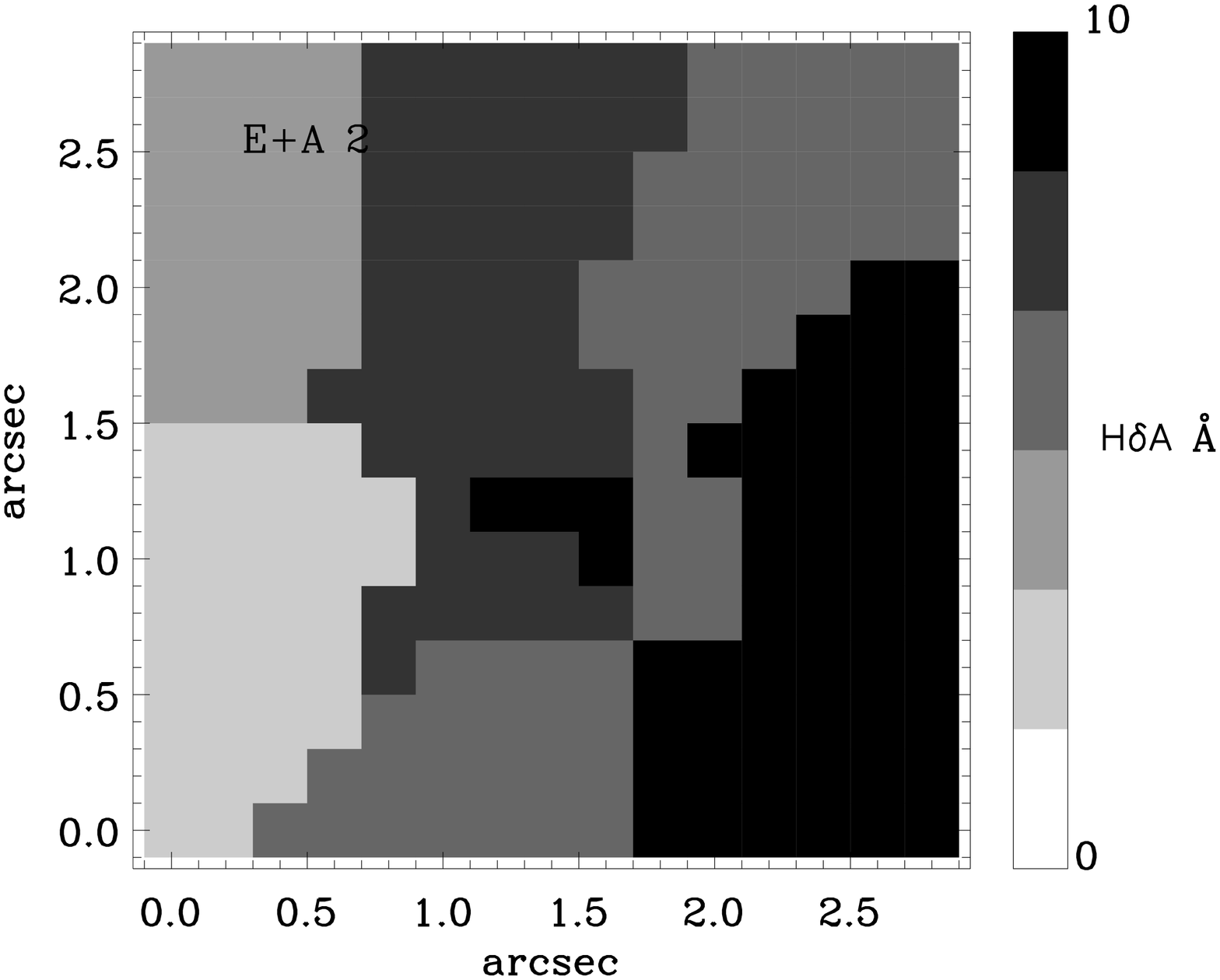}
      \end{minipage}
    \begin{minipage}{0.95\textwidth}
        \includegraphics[width=5.6cm, angle=90, trim=0 0 0 0]{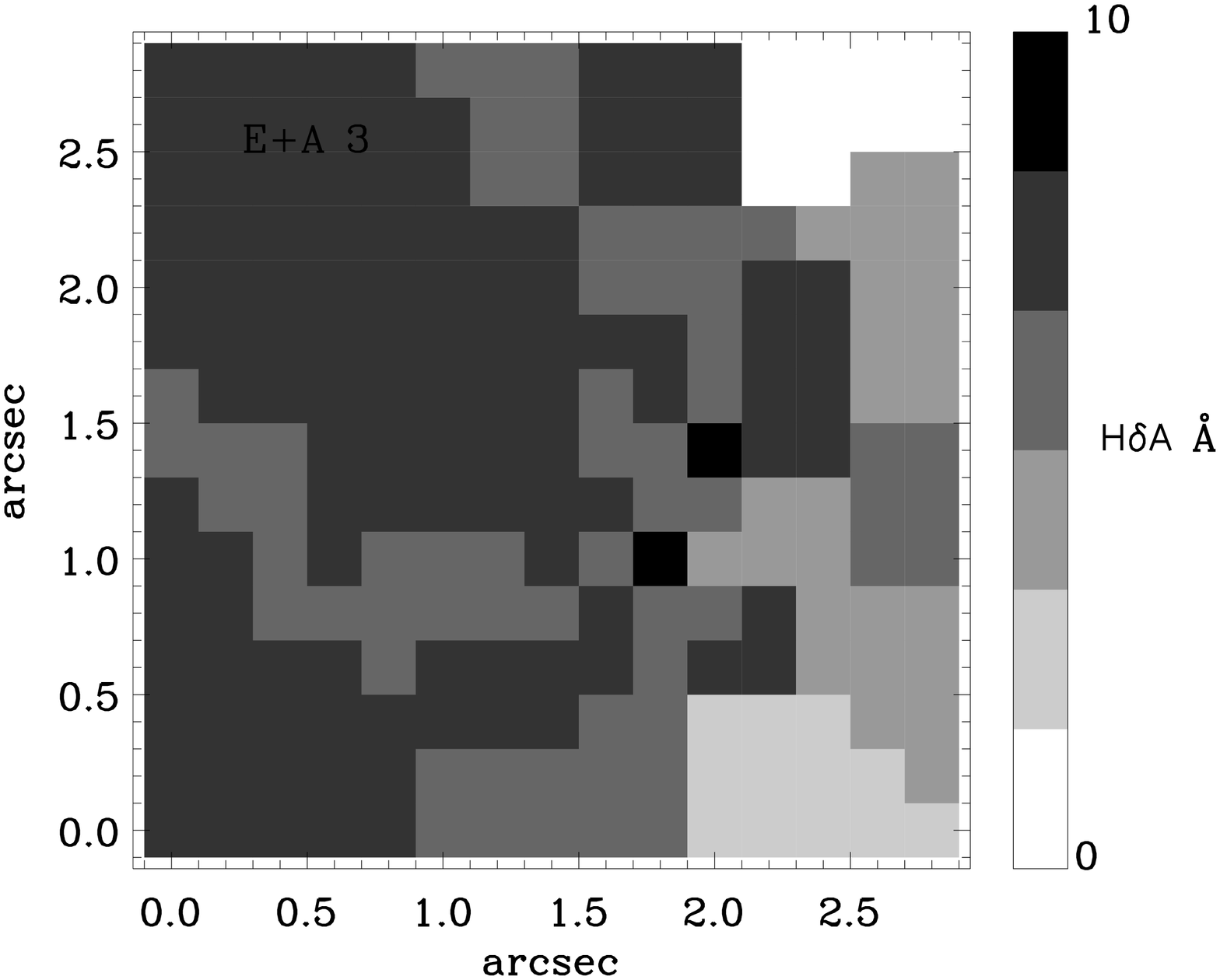}
        \includegraphics[width=5.6cm, angle=90, trim=0 0 0 0]{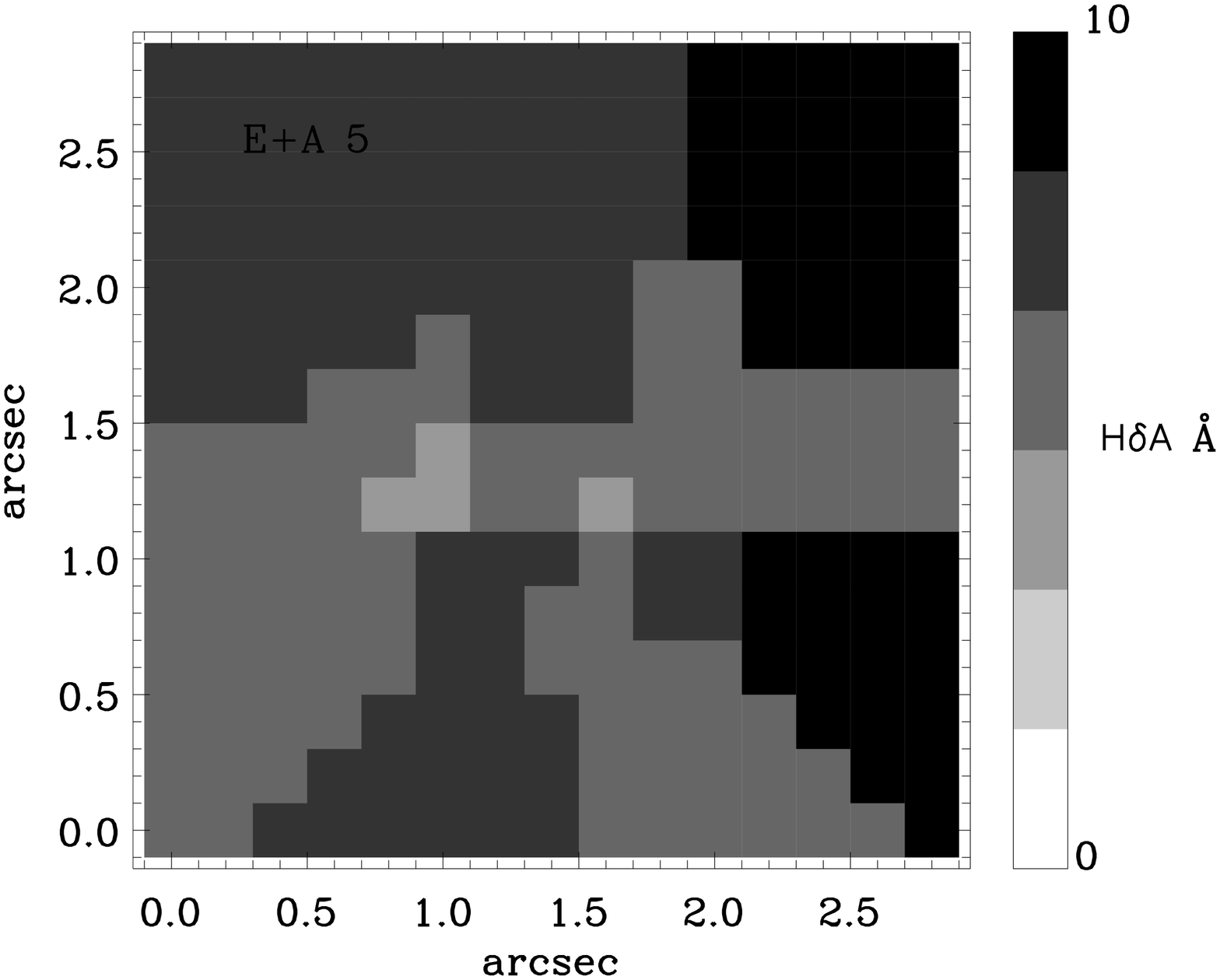}
     \end{minipage}
     \begin{minipage}{0.95\textwidth}
         \includegraphics[width=5.6cm, angle=90, trim=0 0 0 0]{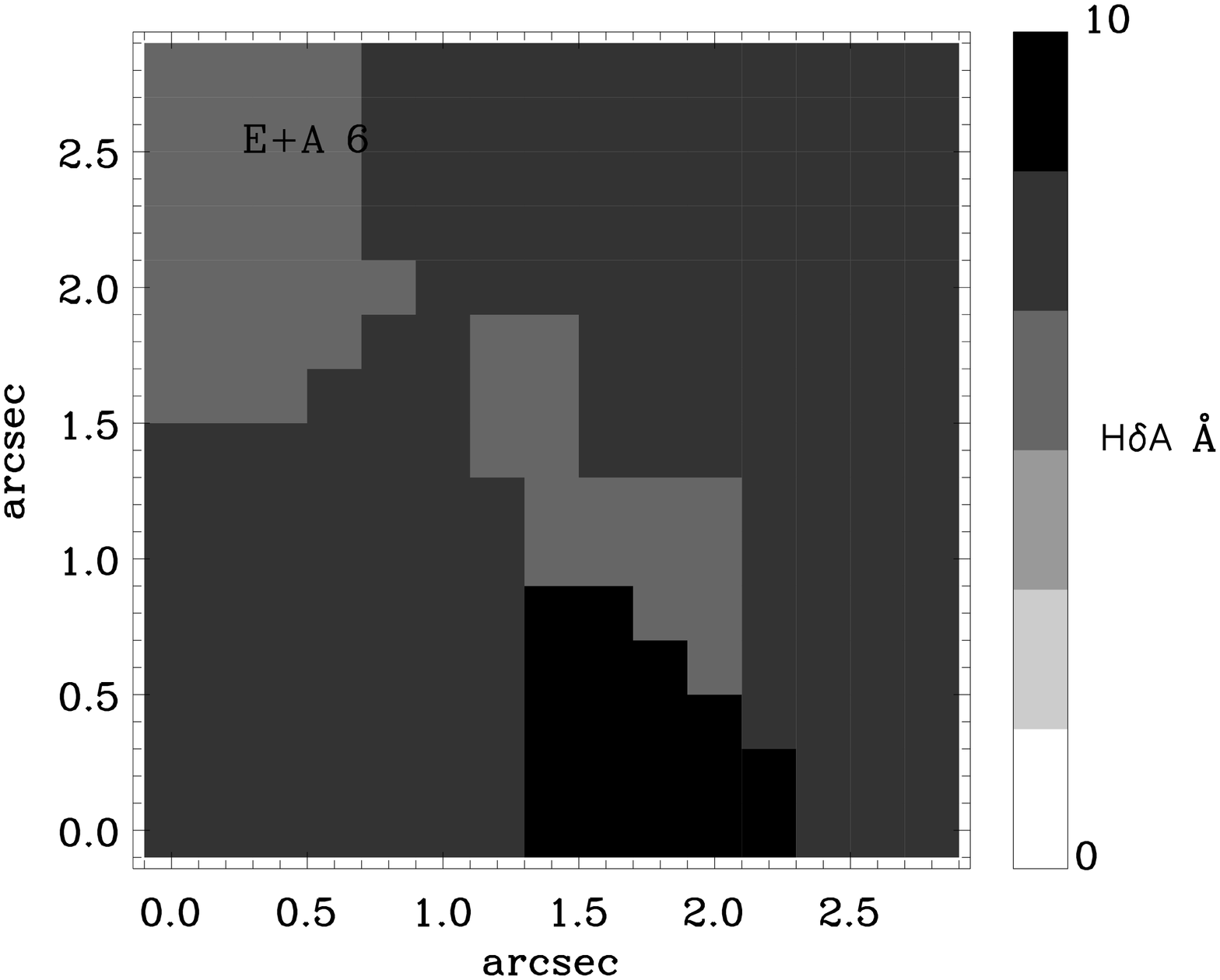}
         \includegraphics[width=5.6cm, angle=90, trim=0 0 0 0]{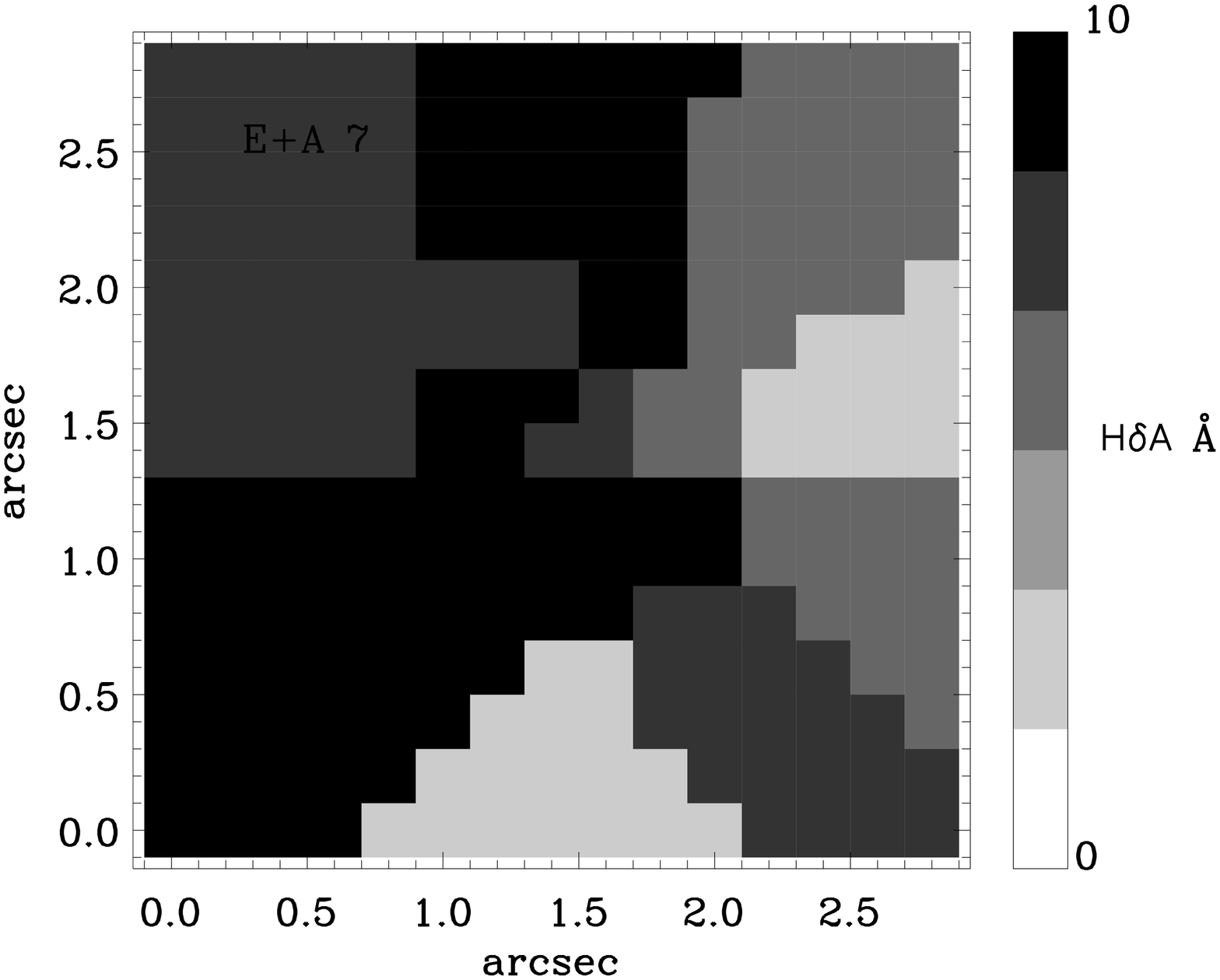}
      \end{minipage}
    \begin{minipage}{0.95\textwidth}
        \includegraphics[width=5.6cm, angle=90, trim=0 0 0 0]{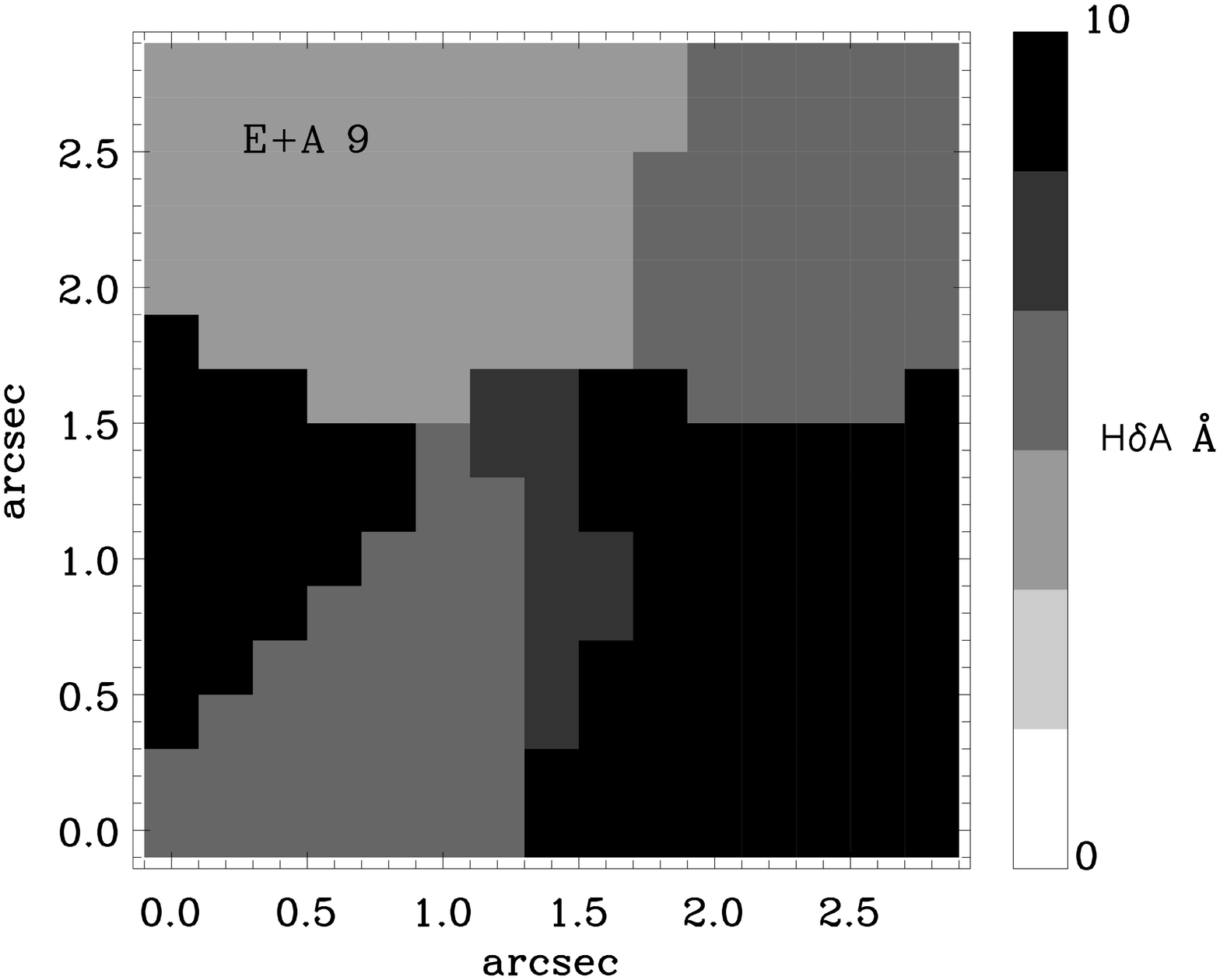}
        \includegraphics[width=5.6cm, angle=90, trim=0 0 0 0]{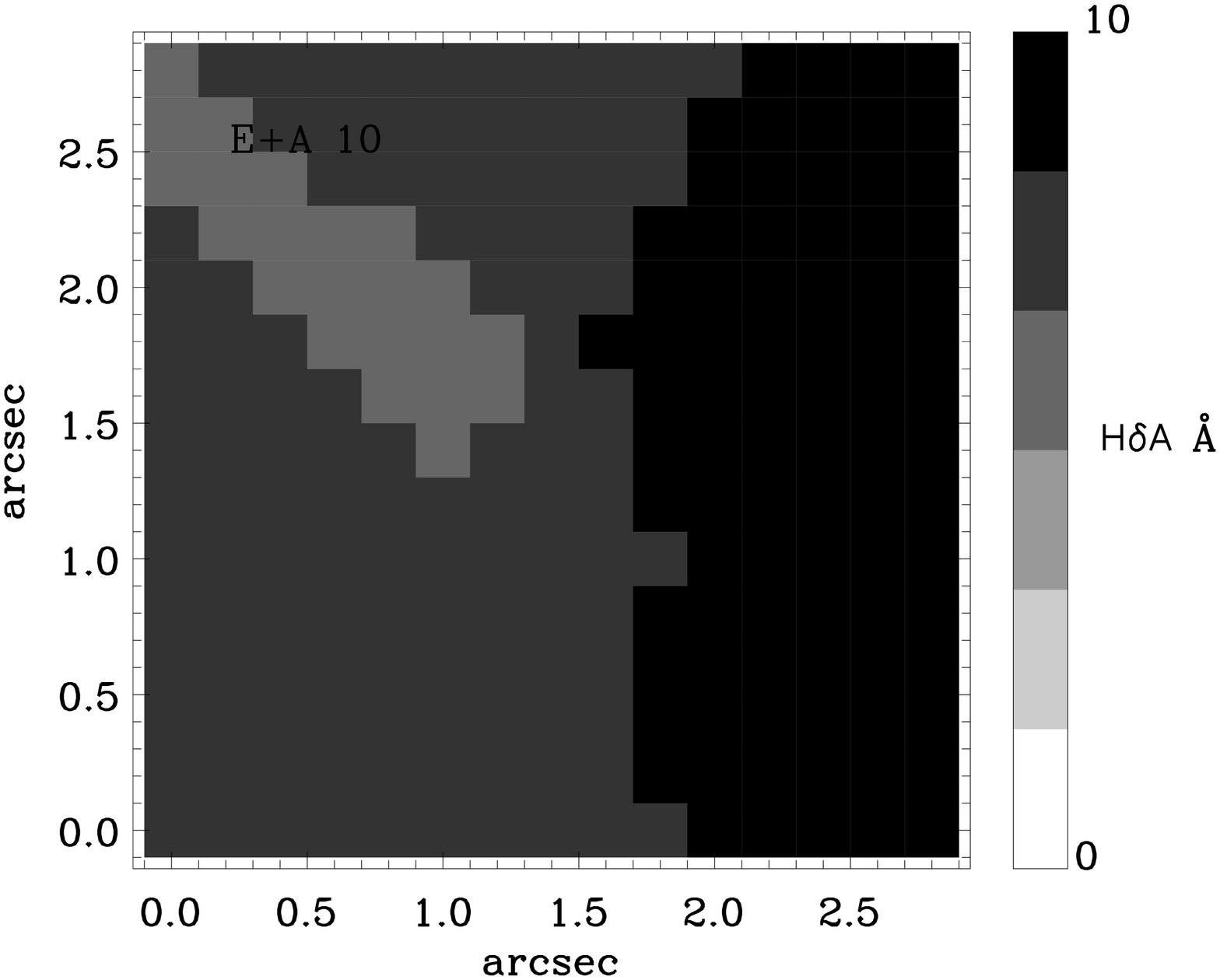}
     \end{minipage}
\end{center}
\caption{\label{fig:hdelta_maps} H$\delta_{\rm A}$ equivalent width maps measured in adaptively binned regions. The sample exhibits uniformly 
strong absorption over the central 3\arcsec.}
\end{figure*}

\begin{figure}
         \includegraphics[width=5.8cm, angle=90, trim=0 0 0 0]{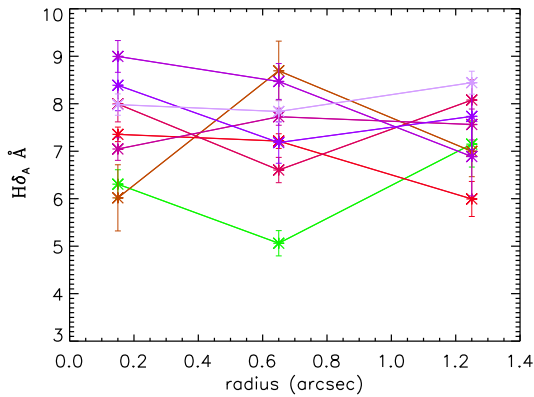}
\caption{\label{fig:hdeltarad} H$\delta$ radial gradients measured from annular binned spectra.}
\end{figure}

We made spatially--resolved line index measurements using the adaptively binned
data.  The H$\delta_{\rm A}$ equivalent width maps for our sample are shown in
Figure \ref{fig:hdelta_maps}. These maps generally reveal a uniformly high
Balmer absorption line equivalent width with little evidence for radial 
gradients present. It should be noted, however, that our data cover only  
the central 3\arcsec\, $\times$ 3\arcsec\, which corresponds to a maximum galacto--centric radius of 
only $\sim 2$ to 5 kiloparsecs. Previous long--slit or photometric studies have claimed cases of E+A galaxies having 
a young population spread over these scales with radial gradients seen only on larger scales \citep{goto08}. 
This scale of a few kiloparsecs is also the scale a merger/interaction-induced starburst is expected to be concentrated on \citep{bekki05}.
Furthermore, the 3\arcsec\, scale is comparable to the expected seeing disk ($\sim 1$\arcsec) which has the effect of 
smoothing out any radial variations. Therefore, while we see uniformly strong  H$\delta$ absorption
across the central few kiloparsecs we cannot rule out the presence of gradients at larger galacto--centric radii.
The lack of significant gradients is more clearly displayed in Figure \ref{fig:hdeltarad} where rather than
use the adaptively binned data we have binned the spectra in three concentric annuli and measured the H$\delta$
line equivalent width. The radial plots reveal no clear radial trends.

\subsection{Template fitting}
Rather than  utilizing just the information obtained by measurement of
discrete line indices, we can use all the information contained in the 
spectroscopic data by fitting the entire spectrum simultaneously. By doing so, 
we can simultaneously gather information on the stellar populations, the streaming 
velocity and velocity dispersion of individual spatial elements of our target 
galaxies. The accuracy of the extraction of kinematics from the 
data is greatly increased since many lines are used simultaneously in
measuring the redshift and line-of-sight velocity dispersion.

As templates for fitting, we use the single--age single--metallicity stellar population synthesis models
of \citet{vazdekis07}. The stellar population spectral energy distributions are constructed from the MILES empirical
 stellar library \citep{sanchez06}. We use a range of templates in the fit ranging in age from 0.1\,Gyr to
13\,Gyrs. The age difference between templates is $\sim 0.1$\,Gyr for ages of
less than 1\,Gyr whilst we use a sparser sampling in age for the older population templates. Solar metallicity 
is assumed for the young templates ($<1$\,Gyr) and a variety of metallicities are allowed for the old stellar 
population templates. The reason a variety of metallicities are not used for the young templates is because the spectra are 
much more sensitive to age than metallicity in this regime, and including more freedom in metallicity does not ultimately improve the fit. 
A list of all templates allowed in the fitting routine is given in Table \ref{tab:templates}.
\begin{table}
\caption{\label{tab:templates} Stellar population templates \citep[from][]{vazdekis07} used for fitting our E+A spectra.}
\begin{tabular}{|c|c|c|} \\ \hline
Template no. & Age (Gyr) & Metallicity \\ \hline
1            &   0.10     & 0.0        \\
2            &   0.20     & 0.0        \\
3            &   0.32     & 0.0        \\
4            &   0.40     & 0.0        \\
5            &   0.50     & 0.0        \\
6            &   0.71     & 0.0        \\
7            &   0.89     & 0.0        \\
8            &   4.47     & -1.68        \\
9            &   4.47     & -1.28        \\
10           &   4.47     & -0.68        \\
11           &   4.47     & -0.38        \\
12           &   4.47     & 0.0        \\
13           &   4.47     & 0.2        \\
14            &   7.98     & -1.68        \\
15            &   7.98     & -1.28        \\
16           &   7.98     & -0.68        \\
17           &   7.98     & -0.38        \\
18           &   7.98     & 0.0        \\
19           &   7.98     & 0.2        \\
20           &   12.59     & -1.68        \\
21            &   12.59     & -1.28        \\
22           &   12.59     & -0.68        \\
23           &   12.59     & -0.38        \\
24           &   12.59     & 0.0        \\
25           &   12.59     & 0.2        \\ \hline
\end{tabular}
\end{table}

The fitting is performed 
using a penalized pixel fitting algorithm \citep{cappellari04} which
fits the spectrum in pixel space using the combination of a number of input user-supplied templates and simultaneously fitting for
redshift and velocity dispersion. Prior to fitting, the science spectra are smoothed from their intrinsic 1.9\,\AA\, spectral
resolution to have the same intrinsic spectral resolution as the template spectra i.e. 2.3\,\AA\, (FWHM).

\begin{figure}
\includegraphics[width=6cm,angle=90]{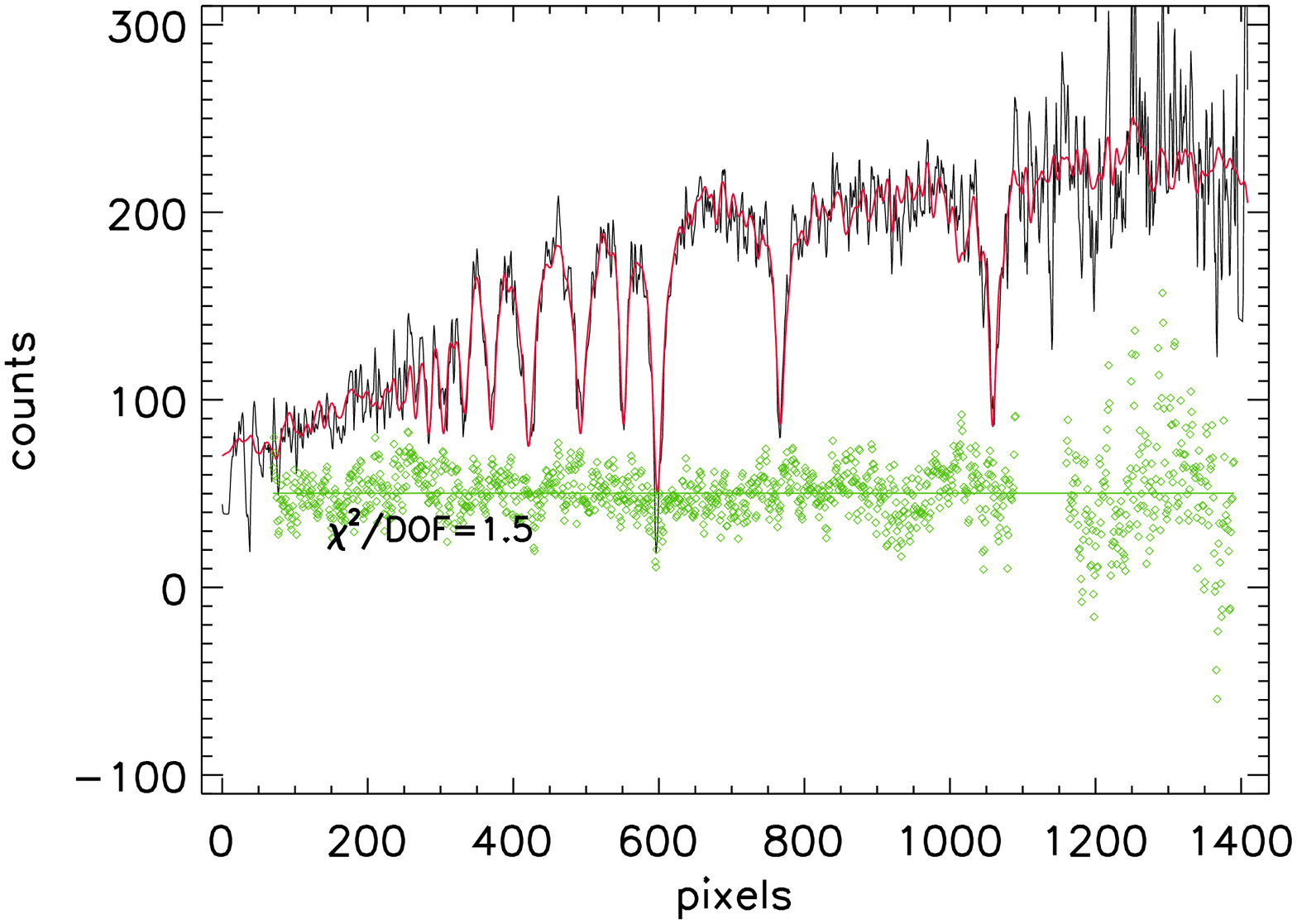}
\includegraphics[width=6cm,angle=90]{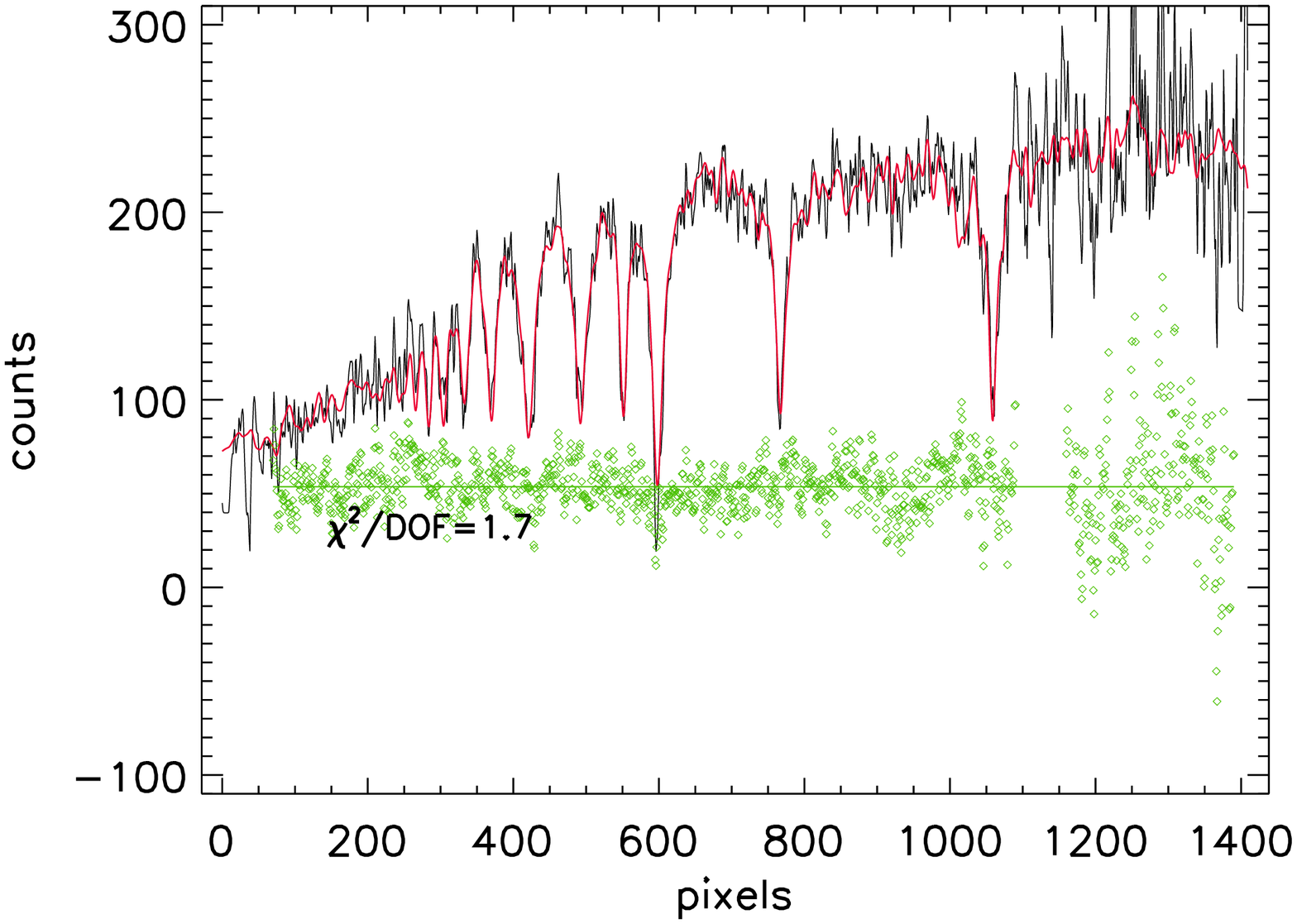}
\caption{\label{fig:specfit} {\it Top panel}: The spectrum of a single adaptively--binned spatial pixel of E+A\_10 where we have
 Voronoi binned to a signal--to--noise ratio of 15. The {\it black line} is the data and the {\it red line} is the best fit using all the templates
in Table \ref{tab:templates}. The residuals
are shown as {\it green points}. The pixel values where no residuals are plotted have been excluded from the fit due to poor
data quality in these regions. {\it Bottom panel}: same as {\it top panel} but using only the templates with ages less than 1\,Gyr in the fit}
\end{figure}

In Figure \ref{fig:specfit} we show an example spectrum of a single, adaptively--binned spatial pixel along
with the best fitting model spectrum produced by the above technique. The observed spectra are generally well fitted by model 
spectra produced in this way and the quality of the data and the goodness of fit shown in Figure \ref{fig:specfit} are fairly typical.

\subsection{Stellar population}
The composite model spectra that best fit our observed E+A spectra are those dominated by young stellar population
templates ($< 1$\,Gyr). In fact, for the most part, inclusion of older stellar templates, whilst often given
small non-zero weighting by the penalized pixel fitting algorithm, does not improve the fit to the data in
a statistically significant way. That is, at the signal--to--noise ratio of the data, we are unable to detect any contribution from
an old stellar population in the spectra. An example of this is shown in Figure \ref{fig:specfit}. In the top panel is
the best fit to the spectrum of one spatial pixel using all the input templates from Table \ref{tab:templates} and in the 
bottom panel is the best fit using only stellar populations synthesis templates with ages of less than 1\,Gyr (i.e. templates 1-7).
There is little difference between the two fits. As an example of the lack of contribution from the old stellar 
templates the mean difference (per spaxel) in the $\chi^2/{\rm DOF}$ values for E+A\_10 
returned by fitting with the different sets of templates is $\sim 0.09$. At this point we caution that our spectral analysis
only really constrains the young population which is the dominant contributor to the stellar light but not the stellar mass.

\begin{figure*}
   \begin{center}
     \begin{minipage}{0.95\textwidth}
         \includegraphics[width=5.6cm, angle=90, trim=0 0 0 0]{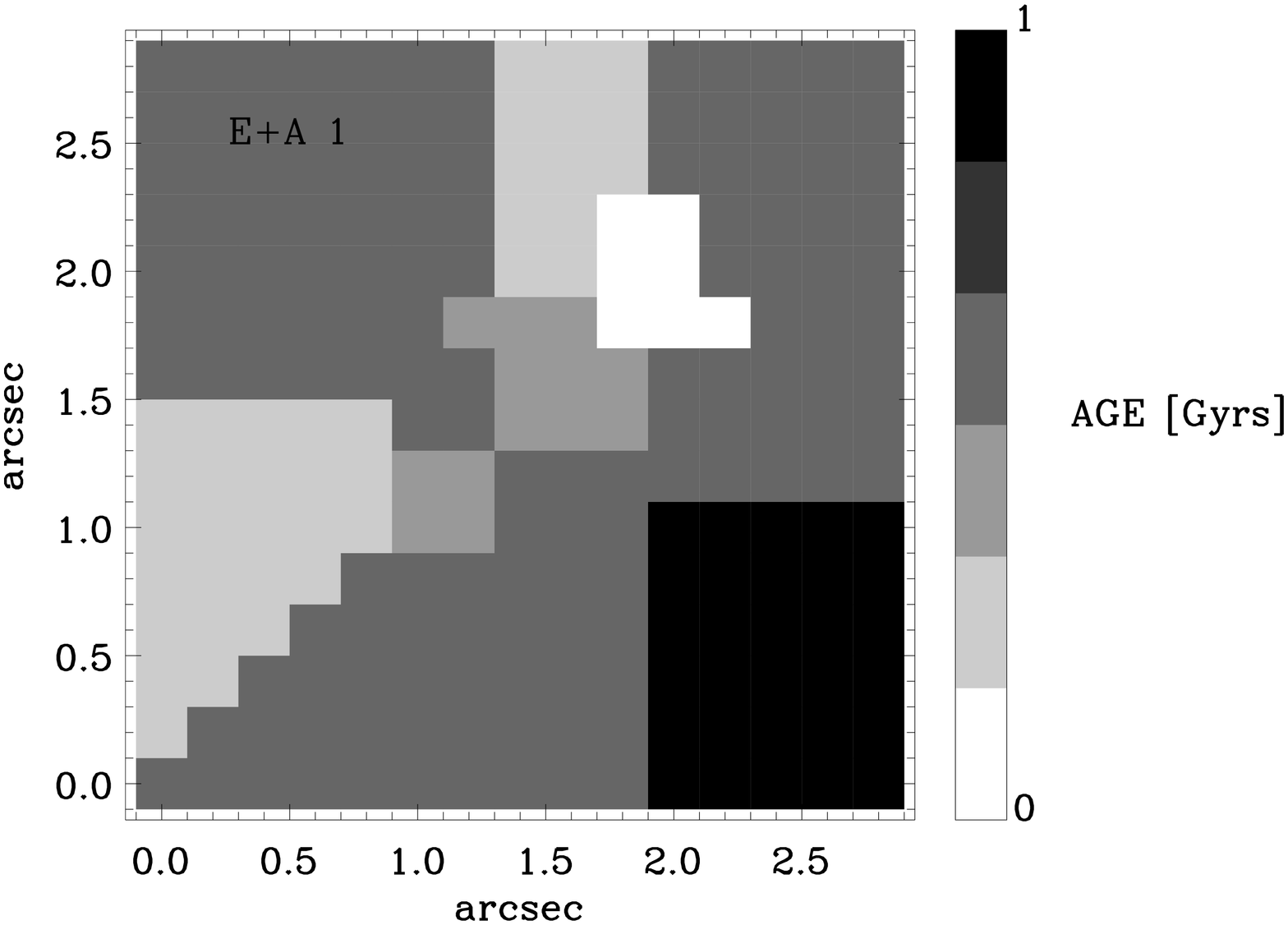}
\hspace{1cm}
         \includegraphics[width=5.6cm, angle=90, trim=0 0 0 0]{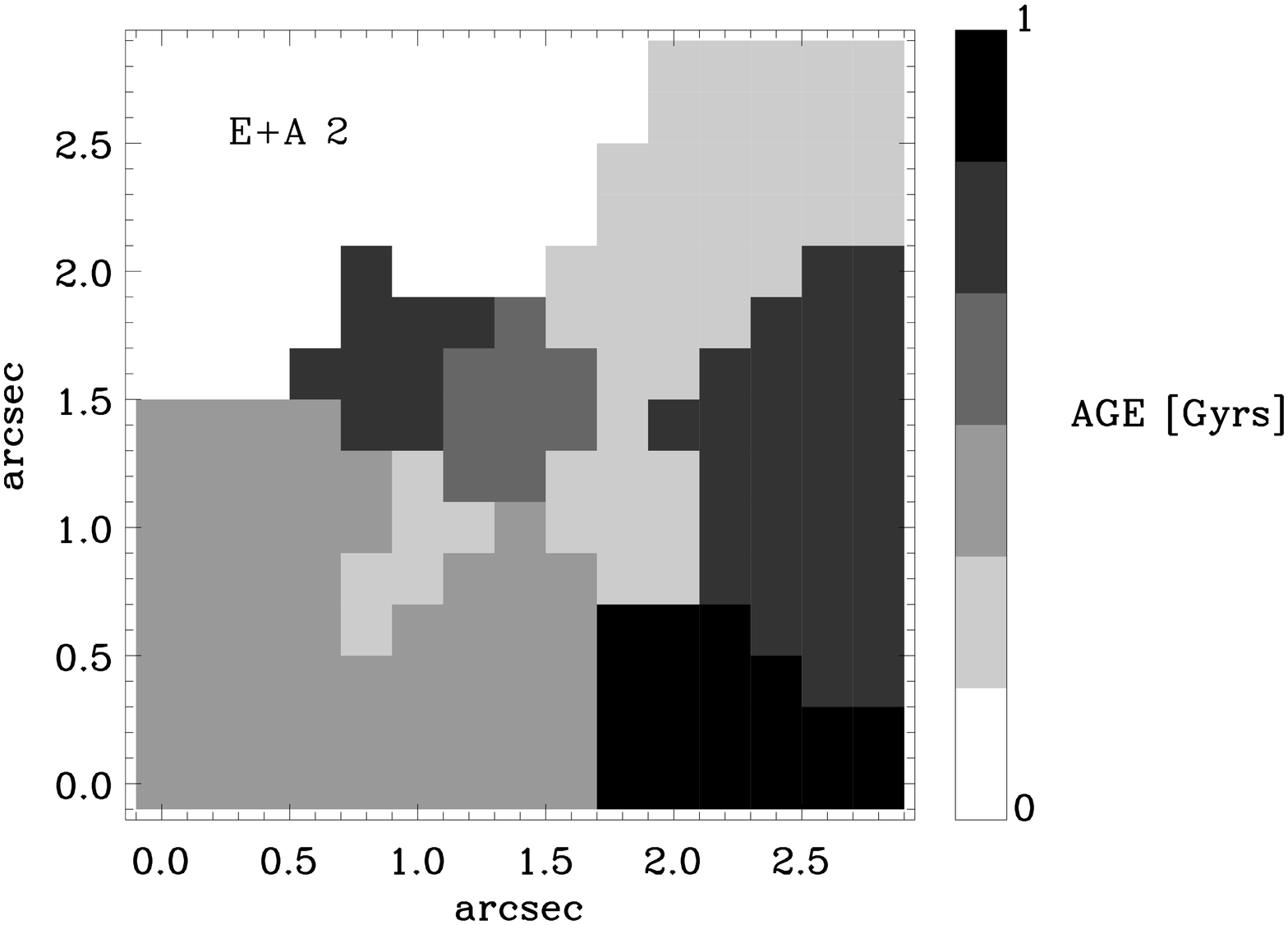}
      \end{minipage}
    \begin{minipage}{0.95\textwidth}
        \includegraphics[width=5.6cm, angle=90, trim=0 0 0 0]{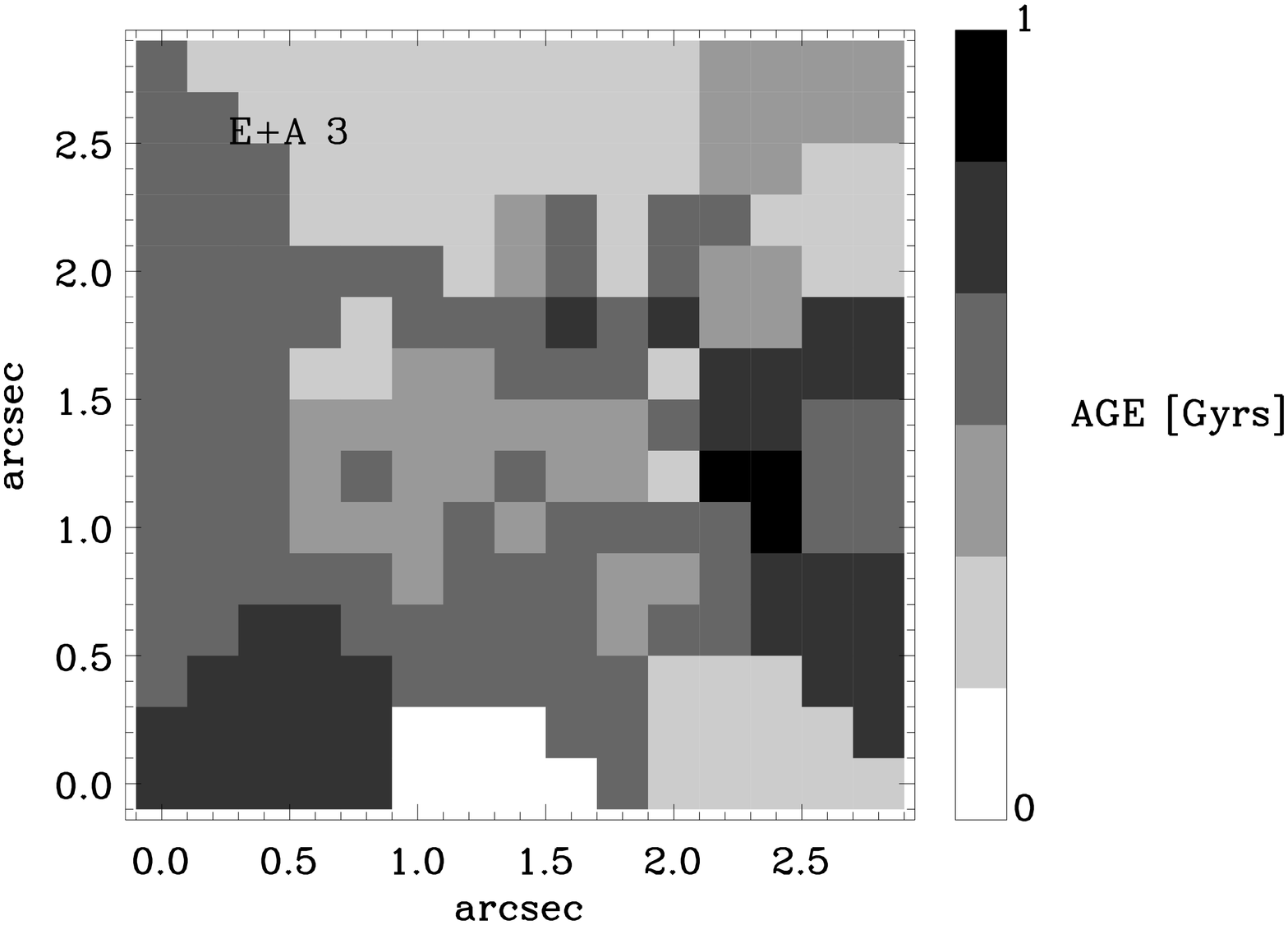}
\hspace{1cm}
        \includegraphics[width=5.6cm, angle=90, trim=0 0 0 0]{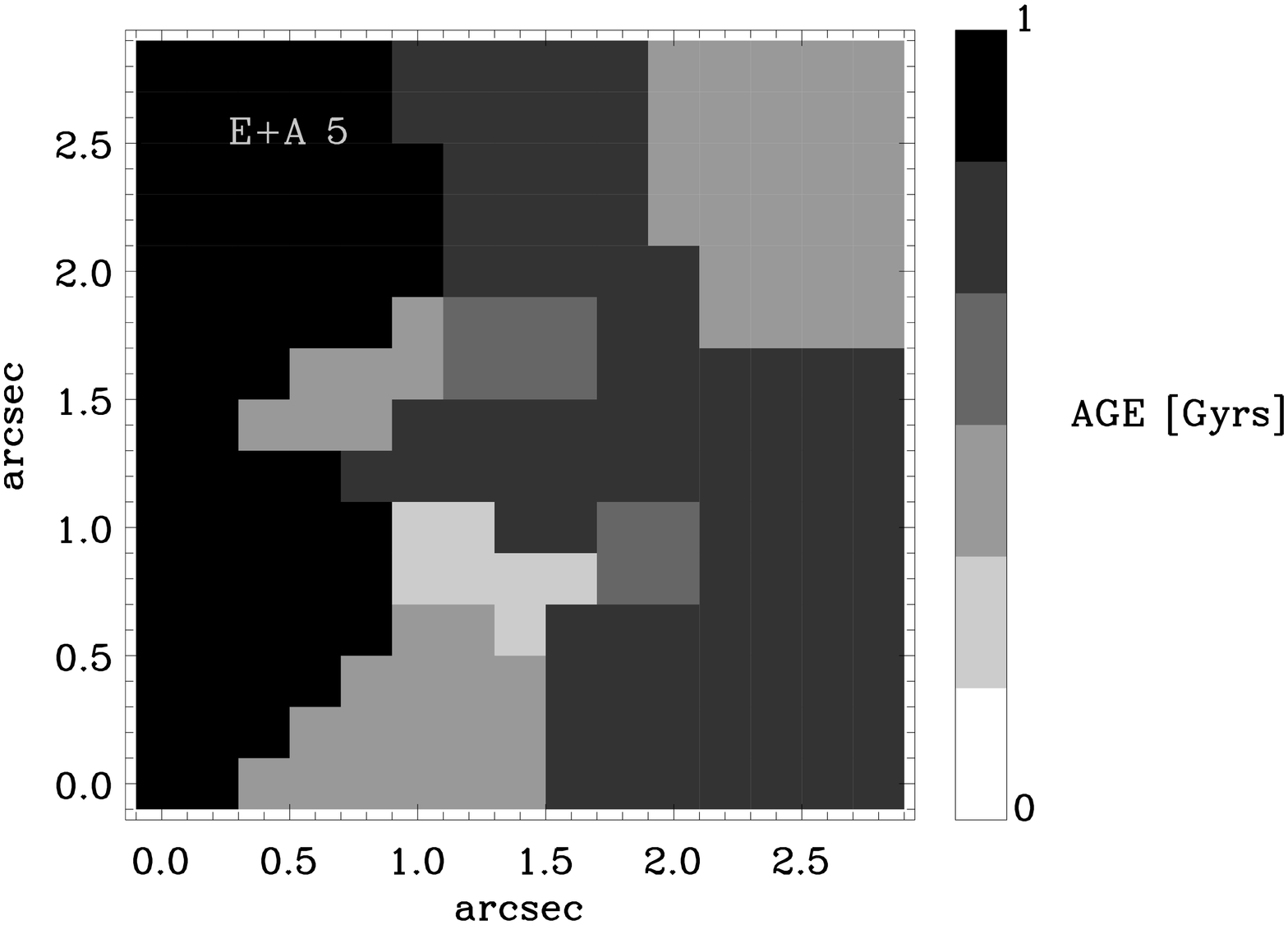}
     \end{minipage}
     \begin{minipage}{0.95\textwidth}
         \includegraphics[width=5.6cm, angle=90, trim=0 0 0 0]{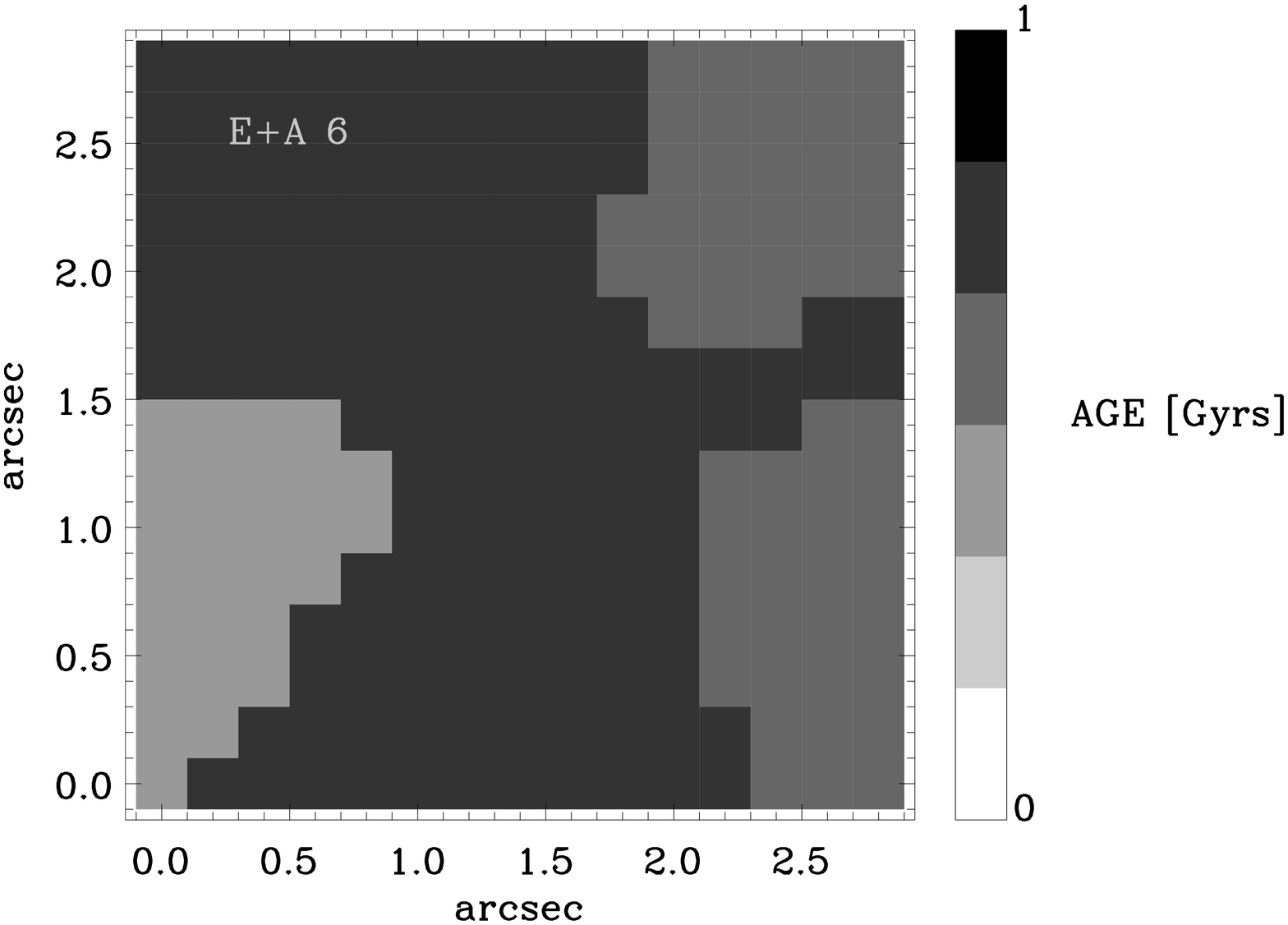}
\hspace{1cm}
         \includegraphics[width=5.6cm, angle=90, trim=0 0 0 0]{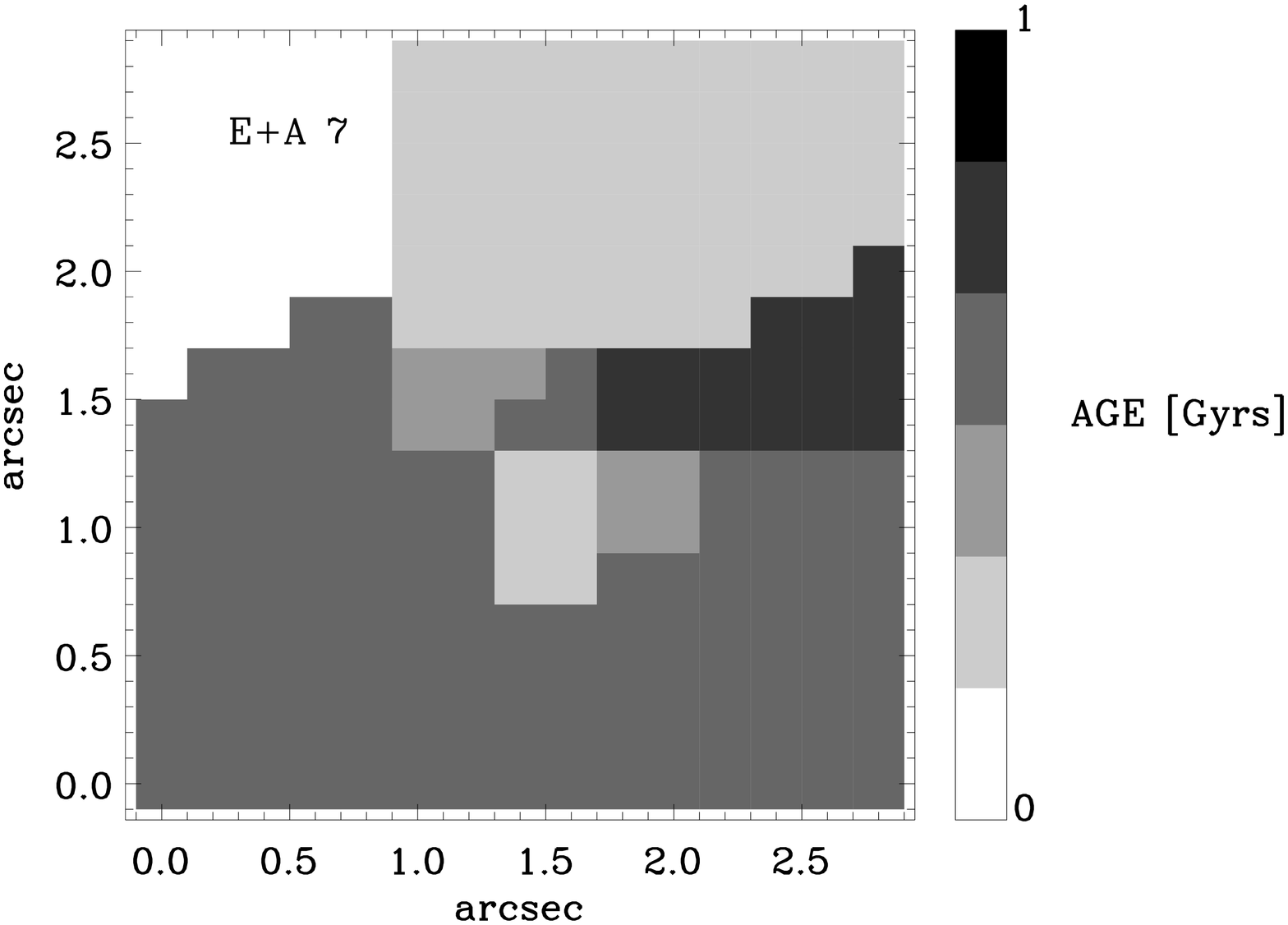}
      \end{minipage}
    \begin{minipage}{0.95\textwidth}
        \includegraphics[width=5.6cm, angle=90, trim=0 0 0 0]{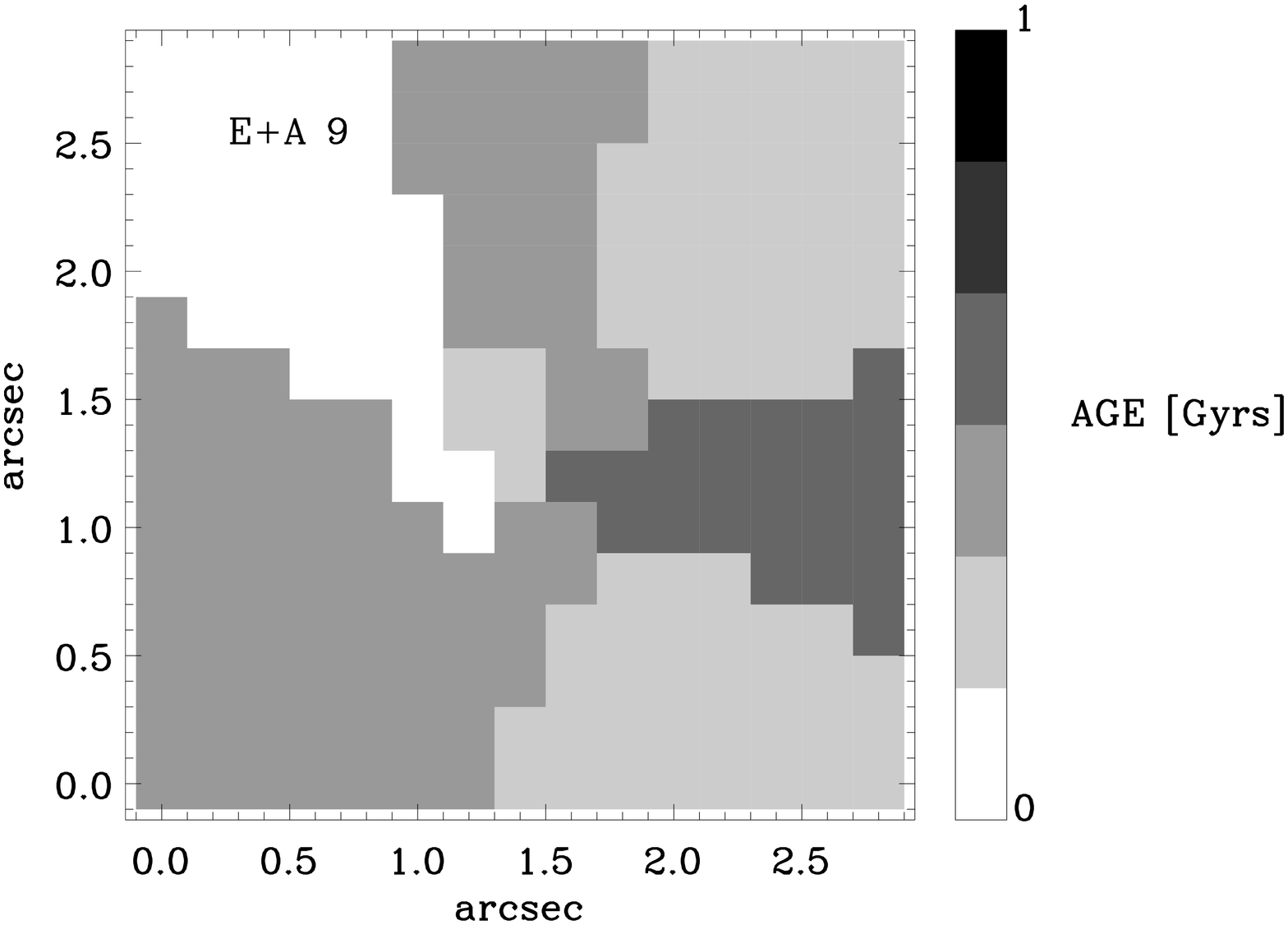}
\hspace{1cm}
        \includegraphics[width=5.6cm, angle=90, trim=0 0 0 0]{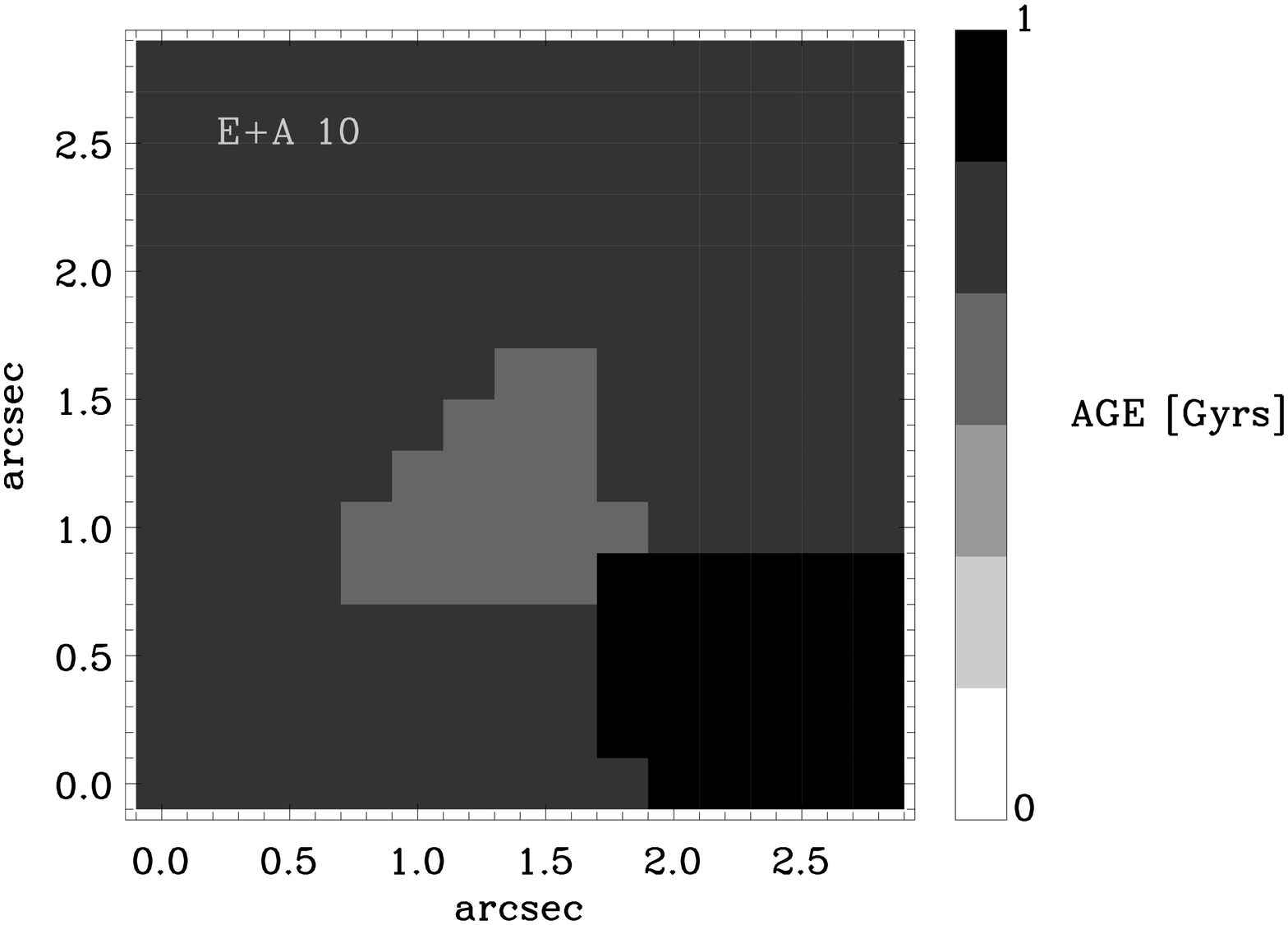}
     \end{minipage}
\end{center}
\caption{\label{fig:agemap} `Age' maps constructed by taking the weighted average age of the best fitting composition
of stellar population models for each spatial element.}
\end{figure*}

\begin{figure}
      \includegraphics[width=5.8cm, angle=90, trim=0 0 0 0]{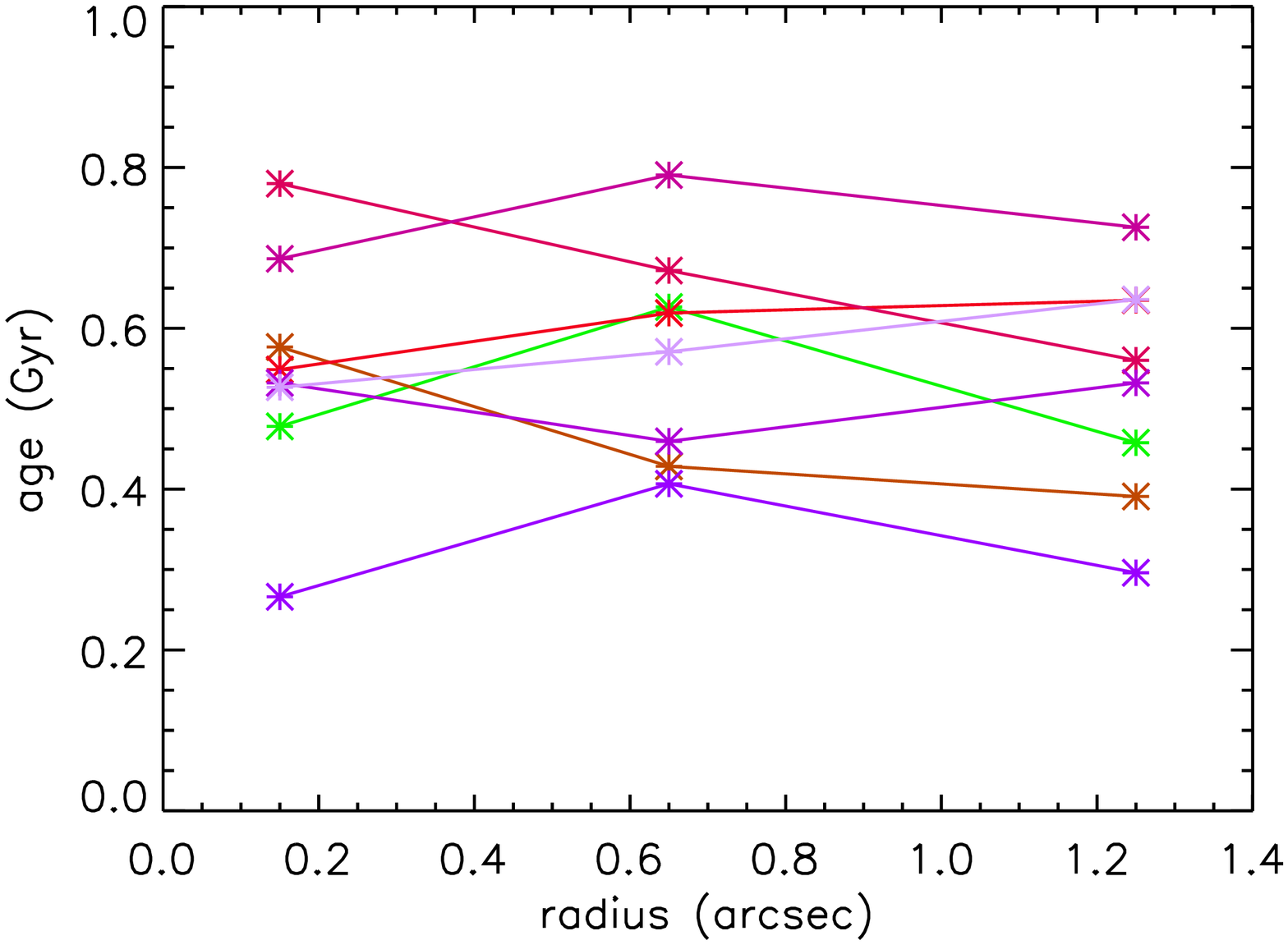}
\caption{\label{fig:agerad} Age estimates versus radius measured from annular binning of the IFU data.}
\end{figure}

We constructed a pseudo `age map' of the young population by taking the weighted age of the best fitting SSP templates from 
each spaxel. These maps are shown in Figure \ref{fig:agemap}. The age maps are somewhat noisy, although there are a couple of 
examples of possible age gradients across the field-of-view (e.g. E+A\_2, E+A\_7). The sample generally have 'weighted ages' of
$\sim 0.5$\,Gyr, perfectly consistent with their E+A status and consistent with the age--metallicity diagnostics of their global 
spectra in Section 4.4.
Radially averaged 'ages' are plotted in Figure \ref{fig:agerad} and there is no sign of significant radial
trends.

\subsection{Internal kinematics}
The stellar kinematics of an E+A galaxy may offer the best clues to what 
represent plausible formation mechanisms. For an equal mass merger 
the remnant should usually be dynamically pressure supported \citep[NGZZ;][]{bekki05}. However, for unequal mass mergers
and tidal interactions, rotation of the young stellar population is expected in most cases \citep{bekki05}. Here the probability
of a rotating remnant increases with increasing mass ratio of the merger progenitors \citep{bournaud08}. Likewise,
if `normal' star--formation in a spiral disk is suddenly truncated then rotation should be present. 

\begin{figure*}
   \begin{center}
     \begin{minipage}{0.95\textwidth}
        \includegraphics[width=5.6cm, angle=90, trim=0 0 0 0]{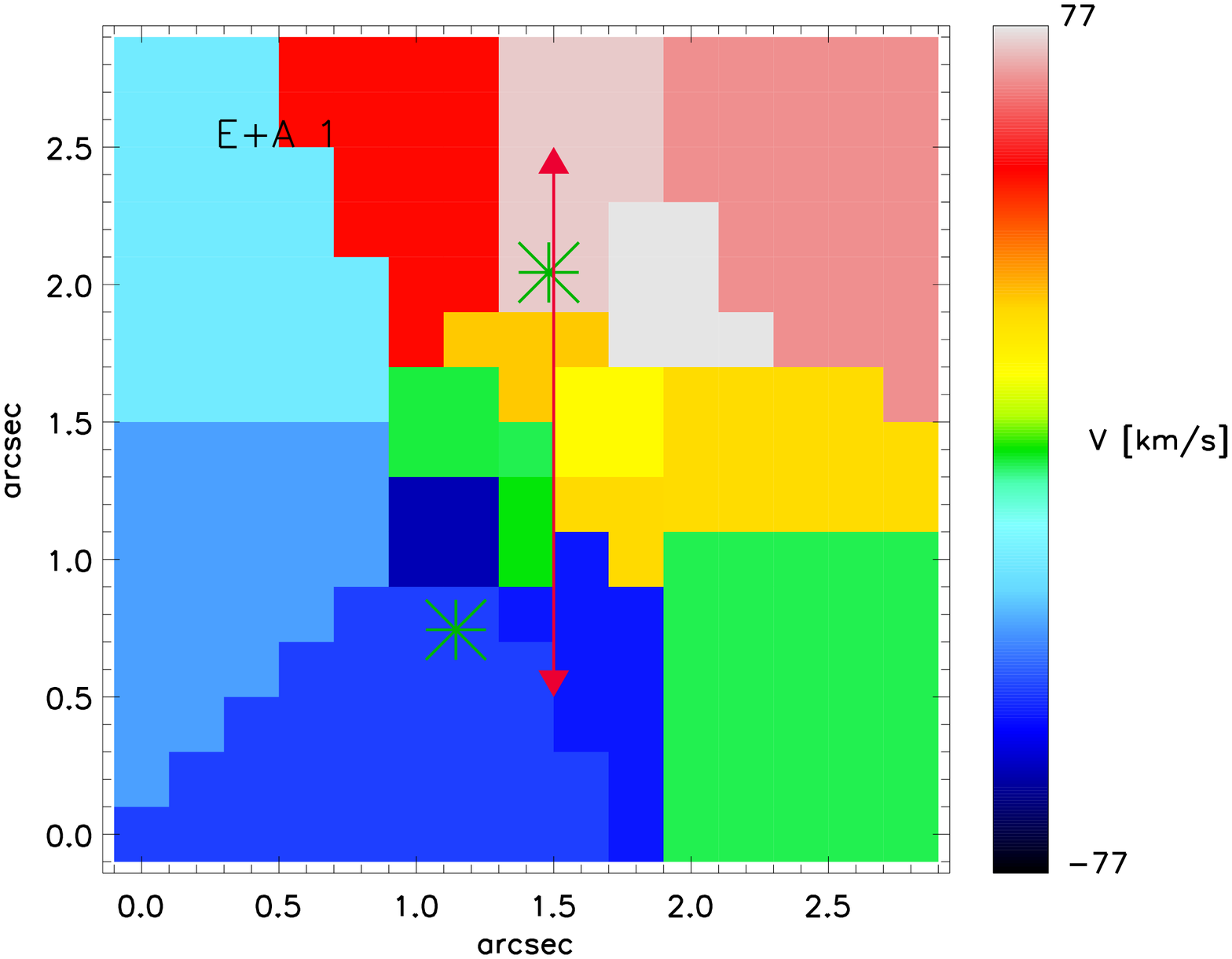}
\hspace{1cm}
         \includegraphics[width=5.6cm, angle=90, trim=0 0 0 0]{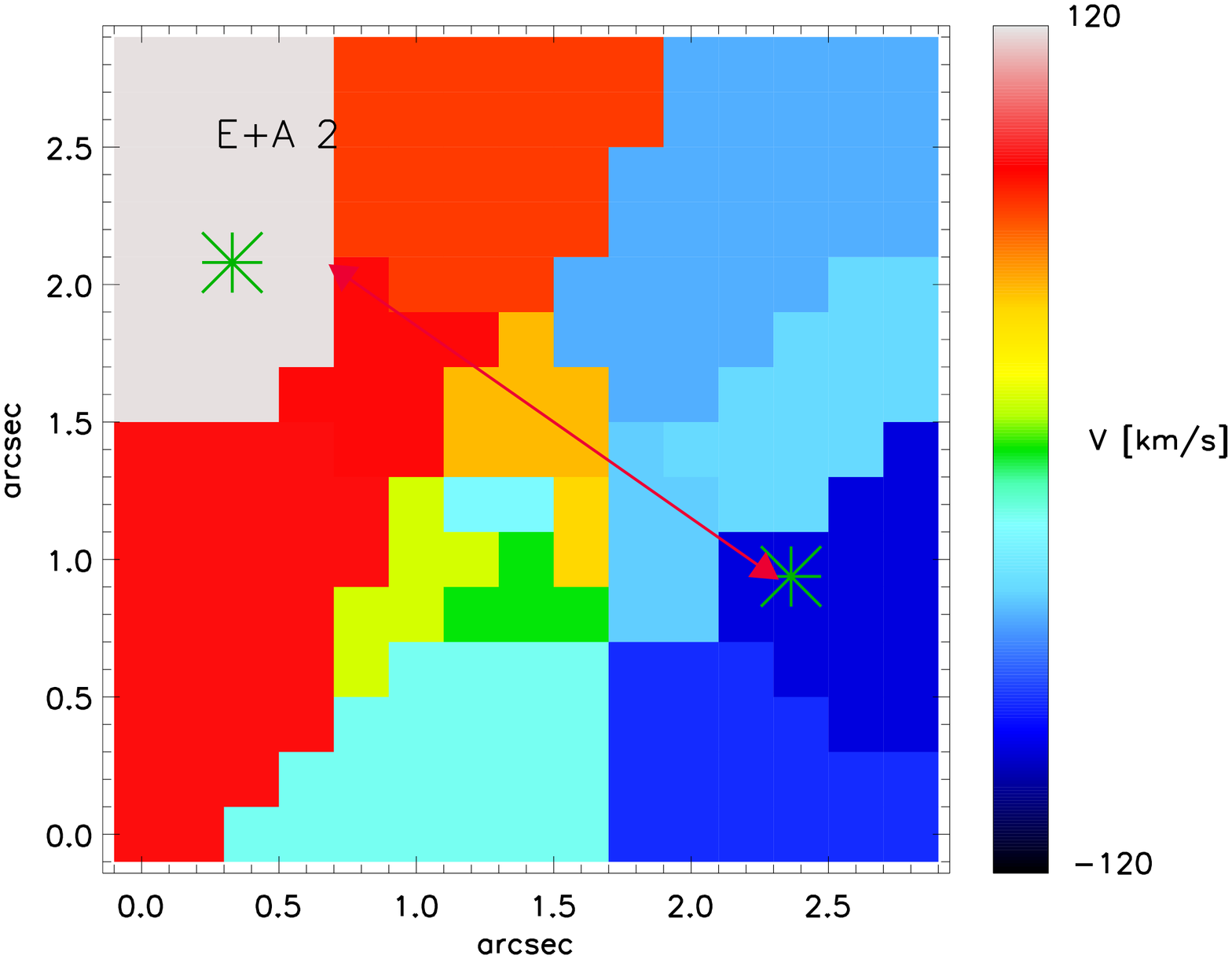}
      \end{minipage}
    \begin{minipage}{0.95\textwidth}
        \includegraphics[width=5.6cm, angle=90, trim=0 0 0 0]{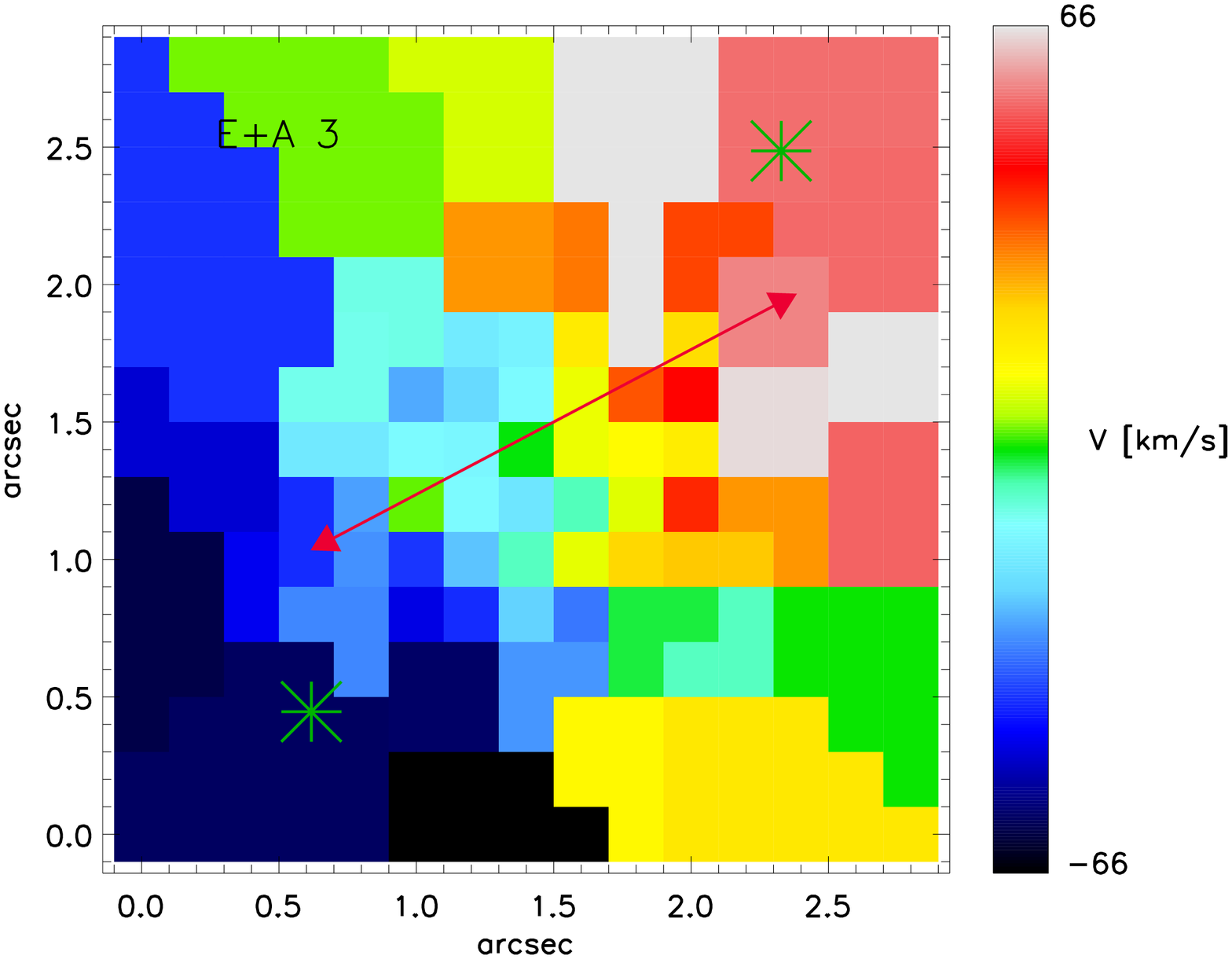}
\hspace{1cm}
        \includegraphics[width=5.6cm, angle=90, trim=0 0 0 0]{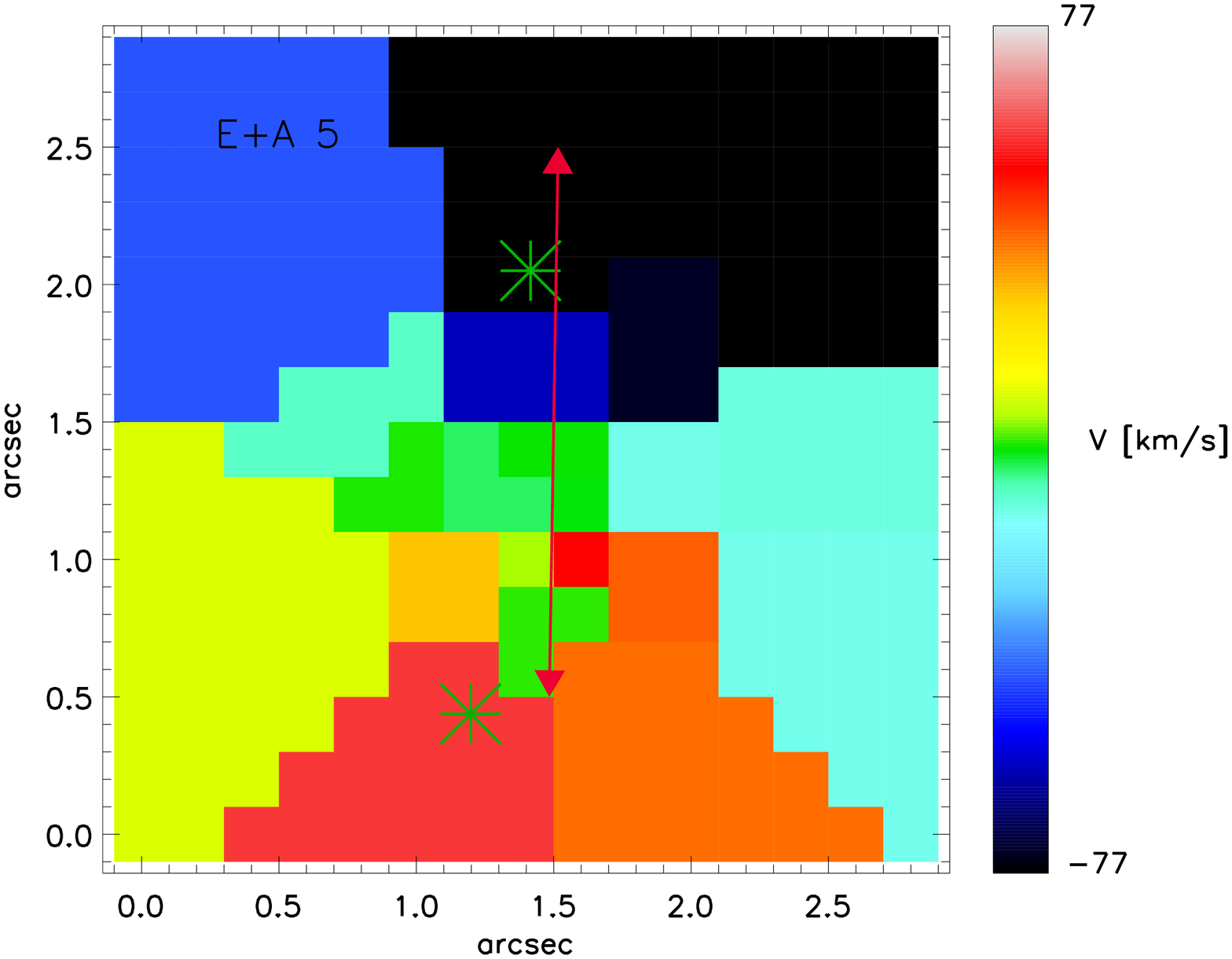}
     \end{minipage}
    \begin{minipage}{0.95\textwidth}
         \includegraphics[width=5.6cm, angle=90, trim=0 0 0 0]{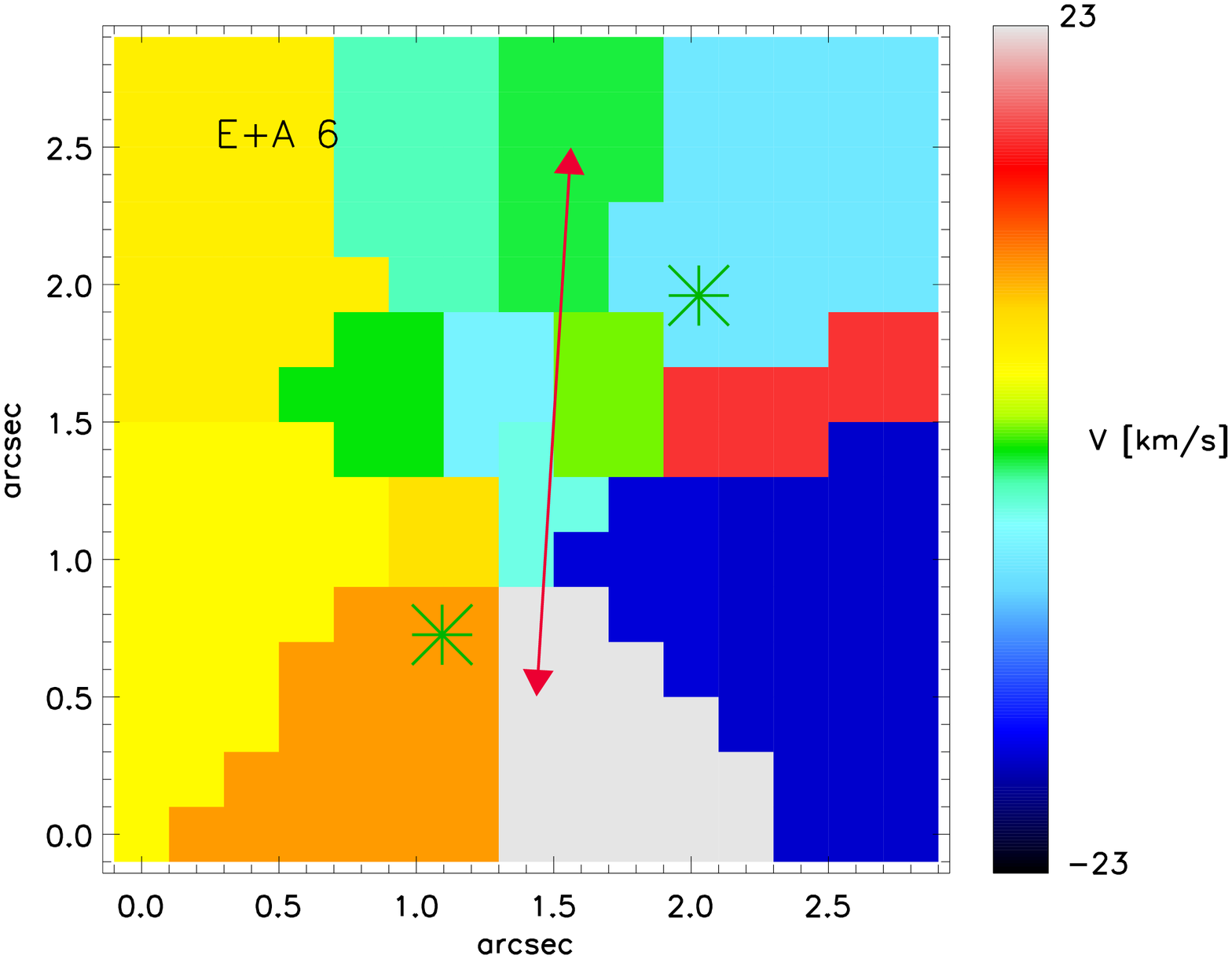}
\hspace{1cm}
         \includegraphics[width=5.6cm, angle=90, trim=0 0 0 0]{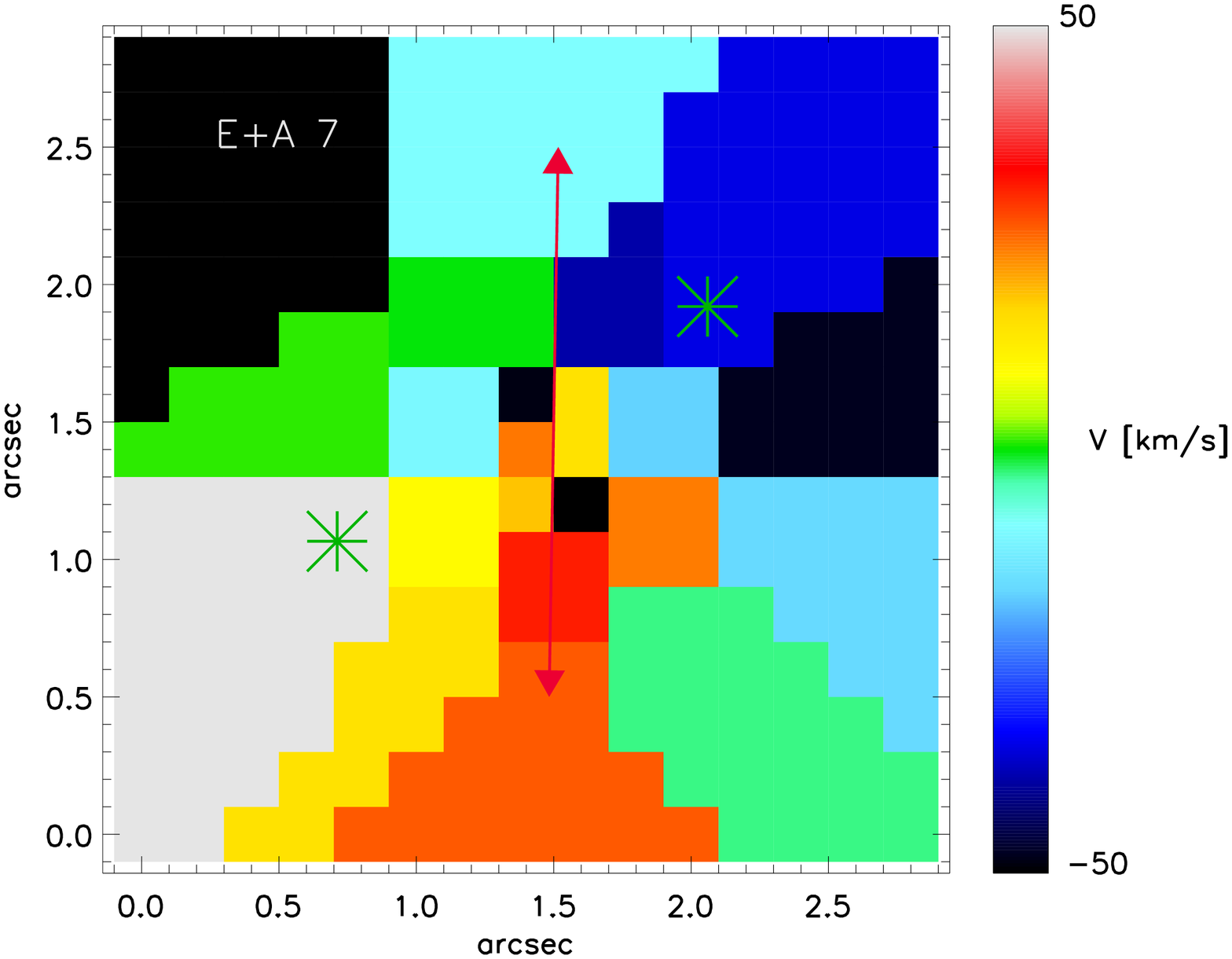}
      \end{minipage}
    \begin{minipage}{0.95\textwidth}
        \includegraphics[width=5.6cm, angle=90, trim=0 0 0 0]{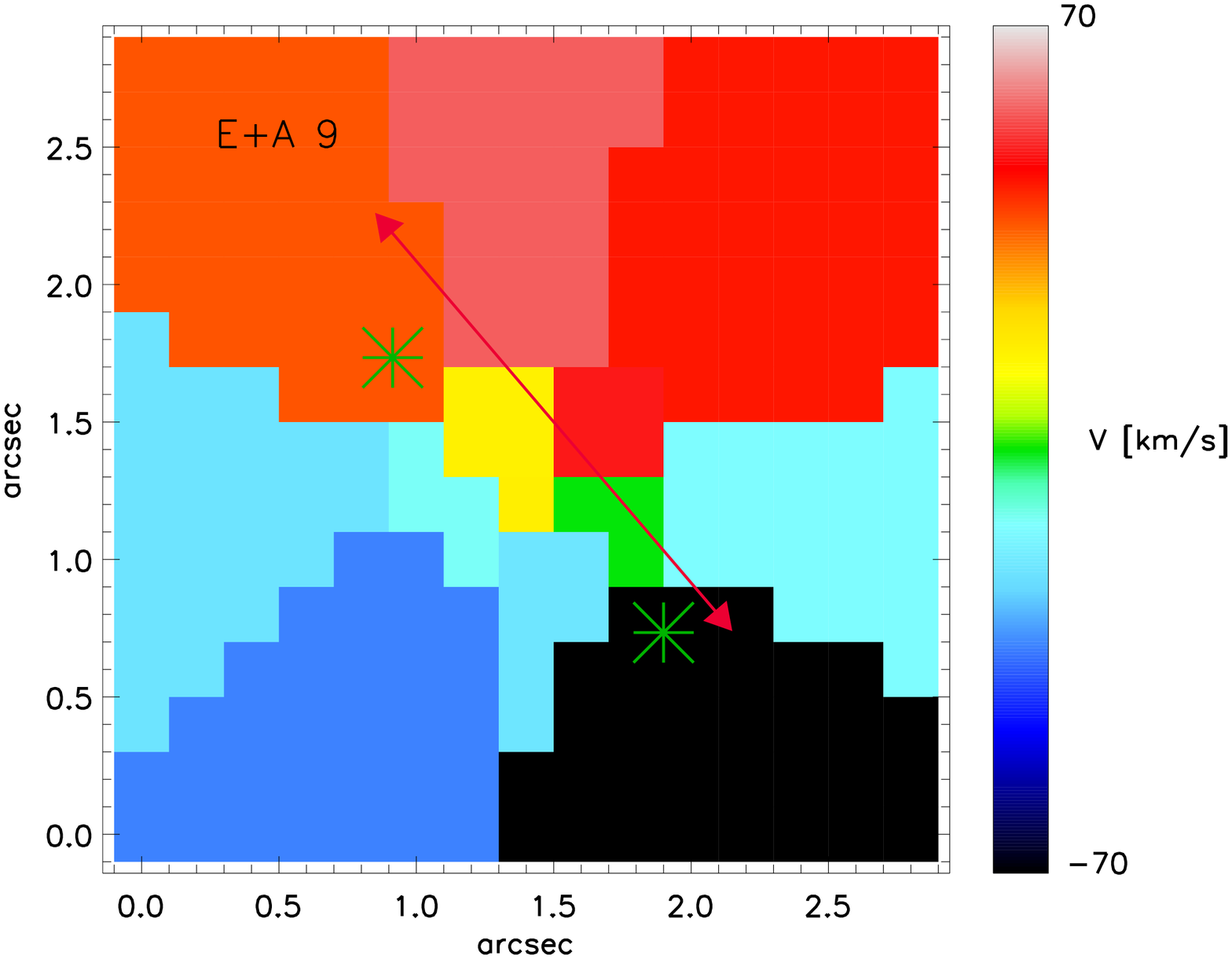}
\hspace{1cm}
        \includegraphics[width=5.6cm, angle=90, trim=0 0 0 0]{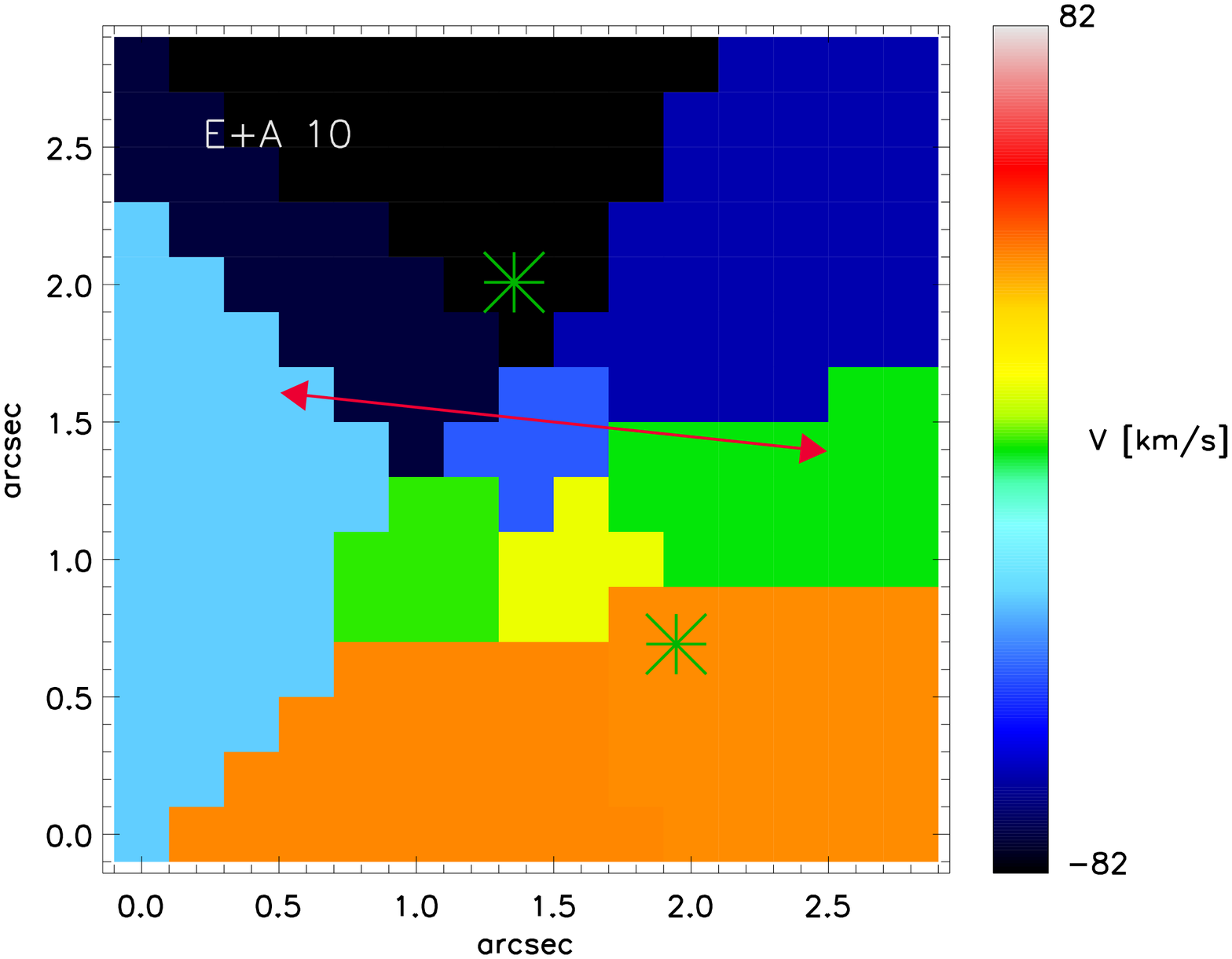}
     \end{minipage}
\end{center}
\caption{\label{fig:velmap} Two dimensional streaming velocity maps for our E+A sample. Rotation is clearly present in all cases. The 
{\it green points} mark the spaxels used to determine the maximum rotation velocity. The {\it red} arrows show the 
angle of the semi--major axes derived from the imaging.}
\end{figure*}

\begin{figure*}
   \begin{center}
     \begin{minipage}{0.95\textwidth}
        \includegraphics[width=5.6cm, angle=90, trim=0 0 0 0]{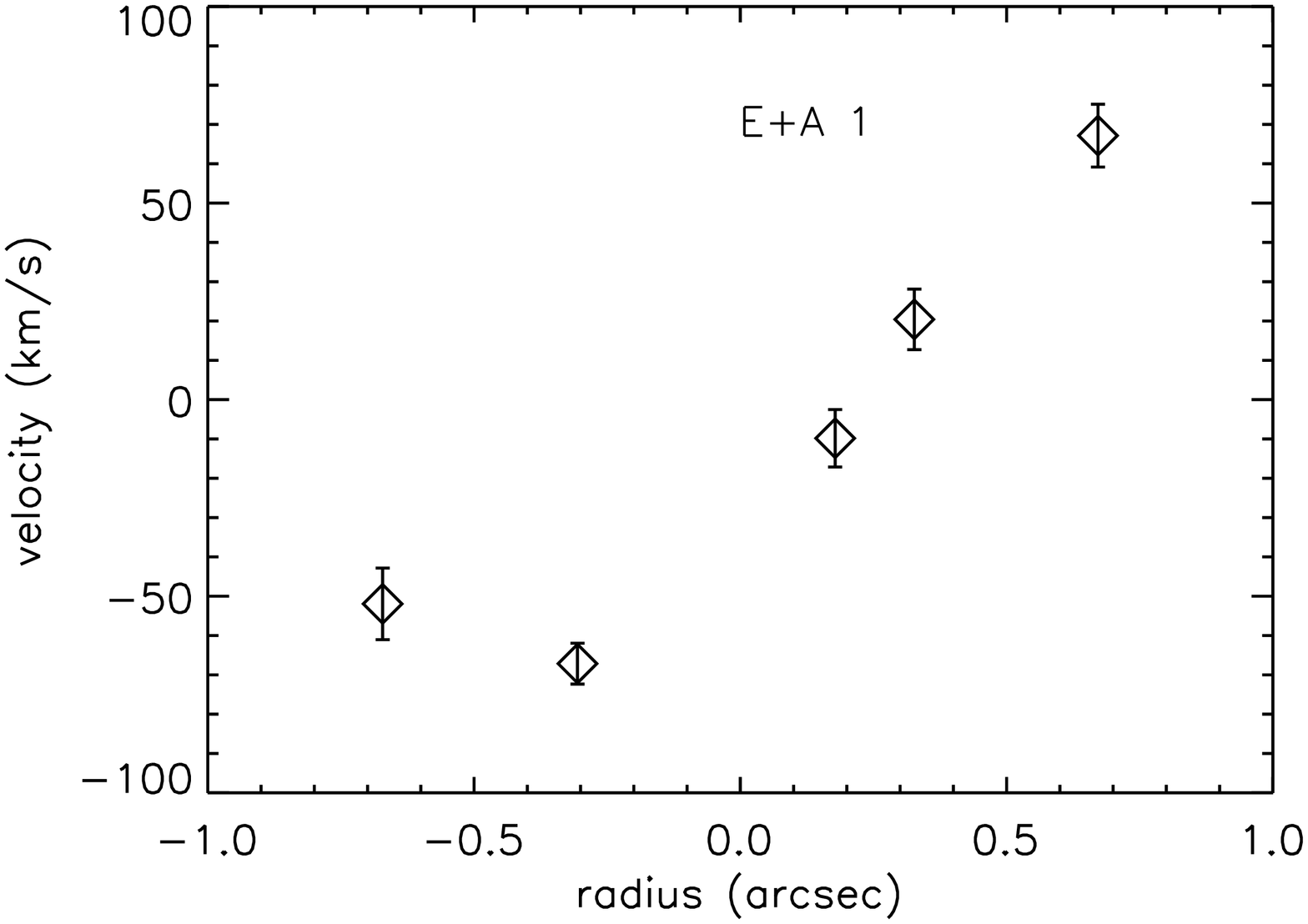}
\hspace{1cm}
         \includegraphics[width=5.6cm, angle=90, trim=0 0 0 0]{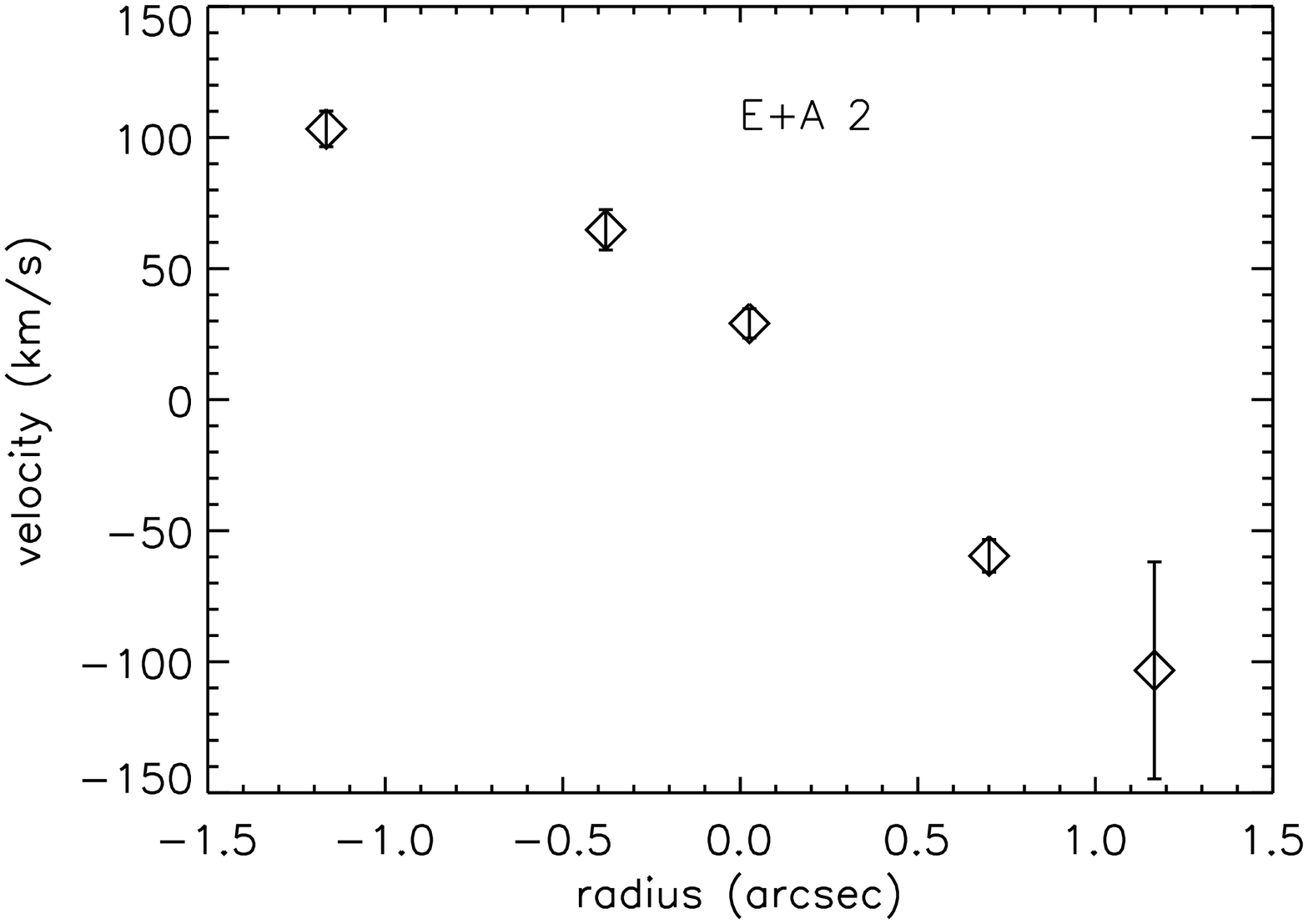}
      \end{minipage}
    \begin{minipage}{0.95\textwidth}
        \includegraphics[width=5.6cm, angle=90, trim=0 0 0 0]{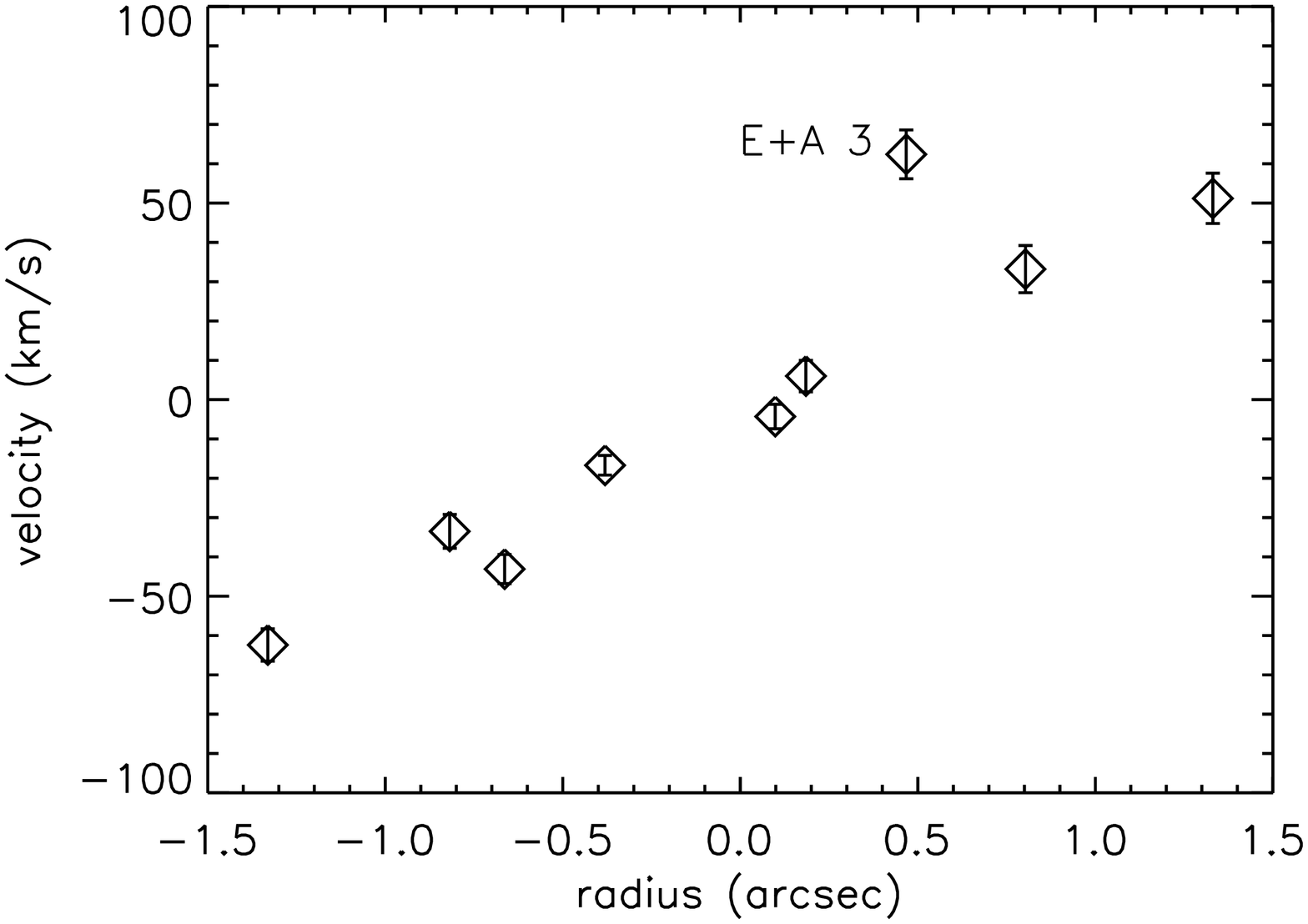}
\hspace{1cm}
        \includegraphics[width=5.6cm, angle=90, trim=0 0 0 0]{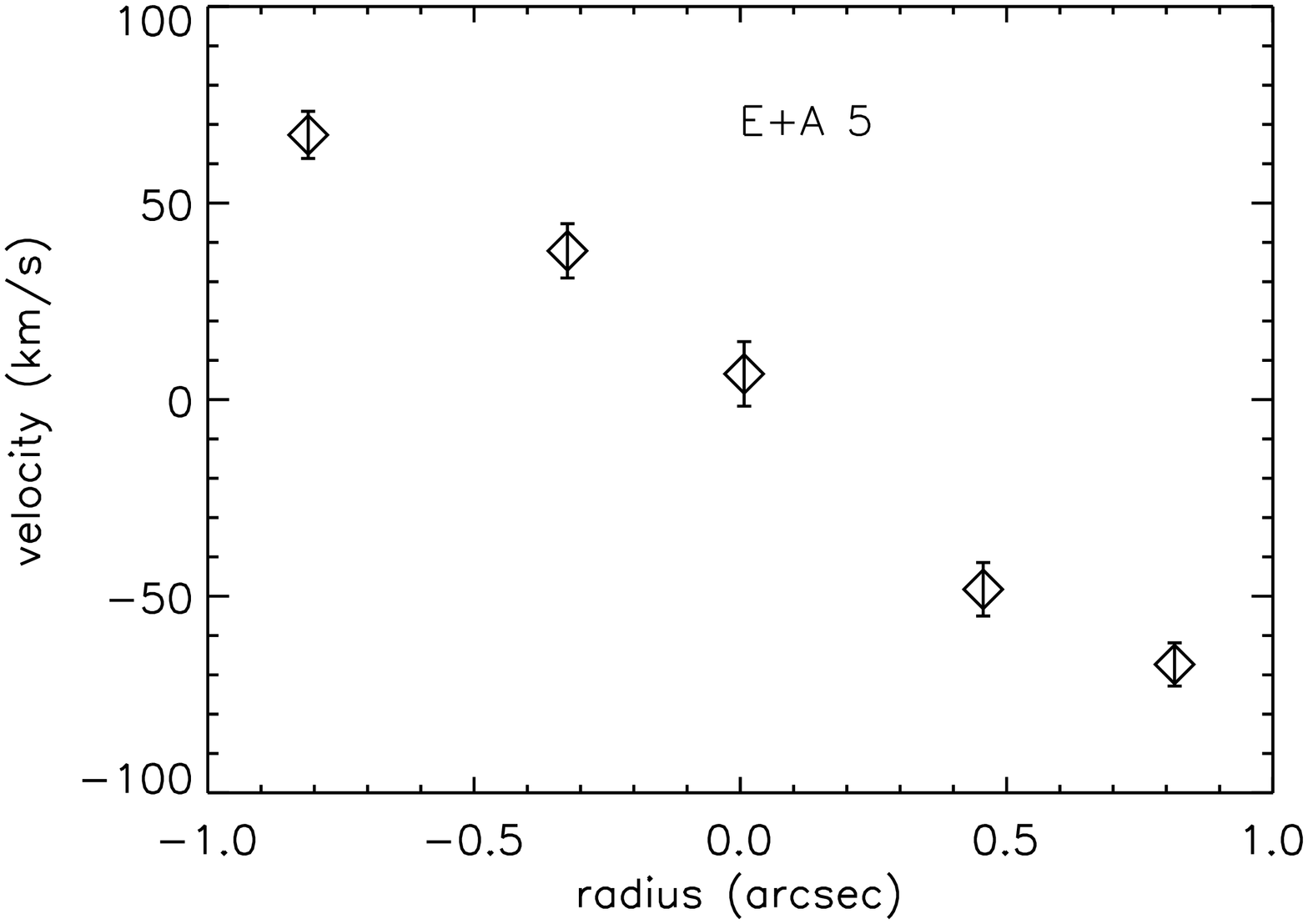}
     \end{minipage}
    \begin{minipage}{0.95\textwidth}
         \includegraphics[width=5.6cm, angle=90, trim=0 0 0 0]{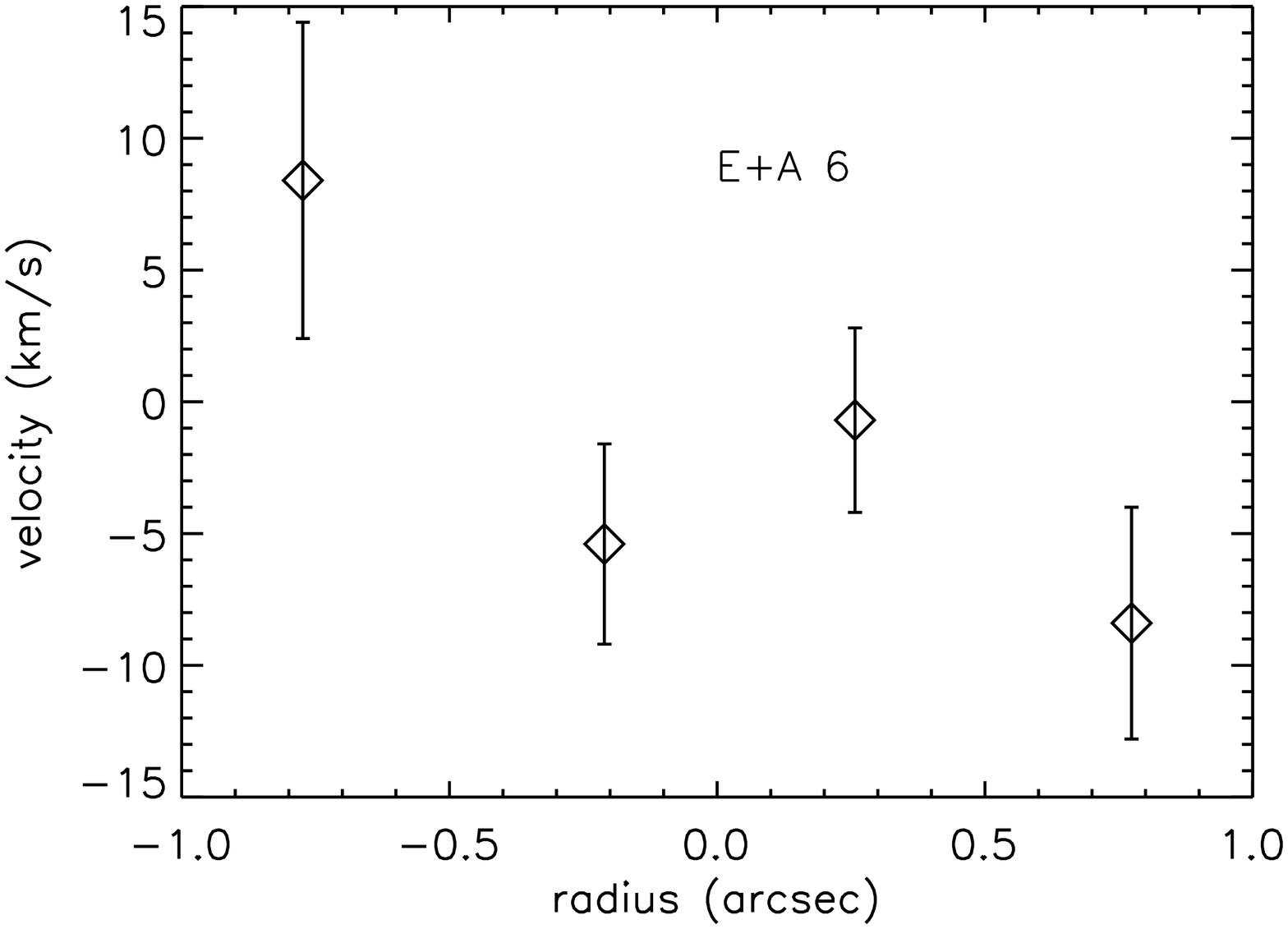}
\hspace{1cm}
         \includegraphics[width=5.6cm, angle=90, trim=0 0 0 0]{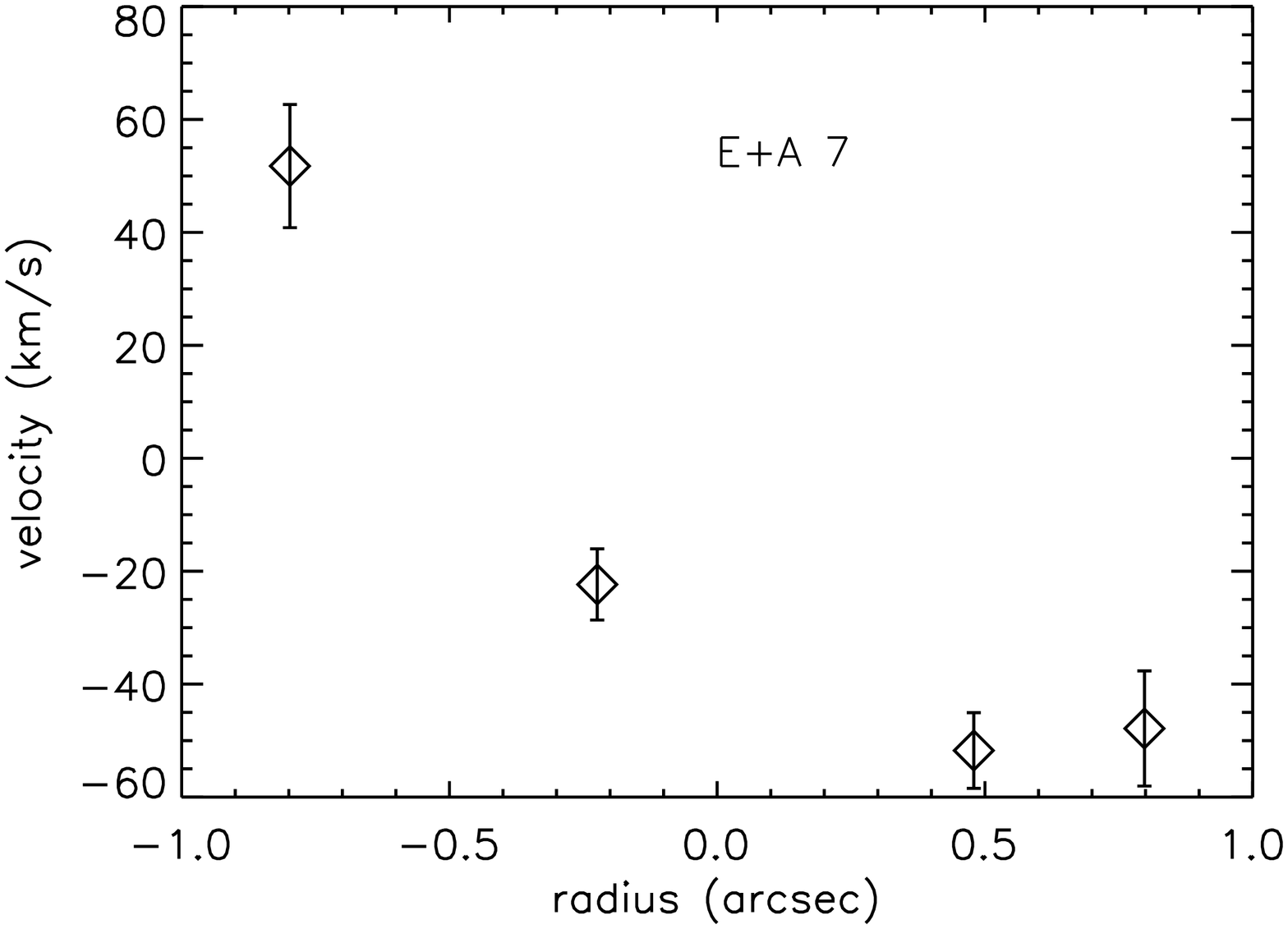}
      \end{minipage}
    \begin{minipage}{0.95\textwidth}
        \includegraphics[width=5.6cm, angle=90, trim=0 0 0 0]{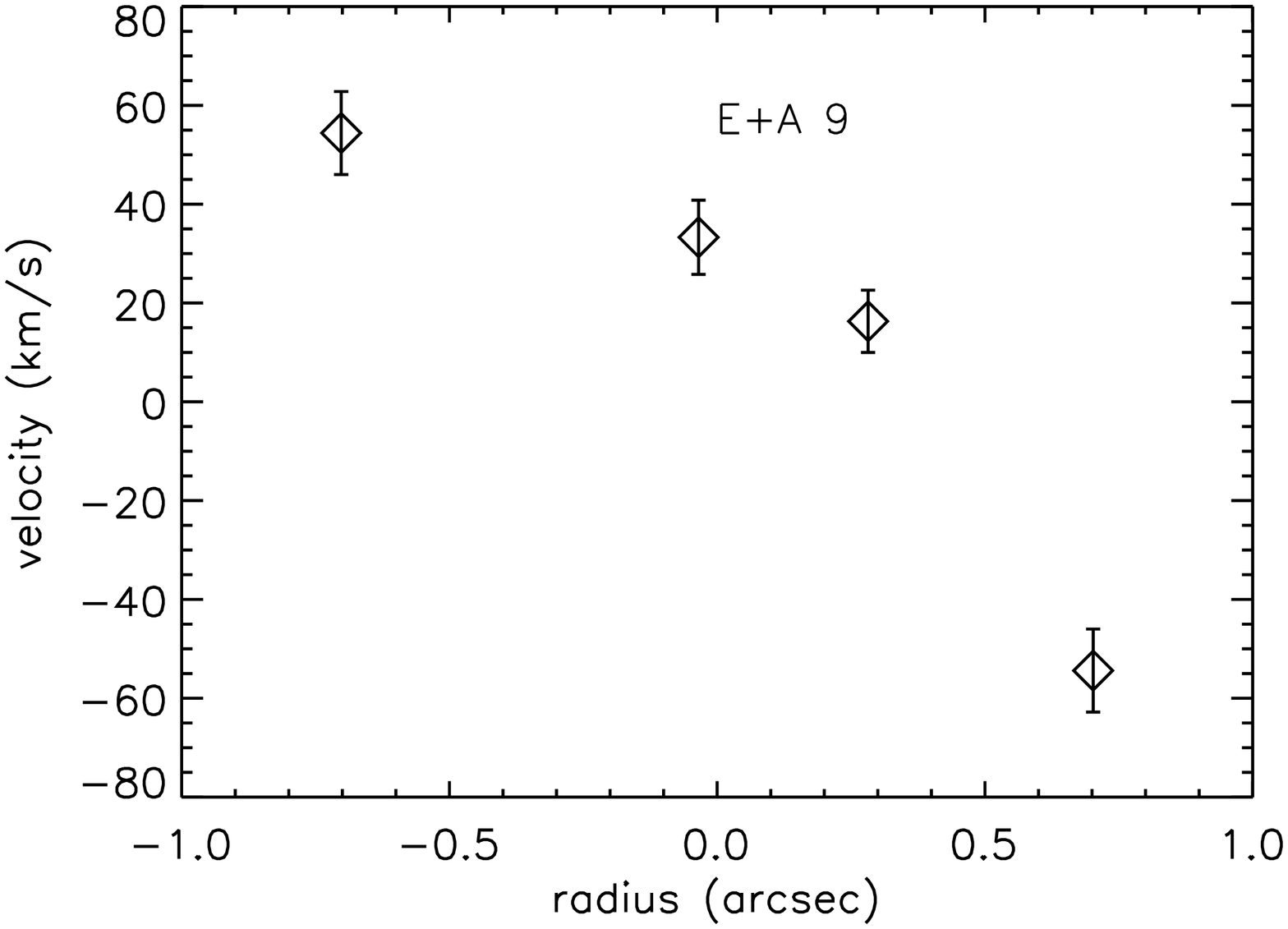}
\hspace{1cm}
        \includegraphics[width=5.6cm, angle=90, trim=0 0 0 0]{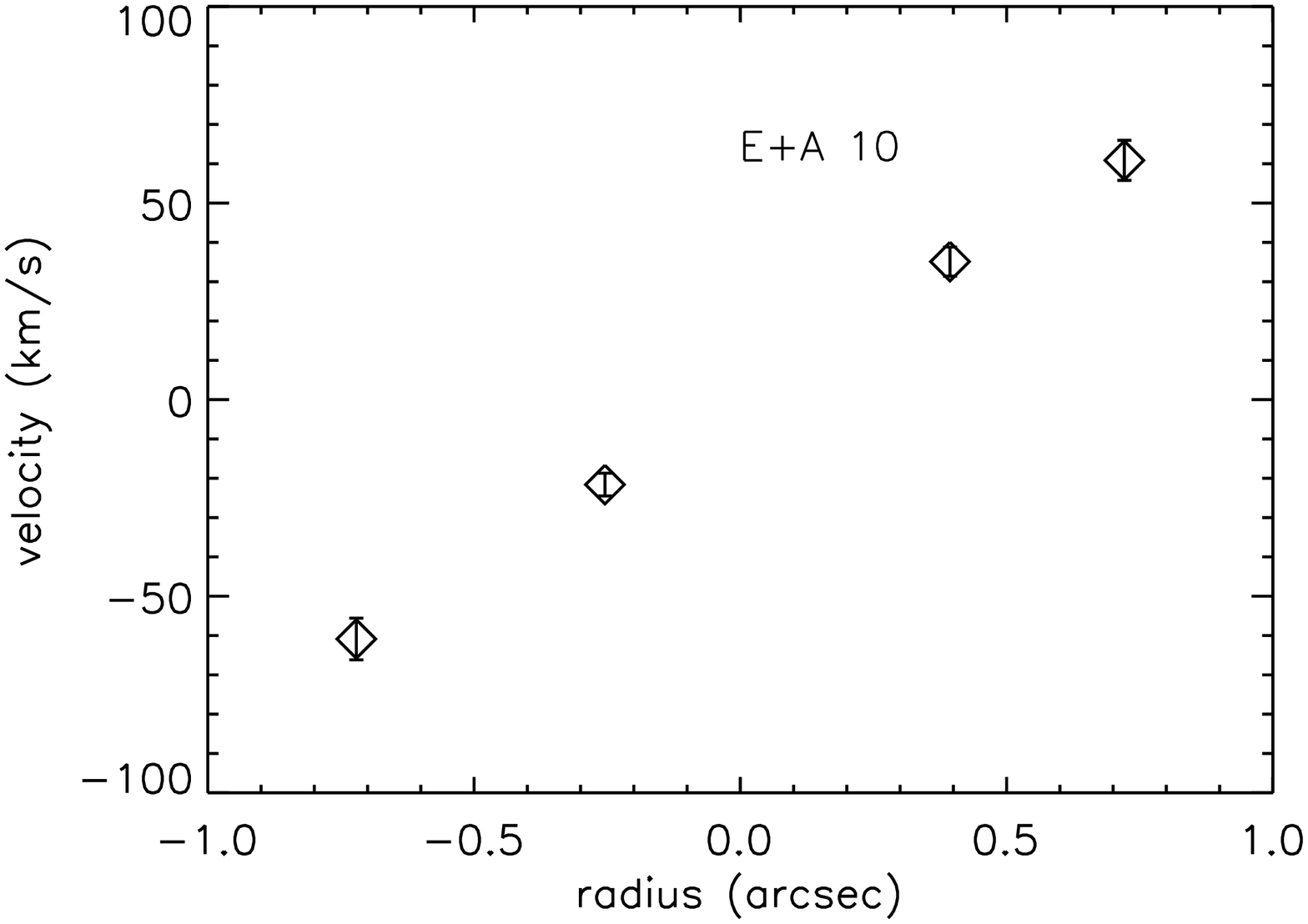}
     \end{minipage}
\end{center}
\caption{\label{fig:velgrad} One dimensional rotation curve for our E+A sample. Rotation is clearly present in all cases.}
\end{figure*}

We have constructed two--dimensional velocity fields for each of the E+As in our
sample that have IFU observations. The recession velocity for each spaxel is a free parameter
in fitting the model templates, described in Section 4.8. The rotation fields are shown in Figure \ref{fig:velmap}. In every
case a clear rotation around the galaxy's center is evident. The observed rotation velocities are generally of
order $V\sin i\sim 60$\,km\,s$^{-1}$, although these are very much lower limits since the data only extend to radii of $\sim 1.5$\arcsec\,
($\sim 2$ to 5\,kpc) and the rotation speed may continue to rise at larger radii. Also the data are smoothed by convolution with the seeing
disk (and also binned spatially) which dampens the amplitude of the rotation (note: the statistical error on 
the velocity determination of each spaxel is generally $\sim 10$\,km\,s$^{-1}$). To further illustrate the `rotation
curve' behaviour seen across the face of the galaxies, the streaming velocities measured for each spaxel
along the line that connects the {\it green stars} in Figure \ref{fig:velmap}, are plotted in 
Figure \ref{fig:velgrad}. Here the {\it green stars} indicate the spaxels identified as having the maximum and
minimum rotation velocities. 

Although several spatially--resolved spectroscopic studies
of E+A galaxies residing in clusters have reported strong rotation \citep{franx93,caldwell96}, the only comprehensive study to date of the internal 
kinematics of nearby field E+A galaxies was performed by \citetalias{norton01}. This study used long--slit spectroscopy 
to obtain spatially resolved spectroscopy of E+A galaxies from the Las Campanas Redshift
Survey \citep{zabludoff96}.
Of the \citetalias{norton01} sample, 14 out of 20 show no evidence for rotation and
two of the remaining 6 show only marginal evidence of rotation. Only 6 of 20 galaxies
in the \citetalias{norton01} sample have a rotational velocity of greater than $v_{rot}\sin i > 40$\,km\,s$^{-1}$
where the average galacto--centric radius of the measured velocity is $\sim 2.8$\,kpc (converted to our assumed cosmology). 
However, \citetalias{norton01} point out that the two galaxies with the most significant rotation are those that 
have the greatest radial coverage. In contrast, 6 out of 8 of the E+A galaxies studied here have 
$v_{rot}\sin i > 40$\,km\,s$^{-1}$, despite being measured over a similar physical radius. Moreover, 
the galaxies in our sample have a similar distribution in absolute magnitude and redshift to those
in NGZZ's sample. Typical large spiral galaxies measured at similar radii have 
$V_{\rm rot}\sin i \sim 70$--140\,km\,s$^{-1}$ \citep{heraudeau99}; our
sample is close to the lower end of this range with most of our E+A galaxies having rotation 
velocities of $\sim 60$\,km\,s$^{-1}$. 

\begin{figure}
         \includegraphics[width=5.8cm, angle=90, trim=0 0 0 0]{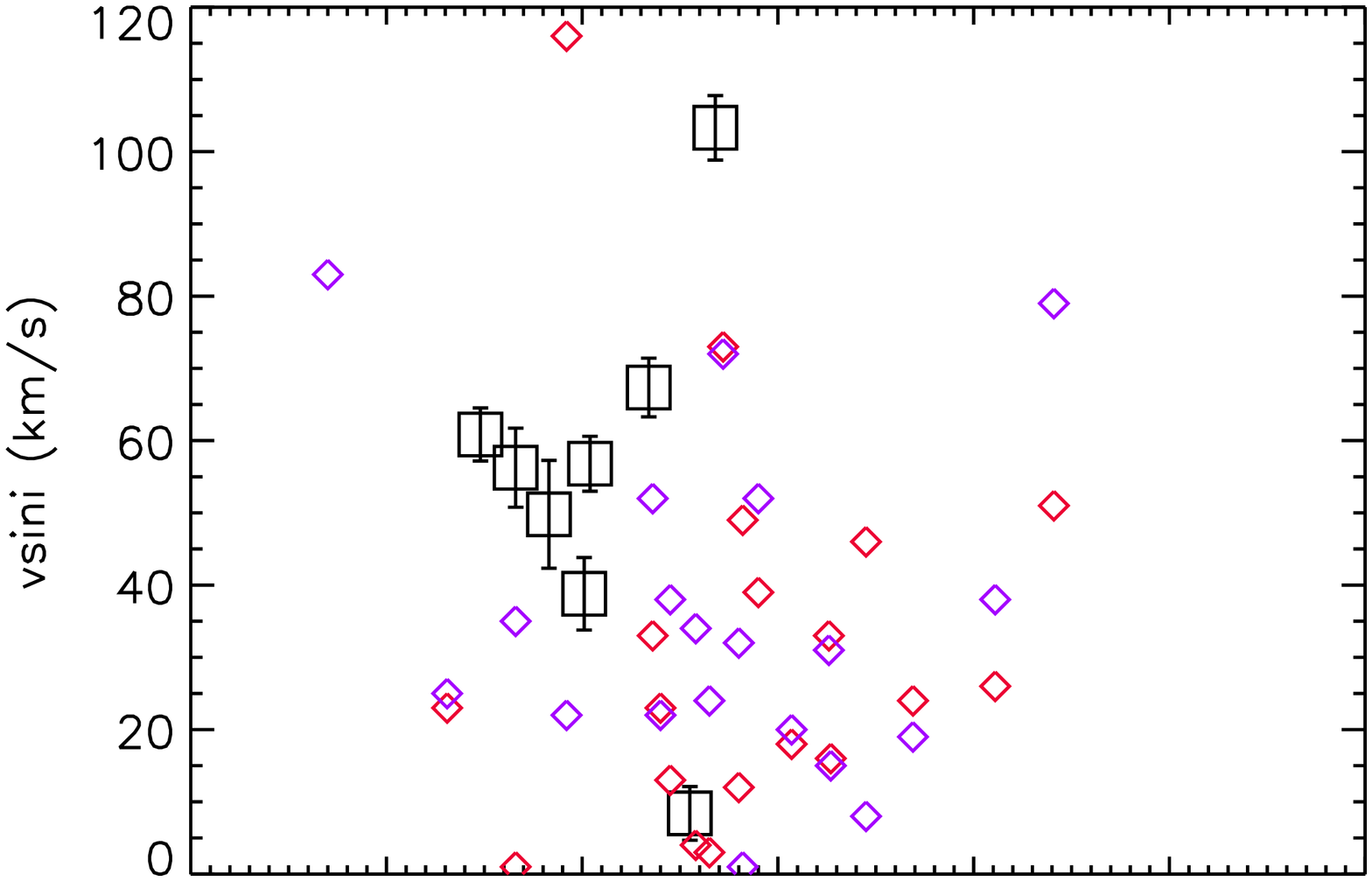}
         \vspace{-1.2cm}\\
        \includegraphics[width=5.8cm, angle=90, trim=0 0 0 0]{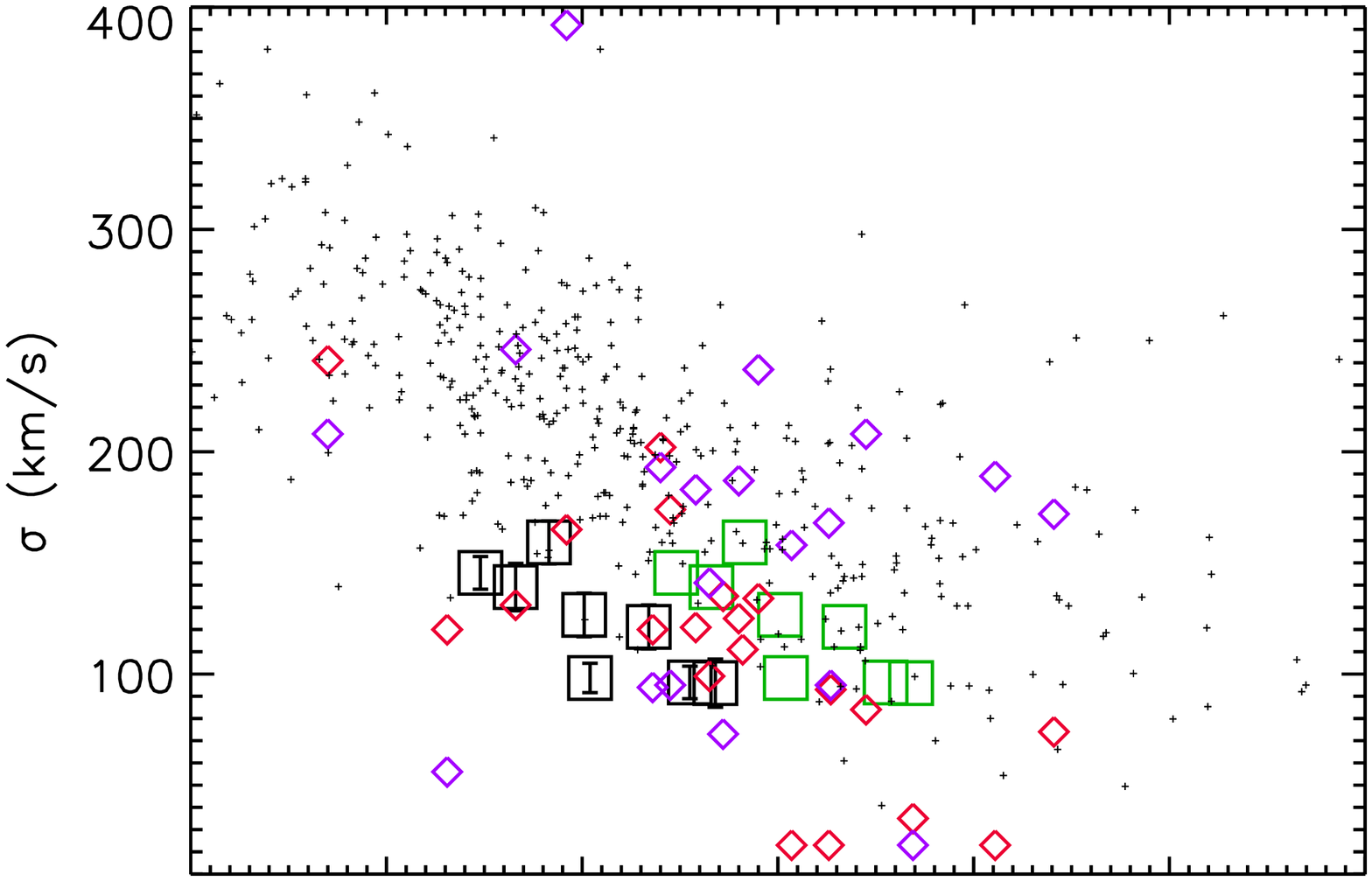}
         \vspace{-1.2cm}\\
      \includegraphics[width=5.8cm, angle=90, trim=0 0 0 0]{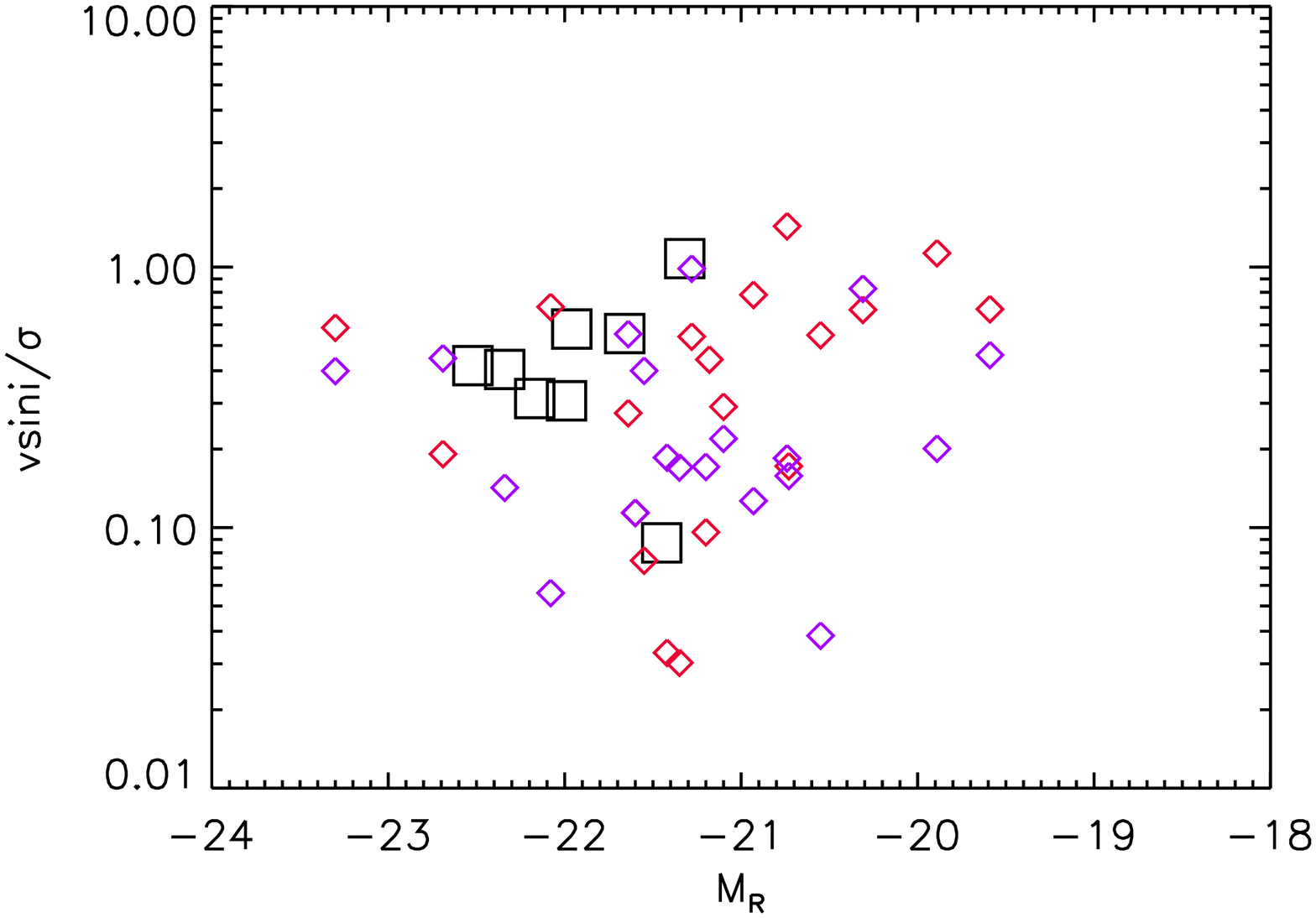}
\caption{\label{fig:nortcomp} {\it Top panel:} The projected rotation velocities of our sample (black squares) compared to
the sample of \citetalias{norton01} ({\it blue and red diamonds}). Note the \citetalias{norton01} fitting technique  simultaneously fit for the kinematics of 
the young and old stellar populations and their results are plotted as {\it blue diamonds} (young component) and {\it red diamonds}
(old component). {\it Middle panel:} Comparison of the central velocity dispersion
of the NGZZ sample ({\it blue and red diamonds}) and our sample {\it black squares}. The {\it small black points} represent the 
Faber-Jackson of normal ellipticals and the {\it green squares} are our data dimmed by 1\,mags which
is the expected brightening due to the A-star light.
{\it bottom panel:} comparison of $v\sin i/\sigma$ of the samples. } 
\end{figure}

In Figure \ref{fig:nortcomp}, we compare the kinematic properties of our sample to the \citetalias{norton01} LCRS sample.
In the top panel we show the measured $v_{\rm rot}\sin i$ values for our sample ({\it black open squares}) along with the $v_{\rm rot}\sin i$
values derived by \citetalias{norton01}. 
We estimate $v_{\rm rot}\sin i$ from the velocity maps in Figure \ref{fig:velmap} by selecting the spaxels which seem to correspond
to the outermost part of the measured rotation curve. 
The \citetalias{norton01} fitting technique simultaneously fitted for the kinematics of 
the young and old stellar populations and their results are plotted as the {\it blue diamonds} (young component) and the {\it red diamonds}
(old component). In principle we should compare to the young component since our spectra are consistent with being `completely'
dominated by a young stellar population, although the two generally span the same parameter space. The rotation velocities measured 
from our sample overlap the high velocity end of NGZZ's distribution, with most of the galaxies in their sample
having significantly lower velocities. In the middle panel of Figure \ref{fig:nortcomp}, we compare the central velocity dispersions,
which were measured over the central 1\arcsec\, ({\it black squares}), 
to those measured by \citetalias{norton01} ({\it red and blue diamonds}). The distribution of central velocity dispersions
are similar in the two samples. Also shown are the velocity dispersions of normal early--type systems ({\it small black points})
from \citet{faber89} where we have converted their $B$-band magnitudes to $R$-band values assuming $B-R=1.25$ \citepalias{norton01}.
Both our data and that of \citetalias{norton01} are too bright to be on the normal early--type relation. However, model predictions of the
expected dimming of E+A galaxies as they evolve suggest they should fade by 0.5--1\,mag over the subsequent 1--2\,Gyrs  
\citep[NGZZ;][]{poggianti99,kelson00}. The {\it green squares} in Figure \ref{fig:nortcomp} show where our E+A galaxies would 
lie once they had faded by this (1.0\,mag) amount, and we see they now overlap the early--type relation.

In the bottom panel of Figure \ref{fig:nortcomp} we compare the $v_{rot} \sin i/\sigma$ values for the two samples. Naturally since our sample has
larger rotation velocities but similar velocity dispersions the $v_{rot} \sin i/\sigma$ distribution of  our sample corresponds
to the high end of the $v \sin i/\sigma$ distribution from \citetalias{norton01}.

\begin{figure}
      \includegraphics[width=5.8cm, angle=90, trim=0 0 0 0]{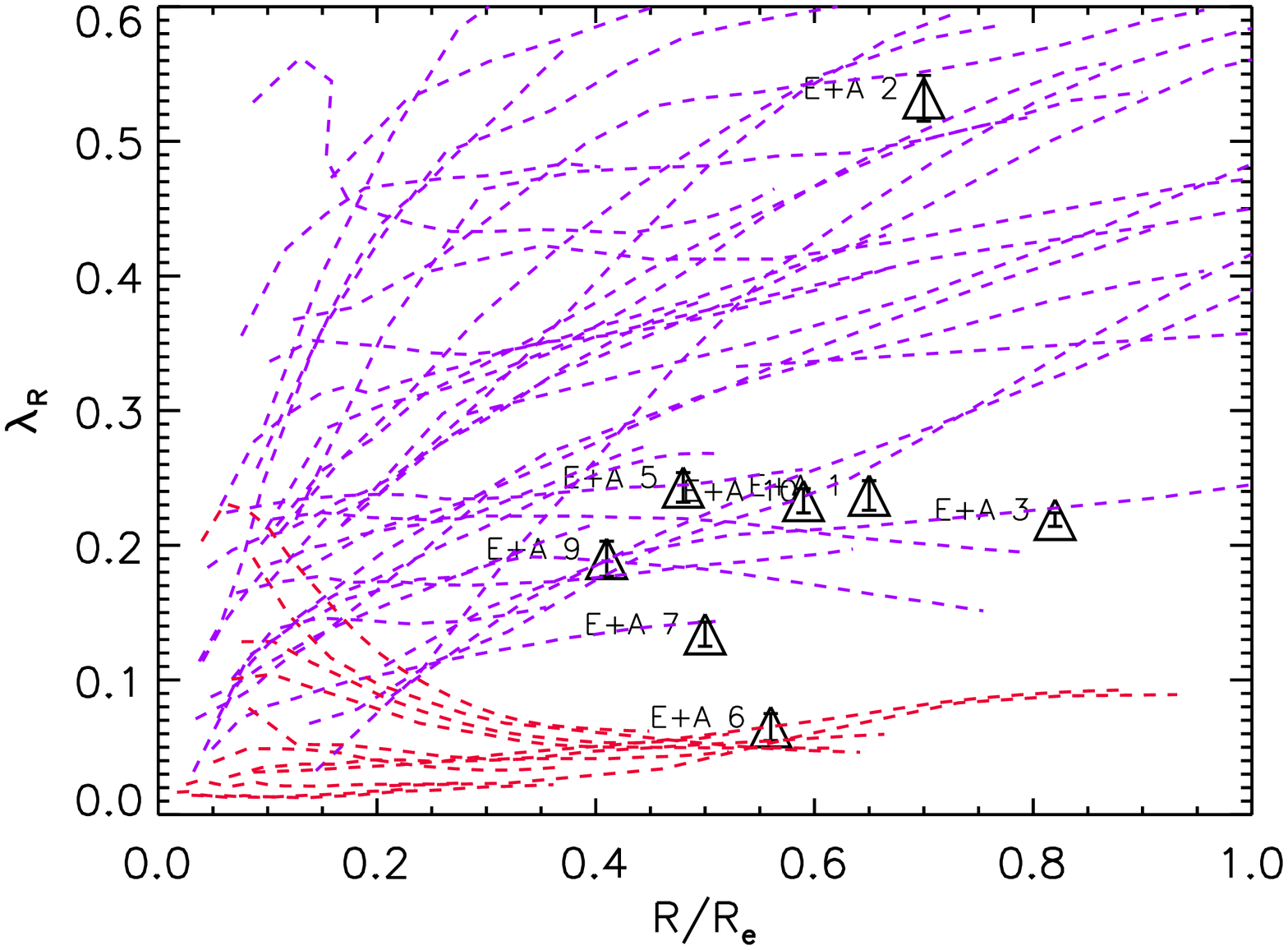}
     \includegraphics[width=5.8cm, angle=90, trim=0 0 0 0]{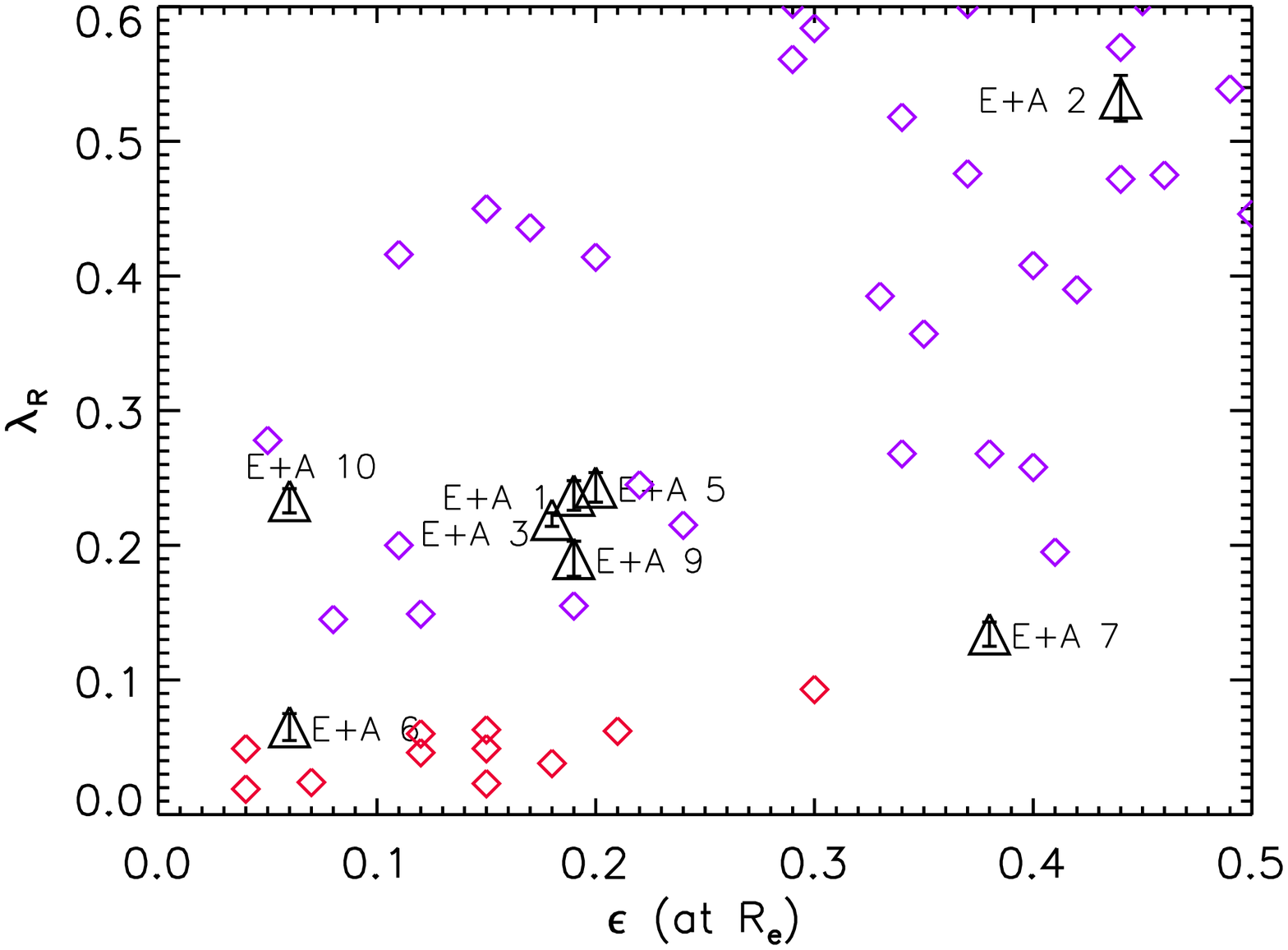}
\caption{\label{fig:lambdar} {\it Top panel:} The $\lambda_{R}$ parameter from \citet{emsellem07} plotted
against the radius at which it is measured (in units of the effective radius) for our E+A sample ({\it black triangles}). The {\it blue dashed lines} are
the tracks for the fast rotators in the SAURON sample \citep{emsellem07} and the {\it red dashed lines} are the tracks for 
the slow rotators. Seven out of eight of our E+As occupy positions in this plane corresponding to fast rotators. E+A\_6 is the
exception lying on the slow rotator tracks. {\it Bottom panel:} The $\lambda_{R}$ parameter plotted against ellipticity for our E+A sample 
({\it black triangles}) where the
$\lambda_{R}$ has been computed over the entire IFU area (i.e. out to the radii shown in the {\it top panel}) and the
ellipticity is calculated at $2R_{e}$. The diamonds show the SAURON sample measured at $1R_{e}$. The blue
symbols are the fast rotators and the red symbols the slow rotators. Again, with the exception of E+A\_6 our sample overlap the fast rotators.
Note: the reason we compute the ellipticity at $2R_{e}$ rather than  $1R_{e}$ is to negate the effect of seeing on the ellipticity.}
\end{figure}

A method for quantifying the rotation of galaxies specifically designed for use with 
two dimentional IFU data was developed by \citet{emsellem07}. Their parameter,
$\lambda_{R}$,  involves luminosity--weighted averages over the two dimensional kinematic field provided
by IFU data and acts as a proxy to the observed projected stellar angular
momentum per unit mass. Using this parameter, \citet{emsellem07} are able to clearly separate the elliptical
galaxy population into two distinct subsets, the `fast rotators' which tend to be relatively low luminosity 
galaxies ($M_{B}>-20.5$), and `slow rotators' which span the entire range in luminosity. 
\citet{emsellem07} found three quarters of the SAURON sample to be fast rotators. In Figure \ref{fig:lambdar} we show the
values of $\lambda_{R}$ measured for the 8 E+A galaxies in our sample for which we have IFU observations.
In the top panel we show the $\lambda_{R}$ parameter measured for each E+A galaxy versus the radius over
which it was measured, and in the bottom panel we plot $\lambda_{R}$ against galaxy ellipticity measured at twice the effective radius. In both plots
we overlay the fast rotators ({\it blue symbols and tracks}) and the slow rotators ({\it red symbols and tracks}) from 
the SAURON sample. Comparison of the $\lambda_{R}$ values found for our E+A galaxies with the tracks for the SAURON sample shows
that 7 out of 8 of the E+As are consistent with being fast rotators and 1 out of 8, E+A\_6, sits on the slow rotator 
tracks. We note, without implication, that E+A\_6 is also the sole blue core object in our sample. 
E+A\_2 has a large $\lambda_{R}$ value in comparision to the other galaxies in the sample and is also the galaxy
which showed the strongest evidence of a substantial disk (see Figures \ref{fig:images} and \ref{fig:iso}). The 
other galaxies classified as fast rotators have $\lambda_{R}$ values at the lower end of the fast rotator range.

There is some minimum value below which $\lambda_{R}$ cannot be reliably measured \citep{emsellem07} 
since noise in the data will always lead to some non--zero value of $\lambda_{R}$ even for an object with zero true rotation. We
can estimate whether or not we are detecting reliable values or $\lambda_{R}$ by simulating this effect. We produced 10,000 simulated velocity fields
for each of our objects by setting the velocity field in the IFU data to zero and adding in a random velocity to each spaxel based on the noise estimates.  
The values of $\lambda_{R}$ produced in this manner were, on average, between 10 and 20\,per cent of the values 
measured from the real data, being as small as 2\,per cent for the high signal-to-noise data obtained for 
E+A\_3 and as high as 53\,per cent for the slow rotator E+A\_6.
This clearly demonstrates that we can reliably measure non--zero  $\lambda_{R}$ values.
\citet{emsellem07} also found that the fast rotator population have well aligned photometric and kinematic axes. In Figure \ref{fig:velmap} we overplot as red 
arrows the position angle of the semi--major axes measured from the imaging. In most cases the photometric axis and kinematic axis are
closely aligned although there are several exceptions. The accuracy of such a comparison is somewhat limited given the generally
low ellipticities of our sample galaxies and the large spaxel size in the velocity data.

\section{discussion}

Our combined photometric and spectroscopic study of our E+A sample has allowed us to
compile a wealth of information on their detailed structure and morphology and the internal
colour distribution and kinematics of their recently formed population of young stars, 
which is spatially resolved on scales of 1-2\,kpc. The primary motivation for acquiring
and assembling these data is to significantly progress our understanding of the physical
mechanisms responsible for E+A formation and their subsequent rapid spectral evolution.
As we now discuss, our study has two important contributions to make in this context: 
(i) to much more firmly establish the key structural and kinematical properties of E+A
galaxies, in particular those that provide the strongest clues as to their formation and 
evolution, and (ii) to cast new light on how E+A galaxies fit within the broader picture
of general galaxy formation and evolution.

A key result to emerge from our study is a very clear picture as to the structural
morphology and generic family of objects E+A galaxies generally belong to. 
The overall surface brightness profiles 
in Figure \ref{fig:iso} show that they are consistent with being early--type systems with 
$r^{1 \over 4}$--like profiles, with only one galaxy showing  clear evidence 
for the presence of an exponential disk. This is further underscored by their visually-determined
morphologies
(which all lie in the narrow range S0-Sa; column 2 of Table 3), and their distribution
within the central velocity dispersion versus magnitude plane (Figure \ref{fig:nortcomp}), 
which is seen to be consistent with the Faber--Jackson relation after taking into account their 
higher transient luminosity resulting from the young A-star population. Both of these findings
further corroborate the previous results of \citet{yang04}, \citet{yamauchi05}, NGZZ, and \citet{yang08} 
thereby making them now well established properties of E+A galaxies.

Another important and common morphological attribute is the high incidence of tidal tails/arms/bridges
and/or disturbed appearance, indicative of tidal interactions and merging. Our deep imaging
shows approximately two-thirds of our sample exhibit features of this kind. This adds strong
support for the conclusions drawn in earlier studies \citep{zabludoff96,yang04,blake04,yang08} that
interaction and merger activity is rife amongst the E+A population. 

While E+A galaxies appear to be relatively homogeneous in terms of their morphology -- predominantly
early--type systems, with a large fraction involved in ongoing or recent mergers and interactions -- the
same cannot be said for their internal colour distributions. As highlighted in columns 5 and 6 of
Table 3, they can have either red, blue or irregular coloured cores, and have negative, flat or positive 
colour gradients. Similarly diverse behaviour in the colour properties has already been reported
by \citet{yamauchi05} in a sample of 22 E+A galaxies from the catalogue of \citet{goto05}. 

A further and related curiosity is the large number of galaxies in our sample that have red cores. This
seems counter to the view that their recent starburst activity was clearly widespread throughout their
central regions, based on the spatial distribution of the E+A spectral signature and hence the young stars that were
formed (as shown by the H$\delta_{A}$ maps in Figure 8). One possible explanation is that we are seeing
a centralized starburst that is dust obscured, which would result in strong absorption lines but redder 
colours. Here the model subtracted images of our galaxies (shown in column 2 of Figure 1) are of 
interest, in particular the residual features very close to their centres, perhaps indicating the
presence of dust lanes and dust obscuration. However, the evidence is at best marginal, with no 
indication that such features preferentially occur in the galaxies with red cores.  
Another explanation might be that the strongest young star contribution is located just outside the core, 
resulting in bluer colours peripheral to the core. Unfortunately, the spatial coverage of our spectroscopy 
is insufficient to test this hypothesis. Our IFU spectroscopy reveals that the H$\delta$ absorption is uniformly 
strong over the central few kiloparsecs but is inadequate in spatial coverage to discern any radial gradients. 
There is limited examples of spatially--resolved spectroscopy of E+A galaxies in the literature
but these have indicated a preference for H$\delta$ strength which is centrally concentrated \citep[NGZZ;][]{goto08} in field E+As. 
In the intermediate redshift cluster AC114, \citet{pracy05} found a dichotomy in H$\delta$ distributions with examples of central concentration
and central deficit.

The last but perhaps most important piece of information provided by our study is that, kinematically, 
our E+A galaxies are very homogeneous in that their populations of recently formed stars in all cases
show unambiguous and significant rotational motion. The only other study of the internal kinematics of 
local field E+A galaxies is that of \citetalias{norton01}. These authors reported much lower levels of rotation 
than found in this work for galaxies that have similar absolute magnitude  and velocity dispersions distributions 
(see Figure \ref{fig:nortcomp}). Also the average radius over which the kinematical measurements were made
is essentially the same in both studies ($\sim 2.8$\,kpc). These very different results could possibly 
be due to technical reasons rather than intrinsic differences between the galaxies targeted in the two
studies. The main difference between the data sets is our use of 2-dimensional spectroscopy 
compared with the long--slit spectroscopy used by \citetalias{norton01}. IFU spectroscopy has the advantage of not requiring
prior knowledge of the axis of rotation and does not suffer from problems associated with the misalignment of the 
rotation axis and the slit. We note that in our sample the rotation and semi-major axis are generally well aligned but
in a subset of objects the differences in positional angle between the two can be substantial.
We also note that strong rotation has been found to be prevalent in E+As in the cluster environment \citep{franx93,caldwell96} 
and \citetalias{norton01} point out that their study is at odds with those previous studies, although selection effects 
made it difficult to draw a firm conclusion in this regard. 

While this ubiquitous rotation seen in our E+A sample is, in itself, a key result of this study, the
strong evidence this provides that E+A galaxies are completely consistent with the `fast rotator'
population of early--type galaxies (as shown in the previous section), is a completely new and hitherto 
unrecognised connection. Importantly, this suggests that E+A galaxies may well be one progenitor
population of the fast rotator galaxies, and hence a feeder population for producing galaxies of this
type\footnote{Although, again, we caution that our observations only constrain the young stellar population which 
will make only a small contribution to the overall light at later times ($> 1$\,Gyr).}. Furthermore, 
there has already been considerable speculation in the literature as to the
physical mechanism(s) responsible for the formation of the fast rotator population, which may well
be applicable to E+A galaxies. Of most relevance here is the work of the SAURON team \citep{emsellem07}, 
who point out that the angular momentum in the inner parts of the fast rotators requires
either gas--rich mergers and the absence of a major dry merger, which would expel angular momentum, 
or the building of a disk-like system via gas accretion. They argue that the specific angular momentum
of fast rotators is mostly built up by either gas-rich minor mergers or other inflow of 
external gas. Very recent simulations by \citet{bournaud08}, which have involved significant
improvements in model resolution, suggest that major (as well as minor) mergers may in some situations 
produce fast rotator galaxies, with a significant amount of angular momentum being retained contrary
to previous expectations.  The dearth of slow rotators in our E+A sample may simply reflect the rarity of 
equal mass galaxy mergers.

The high incidence of the morphological signatures associated with tidal interactions and merging
amongst our E+A galaxies provides a very direct pointer to this being a dominant formation mechanism 
in their case. Very detailed modeling of the expected structural, kinematical and spectrophotometric
properties of an E+A galaxy produced in such an event has been published by \citet{bekki05}. This
was in the specific case of the merger of two gas-rich spirals, whose relative masses were varied
to simulate both major and minor mergers. These simulations clearly demonstrated that a spheroid-dominated
system is formed, which has a positive radial colour gradient and a negative radial H$\delta$ 
absorption strength gradient. They also showed that the projected kinematical and spectroscopic
properties of the simulated E+A galaxies can be remarkably different, depending on the orbital
parameters of the merger, and the young stars formed in the merger should show more rapid rotation
and a smaller central velocity dispersion than the underlying old stellar population. While
some of these predictions are consistent with our observations (e.g., spheroidal morphology, strong
rotation of the young stellar population), and some we cannot test (e.g., negative radial H$\delta$
gradient, kinematics of the old stellar population), the prediction of a positive colour gradient
in the central regions is difficult to reconcile with the large fraction of red cores seen in our sample. 

While such simulation-based models are impressive in their level of detail and continue to improve
\citep[e.g.][]{bournaud08}, some caution needs to be taken in using them to interpret the
observations since they are rather limited in their parameter space coverage, in particular the
type of merger explored. The simulations of \citet{bekki05} focus solely on the merger of 
two gas-rich spirals, and provide no information on what might happen in the merger of a
gas-rich galaxy with a gas-poor galaxy, or if one or both of these galaxies were not spirals.
One of us (K.B.) is currently undertaking a new set of simulations which examine E+A galaxy
formation via the merger of an already gas-poor early--type galaxy and a very gas-rich companion.
While the results from these are very preliminary, they do show considerable promise in 
perhaps addressing the issue of how red cores and hence negative colour gradients might
be formed. In the case of a gas-poor/gas-rich {\it minor} merger, the strong starburst
triggered by the merger does not occur in the centre of the E+A host, leaving its initial
red core intact. Moreover, the starburst components come from the gas-rich companion which is
destroyed during the merging process. The A-type stars formed in the burst are dispersed into
the inner regions of the E+A host, giving rise to the strong central Balmer absorption that
is seen in the observations. Undertaking a full and complete comparison of the structural, 
kinematical and spectrophotometric predictions of these new models with our observational
data set will be an important next step in trying to shed further clues on the details of
E+A galaxy formation.

\section{Conclusion}
We have obtained IFU spectroscopy and deep multi-colour imaging of a sample of
10 nearby E+A galaxies and used them to investigate the spatial distribution of colours, absorption
line strengths, stellar populations and kinematics in these systems. Our main conclusions are as
follows:
\begin{list}{$\bullet$}{\itemsep=0.1cm}
\item Morphologically our E+A galaxies are consistent with being early--type systems based
both on radial surface brightness profiles and visual morphological classification.
\item Our deep imaging reveals a high rate of interactions and tidal features  with 6 out of 10
galaxies displaying some evidence of disturbance. This is consistent with other morphological studies
of E+As.
\item The colour morphologies of our sample are diverse with most galaxies displaying a red core (relative to
the outer parts of the galaxy) but with galaxies also having no core or irregular cores as well as a single galaxy with a distinct blue core. There
is also a diversity in the large scale colour gradients of our sample. The integrated colours of the sample are blue as expected for E+A galaxies.
\item The spatial distribution of A-star light mapped by H$\delta$ absorption or pseudo `age maps' are uniform
over the central few kiloparsec regions studied. But we are unable to spectroscopically probe the interesting
region just beyond these radii where gradients are predicted by models.
\item The young stellar populations in all the E+As in our sample show significant rotation which is inconsistent
with the other major kinematic study of E+A galaxies by \citetalias{norton01} despite similarity in the samples and radial extent probed.
\item The kinematics of our sample are consistent with the overall population of early--type galaxies with 7 out of the
8 being classified as `fast rotators' as defined by \citet{emsellem07}.
\item The high frequency of tidal disturbance and early--type morphologies of our sample argues against simple isolated truncation of a spiral disk.
Our results are consistent with and suggestive of mergers and galaxy--galaxy interactions being a common formation
mechanism for E+As.
\end{list}

\section{Acknowledgements}
This paper is based on observations obtained at the Gemini Observatory, which is operated by the
Association of Universities for Research in Astronomy, Inc., under a cooperative agreement
with the NSF on behalf of the Gemini partnership: the National Science Foundation (United
States), the Science and Technology Facilities Council (United Kingdom), the
National Research Council (Canada), CONICYT (Chile), the Australian Research Council
(Australia), Ministério da Ciência e Tecnologia (Brazil) and SECYT (Argentina).
We would like to thank the staff at the Gemini Observatory and the Australian Gemini
Office for their expert assistance with our observations. 
This research was supported under the Australian Research Council's Discovery Projects funding 
scheme (project number 0559688). We would like to thank the referee, Eric Emsellem, for careful and insightful comments which greatly
improved this paper.
\bibliographystyle{mn2e}
\bibliography{references}

\label{lastpage}
\end{document}